\documentclass[a4paper,fleqn, twocolumn]{article}
\usepackage[a4paper]{geometry}
\usepackage{url}
\usepackage{xcolor}
\usepackage{hyperref}
\usepackage{ulem}
\usepackage{booktabs}
\usepackage{latexsym}
\usepackage{amsmath}
\usepackage{amssymb}
\usepackage{graphicx}
\usepackage[authoryear,longnamesfirst]{natbib}
\usepackage{authblk}
\usepackage{multirow}

\graphicspath{{./plots/}, {./figures/}, {./images/}}

\begin{document}
\let\WriteBookmarks\relax
\def\floatpagepagefraction{1}
\def\textpagefraction{.001}


\title{A Survey of Feature detection methods for localisation of plain sections of Axial Brain Magnetic Resonance Imaging}

\author[1]{Ji\v{r}\'{i} Martin\r{u}}
\author[2,*]{Jan Novotn\'y}
\author[2,3]{Karel Ad\'amek}
\author[1,2]{Petr \v{C}erm\'ak}
\author[4]{Ji\v{r}\'{i} Kozel}
\author[4]{David \v{S}koloud\'{i}k}
\affil[1]{Institute of Computer Science, Silesian University in Opava, Silesian University in Opava, Czech Republic}
\affil[2]{Research Centre for Theoretical Physics and Astrophysics, Institute of Physics, Silesian University in Opava, Czech Republic}
\affil[3]{Oxford e-Research Centre, Department of Engineering Science, University of Oxford, United Kingdom}
\affil[4]{Center for Health Research, Ostrava University Medical Faculty, Czech Republic}
\affil[*]{Corresponding author: Jan Novotn\'y, jan.novotny@physics.slu.cz}




\maketitle
\begin{abstract}
Matching MRI brain images between patients or mapping patients' MRI slices to the simulated atlas of a brain is key to the automatic registration of MRI of a brain.
The ability to match MRI images would also enable such applications as indexing and searching MRI images among multiple patients or selecting images from the region of interest. 
In this work, we have introduced robustness, accuracy and cumulative distance metrics and methodology that allows us to compare different techniques and approaches in matching brain MRI of different patients or matching MRI brain slice to a position in the brain atlas. 
To that end, we have used feature detection methods AGAST, AKAZE, BRISK, GFTT, HardNet, and ORB, which are established methods in image processing, and compared them on their resistance to image degradation and their ability to match the same brain MRI slice of different patients.
We have demonstrated that some of these techniques can correctly match most of the brain MRI slices of different patients. When matching is performed with the atlas of the human brain, their performance is significantly lower. 
The best performing feature detection method was a combination of SIFT detector and HardNet descriptor that achieved 93\% accuracy in matching images with other patients and only 52\% accurately matched images when compared to atlas.
\end{abstract}




\section{Introduction}
\label{sec:Introduction}
Image matching is a feature extraction technique, which seeks to establish a correspondence between two or more images using keypoints and descriptors. It is an active area of research used in many fields such as robotics \citep{book:San:2014:, Nak-Tom-She:2019:AKAZE_visual_odometry}, object detection, object tracking, security \citep{Deng:2020, Sun:2004, Moore:2017}, and medicine \citep{Cho-Acha:2020:, kei-Lip-Jai:2018:, Jos-Pha:2012:MRI}. However, not all methods applicable to images of everyday environments (rooms, landscapes, etc.) are suitable to be used in medicine. Specifically for use on images from various modalities such as computed tomography (CT), magnetic resonance imaging (MRI), ultrasound, thermograph, electroencephalography, nuclear medicine functional imaging, etc. It was found that although the currently used open-source methods for keypoint search and image similarity matching work well for general images, they are often unsatisfactory for medical images from modalities such as CT, MRI and ultrasound~\citep{Noborio:2019}.

Diagnosis of patients is currently labour intensive process limited to only a few experts, who are often overloaded. Even with recent advantages in the development of diagnostic tools~\citep{kei-Lip-Jai:2018:, Deng:2020, Lit-etal:2017:Survey_deep_learning}, this trend may not improve in the future. Training personnel, although vital, is a slow and challenging process as the skills required are complex. Among those is the spatial orientation needed to diagnose patients based on the MRI images. A visual aid that would locate MRI images within the atlas of a human brain would help experts diagnose patients faster by pre-selecting in MRI images only regions of interest (ROI). It could also help in training new medical personnel. Such a tool can also register images according to the correct shift in MRI images automatically and can serve as a filter for further processing. To enable this, we need to find corresponding brain slices from an atlas (or another patient) with a brain slice of interest and rank them based on the position within the atlas.  However, there is a lack of suitable methodology that could be used to evaluate such tools.

In~\citet{Noborio:2019} the authors addressed the comparison of brain images that changed due to gravitational forces during brain surgery and compared the suitability of using SIFT, KAZE, AKAZE and ORB methods. They found that these methods were not suitable for detecting key points inside the skull and proposed an image pre-processing method using Sobel filter and Canny edge operator to improve their performance, with which they achieved significantly better results. According to our own research, this approach is satisfactory in the case of the same patient and the same data-set (created at the same time on the same device), but not to compare images (slices) between a reference (atlas, phantom) and different patients.      

\citet{Tareen:2018} performed a comparison of the SIFT, SURF, KAZE, AKAZE, ORB and BRISK methods. However, this comparison was performed on common types of images (mountains, graffiti, terrain, buildings, bricks) containing a large number of edges and corners, which are used by these methods for keypoint detection. \citet{Survey:Boj-Bar} conducted a wide-ranging study of different combinations of classical detectors and descriptors. Based on this study, we have selected a detector-descriptor combination of AGAST + SIFT (referred in the text as AGAST) and GFTT + SIFT (GFTT).

Besides hand-crafted keypoint detectors and descriptors, alternative deep-learning detectors and descriptors are also being actively researched. A large survey of deep-learning-based methods was performed by \citet{2020arXiv200301587J}. \citet{2020arXiv200301587J} concluded that the performance of the deep-learning methods is on par with the hand-crafted method. Still, when the parameters of classical methods are tuned appropriately, these methods outperform deep-learning approaches. Among the best deep-learning models from this survey were HardNet by \citet{NIPS2017_831caa1b} and SOSNet by \citet{2019arXiv190405019T}. In our comparison, we have used HardNet as it is a part of the end-to-end pipeline which is ready to use. 

A comprehensive survey and detailed description of available feature detection (FD), both handcrafted and deep-learning methods and techniques were performed by \citet{Ma:etal:2021:matching_explain}. For further details on methods used in this work we refer interested readers to this survey. This survey however does not consider medical images in any detail.

In practice, MRI images from different patients or the same patient but taken at different times or on different equipment can be affected by noise, different scaling, and rotation. Therefore, to compare them and find the most similar slice, is necessary to optimise methods to be more robust and invariant to these transformations and noise or to use appropriate pre-processing steps.  

The novelty of this work and its contribution is the definition of a methodology for performance comparison of the different image processing techniques in searching, registration, and selection of MRI brain scans in the region of interest. Furthermore, using our methodology, we have demonstrated that it is possible to automatically register MRI brain scans and select images from the region of interest using state-of-the-art FD techniques (AGAST, AKAZE, BRISK, GFTT, HardNet, and ORB). We evaluate their effectiveness and sensitivity to the noise and geometrical deformations of MRI brain scans.

In section~\ref{sec:methods} we defined three metrics: robustness $R$, accuracy $A$, and cumulative distance $C$ and described used methods, datasets, and experimental setup. In section~\ref{sec:robustness} we test the FD methods invariability on a set of images from the same patient affected by selected image degradations (rotation, noise, scaling). In section~\ref{sec:accuracy} we investigate the ability of FD methods to match MRI images of different patients and how different image preprocessing steps can improve the results. In section~\ref{sec:atlas} and determine how well the FD methods can locate specific MRI slices within the simulated atlas. Lastly, we conclude obtained results in section~\ref{sec:Conclusion}. 

\section{Experimental setup, datasets and methods}
\label{sec:methods}
The image preprocessing for medical images is an important and challenging task \citep{Vas-etal:2017:}, especially when a diagnosis is in consideration (e.g. tumour segmentation) \citep{Pat-Udu:2012}. Contrast and Image quality are in general the major problems in medical imagery. Also, when we want to interpret images of the human brain with different data sets or atlases often an initial alignment (also called registration) of the images is needed \citep{Colling:1994:Registering}. The various image processing applications are for example provided by The McConnell Brain Imaging Centre (e.g. BEaST -- Brain Extraction based on non-local Segmentation Technique \citep{BEAST}, INSECT -- method to separate a structural MRI). Part of our investigation is to study and define the effects of an image enhancement on the ability of FD methods to match MRI images of healthy patients. 

For our tests we downloaded a dataset from open-science neuroinformatics database OpenNeuro (formerly known as OpenfMRI) maintained by a research group around Russell Poldrack \citep{Pol-etal:2013:OpenNeuro, Pol:2017:OpenfMRI}. We used a dataset CEREBRuM: a 3T MRI segmentation tool \citep{dataset:cerebrum} used in the design and testing of an optimised end-to-end convolutional neural network (CNN) architecture \citep{bontempi:2019:cerebrum}. The dataset consists of 947 off-scanner MR images (3 Tesla T1-weighted 1~mm isotropic MP-RAGE 3D sequences scanned at the Centre for Cognitive Neuroimaging at the Institute of Neuroscience and Psychology, University of Glasgow), which of 7 are publicly available as Neuroimaging Informatics Technology Initiative file format (NIfTI). For simplicity this work is limited to axial MRI slices which were exported into Portable Network Graphics (PNG) files by using \texttt{nii2png} software\footnote{\url{https://github.com/alexlaurence/NIfTI-Image-Converter}} from original NIfTI files. Also, the FD methods are not well suited to work with NIfTI formats. Following this process we obtain a series of 170~PNG MRI brain axial slices for each patient.

Comparing different FD methods is not trivial. These methods tend to produce a different number of keypoints per image and to complicate the comparison further, each FD method may associate different weights to the matched keypoints. Therefore, to rank different methods, we need to depart from absolute number of keypoints and base our comparison on a more general metric. For this purpose, we have devised four metrics, each describing a different aspect of the FD method's ability to match MRI slices. Those are signal-to-noise ratio (SNR), which conveys confidence in matching the two MRI slices. An \textit{accuracy}, which expresses how many MRI slices are correctly matched within a given error, and a \textit{cumulative distance}, which is a total error measure in the number of slices for both correctly matched and mismatched slices. Furthermore, we have defined \textit{robustness} to evaluate the sensitivity of different FD methods to the amount of noise, change in scale and rotation.

The signal-to-noise ratio (SNR) is a standard metric used in many areas of signal processing. In this work, we have used it because it allows us to compare FD methods that produce a different number of keypoints. The SNR is calculated as follows
\begin{equation}
\label{eq:snr}
\mathrm{SNR}_{ij} = \frac{x_{ij} - \mu(k)}{\sigma(k)}\,,
\end{equation}
where $x_{ij}$ is the number of matched keypoints between reference image $j$ and image $i$. The mean $\mu(k)$ and the standard deviation $\sigma(k)$ are determined based on the number of matched keypoints between image $j$ with a set of all images from patient $k$, which also contains image $i$ used in the comparison. Image $j$ could belong to a different patient or the same patient.

First, we have investigated the sensitivity of the selected methods (AGAST, AKAZE, BRISK, GFTT, HardNet, ORB) to the degradation imposed on the MRI images, such as the scaling factor, different noise levels, rotation, alignment and changes in contrast. To measure the invariability of a method to MRI image degradation, we used a set of MRI images from the same patient. We have calculated robustness $R$ as
\begin{equation}
\label{eq:robustness}
R_x = \frac{\mathrm{SNR}_{ii,x}}{\mathrm{SNR}_{ii}}\,,
\end{equation}
where the $\mathrm{SNR}_{ii}$ corresponds to the SNR of the match between non-degraded image $i$ with itself, and $\mathrm{SNR}_{ii,x}$ is the SNR of the match between degraded image $i$ and non-degraded image $i$. The type of degradation $x$ used might be rotation (r), noise (N) and scaling (s). For simplicity, we choose the rotation of 5${^\circ}$, upscaling by 5\%, Gaussian noise with 10, 20 and 30 standard deviations (STD). The 5${^\circ}$ angle was chosen based on recent study by \cite{prabhu2021tilt} which shows that observed in the T1- and T2-weighted MR images obtained is in range from 2${^\circ}$--6${^\circ}$. The choice of upscaling of 5\% is selected as a potential accepted difference for different age ranges and gender. In MRI images, the typical types of noise are Rician noise, Gaussian Noise and Rayleigh noise \citep{Goy-etal:2018:noise}. However, in most of the clinical applications prevails the Gaussian noise \citep{Bha-etal:1995:mri_noise}.

To extend our study of FD methods to the comparison of two different patients or a patient to the atlas, we have introduced two more metrics: accuracy, and cumulative distance. In such a case the correct alignment (rotation, scaling) and contrast (noise) are essential, especially when the selected FD method is sensitive to one or more of these factors. This process is commonly known as ``registering data'', e.g., transforming the medical data to the Talairach coordinate system~\citep{Colling:1994:Registering}.

Accuracy is an aggregated metric and it tells us how many MRI slices were correctly matched, that is when normalised, it tells us a ratio of correctly matched images in percentage. Accuracy is calculated as follows
\begin{equation}
\label{eq:accuracy}
    A_{d,c} = \sum_{i=1}^{n} x_i/n\,,\quad x_i = \begin{cases} 1\, \mathrm{if,}\, |y_e - y_b| \leq d \\ 0,\,\mathrm{otherwise}
    \end{cases}\,,
\end{equation}
where $y_e$ is the correct index of the image which should be selected by the FD method, $y_b$ is the index of the image which is the best match of the source image $i$, and $n$ is the total number of images we took into account. Since MRI slices that are next to each other are similar (depending on resolution and MRI slice step) the interval parameter $d$ is the neighbourhood of image indices where the image match is considered correct. Lastly, $c$ indicates preprocessing step(s) applied to the data.

For the general overview of how compact or spread out the predictions of the selected FD method and preprocessing is, we also introduce the cumulative distance as
\begin{equation}
\label{eq:cum_distance}
    C_{c} = \sum_{i=1}^{n} |y_e - y_b|\,.
\end{equation}

When combined the three metrics allow us to describe the behaviour of the FD method. High accuracy together with high cumulative distance means that images are mostly correctly matched with some mismatched images with a large distance from the correct index. A typical mismatched image has a low or high image index. On the other hand low accuracy with a small cumulative distance means that results are centred around correct values but have a larger spread. In this case, increasing $d$ would increase the accuracy. The SNR role is to indicate how the other two metrics are reliable as it expresses confidence in the match. A high SNR value indicates that matching two images is probably not due to noise while low SNR indicates that selecting which images match was affected by the noise thus accuracy and cumulative distance may be affected.

\begin{figure}[htbp]
    \centering
    \includegraphics[width=\linewidth]{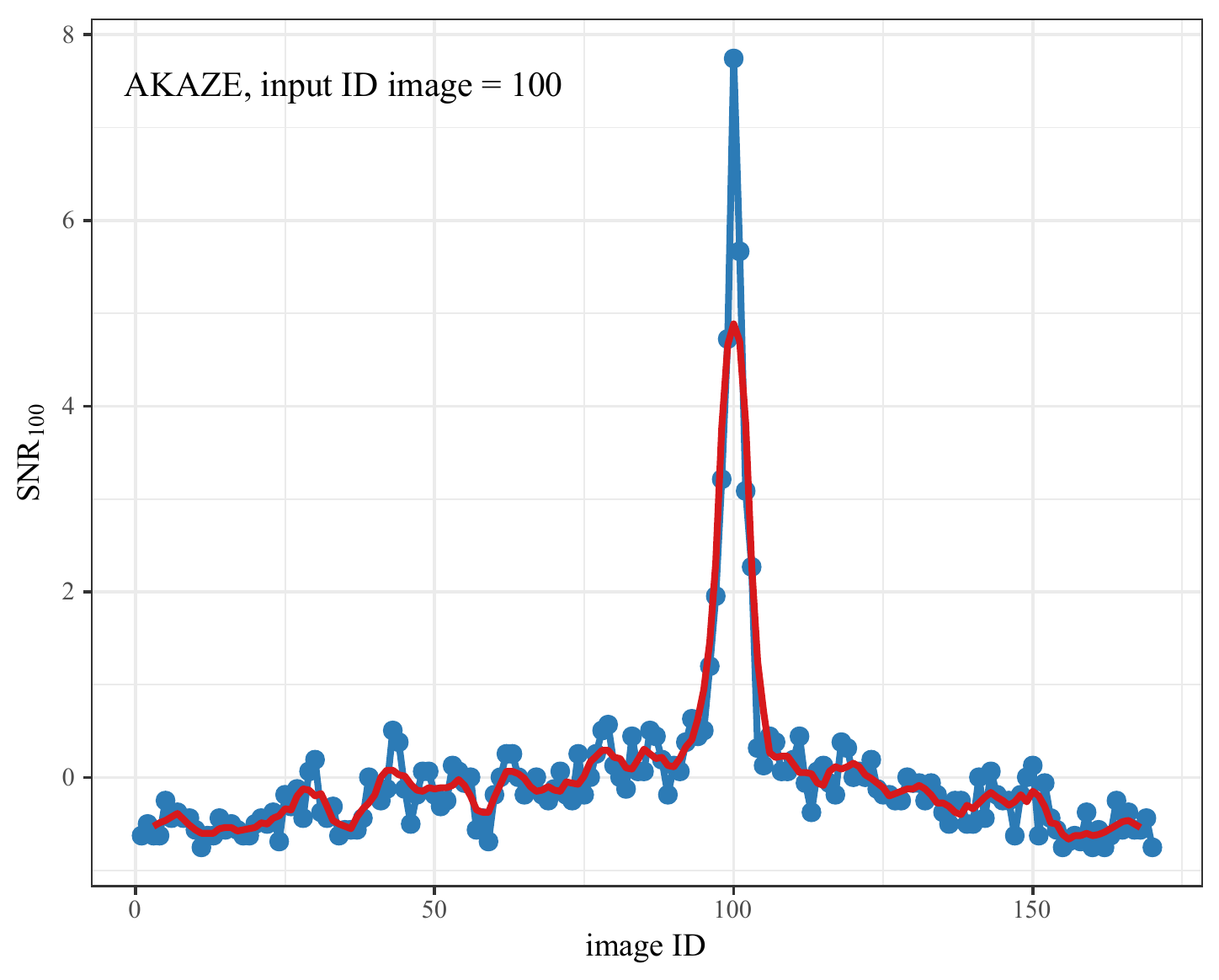}
    \caption{Example of computed series of SNR computed by using Eq.~\eqref{eq:snr} on the sets of same subject. The blue line with points represents SNR in the case of input image $i=100$, the red line corresponds to the moving average.}
    \label{fig:moving_average}
\end{figure}

To find a match for an image $i$ of a patient, we have to compare the image using one of FD methods with all images from the reference (another patient or the atlas). This involves calculating the number of matched keypoints for all images, respectively calculating SNR for each one and then selecting the highest SNR to get the image slice that matches the source image the most. In the end of this process, we have a series of SNR values, which could be affected by noise. To reduce the effect of the noise on the selection of the best matching image, we use a moving average with a window size of 7 samples. The optimal size of the window depends on the slice step. For larger slice steps shorter window is more appropriate. An example of the SNR series for input image $i=100$ with the computed moving average is shown in Fig.~\ref{fig:moving_average}.

For the all mentioned FD methods and feature matching methods (AGAST, AKAZE, BRICK, GFTT, HardNet, ORB), we wrote a few simple python scripts (version 3.9) using the Open Source Computer Vision Library (OpenCV -- version 4.5.1.48) \citep{opencv_library} and Kornia library (version -- 0.6.4) \citep{eriba2019kornia}. For the data analysis, we used the R programming language (version 3.6.1). All scripts are available at \url{https://github.com/jan2nov/FDMinM}. The criteria and parameters used for matching keypoints are for the simple Low ratio test, we used the limit around 0.75, knnMatch with the out-of-best matches found per each query descriptor equal to 2 or as in the case of HardNet mutual nearest neighbour function with threshold distance 0.95.

\begin{figure}[htbp]
    \centering
    \includegraphics[width=\linewidth]{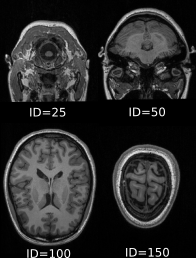}
    \caption{Preview of the axial MRI planes (images) highlighted in some of our tests.}
    \label{fig:select_images}
\end{figure}

\begin{figure*}
    \centering
    \includegraphics[width=.98\linewidth]{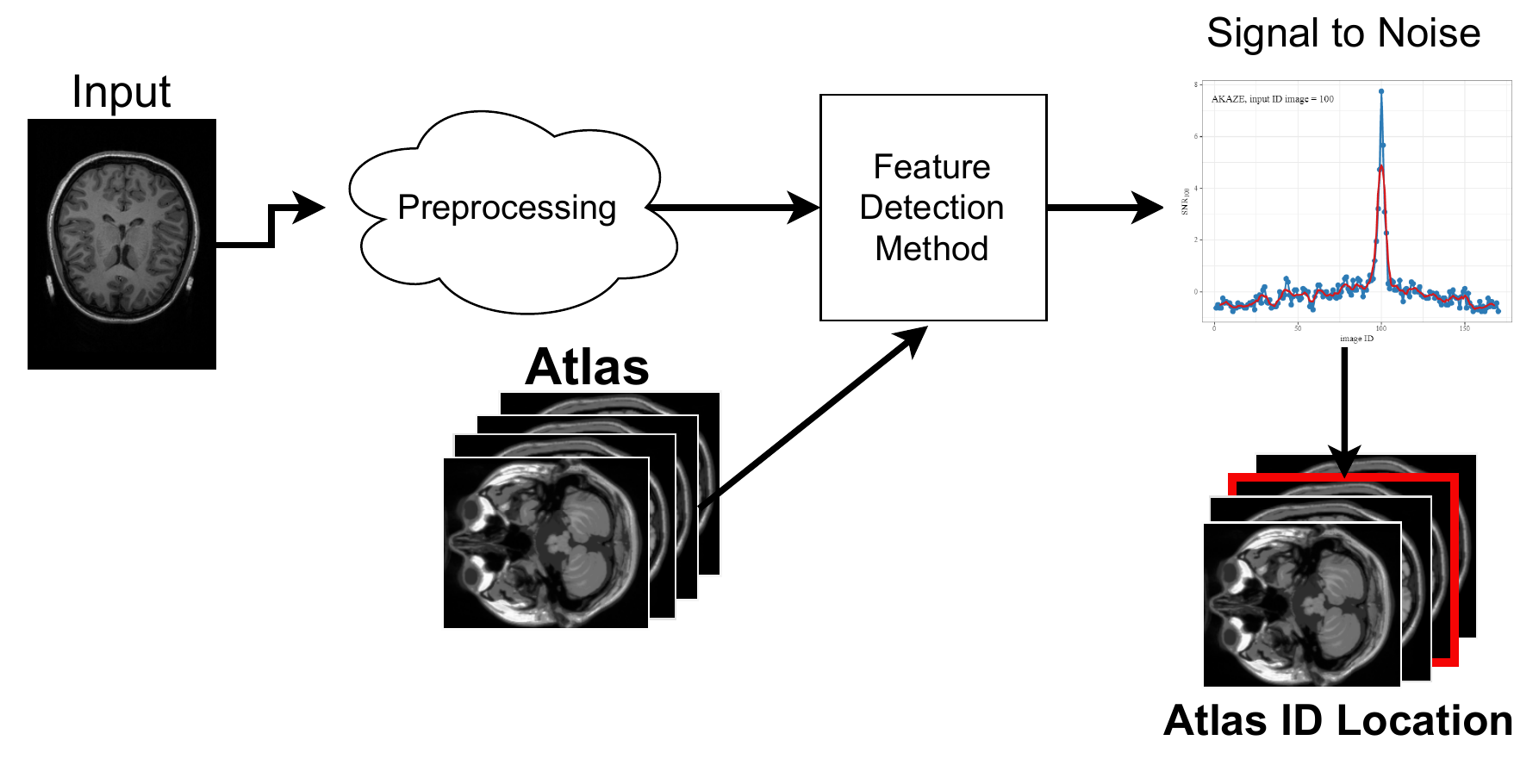}
    \caption{\label{fig:flowchart}Schematic diagram of the process of selecting the most similar MRI image relative in the atlas in relative to the input MRI image.}
\end{figure*}

To illustrate the ability of different FD methods to match different MRI slices, we have chosen representative samples that are shown in Fig.~\ref{fig:select_images}. In Figure~\ref{fig:flowchart} we illustrate a flowchart of the process of selecting the most similar MRI image in the atlas (or in some of our tests a patient) relative to the input ID image. In the preprocessing part, it is then possible to apply various corrections to the image (rotation, scaling, contrast), including possible other corrections. Everything is then passed through the selected FD method and computed a SNR distribution function for each atlas image is created. The one with the largest SNR value then indicates the position of the most similar input image from the atlas.

\section{Results and Discussion}
\label{sec:Results}
\subsection{Relation between the signal to noise ratio with image degradation}
\begin{figure*}
    \centering
    \includegraphics[width=.33\linewidth]{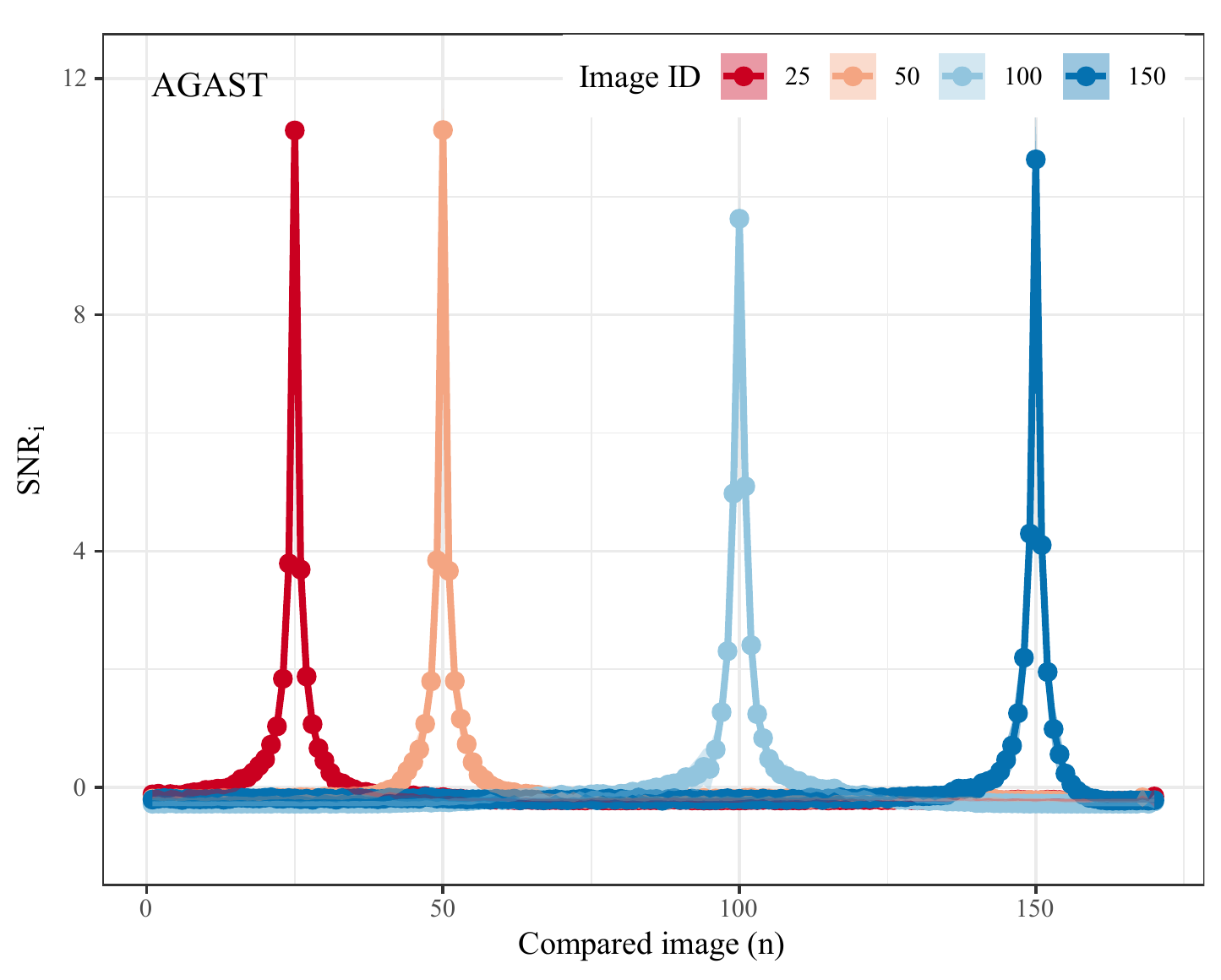}\hfill
    \includegraphics[width=.33\linewidth]{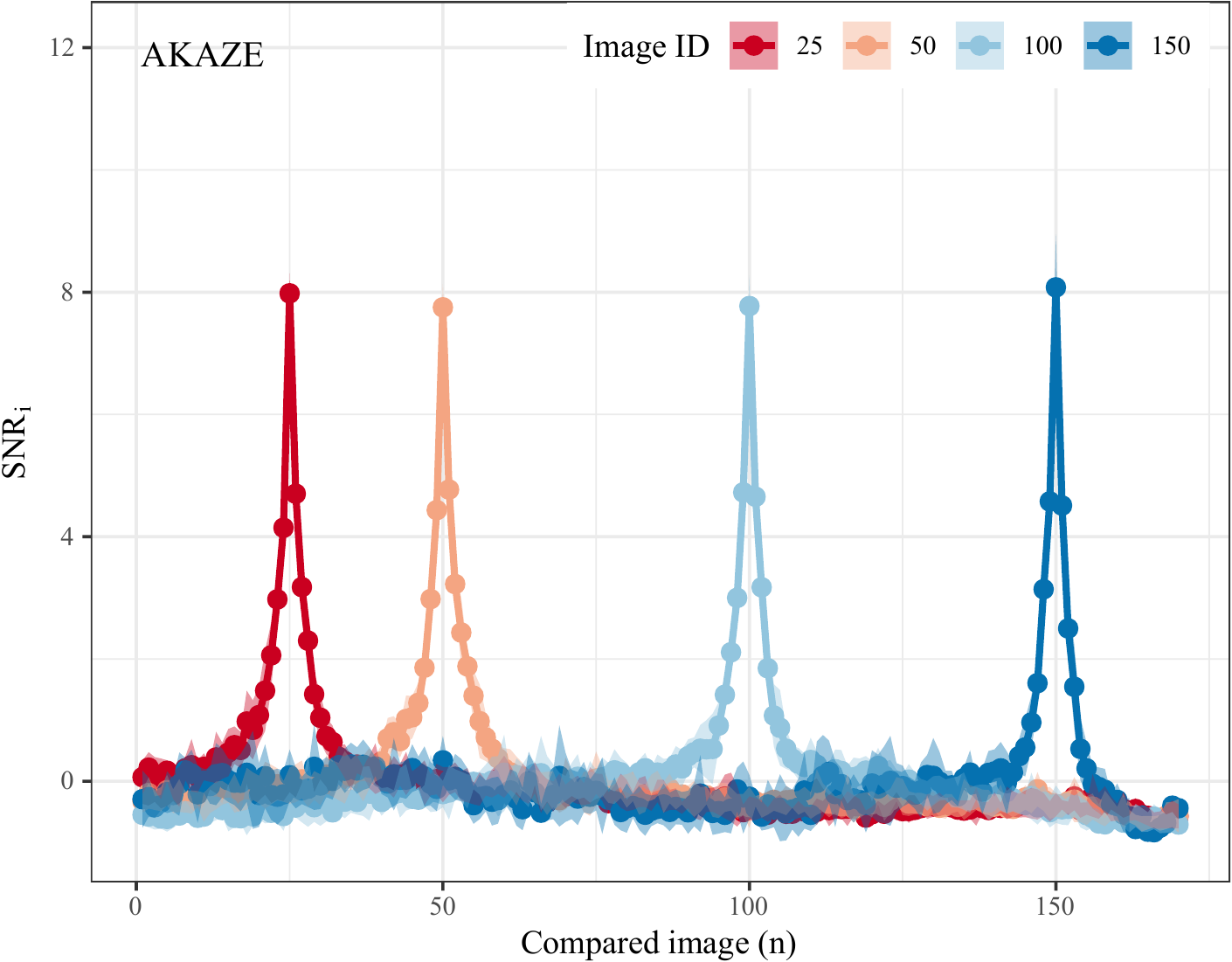}\hfill
    \includegraphics[width=.33\linewidth]{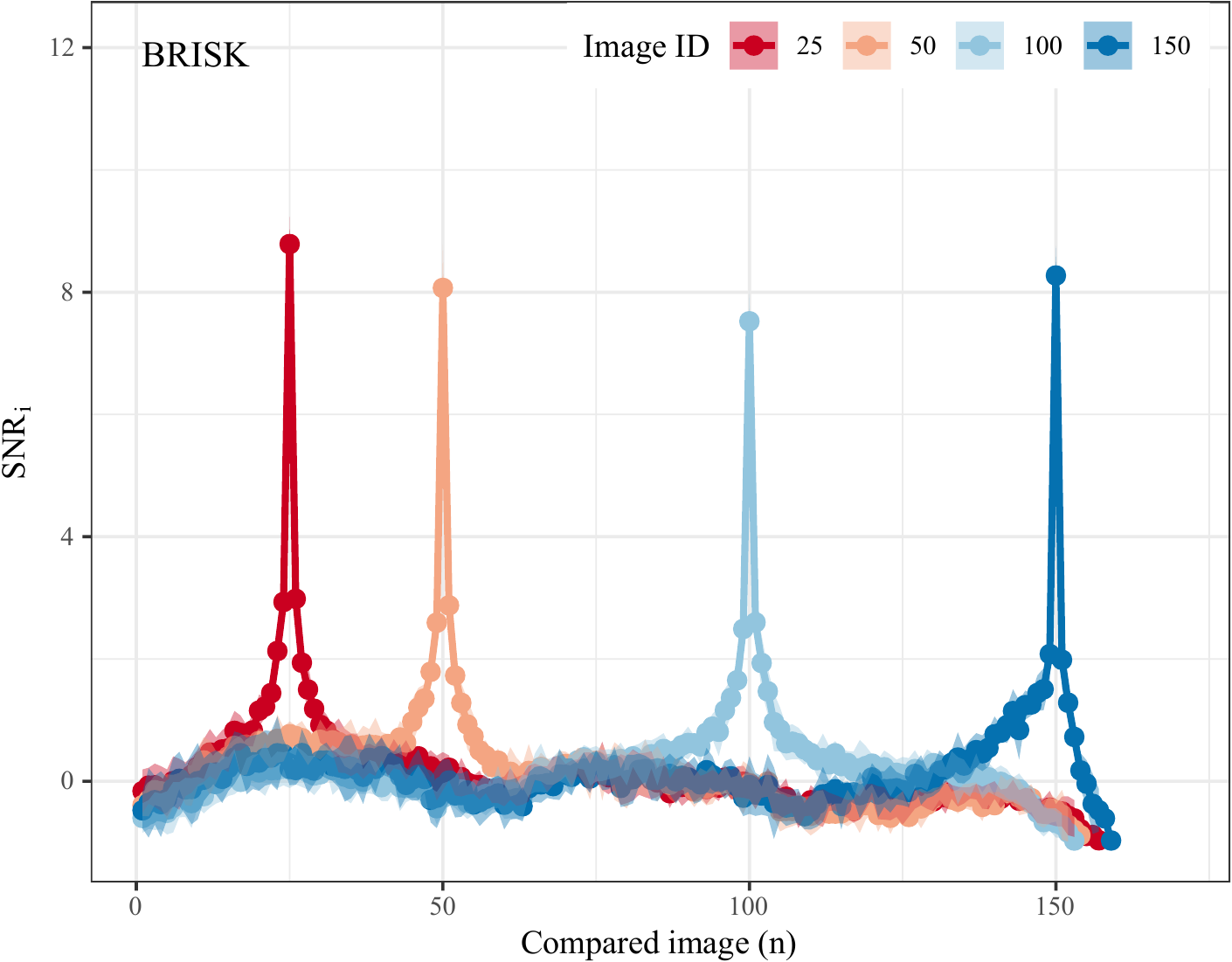}\\
    \includegraphics[width=.33\linewidth]{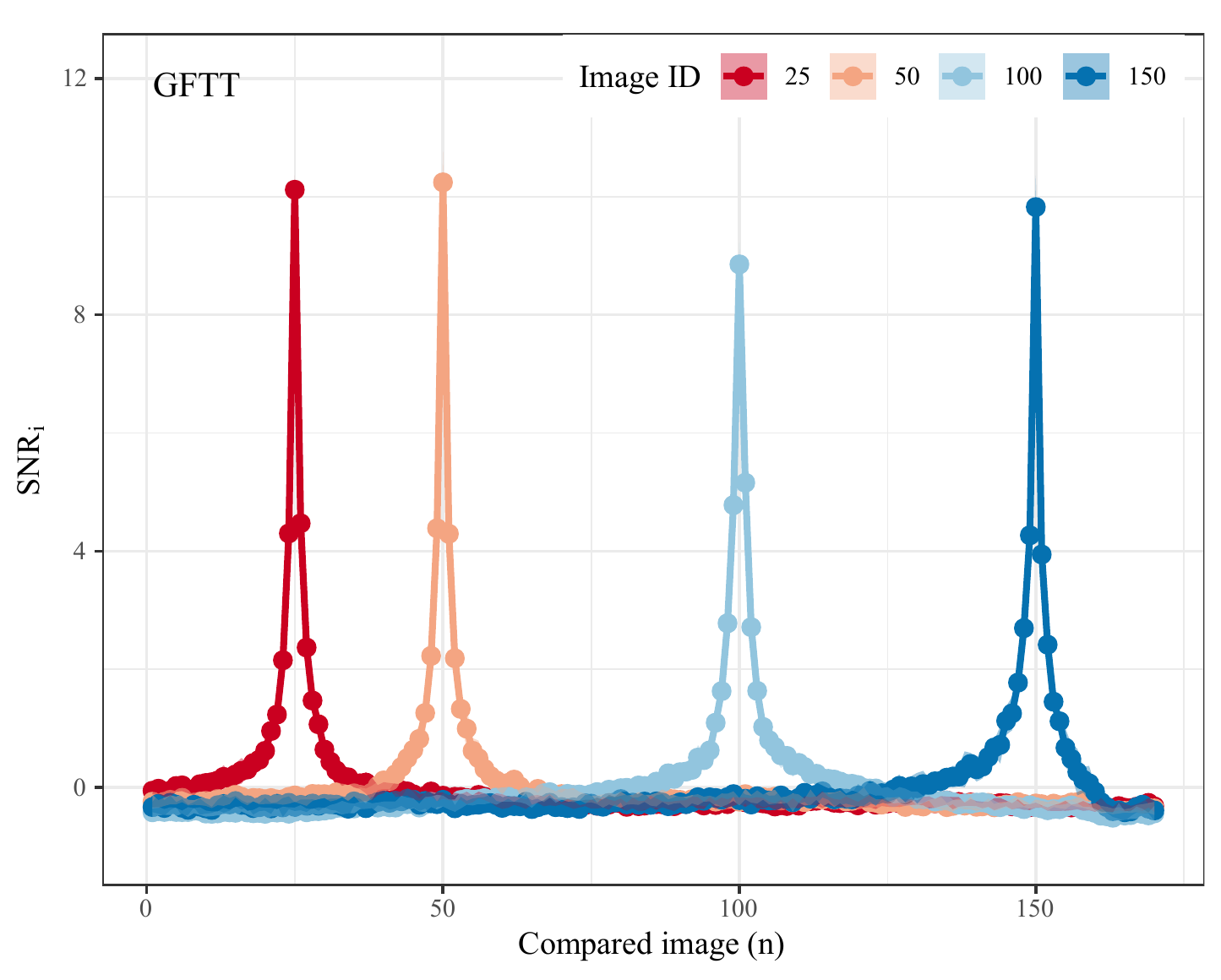}\hfill
    \includegraphics[width=.33\linewidth]{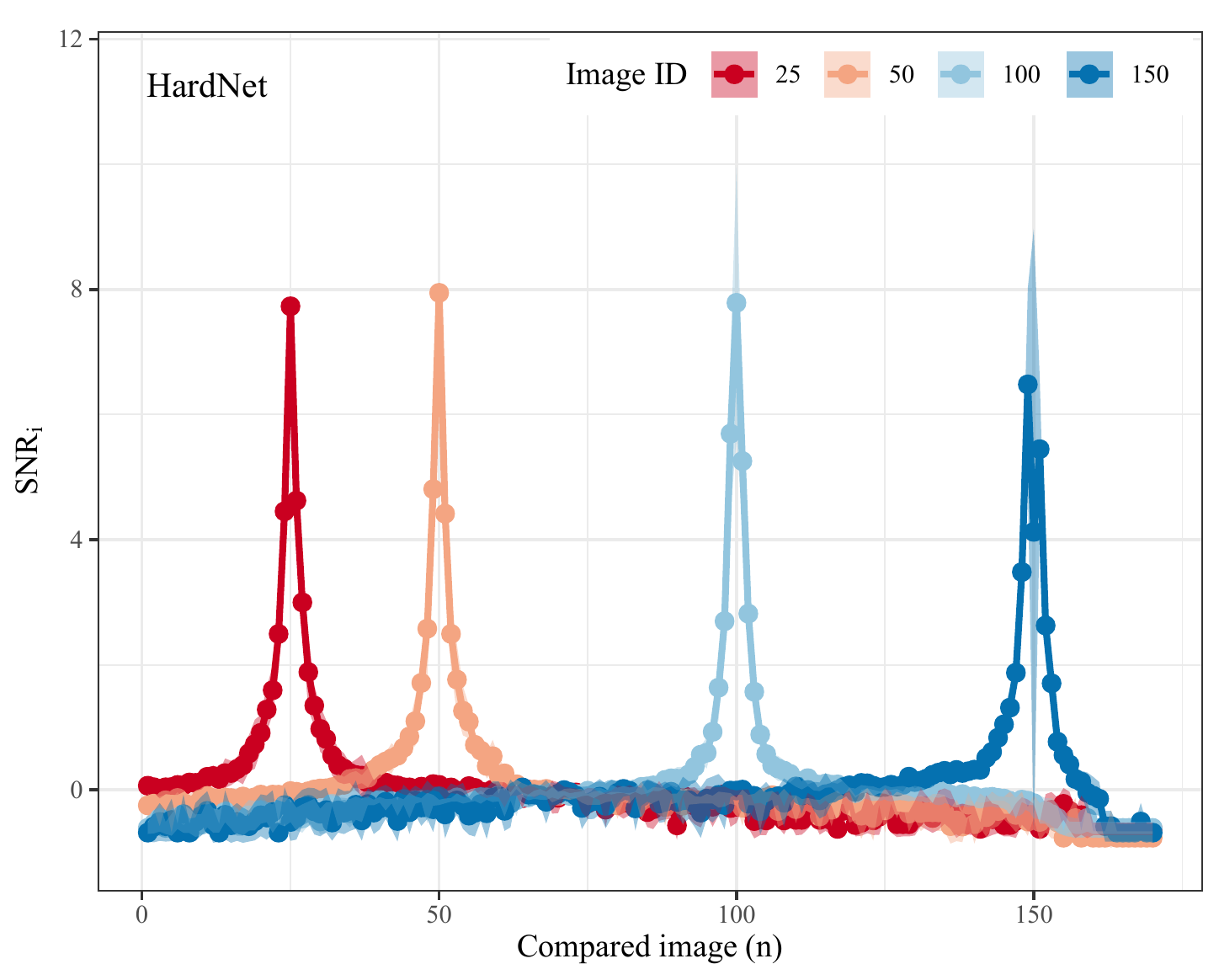}\hfill
    \includegraphics[width=.33\linewidth]{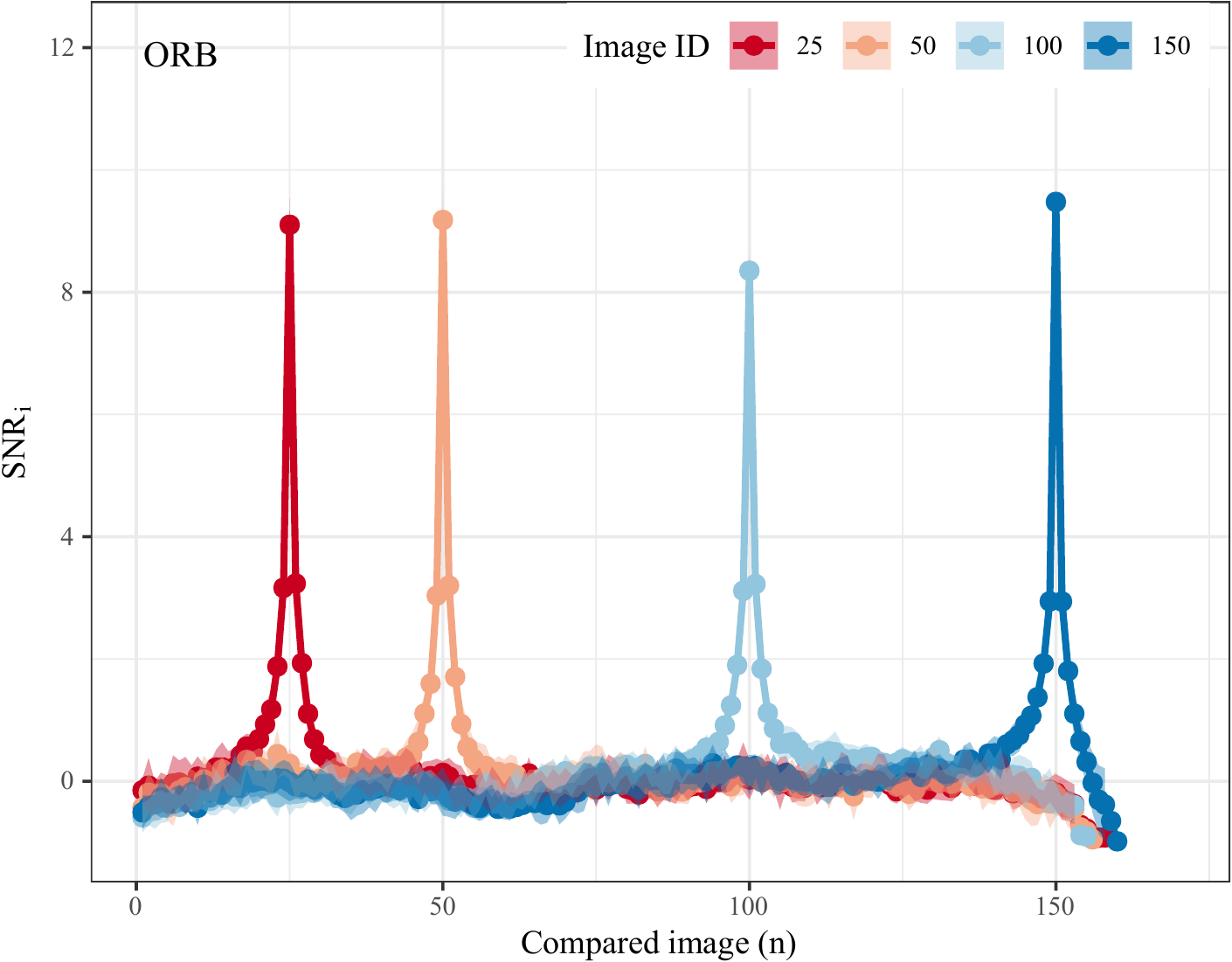}
    \caption{Behaviour of the SNR (y-axis) defined by Eq.~\eqref{eq:snr} for selected input images (25, 50, 100, 150) in case of used FD methods: AGAST (first row left), AKAZE (first row center), BRISK (first row right), GFTT (second row left), HardNet (second row center) and ORB (second row right). The solid lines correspond to an average computed from a series of SNR of each subject, whilst the coloured area to the standard error.}
    \label{fig:snr_identity}
\end{figure*}

To determine how each FD method is behaving with each image degradation we first present the case of SNR values (Eq.~\eqref{eq:snr}) computed on the same set of non-enhanced, non-degraded images as the input images (``identity''). The results of this test for a few selected ID images (25, 50, 100, 150) are shown in Fig.~\ref{fig:snr_identity} (higher value is better). As we can see, most of the FD methods are able to find the right axial image with high certainty, respectively on average the SNR values for all input images from best to worst FD methods are AGAST ($\sim$10.6), GFTT ($\sim$9.7), ORB ($\sim$8.9), BRISK ($\sim$8.1), AKAZE ($\sim$7.7), and HardNet ($\sim$7.4). Only HardNet in a few cases (for example the input image ID=150) noted the neighbouring images as better matches than the input image itself.

\begin{figure*}
    \centering
    \includegraphics[width=.33\linewidth]{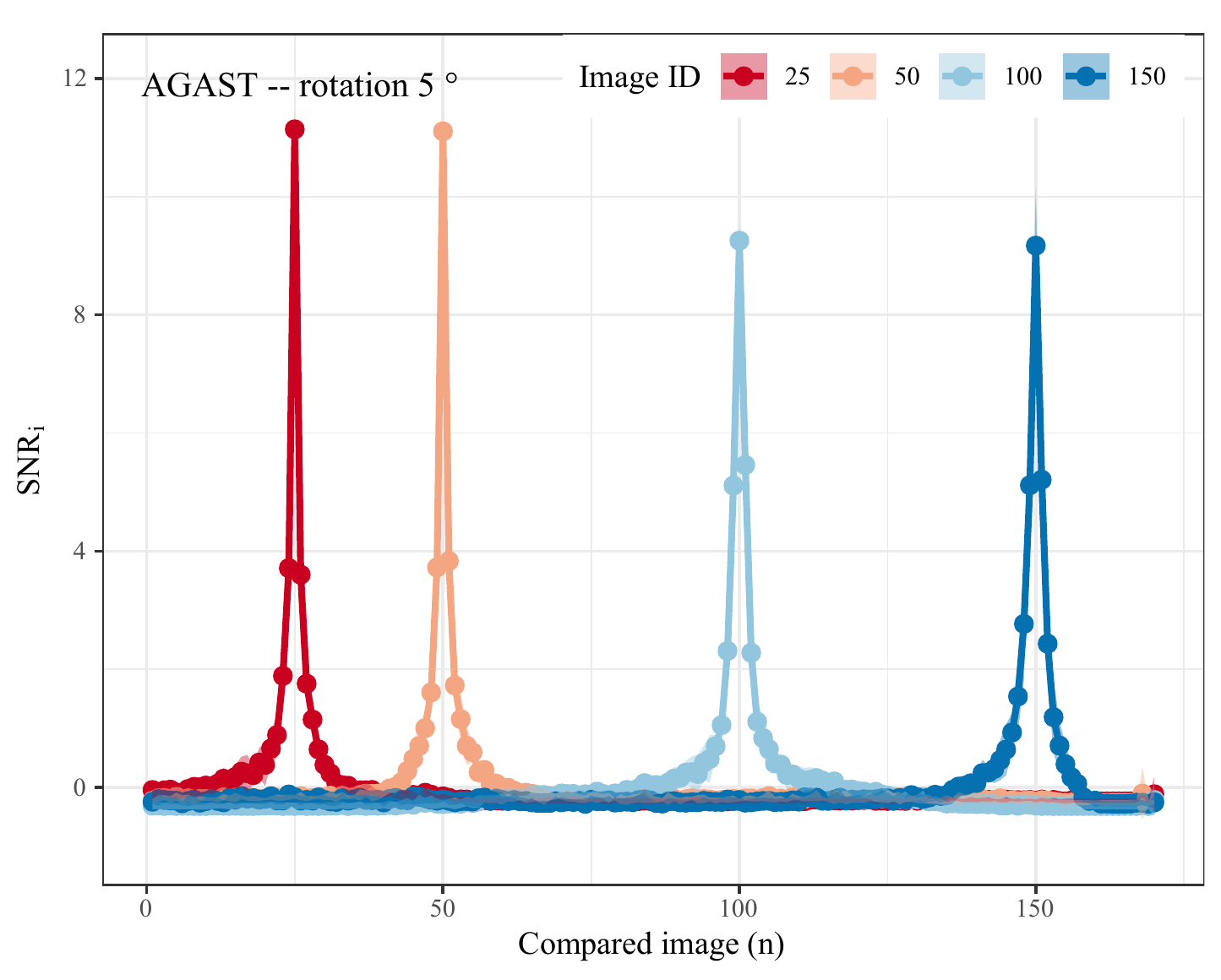}\hfill
    \includegraphics[width=.33\linewidth]{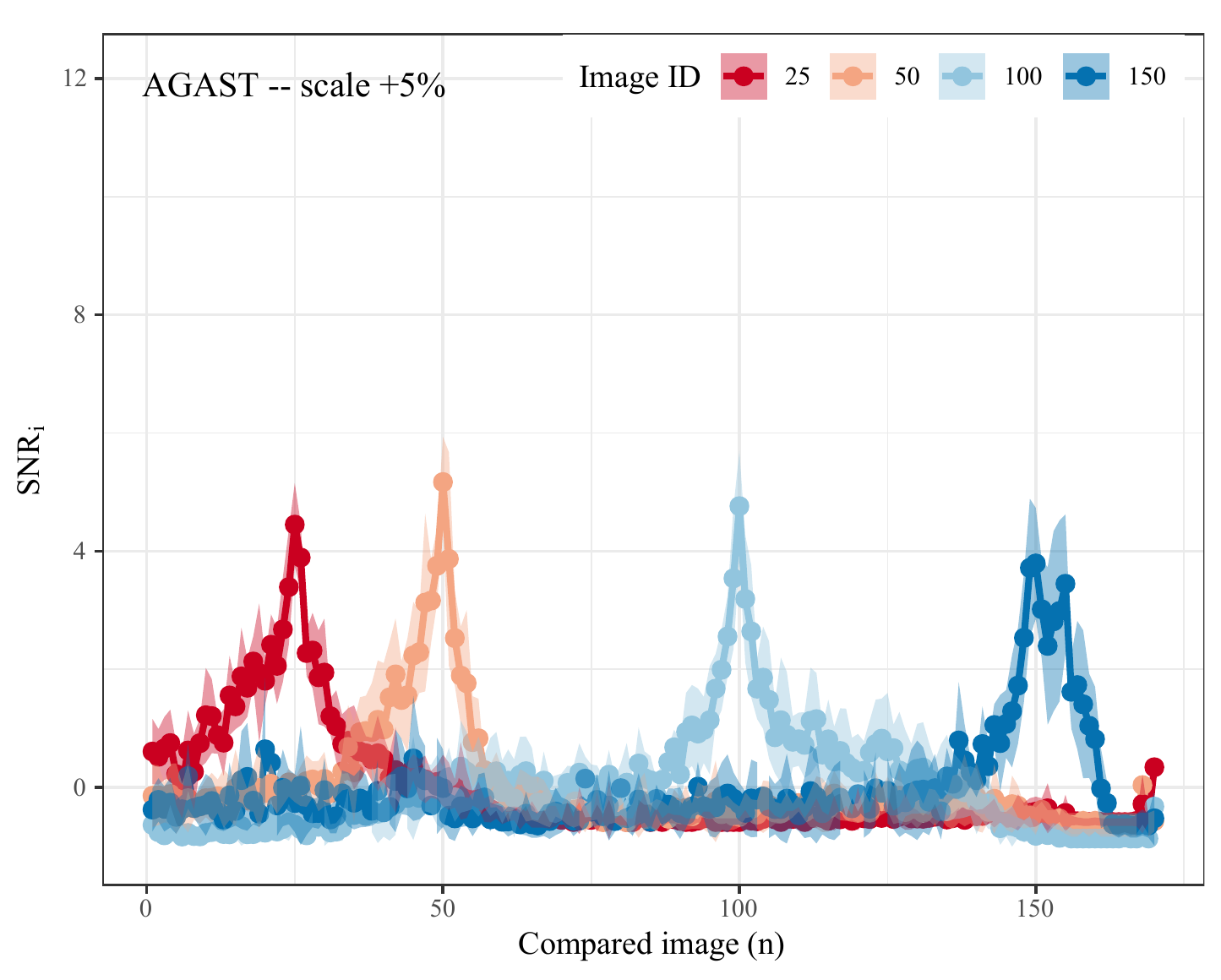}\hfill
    \includegraphics[width=.33\linewidth]{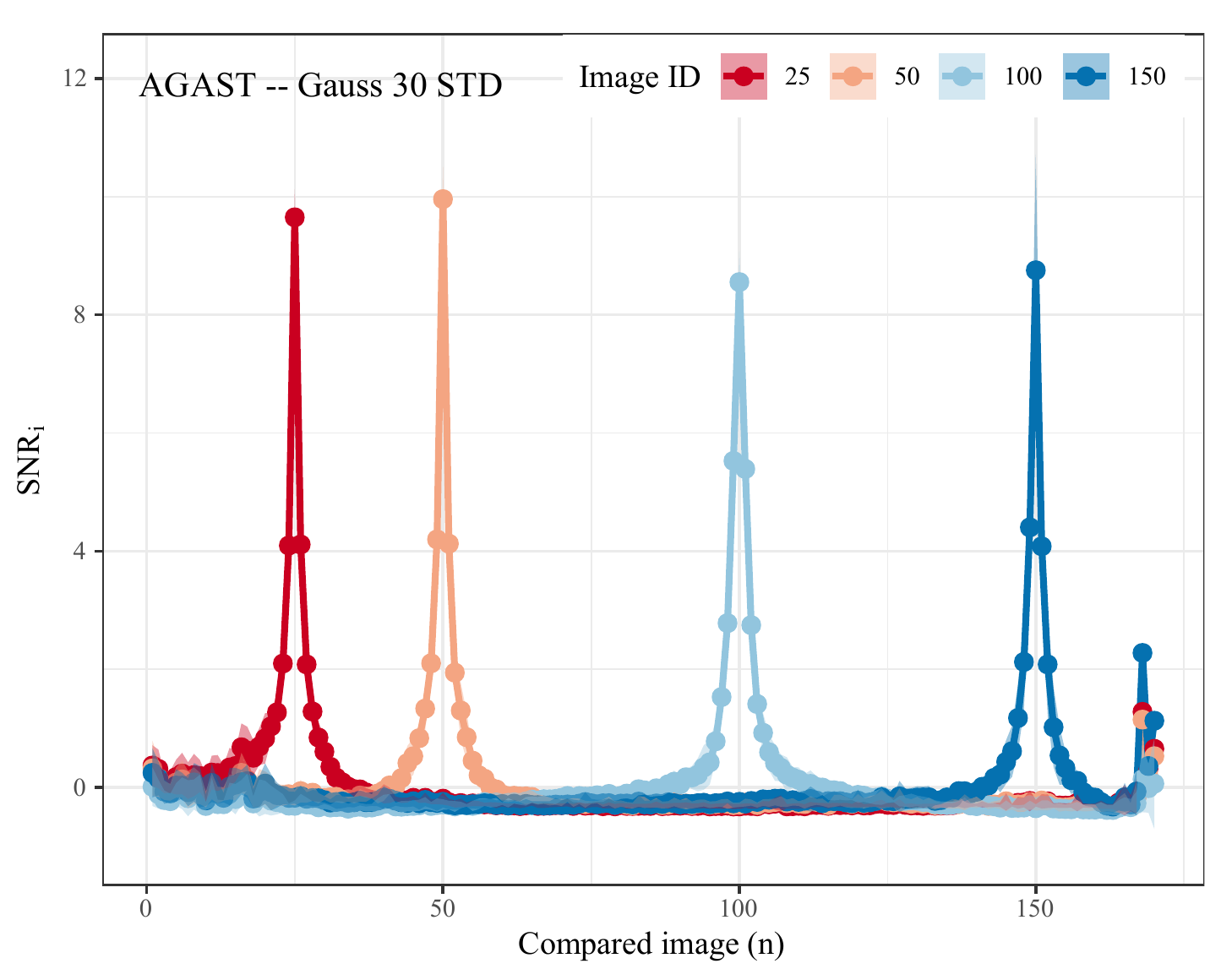}\\
    \includegraphics[width=.33\linewidth]{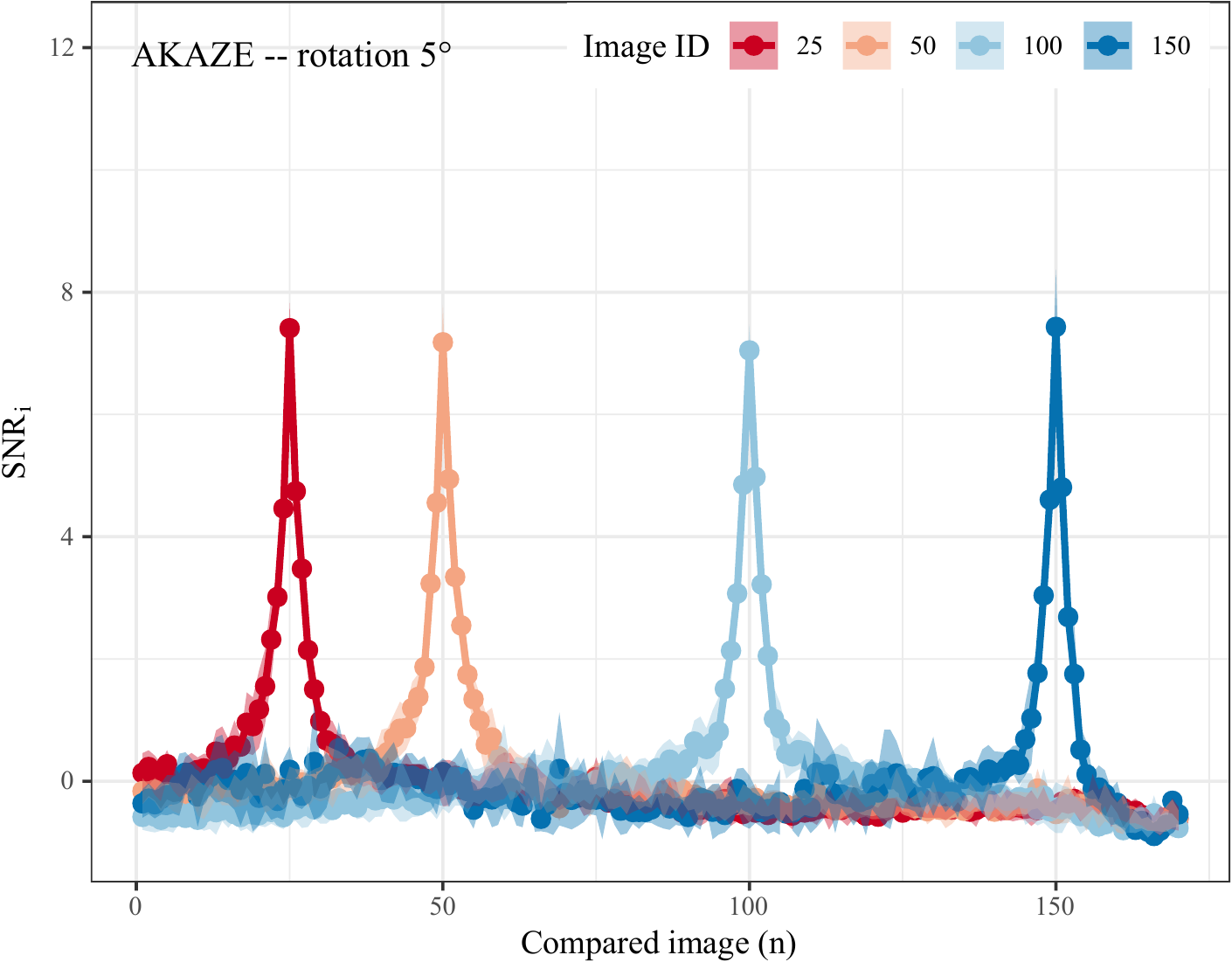}\hfill
    \includegraphics[width=.33\linewidth]{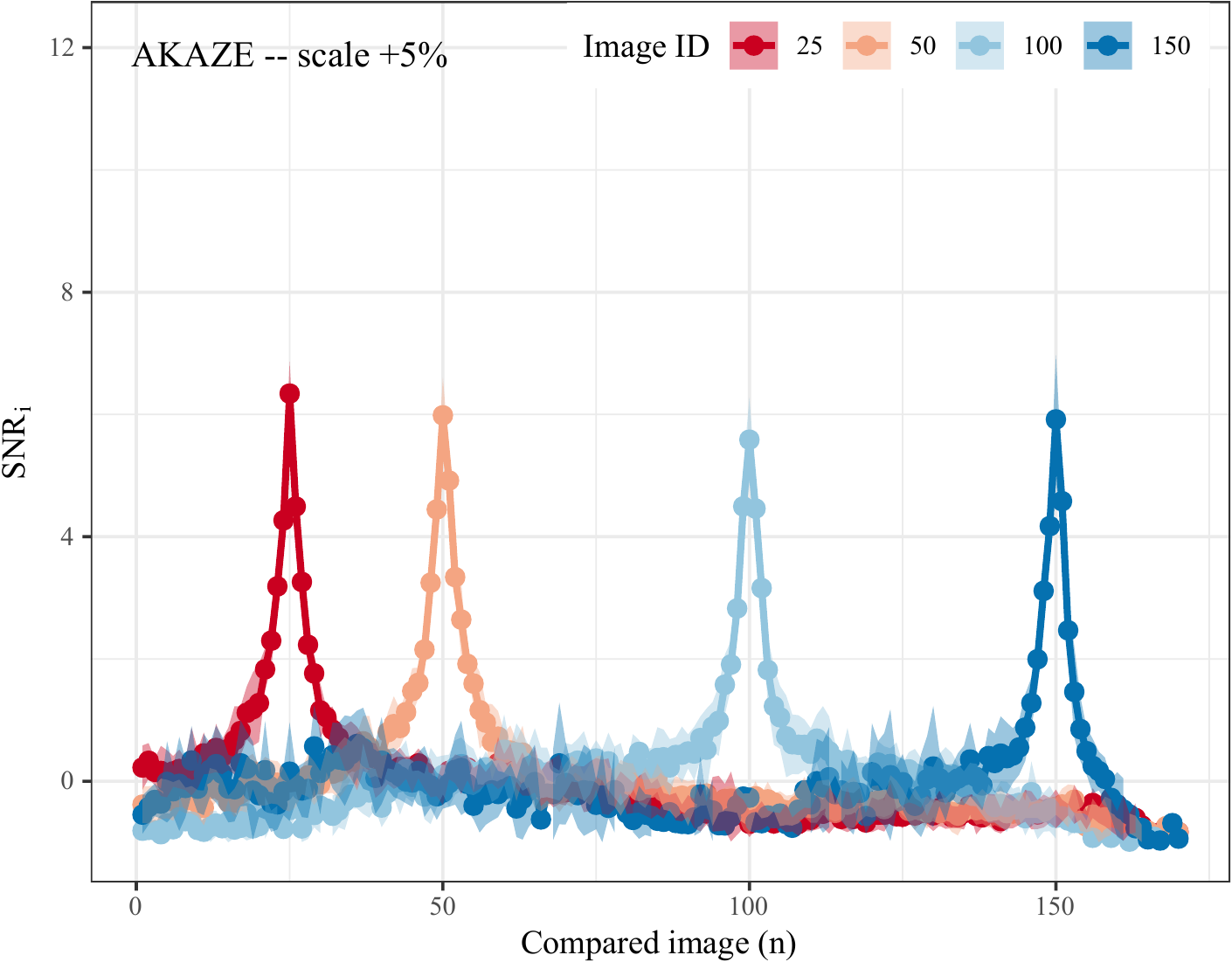}\hfill
    \includegraphics[width=.33\linewidth]{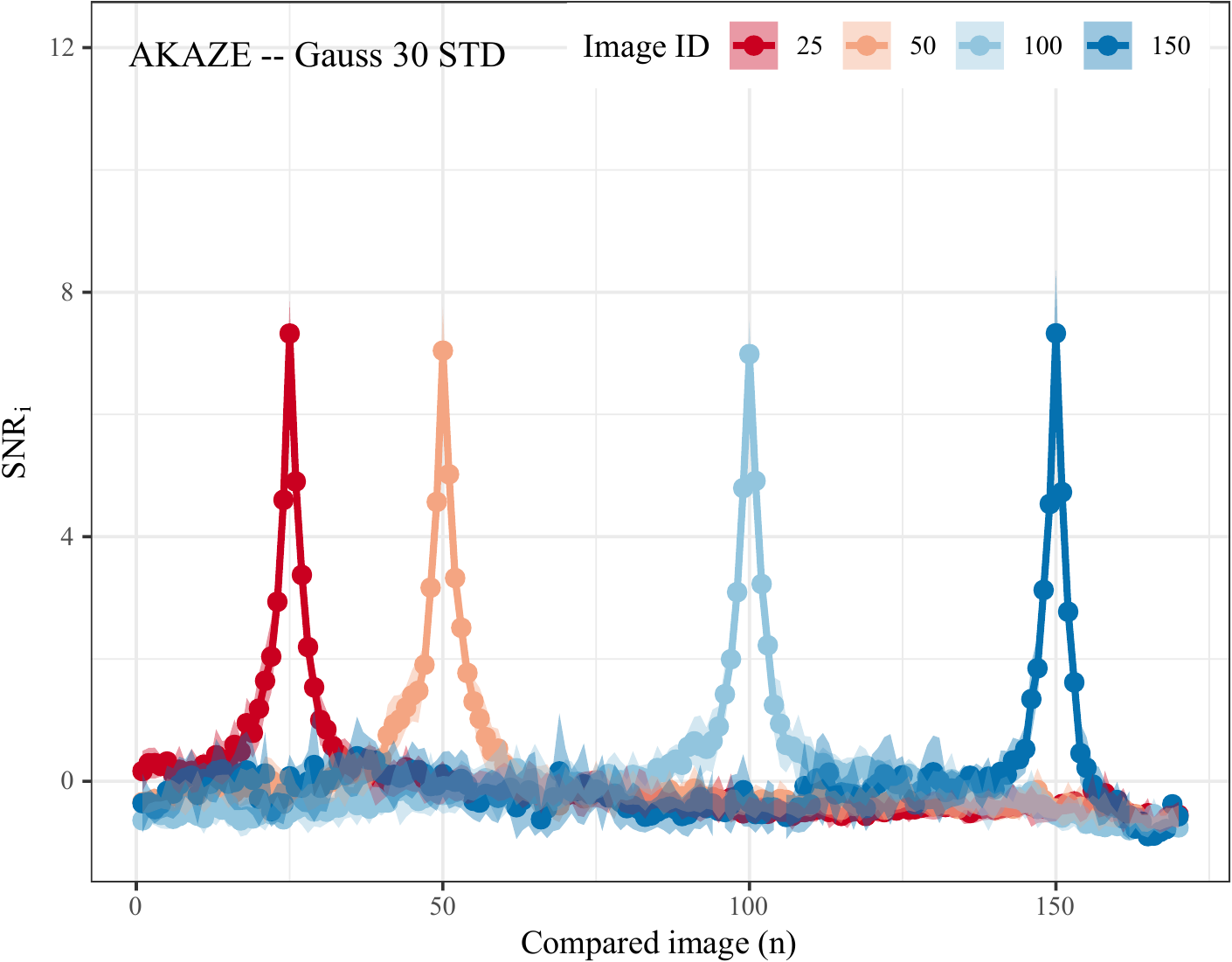}\\
    \includegraphics[width=.33\linewidth]{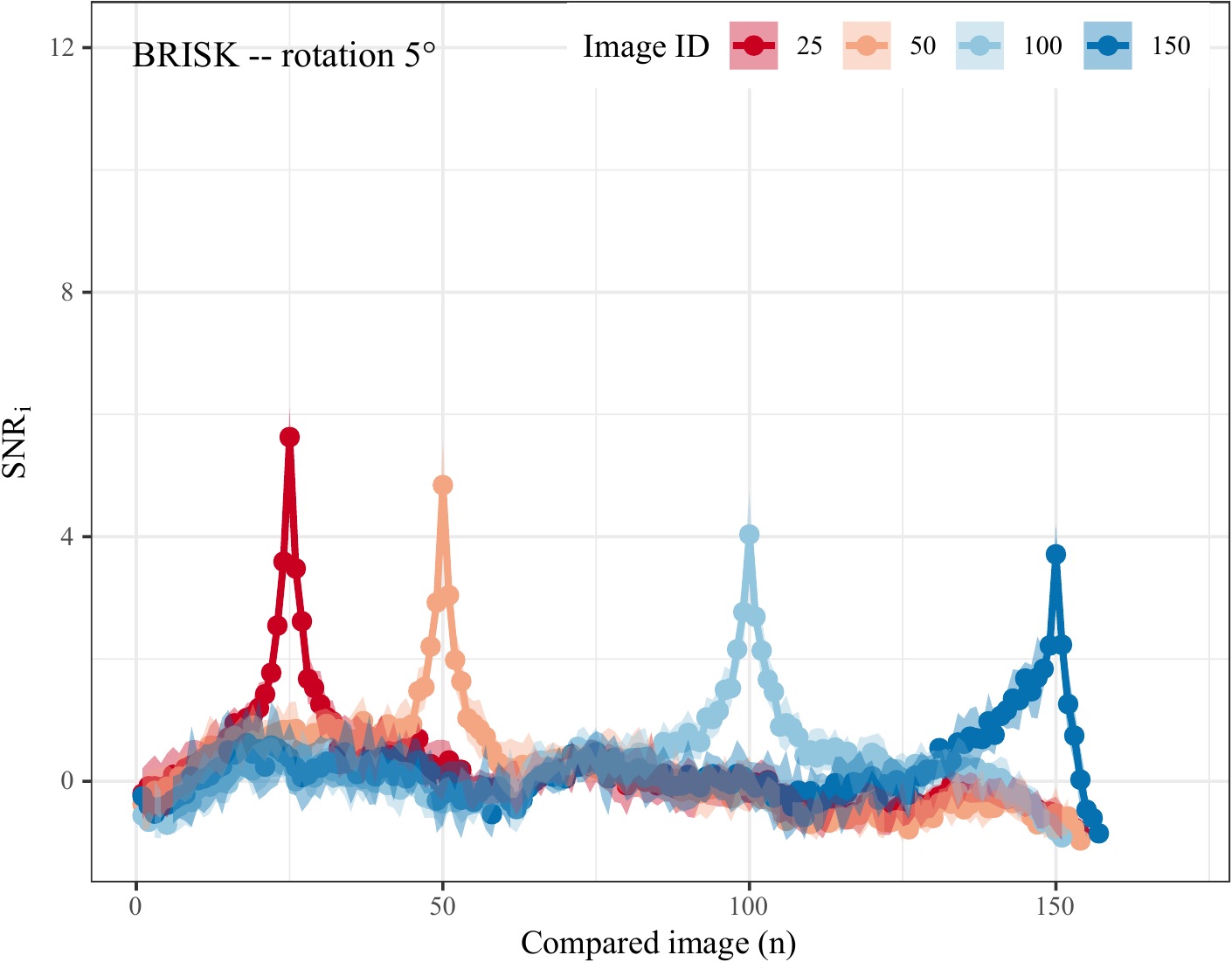}\hfill
    \includegraphics[width=.33\linewidth]{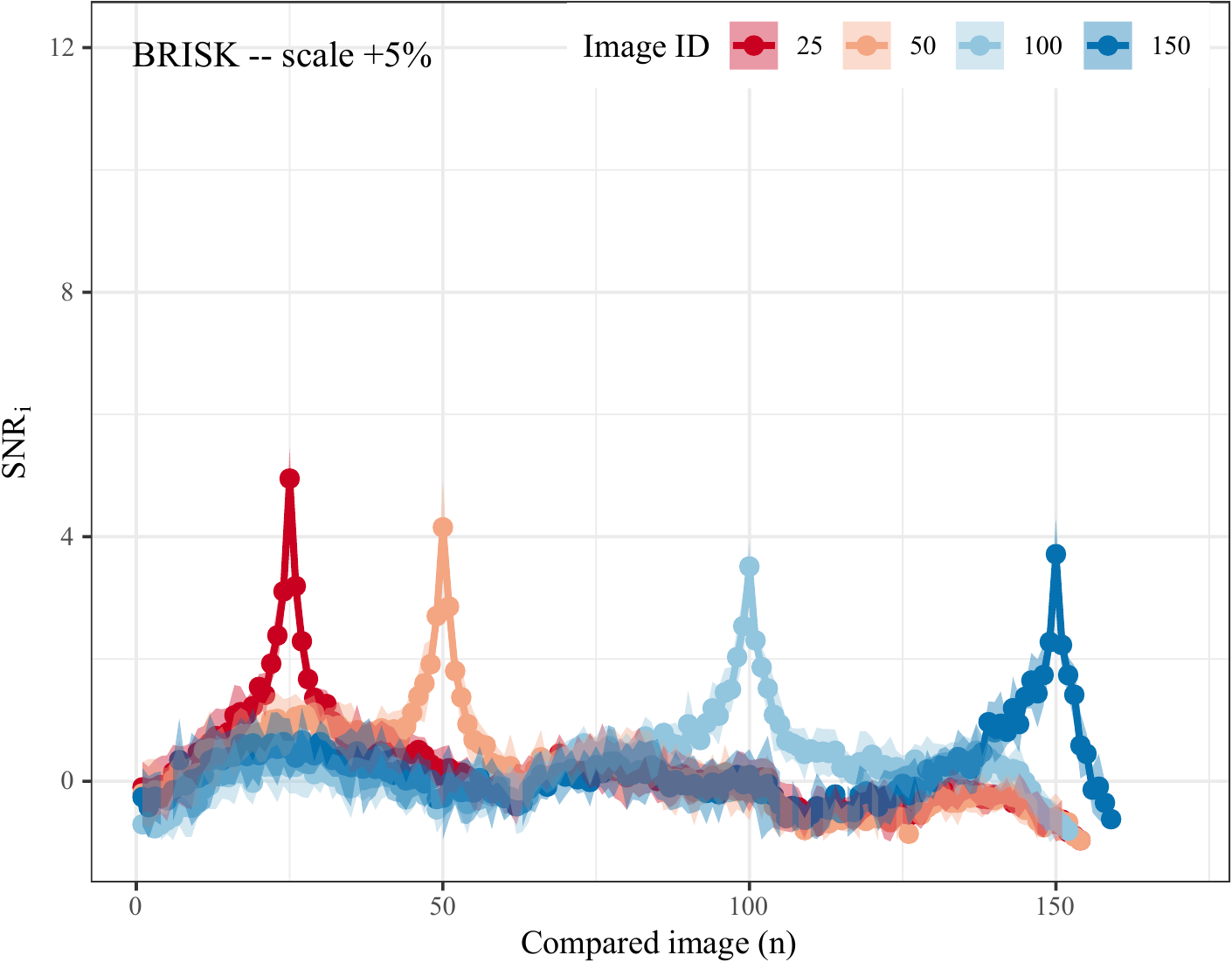}\hfill
    \includegraphics[width=.33\linewidth]{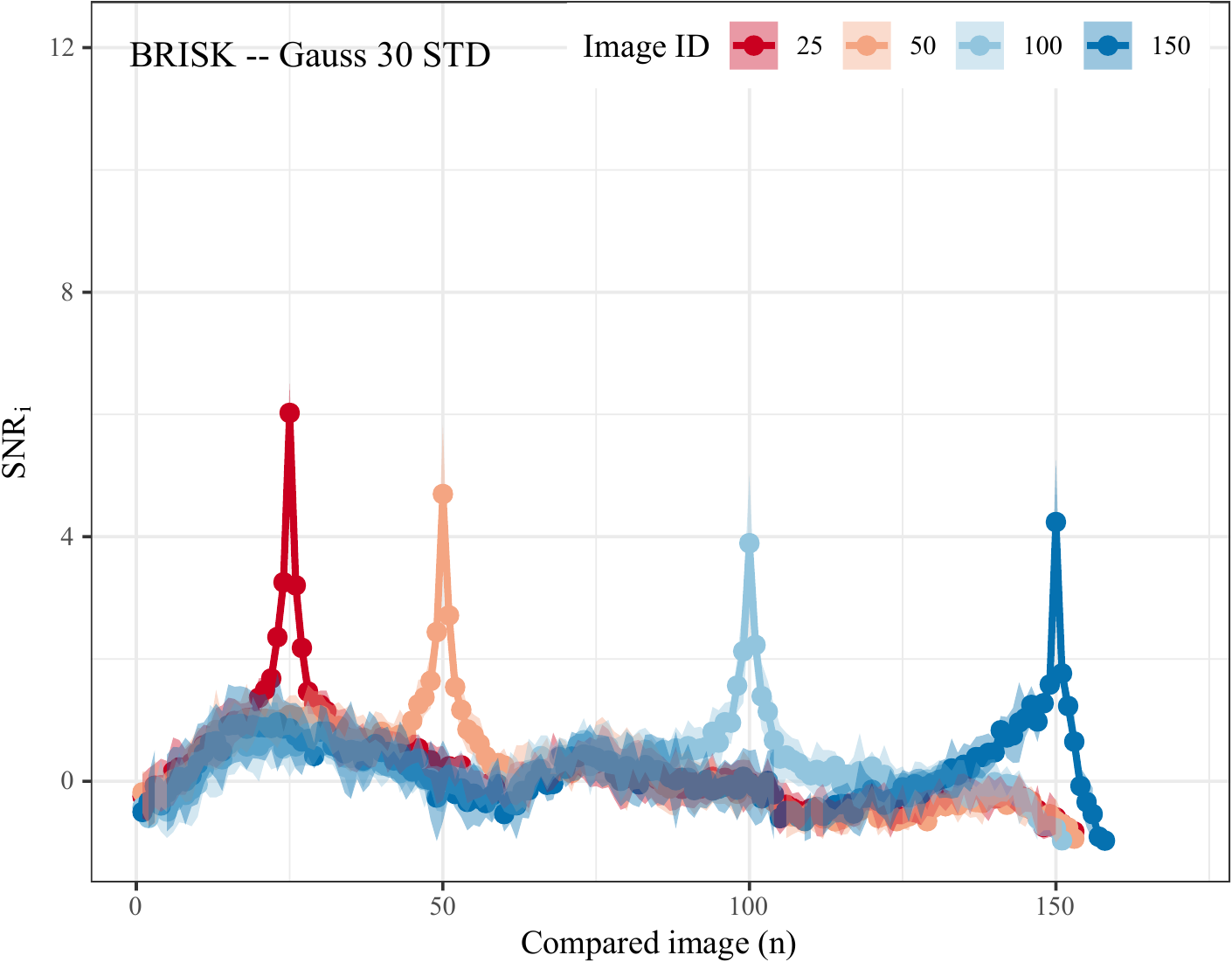}
    \caption{\label{fig:snr_degradation}The values of SNR (y-axis) computed by using Eq.~\eqref{eq:snr} (higher values is better) for selected input images (25, 50, 100, 150) for the first three used FD methods: AGAST (first row), AKAZE (second row) and BRISK (third row). The first column corresponds to an input image degradation with rotation by 5$^\circ$, the second column to up-scaling the image by 5\%, and the third column to the case when Gaussian noise of standard deviation 30 is added. The colour lines and the coloured region are following the same description as in Fig.\ref{fig:snr_identity}.}
    
\end{figure*}

\begin{figure*}
    \centering
    \includegraphics[width=.33\linewidth]{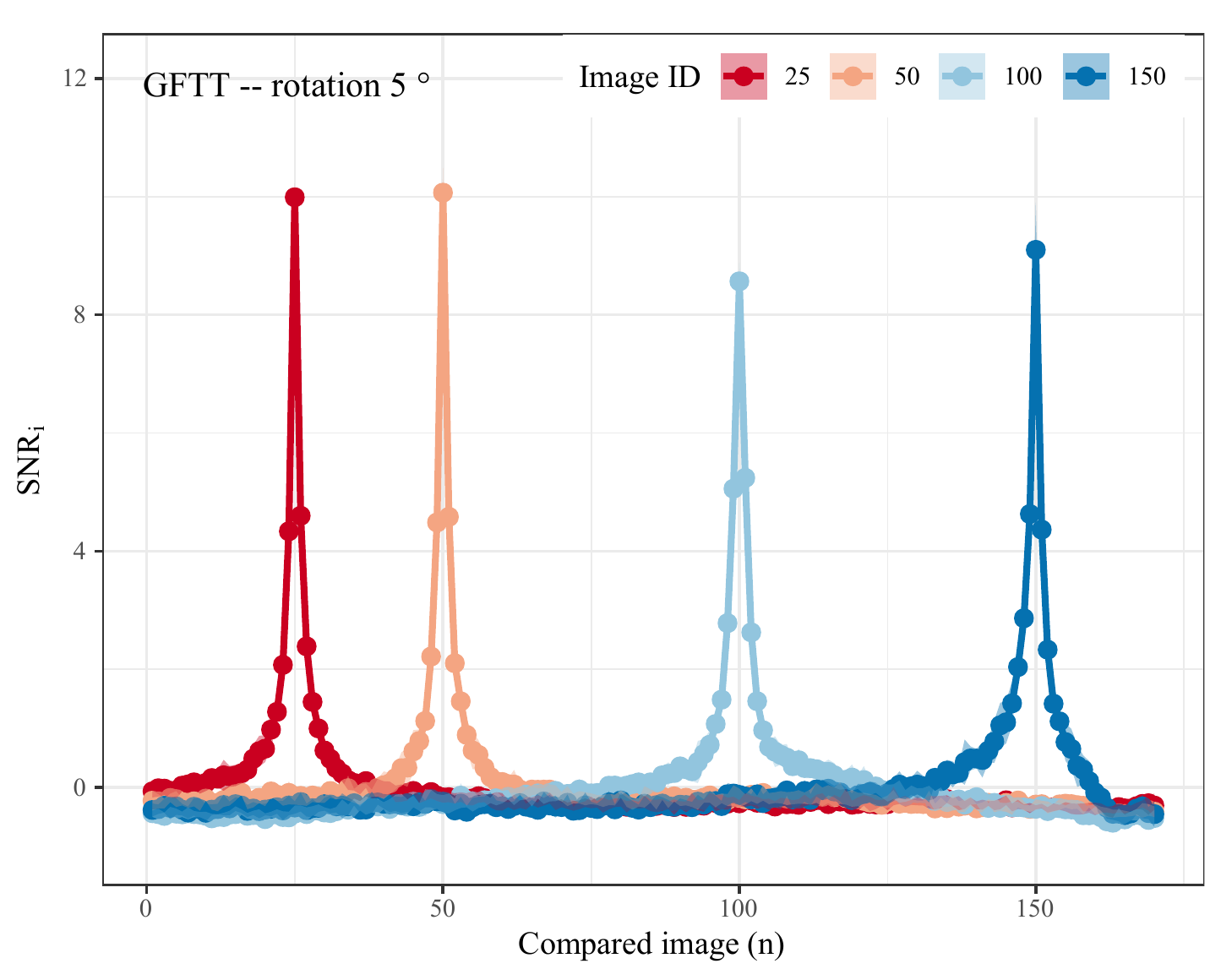}\hfill
    \includegraphics[width=.33\linewidth]{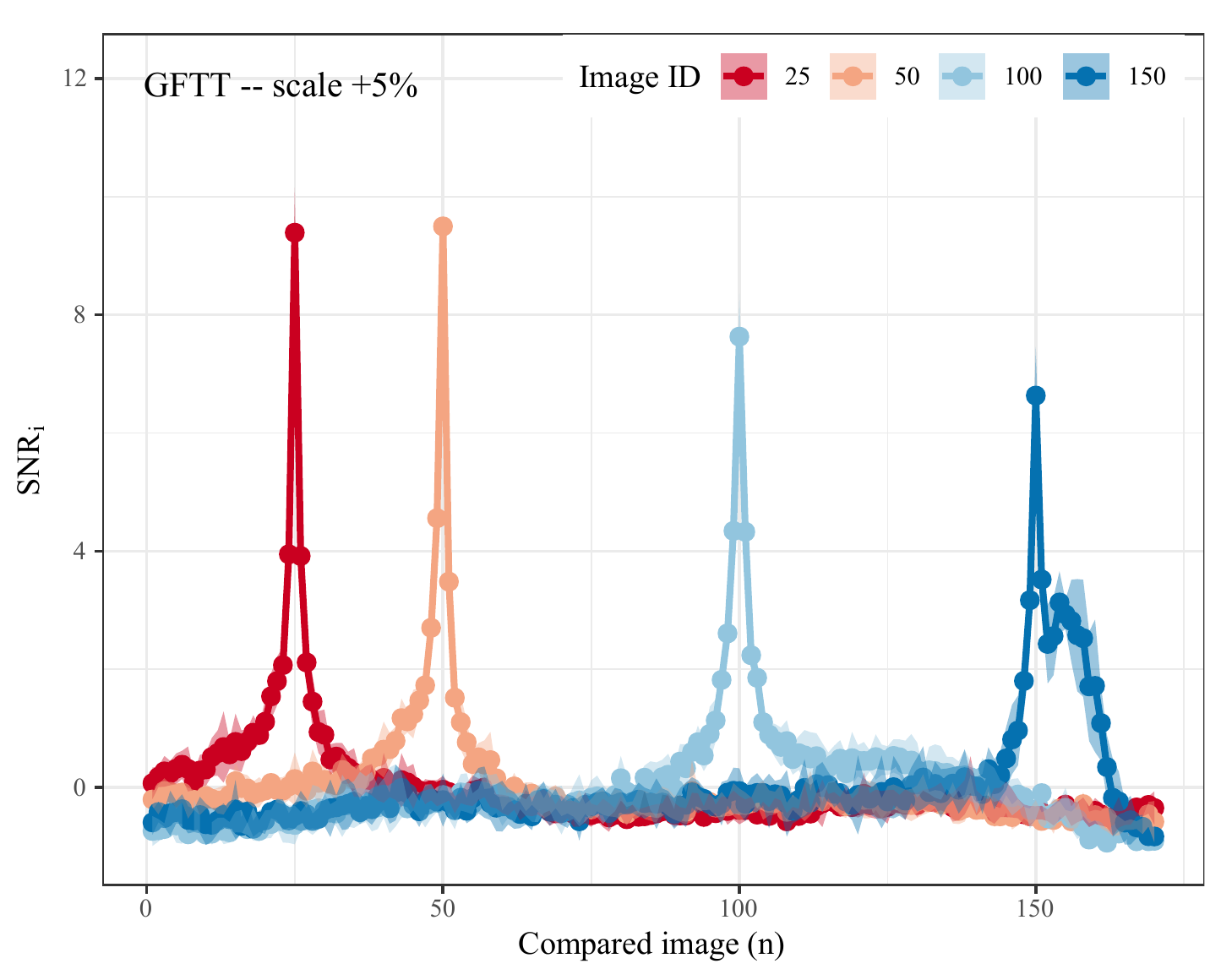}\hfill
    \includegraphics[width=.33\linewidth]{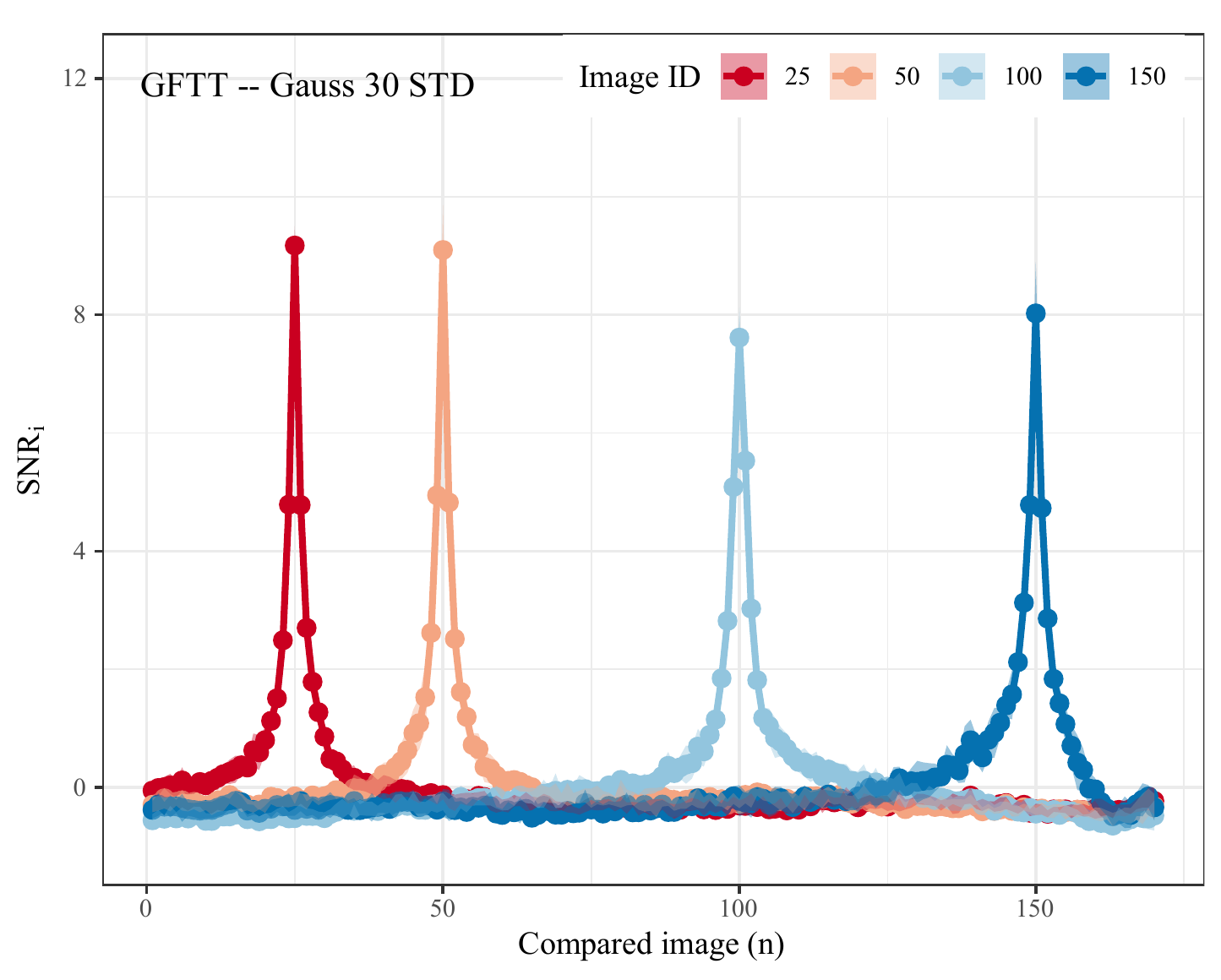}\\
    \includegraphics[width=.33\linewidth]{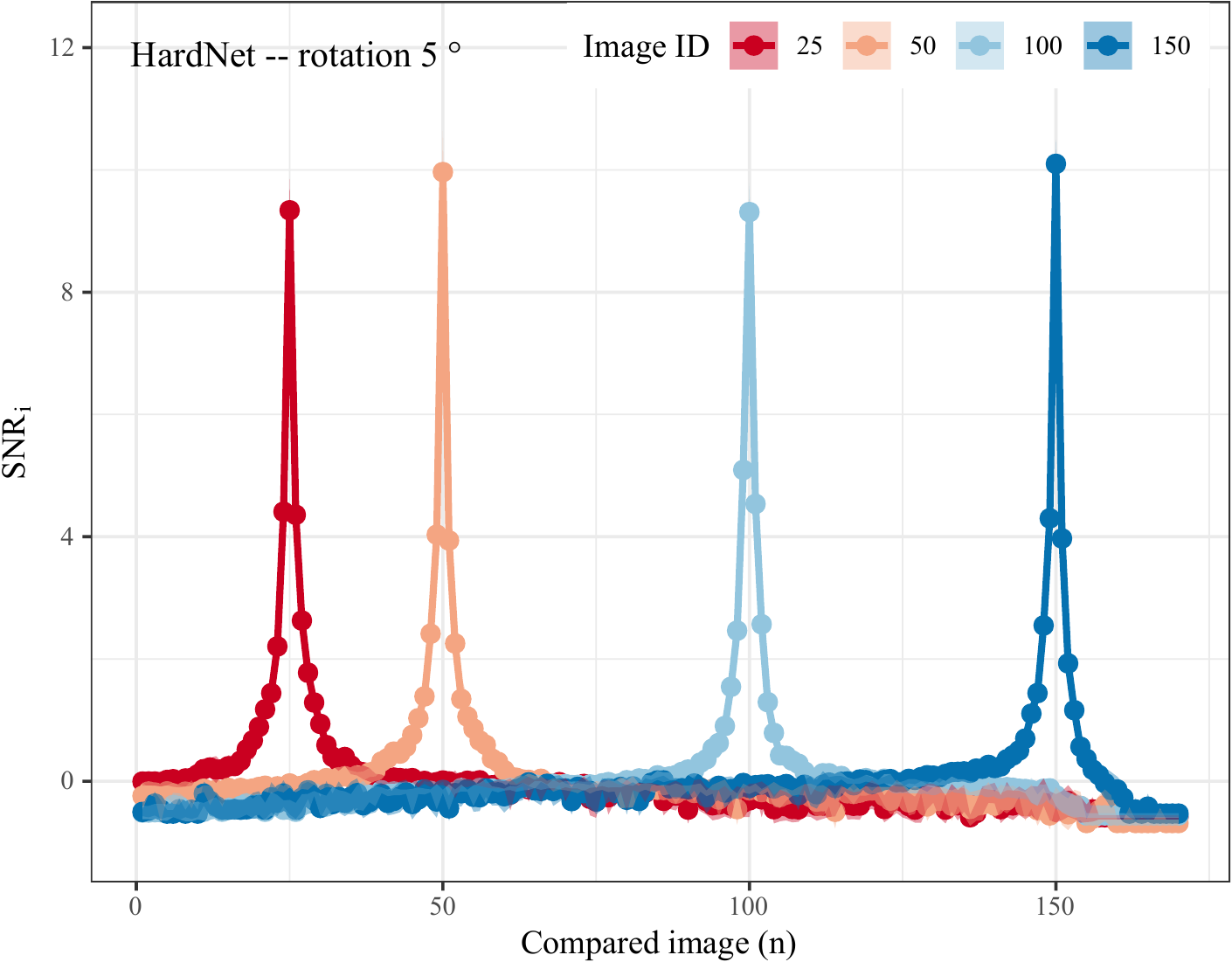}\hfill
    \includegraphics[width=.33\linewidth]{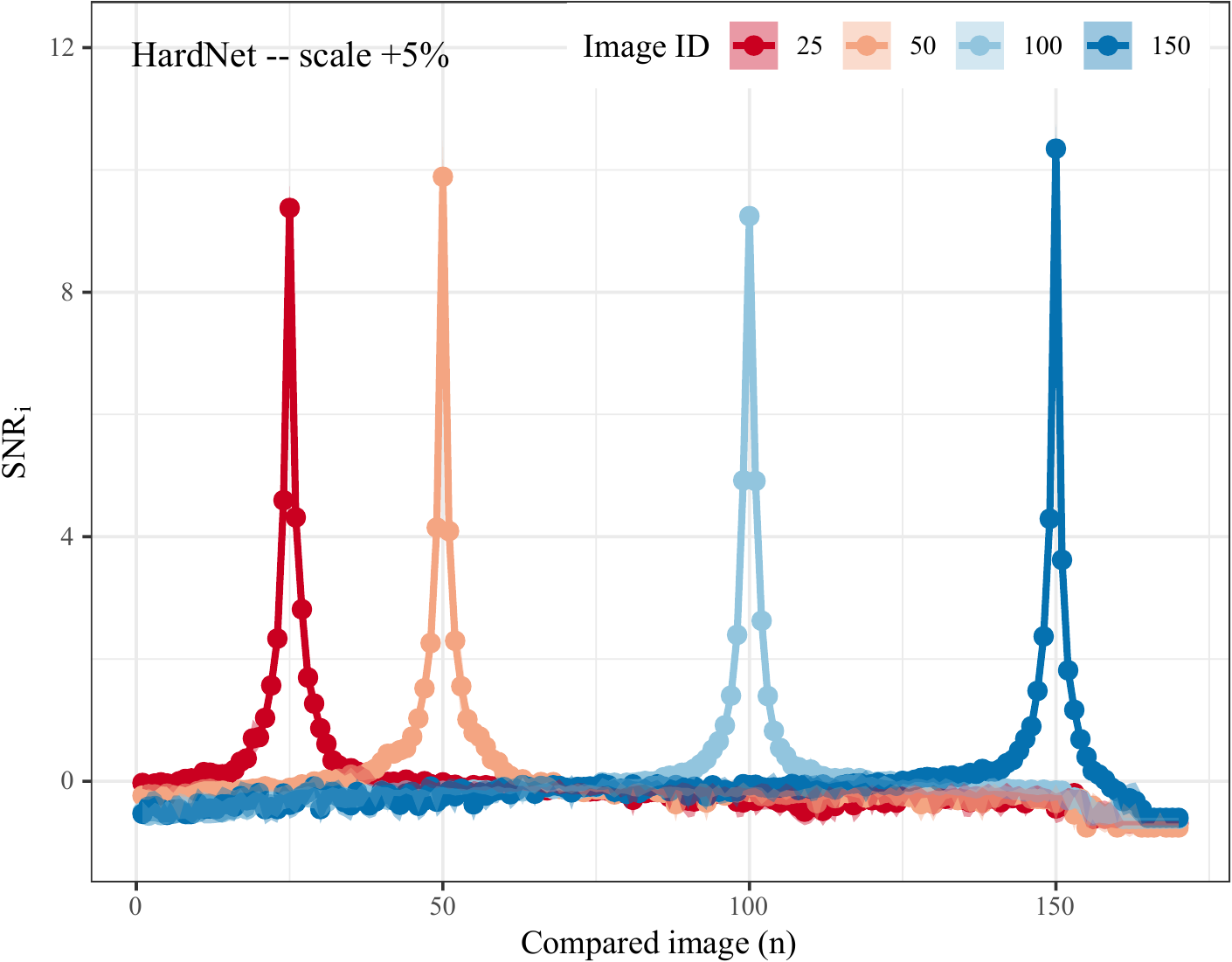}\hfill
    \includegraphics[width=.33\linewidth]{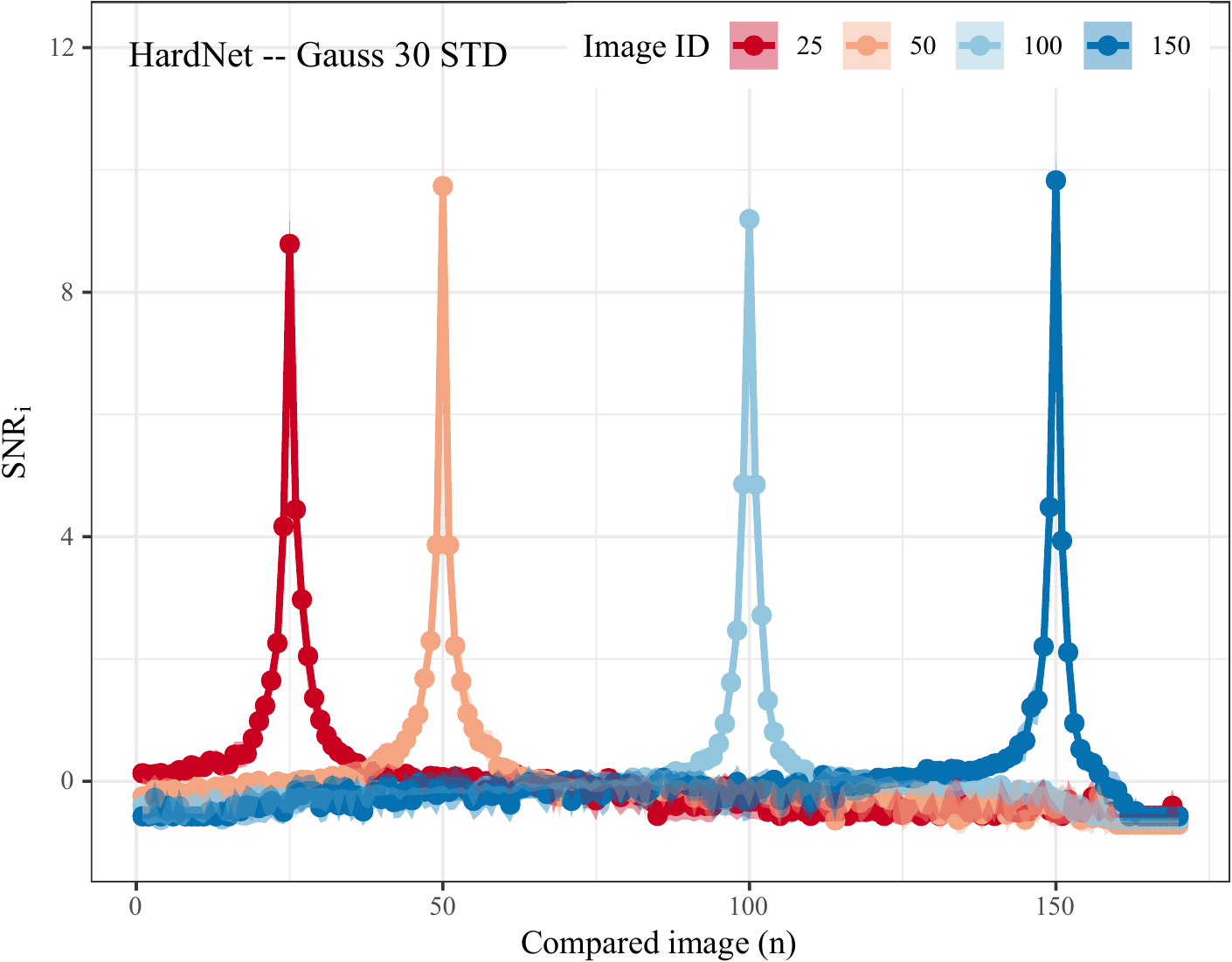}\\
    \includegraphics[width=.33\linewidth]{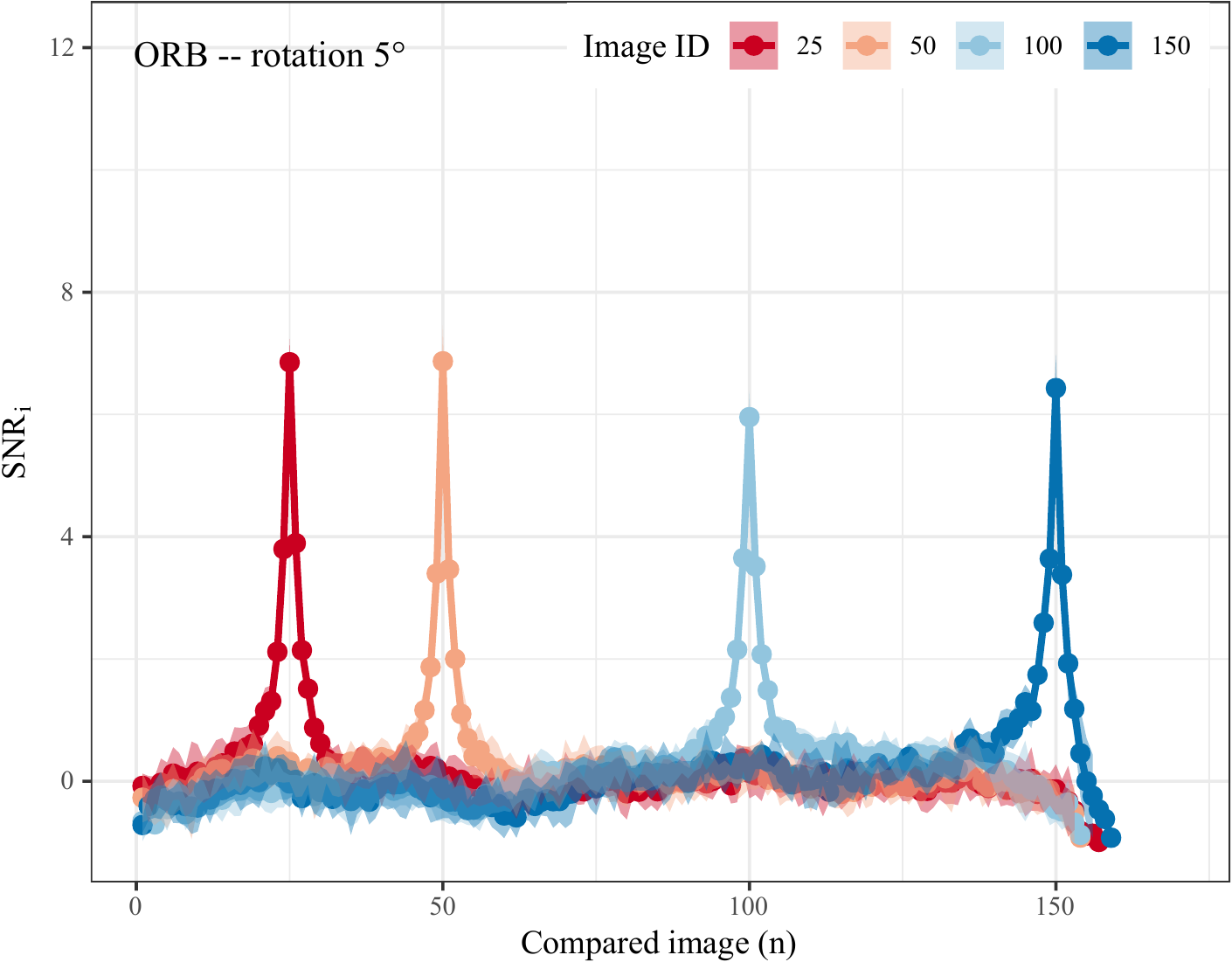}\hfill
    \includegraphics[width=.33\linewidth]{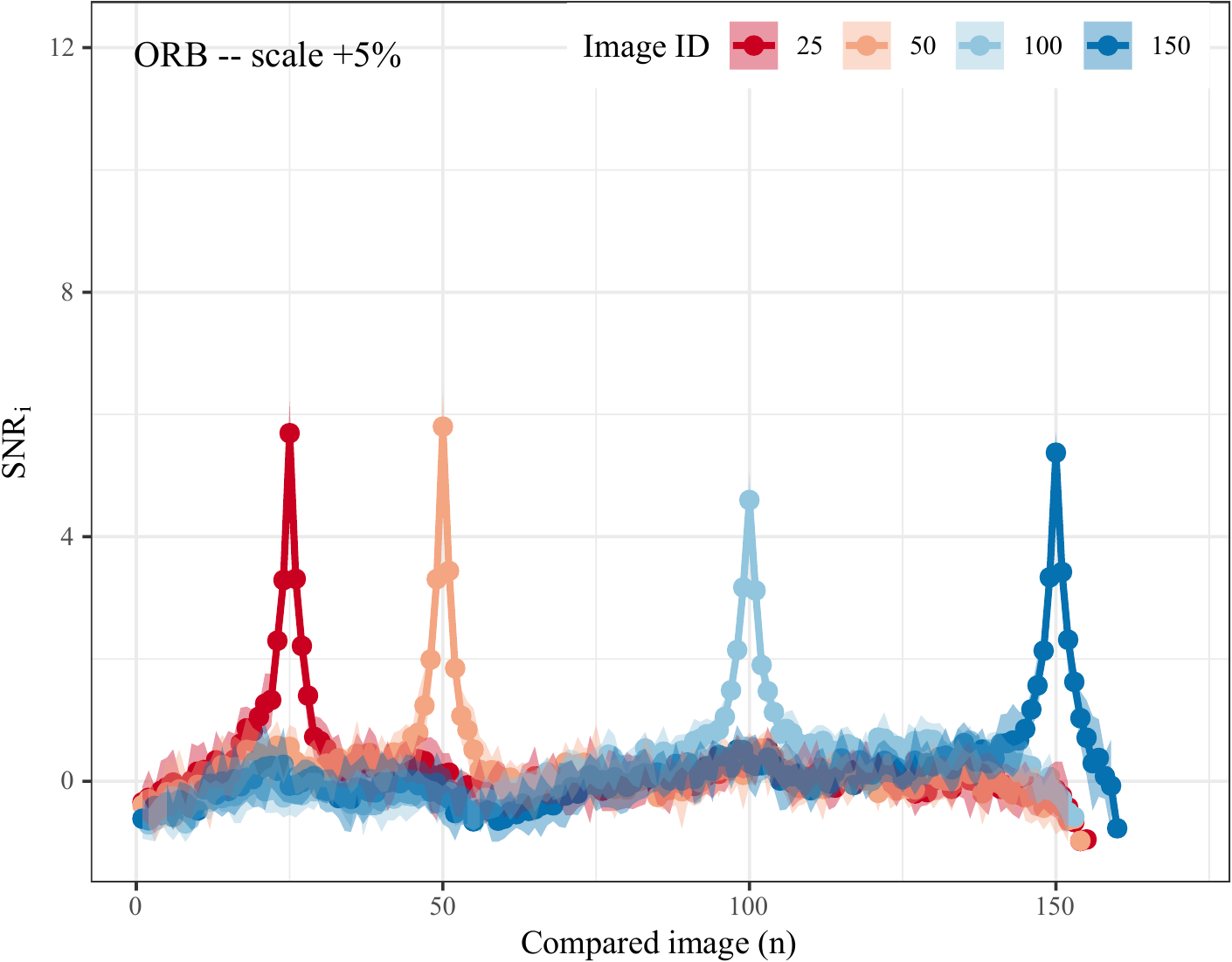}\hfill
    \includegraphics[width=.33\linewidth]{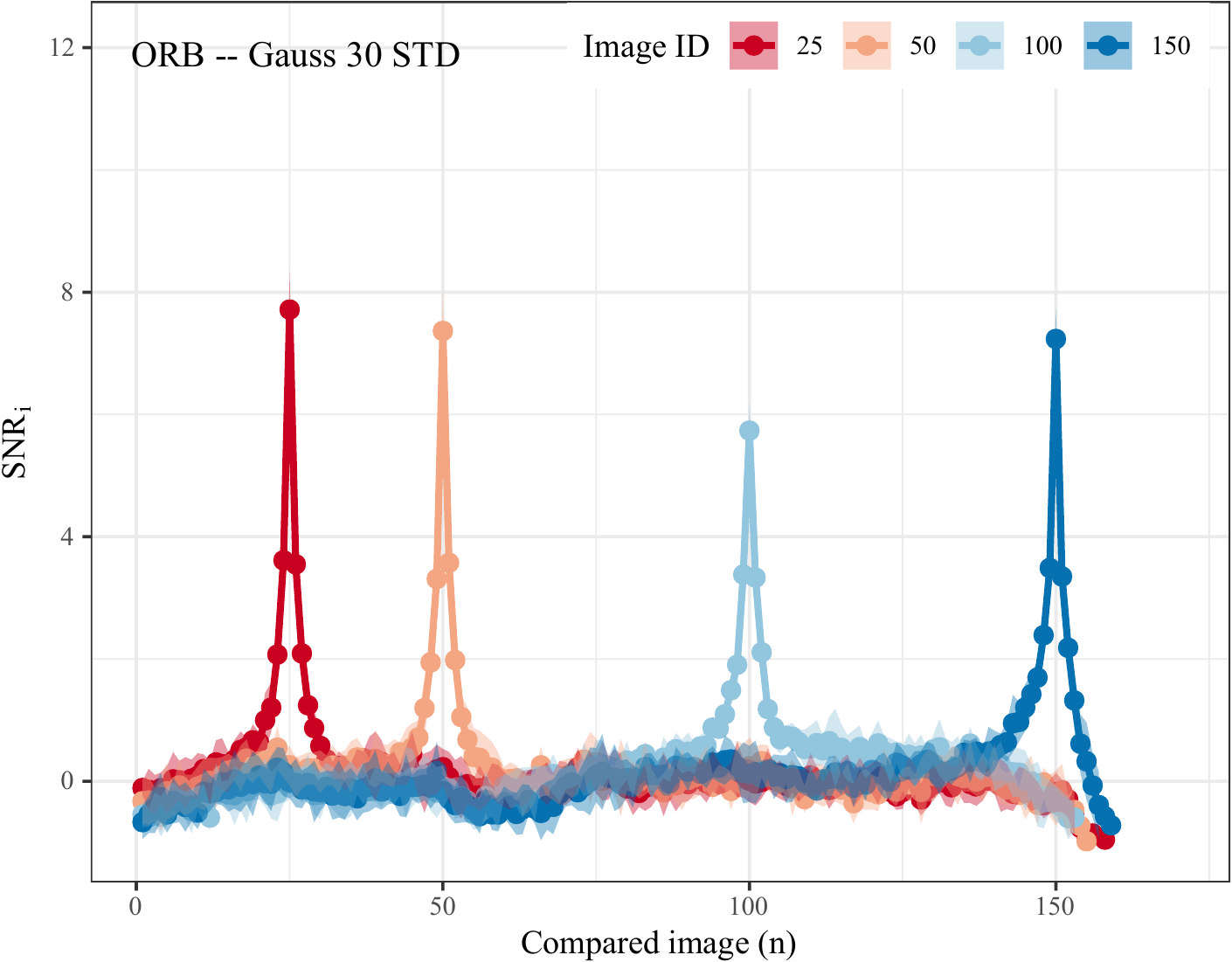}
    \caption{\label{fig:snr_degradation2}Continuation of Figure~\ref{fig:snr_degradation} showing the value of SNR (y-axis) computed by Eq.~\ref{eq:snr} with the next three FD methods (in rows): GFTT, HardNet, and ORB, for each image degradation (columns: rotation, scale, noise).}
\end{figure*}

In Figures~\ref{fig:snr_degradation} and \ref{fig:snr_degradation2}, similarly as in the previous case in Figure~\ref{fig:snr_identity} for image IDs 25, 50, 100 and 150, we show the behaviour of the SNR values when the image degradation to the input images are applied. Namely, in the first columns a clock-wise rotation of 5$^\circ$ is used, in the second columns images were up-scaled by 5\%, and in the third columns a Gaussian noise with STD=30 is used. Please note, that we also tested the Gaussian noise with 10 and 20 STD and found out, that in general it slowly decreases the SNR values overall. For this reason we included only the results with STD 30 as the extreme case. In general, the scaling factor is the one where the SNR values are affected the most, followed by rotation, and lastly the image noise (see the values of SNR). 

From the perspective of the FD methods the GFTT and AKAZE are the most invariant in all image degradation factors, especially in the case of Gaussian noise, where the loss in the values of SNR is almost negligible. AGAST is the most invariant in the case of rotation degradation. The ORB and BRISK methods are affected by the changes in the input images and we can see a $\sim$40\% loss in the SNR values. HardNet is benefiting from the image degradations of the neighbouring images, respectively the decreasing number of matching points, and thus improving the SNR. In the Table~\ref{tab:snr_loss} we summarise the average of obtained SNR for all input images when related degradation is used.

\begin{figure*}[htbp]
    \centering
    \includegraphics[width=.33\linewidth]{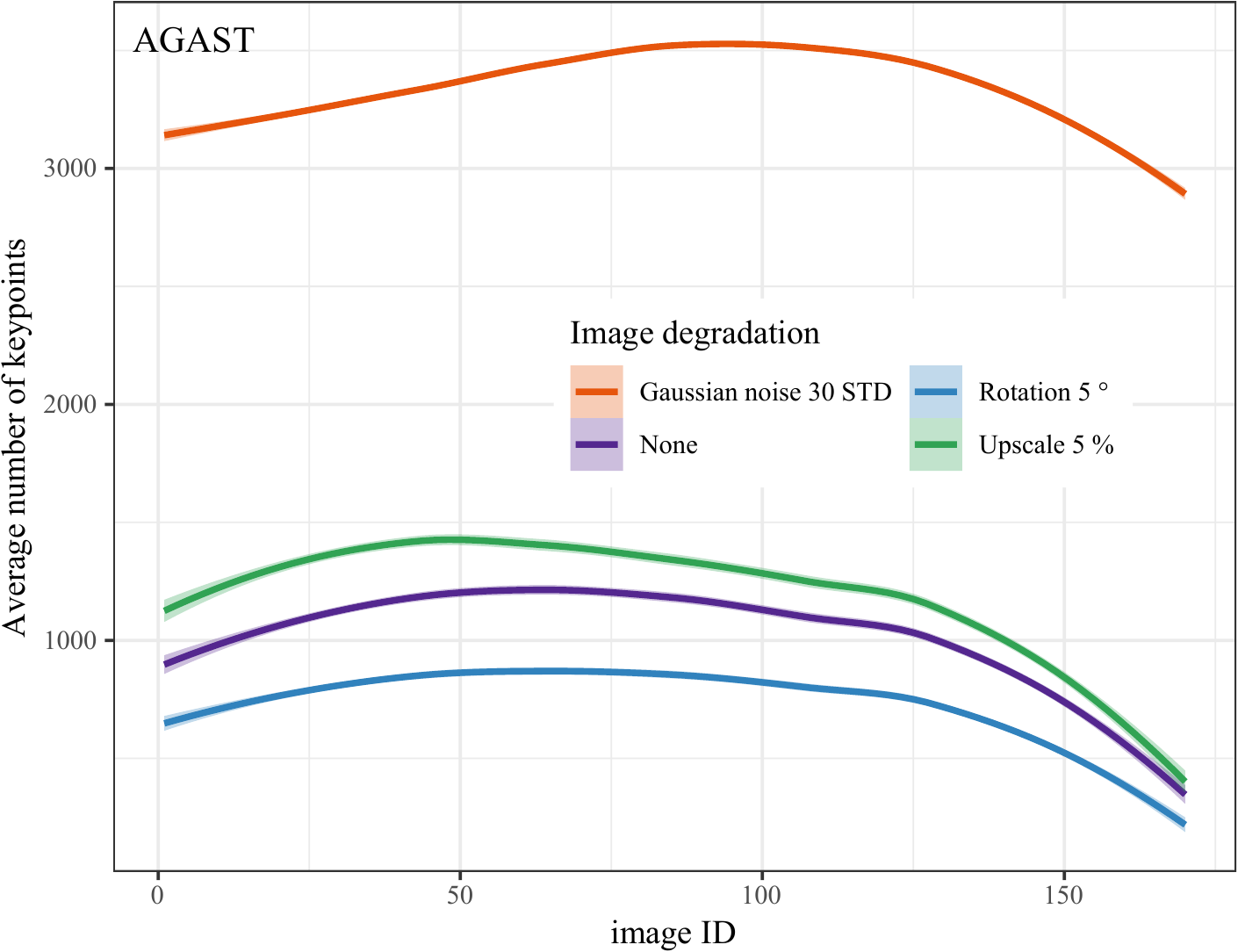}\hfill
    \includegraphics[width=.33\linewidth]{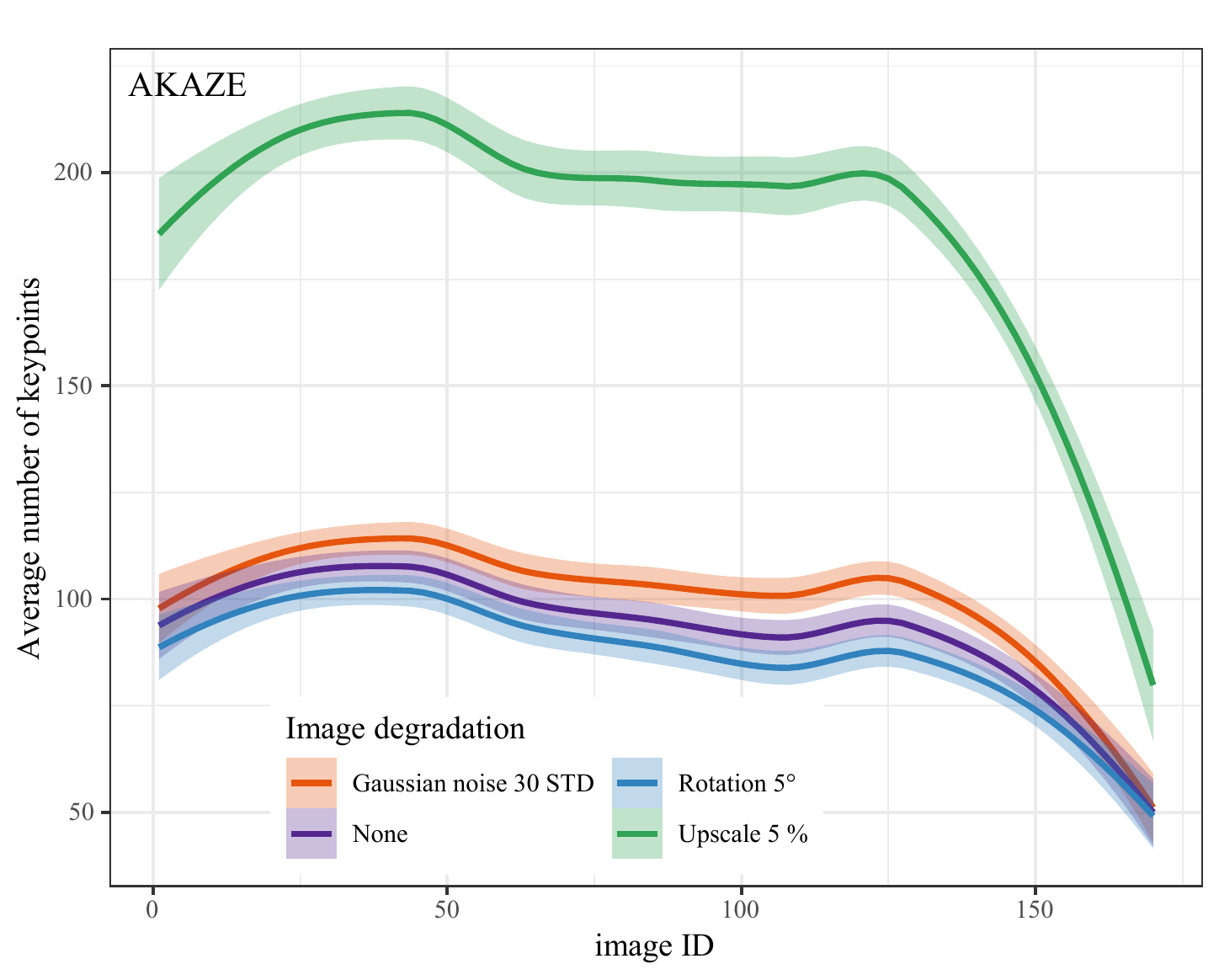}\hfill
    \includegraphics[width=.33\linewidth]{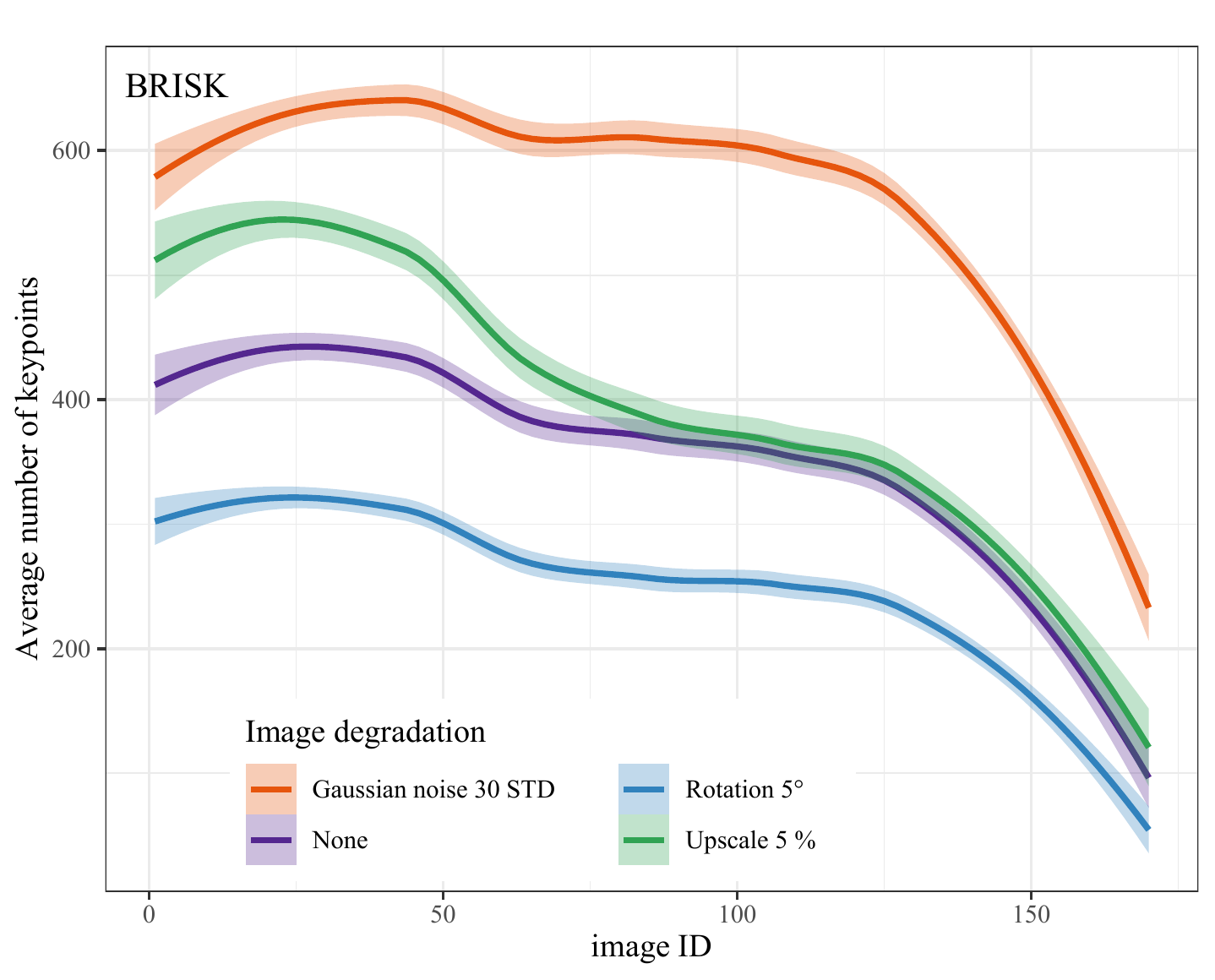}\\
    \includegraphics[width=.33\linewidth]{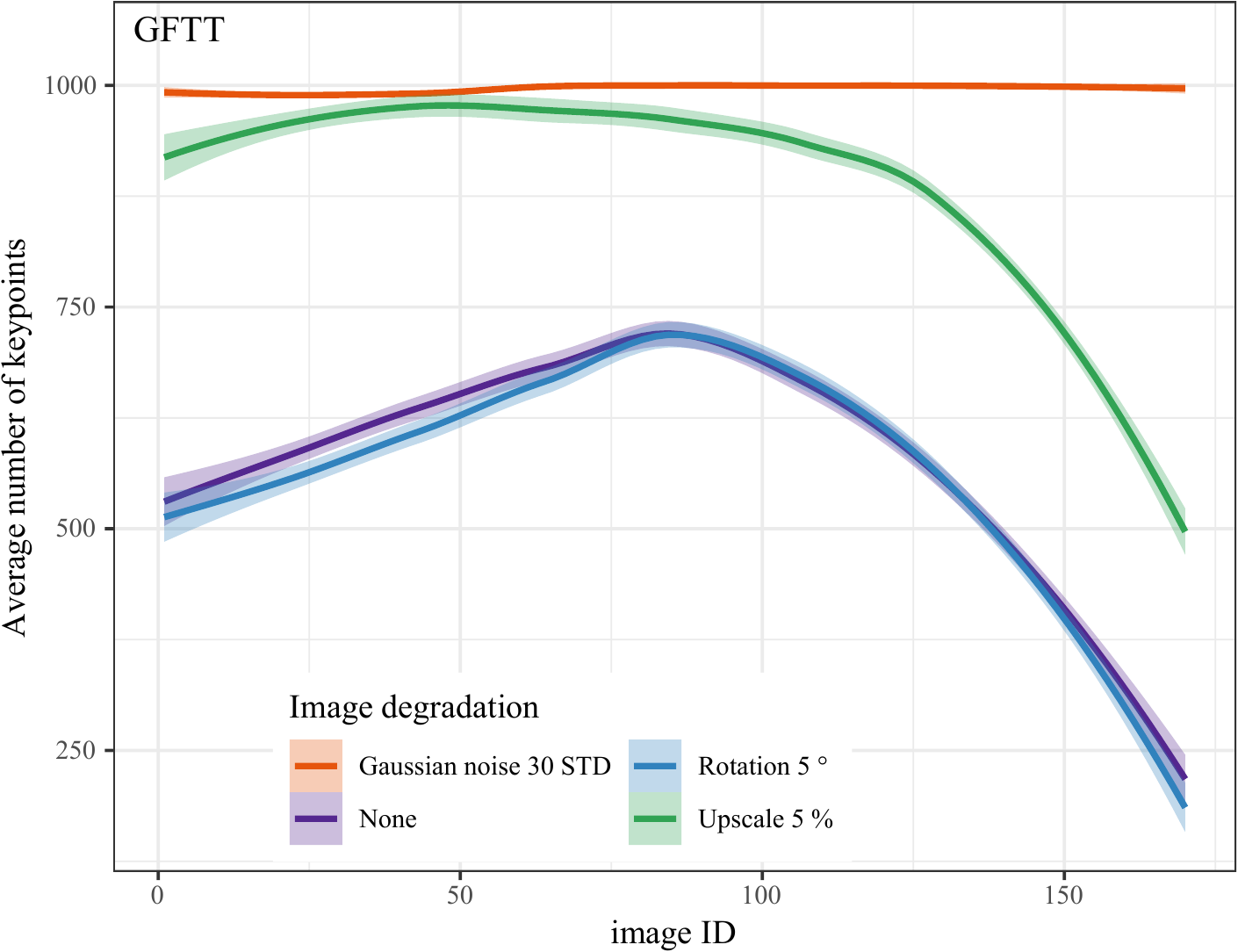}\hfill
    \includegraphics[width=.33\linewidth]{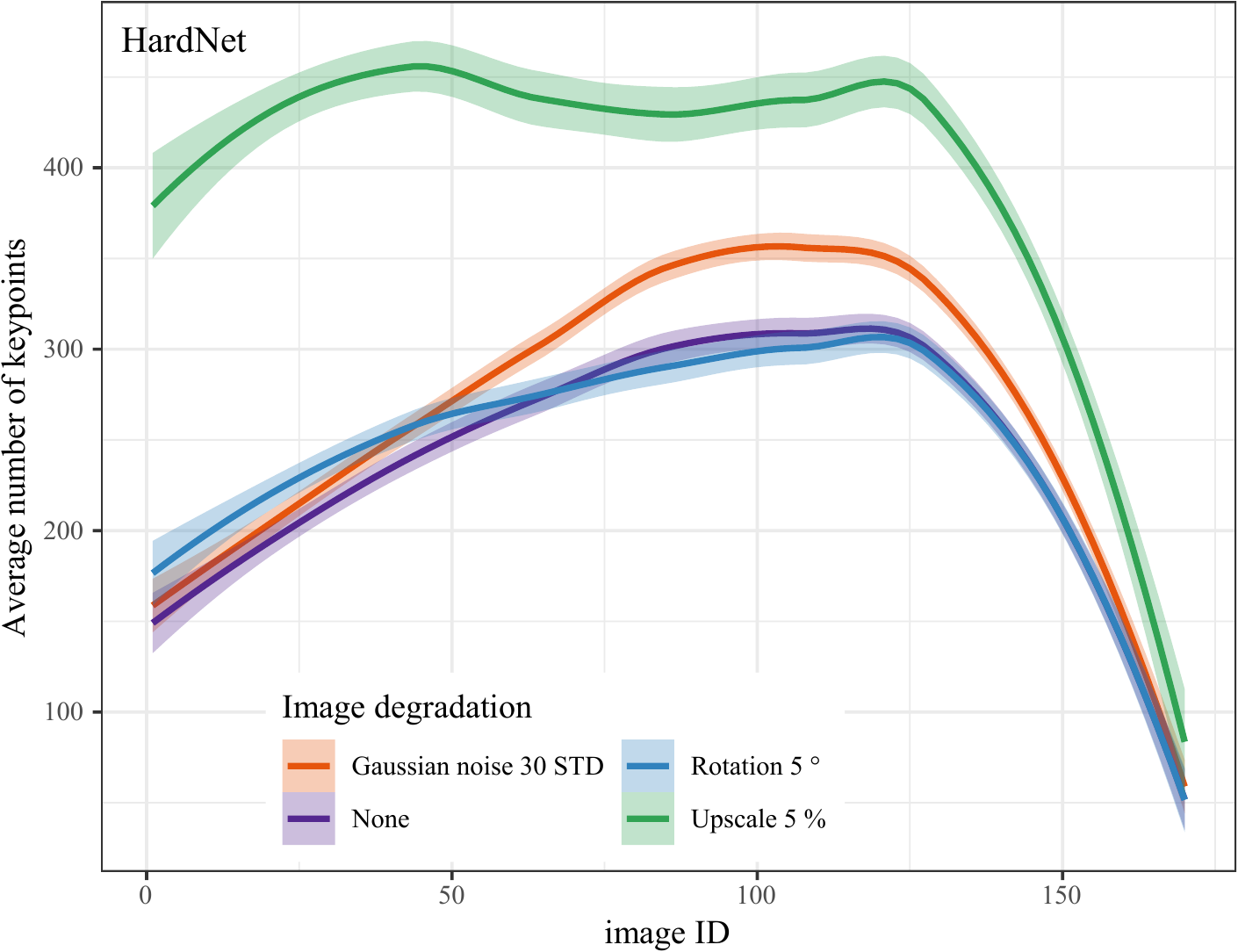}\hfill
    \includegraphics[width=.33\linewidth]{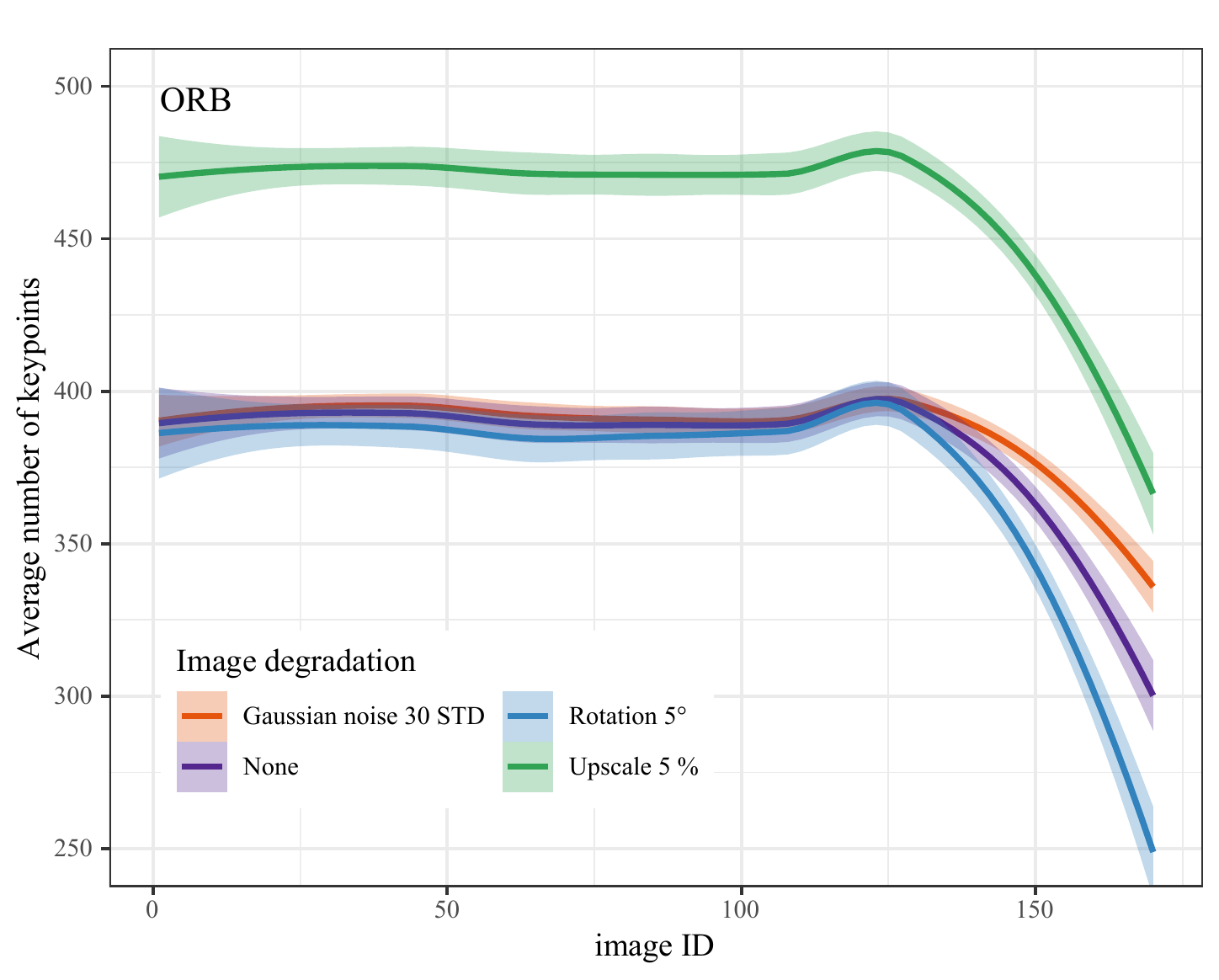}\\
    \caption{The average number of keypoints for each input image found by FD methods AGAST (the first row left), AKAZE (first row middle), BRISK (first row right), GFTT (the second row left), HardNet (second row middle), and ORB (second row right) with and without image degradation. The solid line represents the mean value computed from the 7 subjects and the coloured area is the standard deviation.}
    \label{fig:nkeypoints_degradated}
\end{figure*}

\begin{table*}[htbp]
\begin{center}
\caption{\label{tab:snr_loss} Summary of the mean SNR for AGAST, AKAZE, BRISK, GFTT, HardNet, and ORB methods for cases when the non-degenerated and degenerated input images are taken into account.}
\begin{tabular}{@{}l|cccc@{}}
\toprule
\multicolumn{1}{l|}{Feature detection} & \multicolumn{4}{c}{Image degradation factor:} \\
\multicolumn{1}{l|}{method} & None & Rotation & Upscaling & Noise \\ \midrule
AGAST          & 10.60 & 10.31     & 4.70      & 9.15    \\
AKAZE          &  7.70 &  7.02     & 5.78      & 6.92        \\
BRISK          &  8.06 &  4.53     & 3.87      & 4.36       \\
GFTT           &  9.66 &  9.22     & 8.28      & 8.47    \\
HardNet        &  7.41 &  9.43     & 9.39      & 9.19    \\
ORB            &  8.93 &  6.47     & 5.24      & 6.87        \\
\bottomrule
\end{tabular}
\end{center}
\end{table*}
In terms of competitive comparisons, AGAST found the highest number ($\sim$2000) of keypoints (also called interest points) in most input slices of the MRI images, followed by GFTT ($\sim$600), and then by BRISK ($\sim$450). When we applied the rotation the number of keypoints dropped or remained the same for all used methods. All other image degradations (upscaling and Gaussian noise) behave slightly differently for each feature detector tested. For AKAZE, ORB, HardNet (SIFT) and GFTT, the upscaling factor show the highest increase in the number of interest points, whilst for BRISK and AGAST it is the Gaussian noise. Please note, that from image ID=150 the number of keypoints rapidly decreases. This can be expected as input images approach the top of the head, where there is less and less usable information not only from the medical point of view, but also from the computer vision point of view. However, for the completeness of our survey, we present the results in these cases as well. The average number of found keypoints for each used method and image degradation are shown in Fig.~\ref{fig:nkeypoints_degradated}. From the point of computational performance, the order is from the fastest to the slowest GFTT, AGAST, HardNet, AKAZE, ORB and BRISK (0.2, 0.25, 1, 5, 8 and 15 minutes) to compare 1 input image with 170 others.

\subsection{Robustness}
\label{sec:robustness}
\begin{figure*}[htbp]
    \centering
    \includegraphics[width=.33\linewidth]{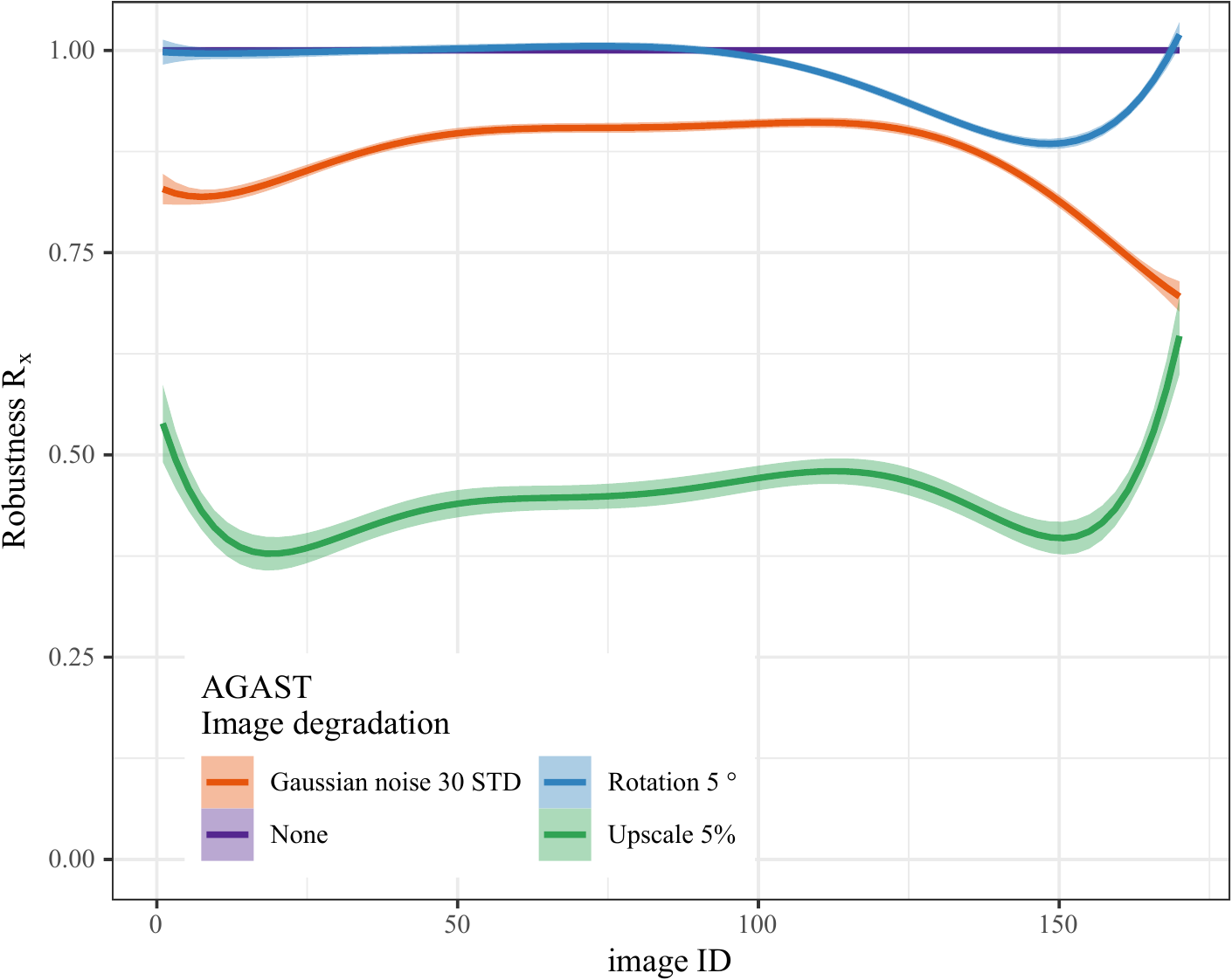}\hfill
    \includegraphics[width=.33\linewidth]{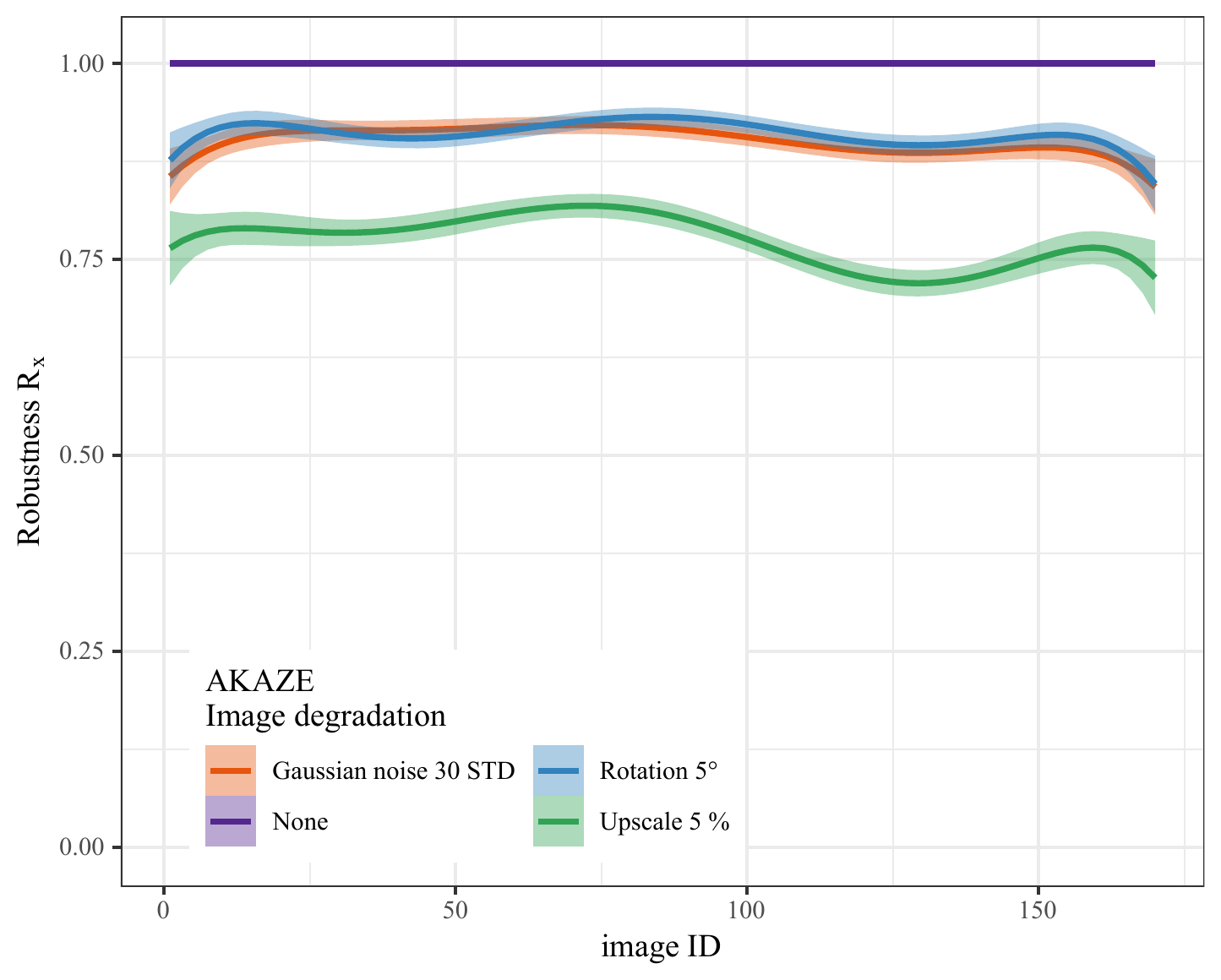}\hfill
    \includegraphics[width=.33\linewidth]{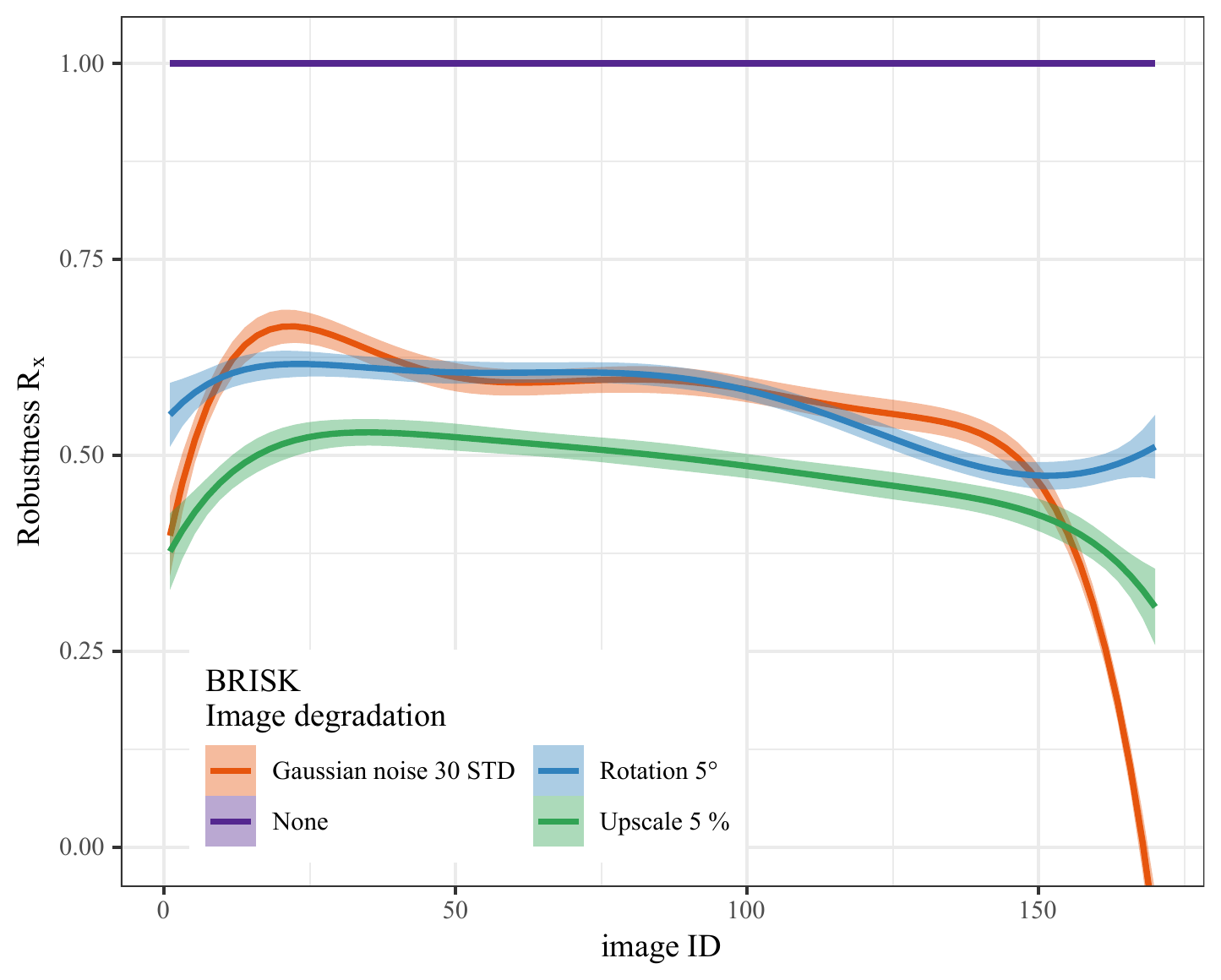}\\
    \includegraphics[width=.33\linewidth]{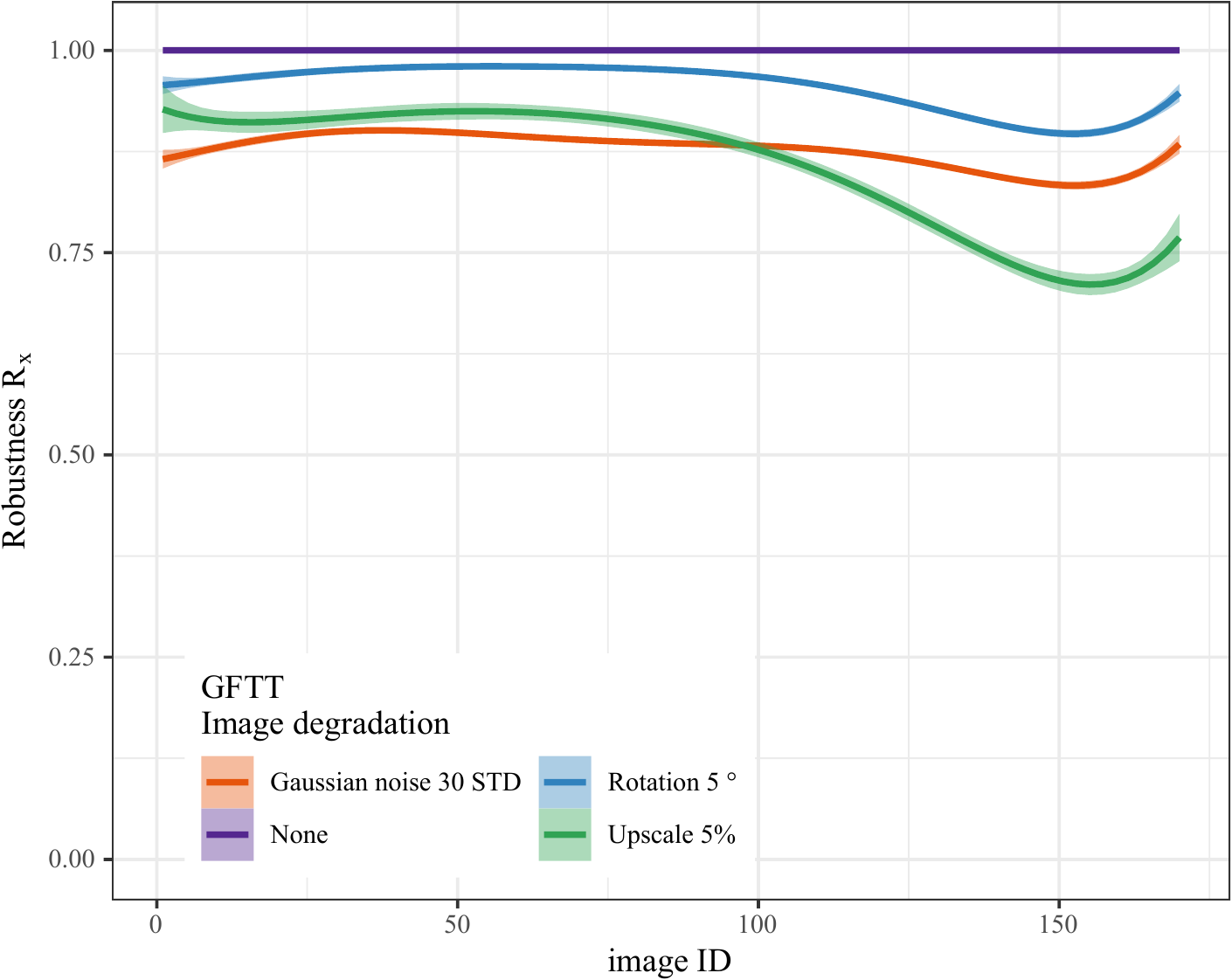}\hfill
    \includegraphics[width=.33\linewidth]{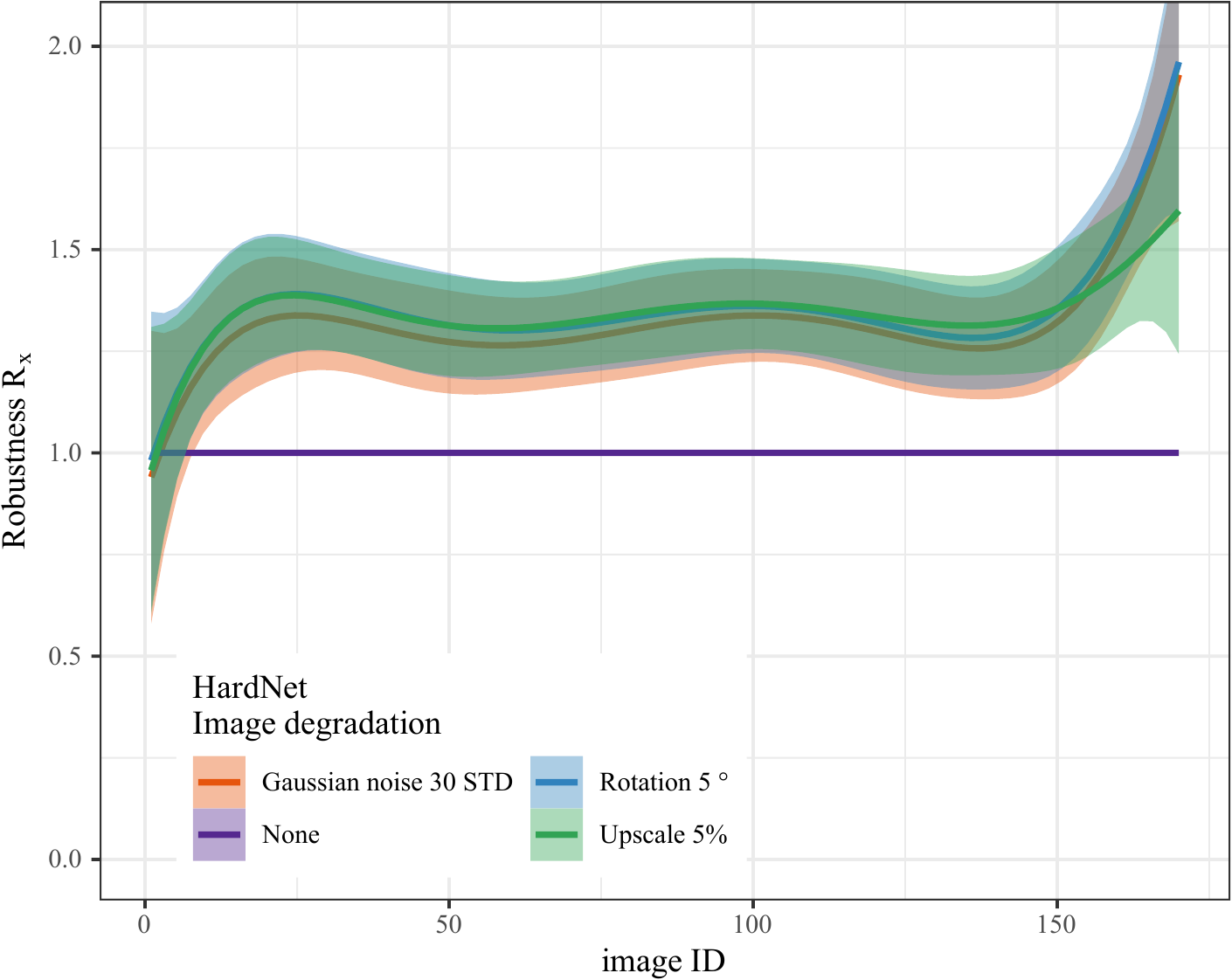}\hfill
    \includegraphics[width=.33\linewidth]{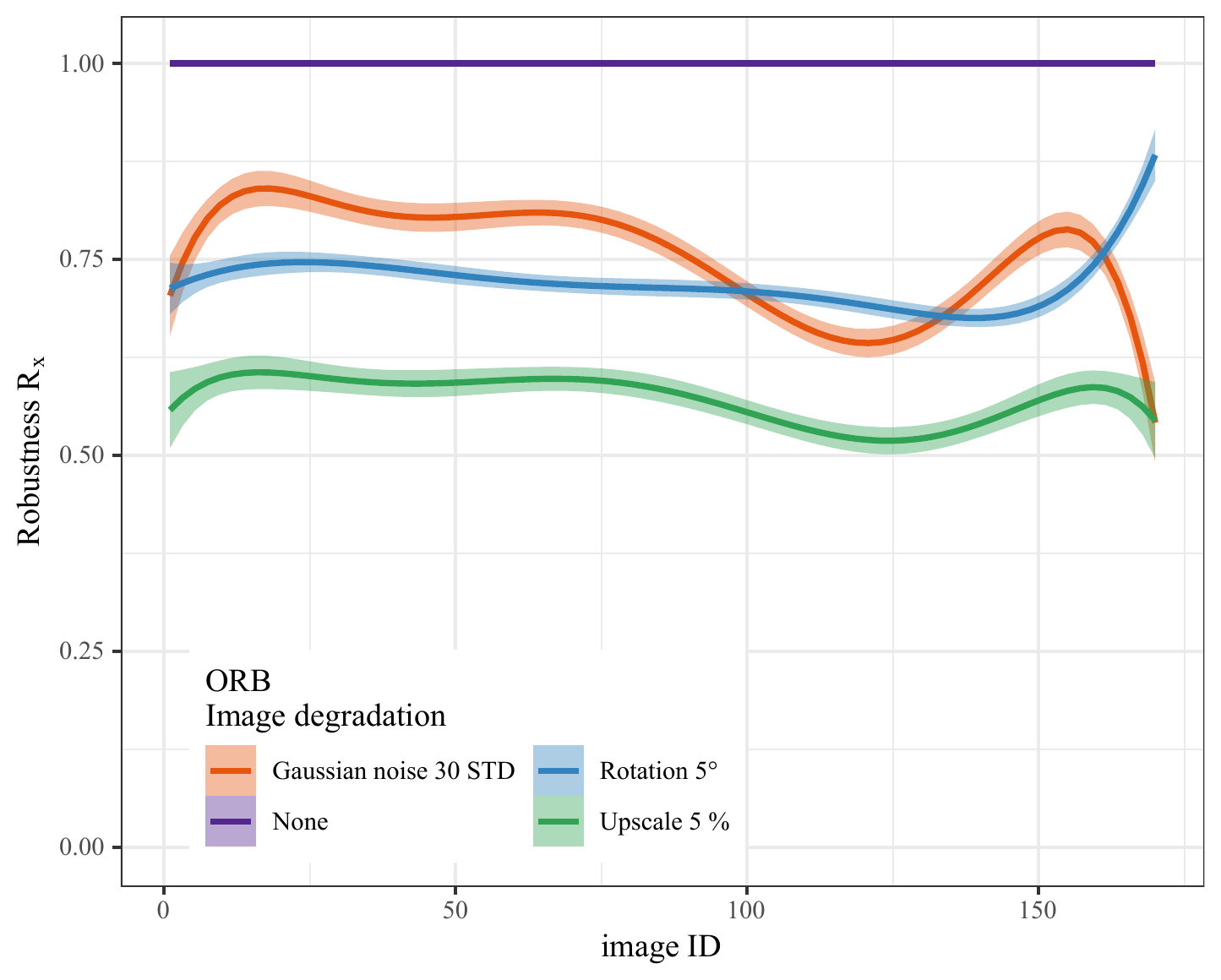}\\
    \caption{The robustness $R_x$ (y-axis) computed by the Eq.~\ref{eq:robustness} for each input image for AGAST (first row left), AKAZE (first row center), BRISK (first row right), GFTT (second row left), HardNet (second row center), and ORB (second row right). The solid line represents the mean value computed from the 7 subjects, their colour to the appropriate image degradation (rotation, upscaling, noising) and the coloured area to the standard deviation.}
    \label{fig:robustness}
\end{figure*}

These tests serve especially for a better estimation of the problems with the selected FD methods and to compare their invariability to typical image degradation in medical science, respectively in our case for MRI. In Figure~\ref{fig:robustness} we show the robustness $R_x$ calculated by Eq.~\ref{eq:robustness} for each method and image degradation as a function of the input image IDs. Since the robustness $R_x$ corresponds to the ratio between the SNR affected by the image degradation and the unaffected, a higher value of the robustness $R_x$ reflects the behaviour of the FD method invariant behaviour invariant to the selected image degradation.

The most problematic image degradation type for selected FD methods is the upscale factor, where AGAST, ORB and BRISK show the loss of SNR around 50\,\%. The rotation and the Gaussian noise demonstrate similar changes in robustness for all FD methods, with the greatest influence in the case of the BRISK and the least in the case of AGAST and AKAZE. On the other hand HardNet show for all image degradation an increase in robustness, respectively value higher than 1 (see the second row, the middle column of Fig.~\ref{fig:robustness}). This is due to the fact that the SNR of the neighbouring image IDs (``the noise'') decreases (see Figure~\ref{fig:snr_degradation}), and thus the input image ID SNR increase.   

The most invariant FD methods are GFTT and AKAZE, followed by AGAST, ORB and BRISK. The GFTT and AKAZE method shows in general a loss of  10 and 14\,\% SNR compared with their original SNR ($R_x{\sim}$0.86--0.90), AGAST has, except for the rotation degradation where is the most invariant from all methods, around 20\,\% ($R_x{\sim}$0.60), ORB has around 32\,\% ($R_x{\sim}0.68$), and BRISK 47\,\% ($R_x{\sim}0.53$). Instead, HardNet shows a gain of 10\,\% ($R_x{\sim}1.1$).

\subsection{Accuracy}
\label{sec:accuracy}
As mentioned in the previous section the accuracy $A_{d,c}$ (defined by Eq.~\ref{eq:accuracy}) gives us the general overview of how the selected FD methods behave when a patient is compared to another patient or an atlas. For our test we selected patient number 7 to perform as an atlas. In this section we compare data taken from the same machine, i.e., with similar image properties. In the next section we provide the comparison with a~selected atlas provided by the International Consortium for Brain Mapping (ICBM).

\begin{figure*}[p]
    \centering
    \includegraphics[height=.125\textheight]{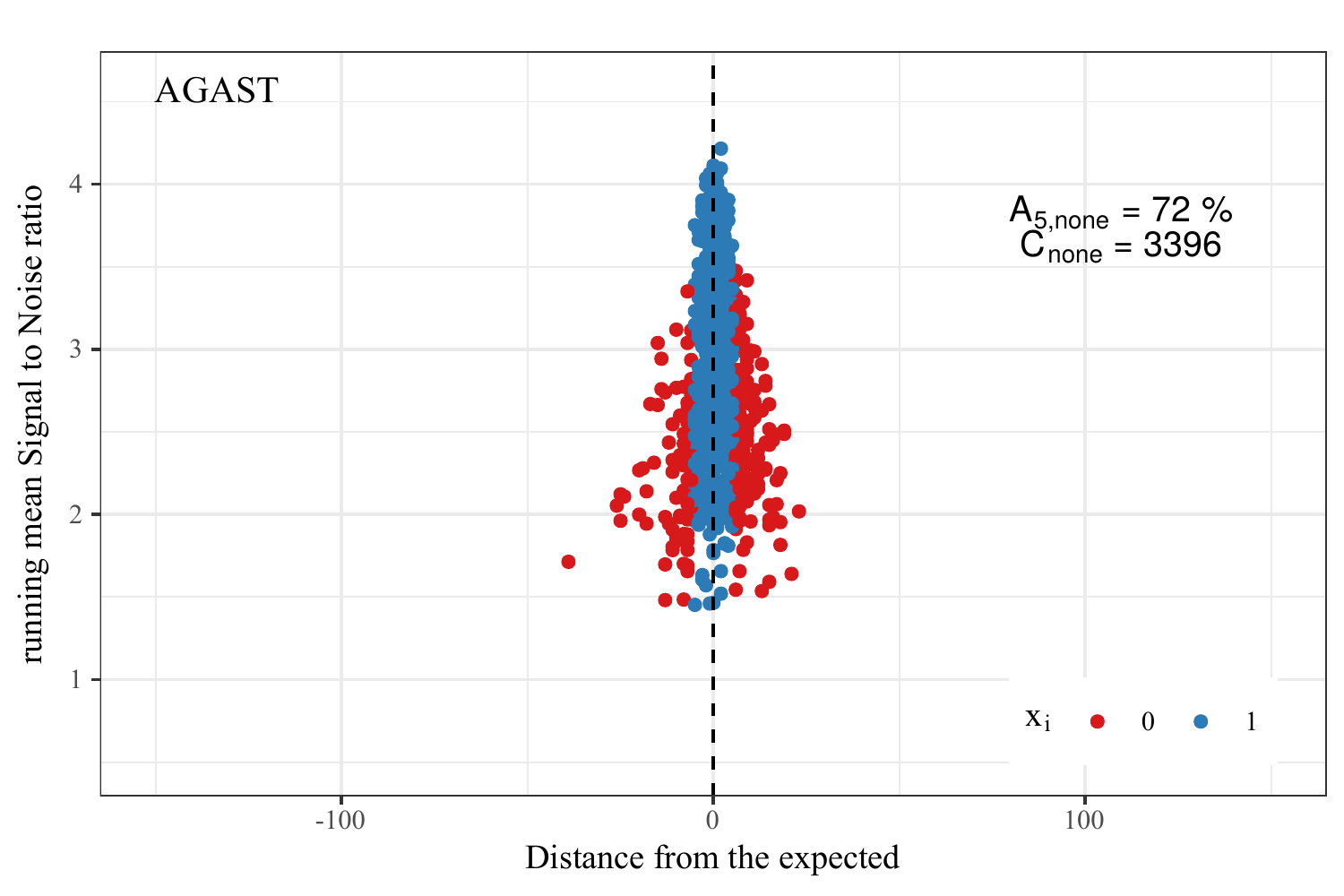}
    \includegraphics[height=.125\textheight]{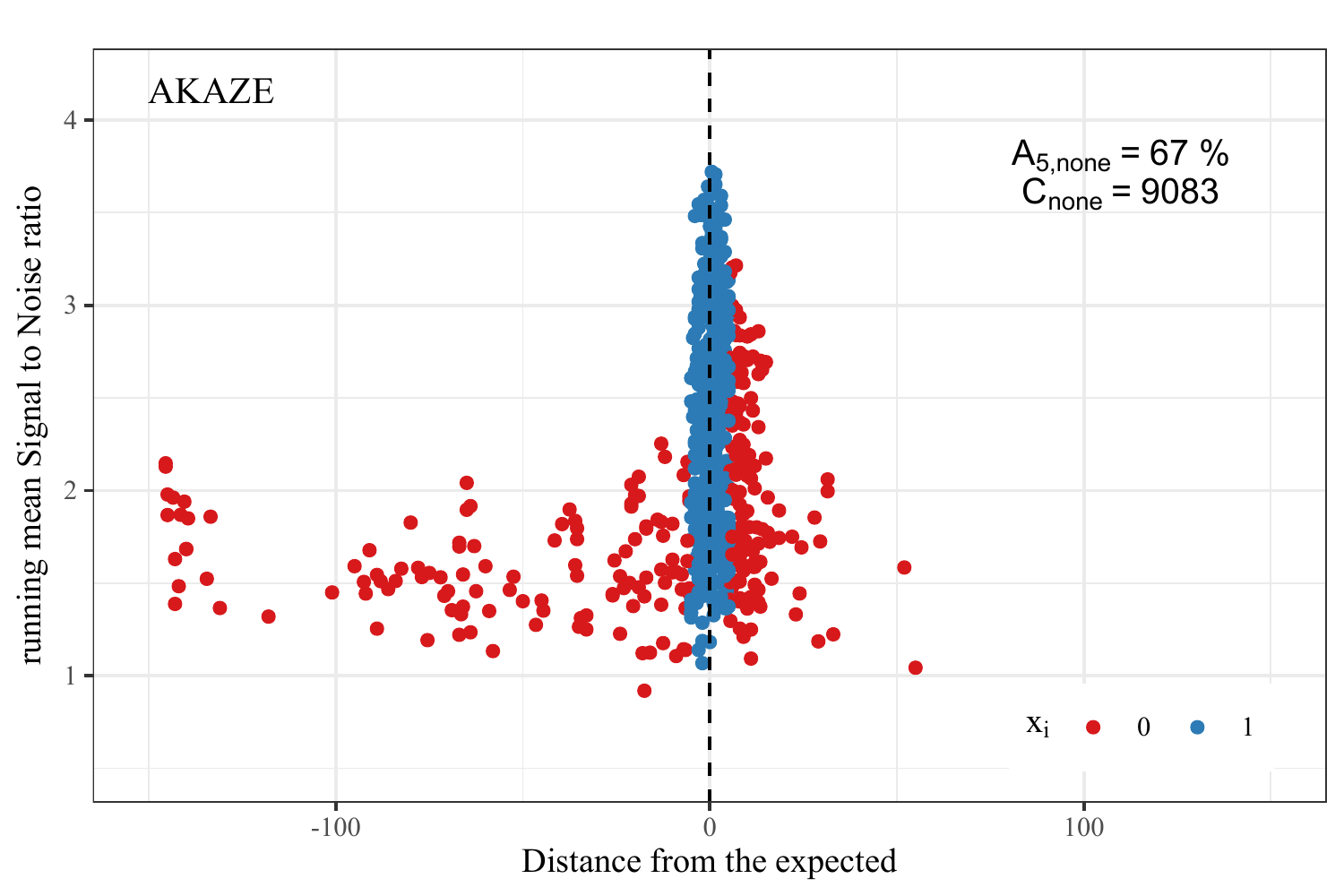}
    \includegraphics[height=.125\textheight]{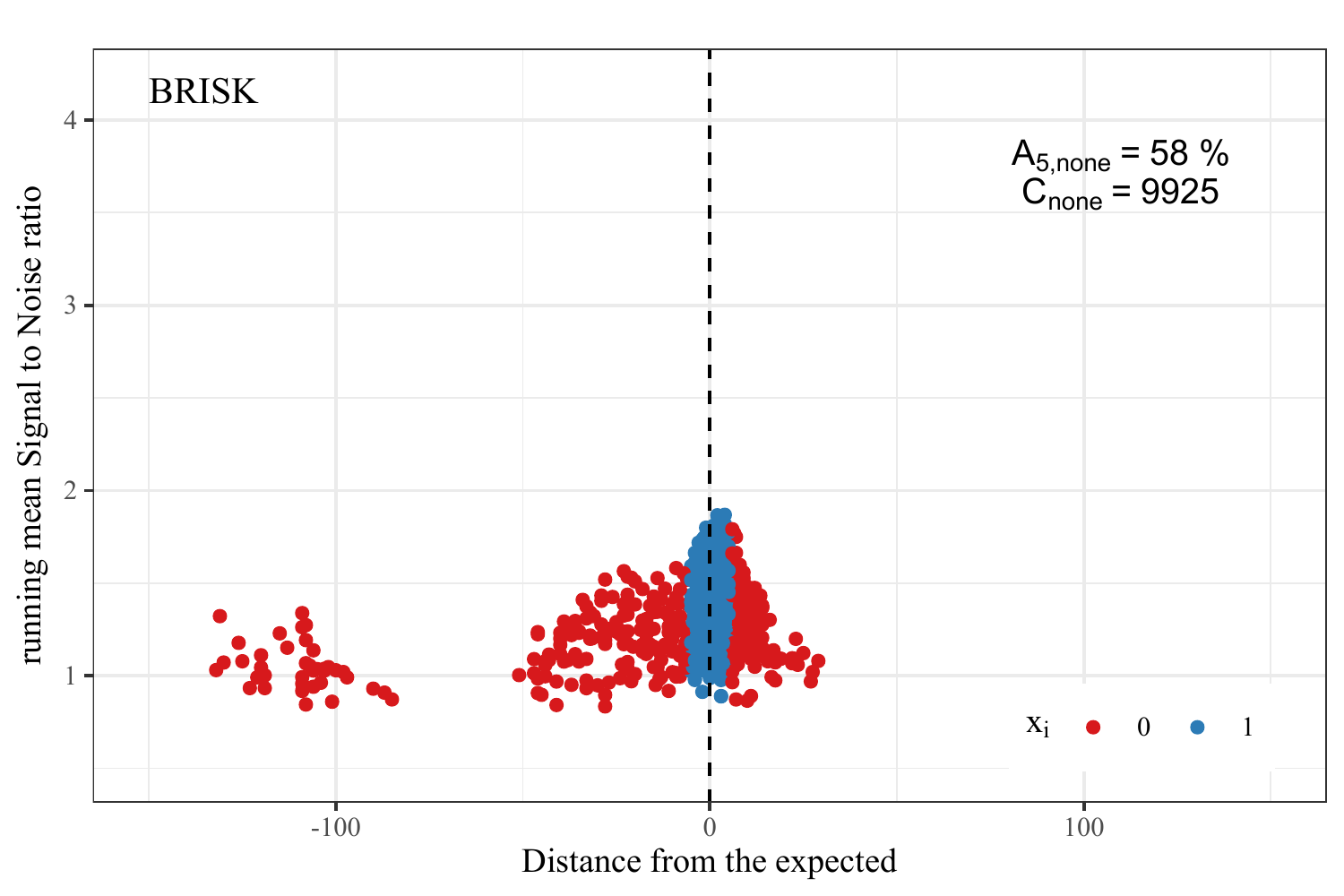}\\
    \includegraphics[height=.125\textheight]{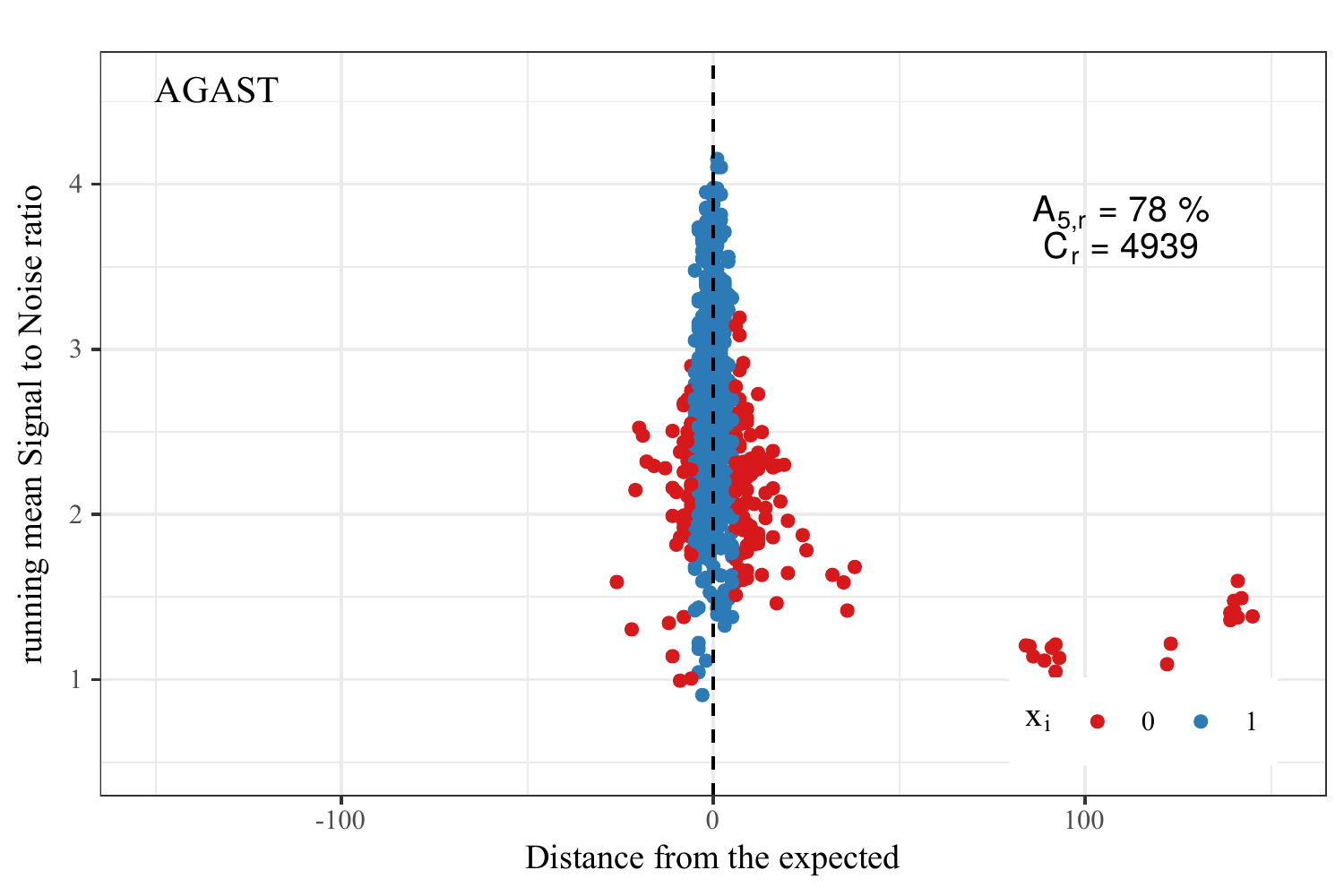}
    \includegraphics[height=.125\textheight]{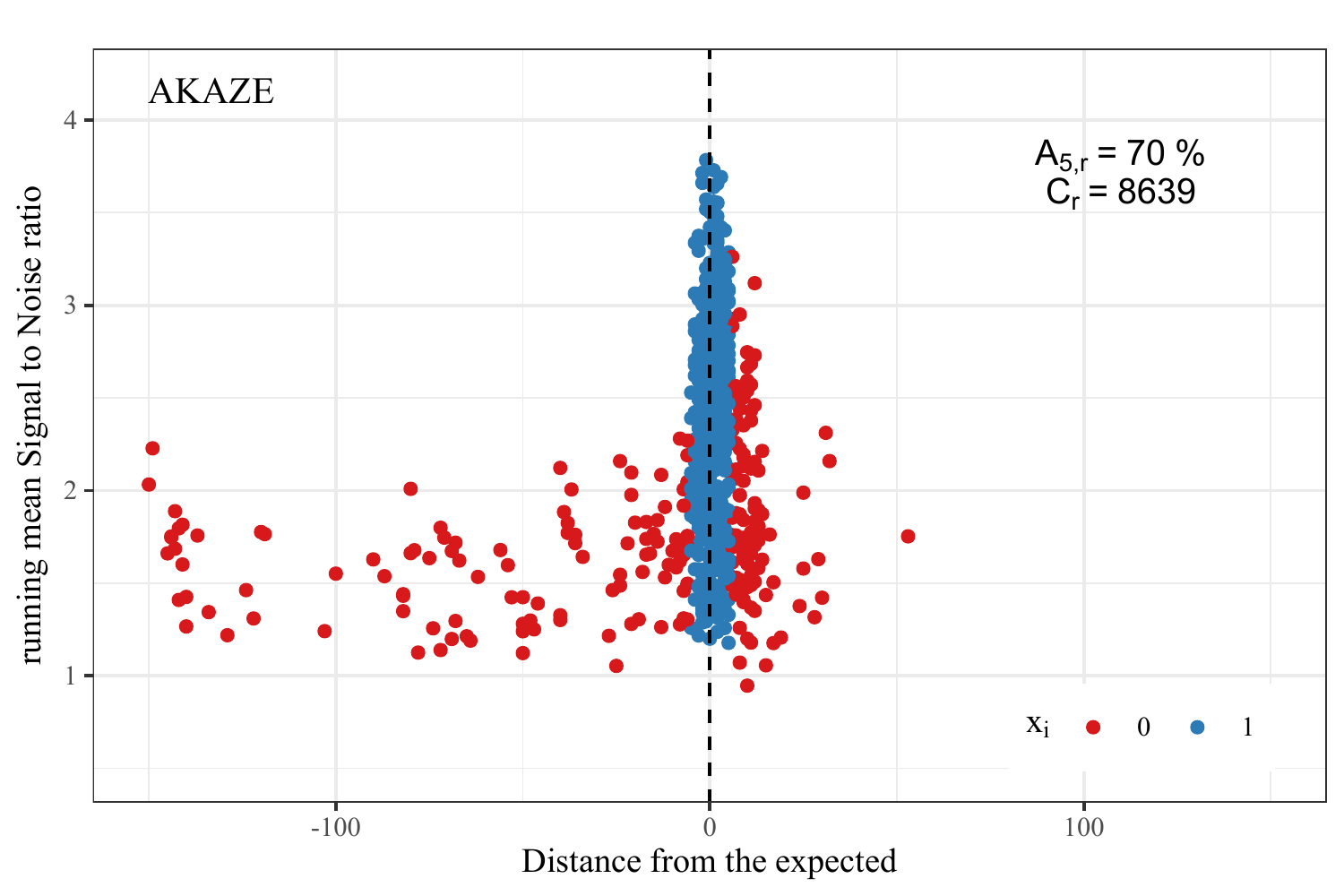}
    \includegraphics[height=.125\textheight]{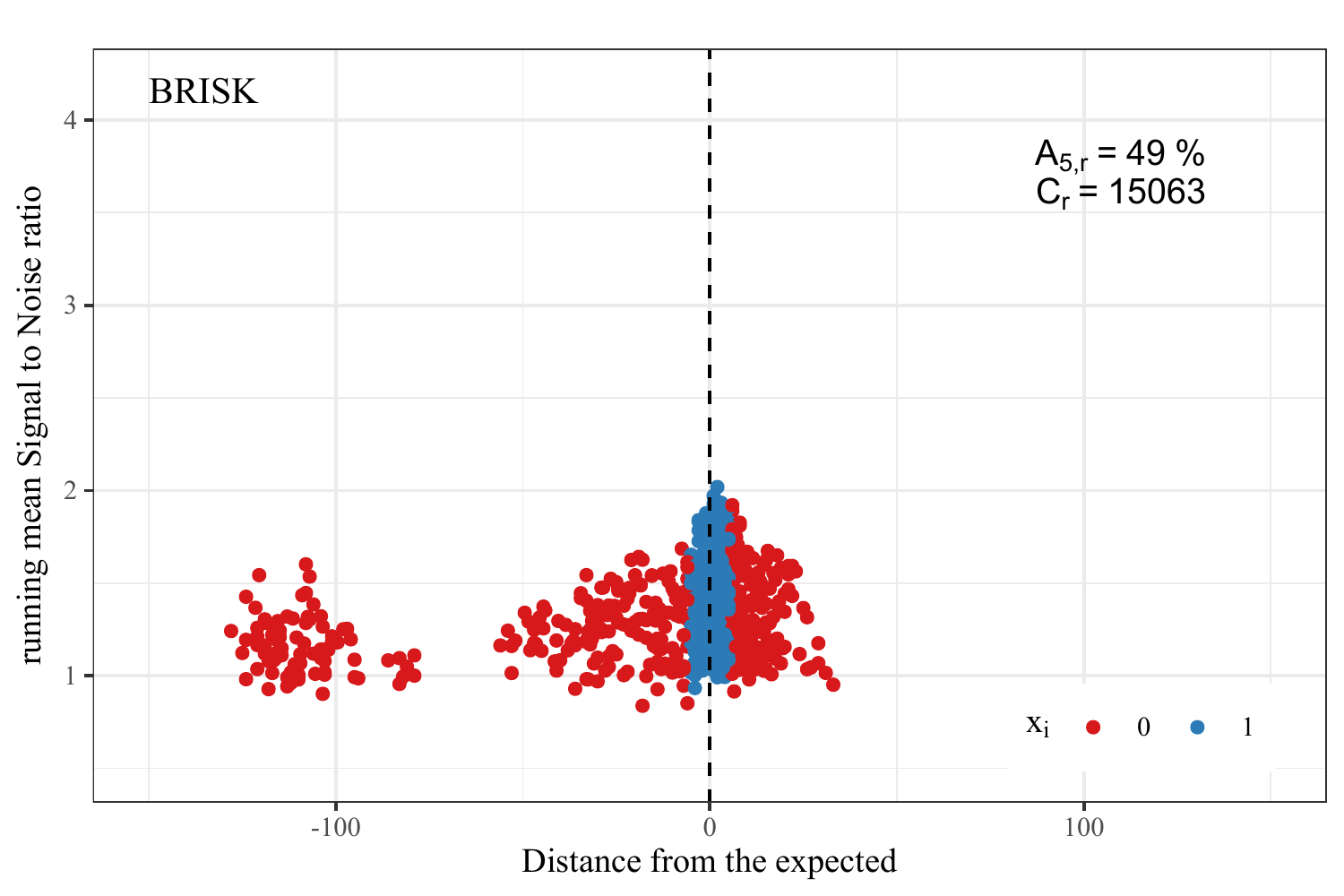}\\
    \includegraphics[height=.125\textheight]{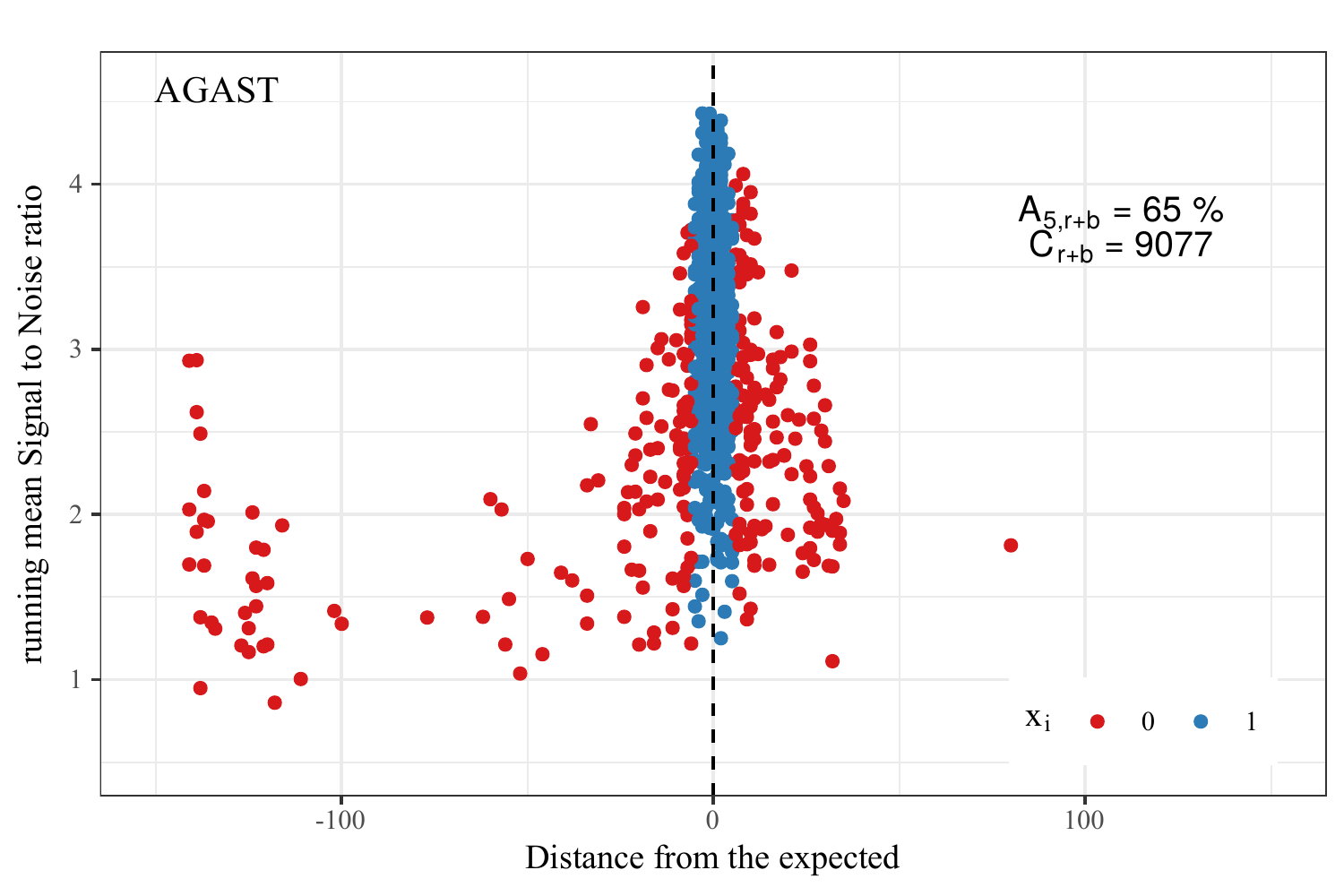}
    \includegraphics[height=.125\textheight]{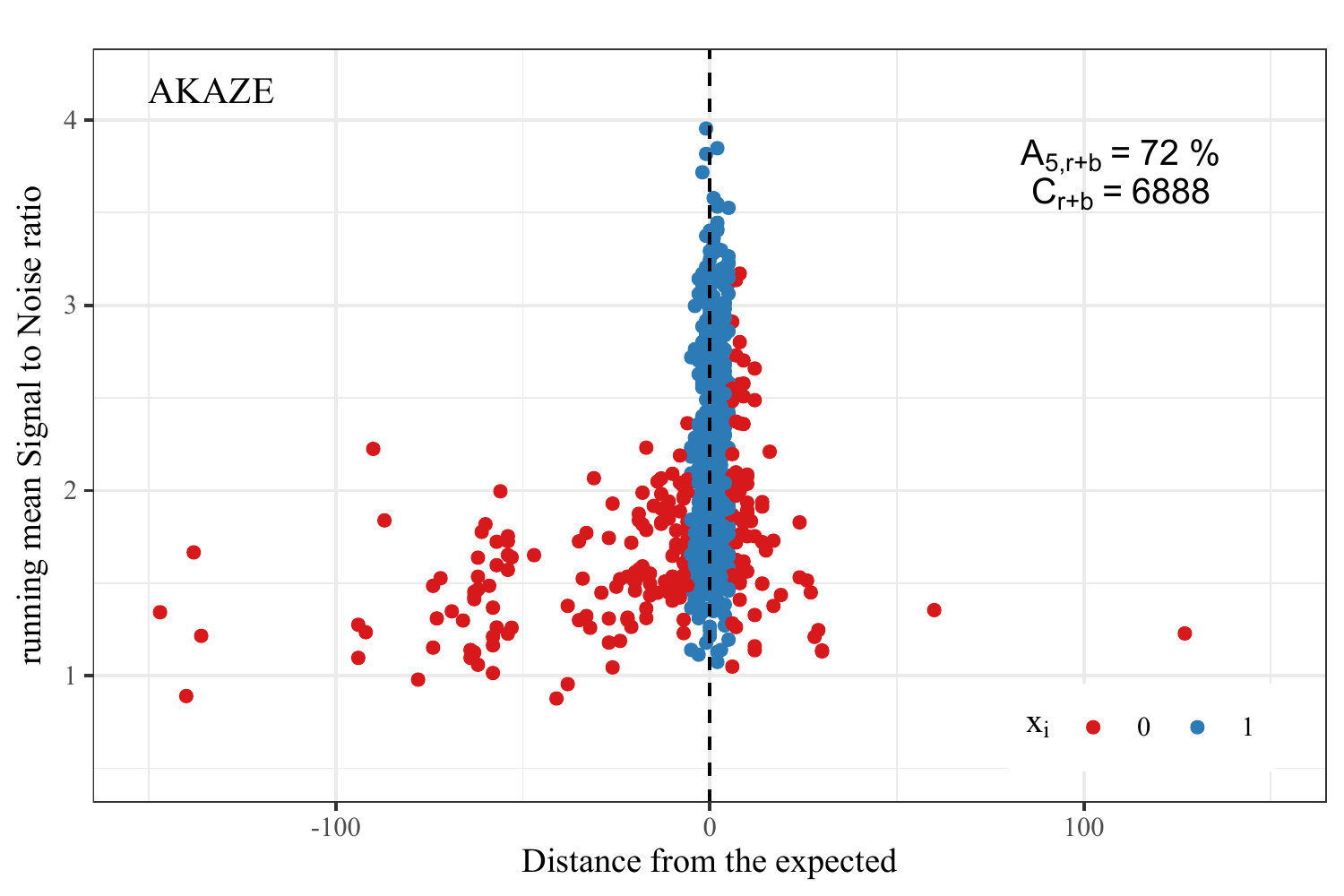}
    \includegraphics[height=.125\textheight]{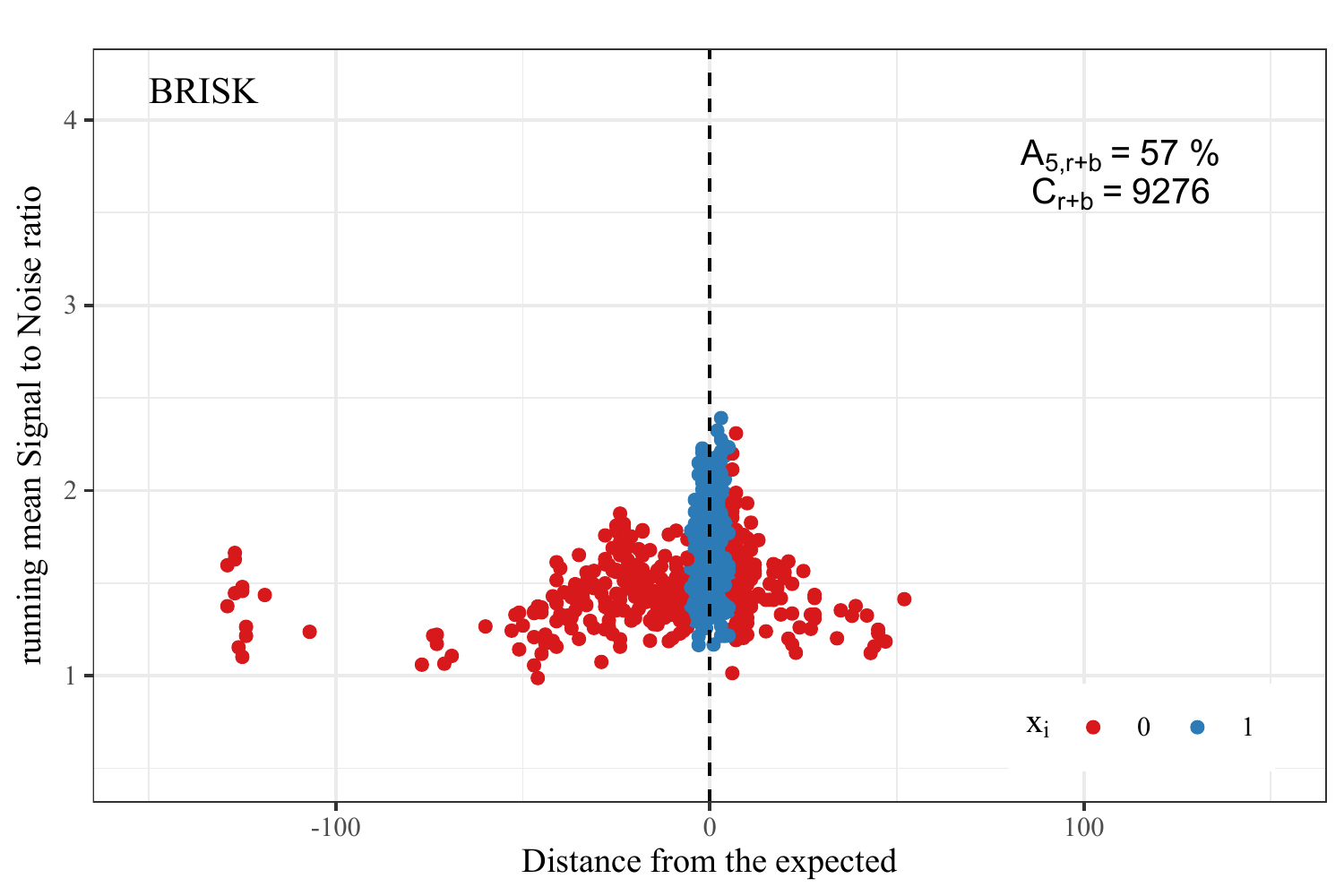}\\
    \includegraphics[height=.125\textheight]{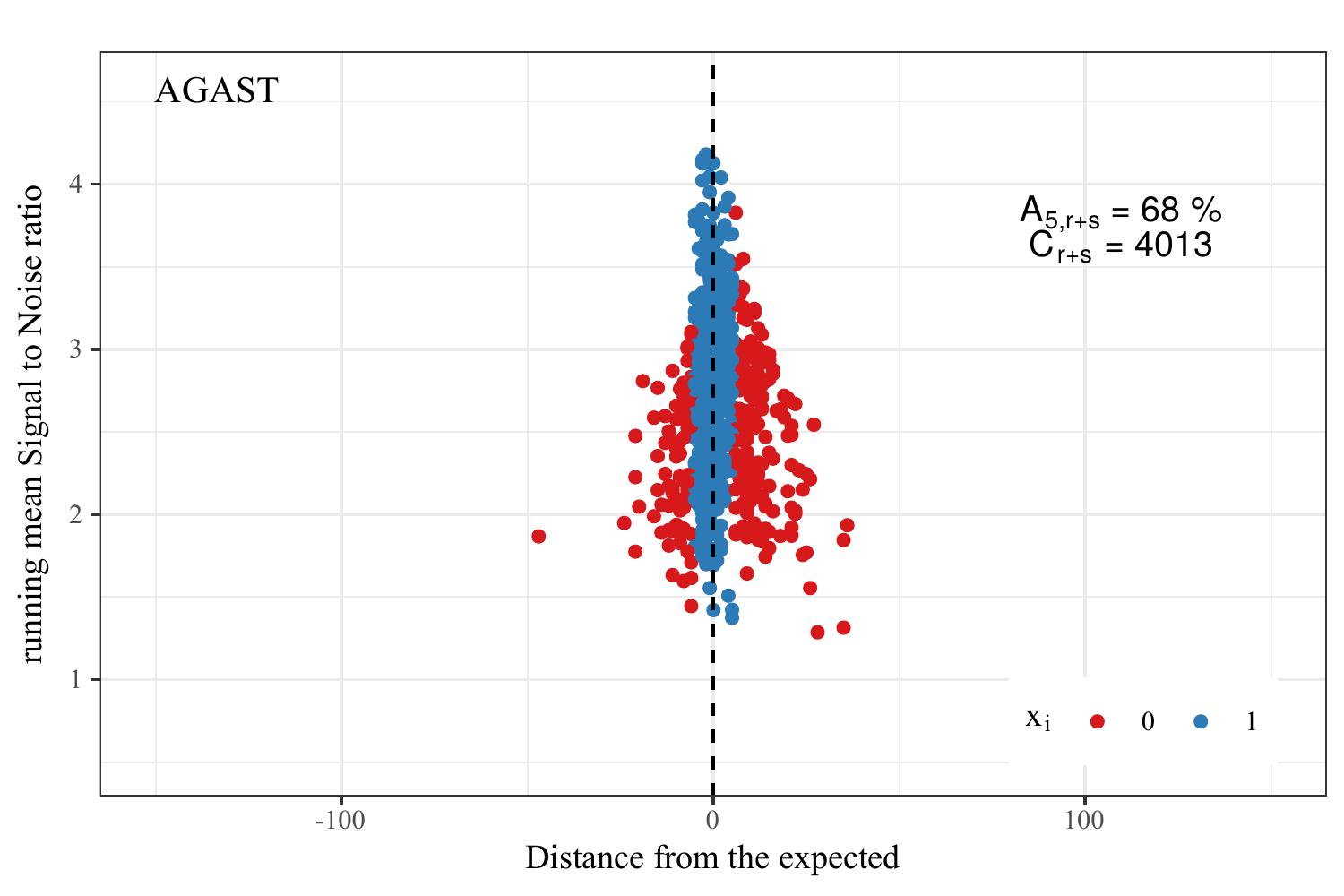}
    \includegraphics[height=.125\textheight]{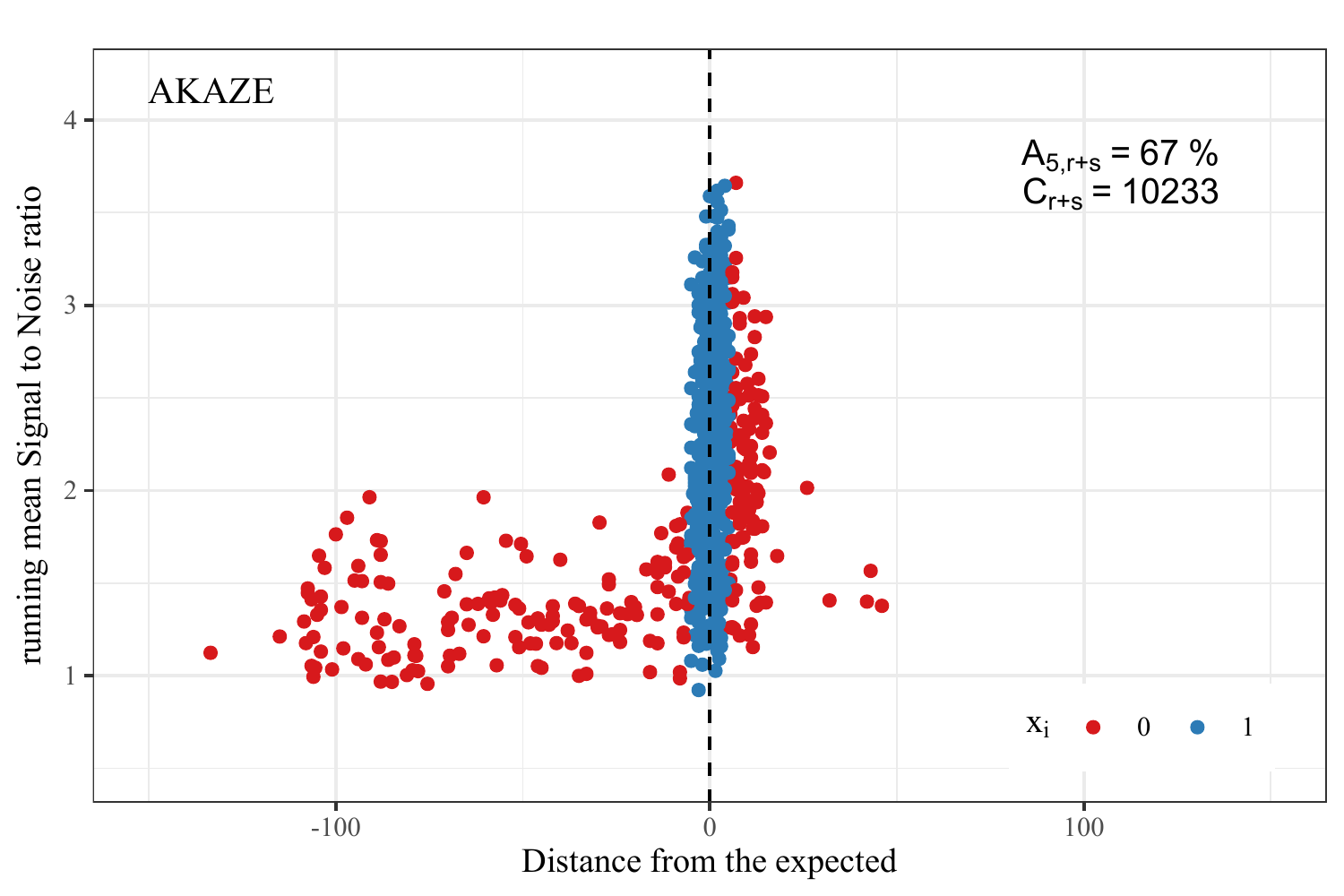}
    \includegraphics[height=.125\textheight]{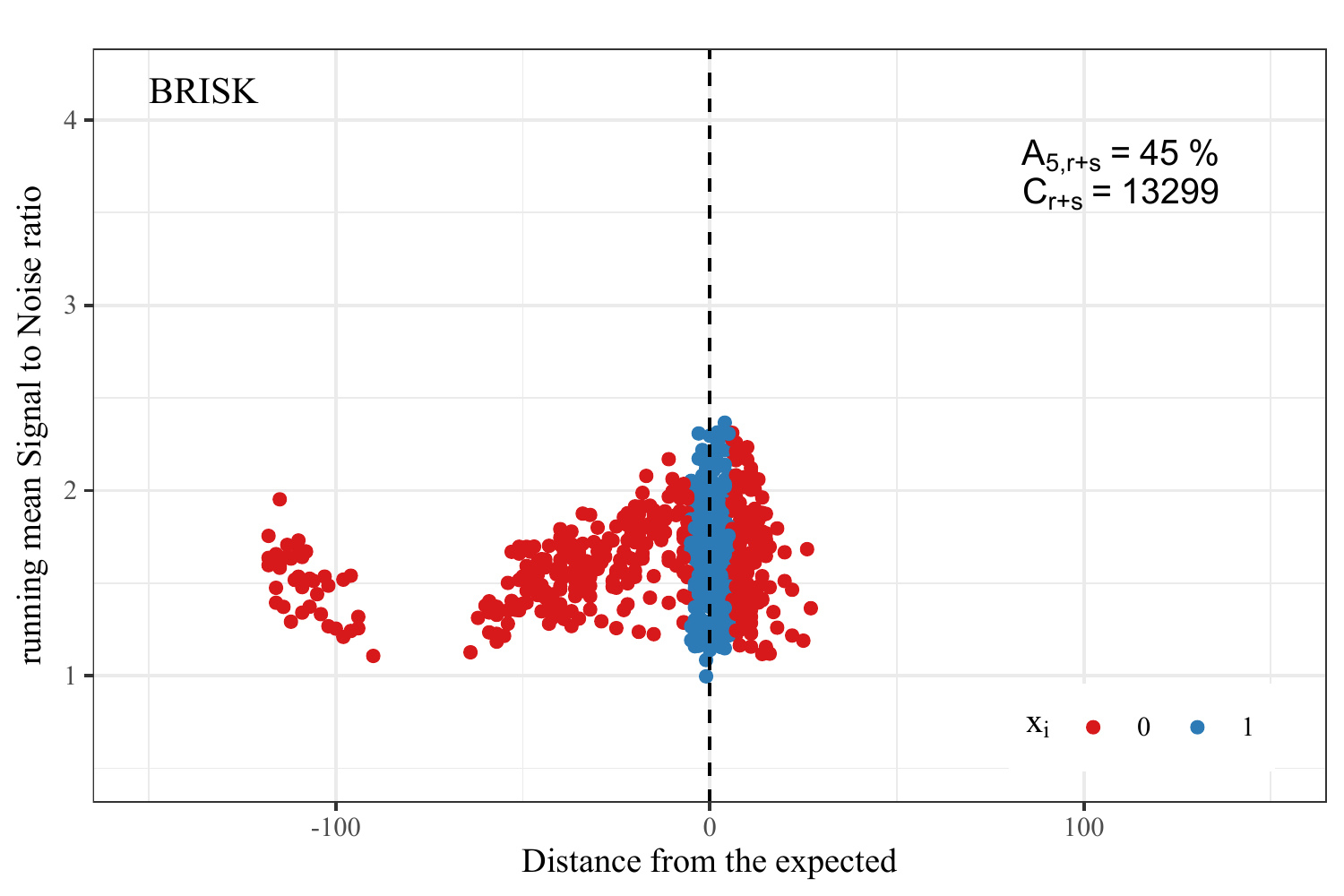}\\
    \includegraphics[height=.125\textheight]{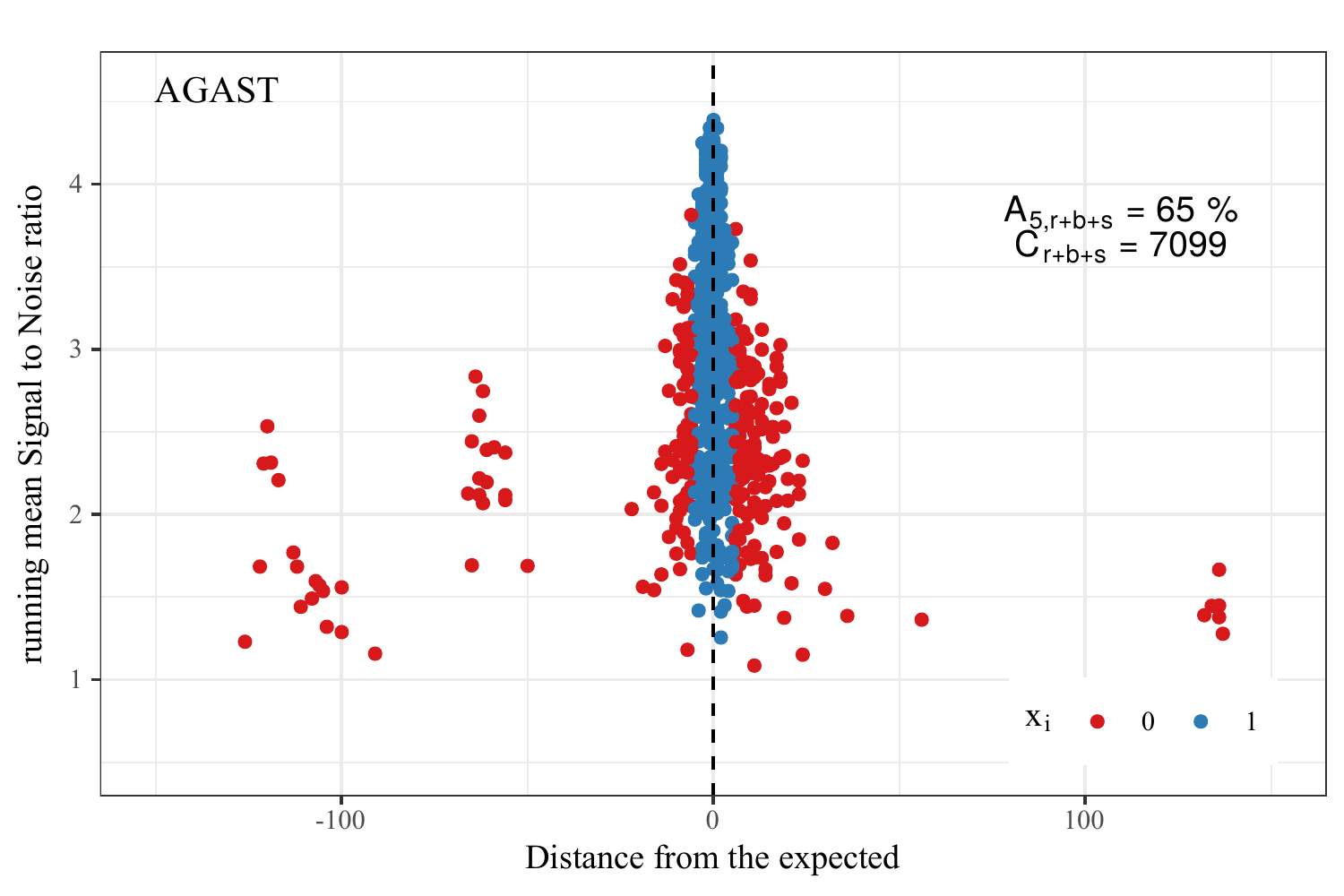}
    \includegraphics[height=.125\textheight]{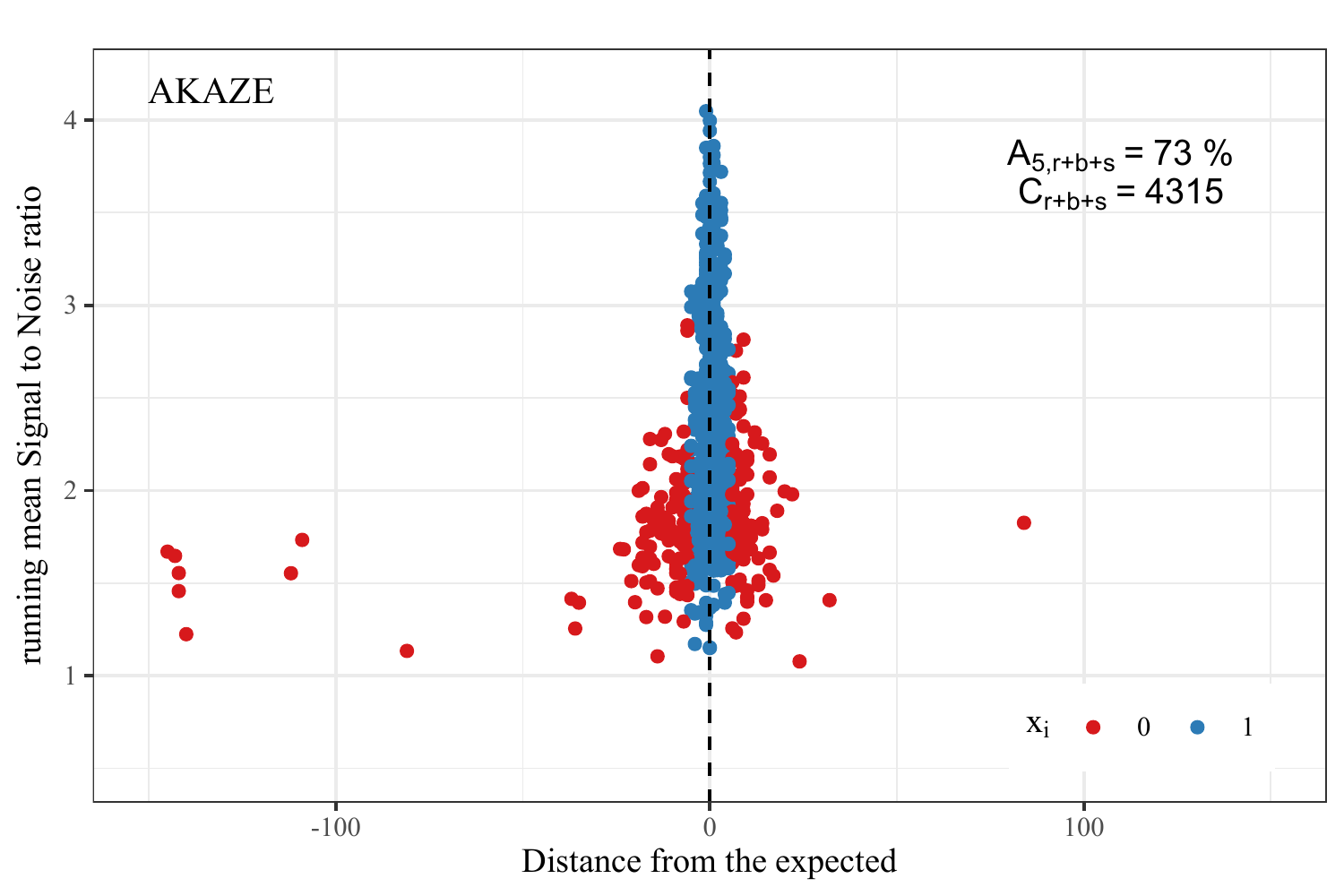}
    \includegraphics[height=.125\textheight]{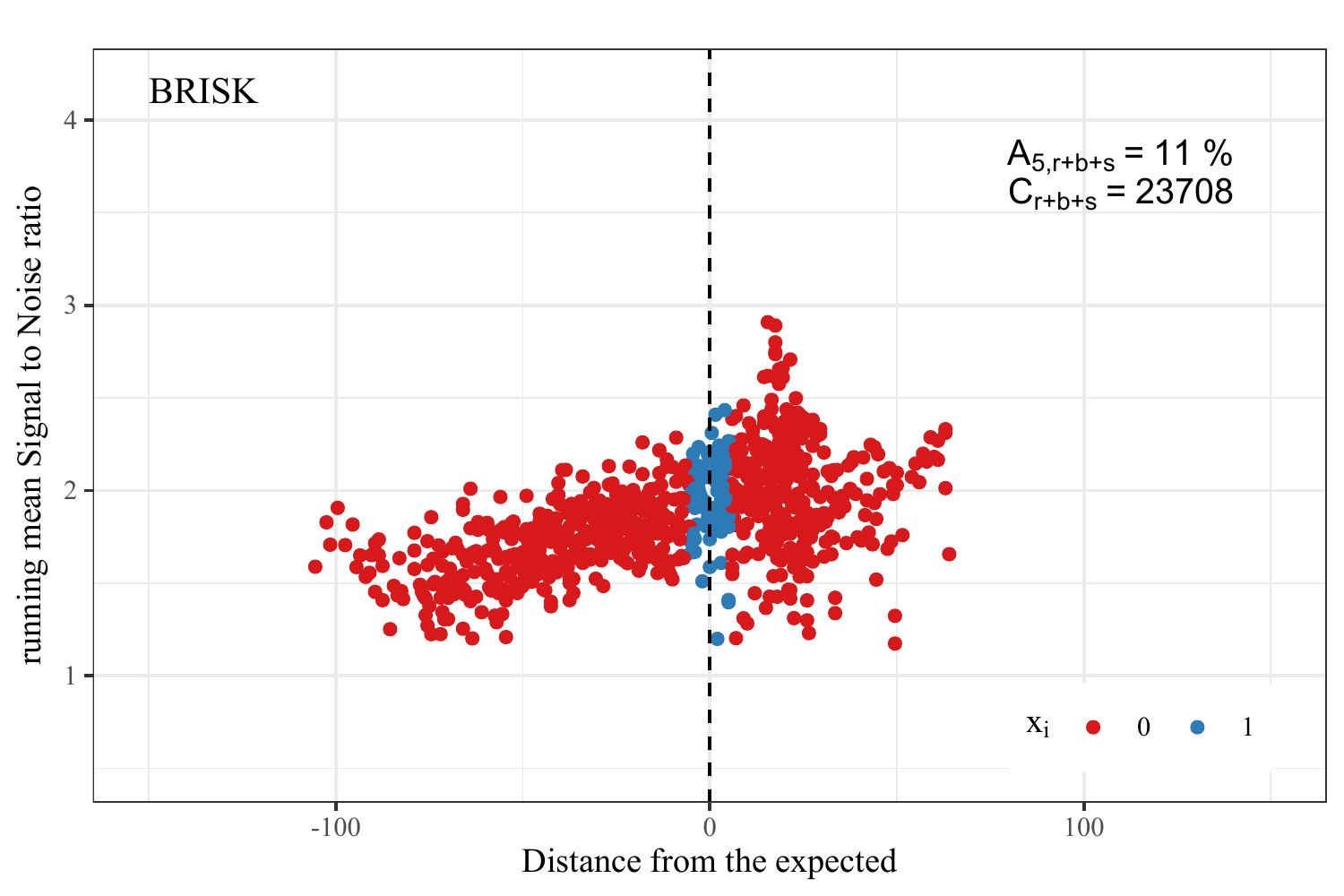}\\
    \includegraphics[height=.125\textheight]{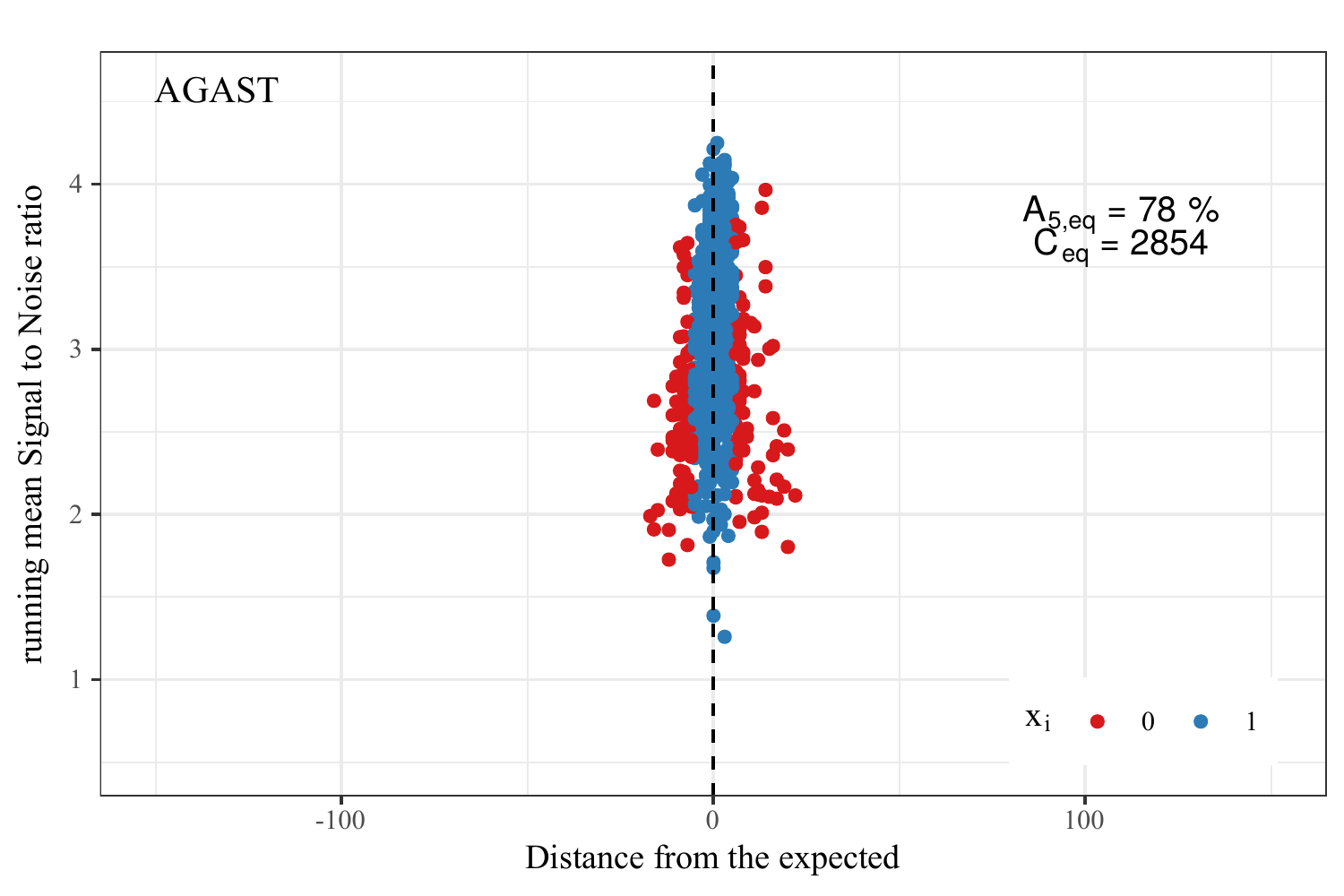}
    \includegraphics[height=.125\textheight]{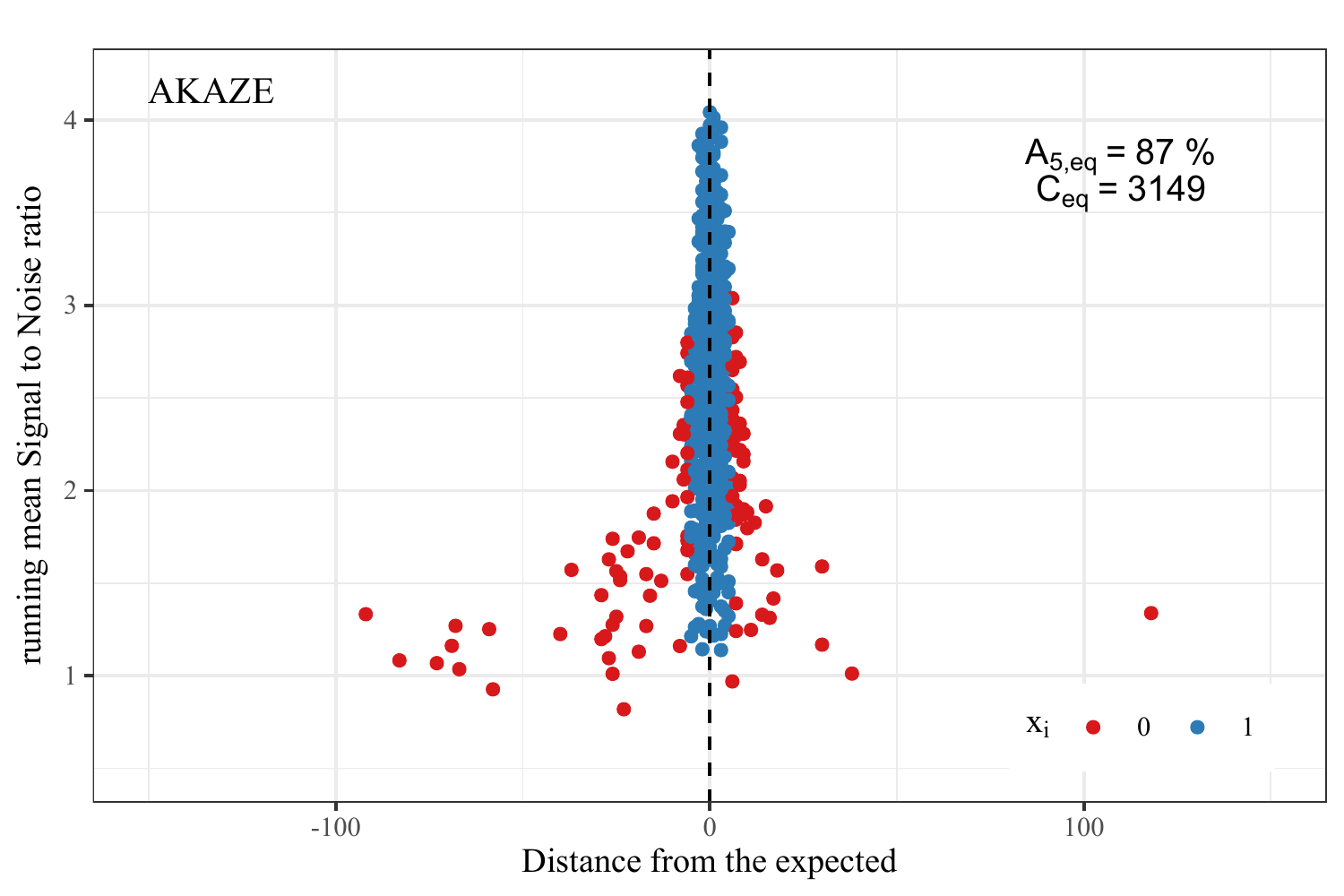}
    \includegraphics[height=.125\textheight]{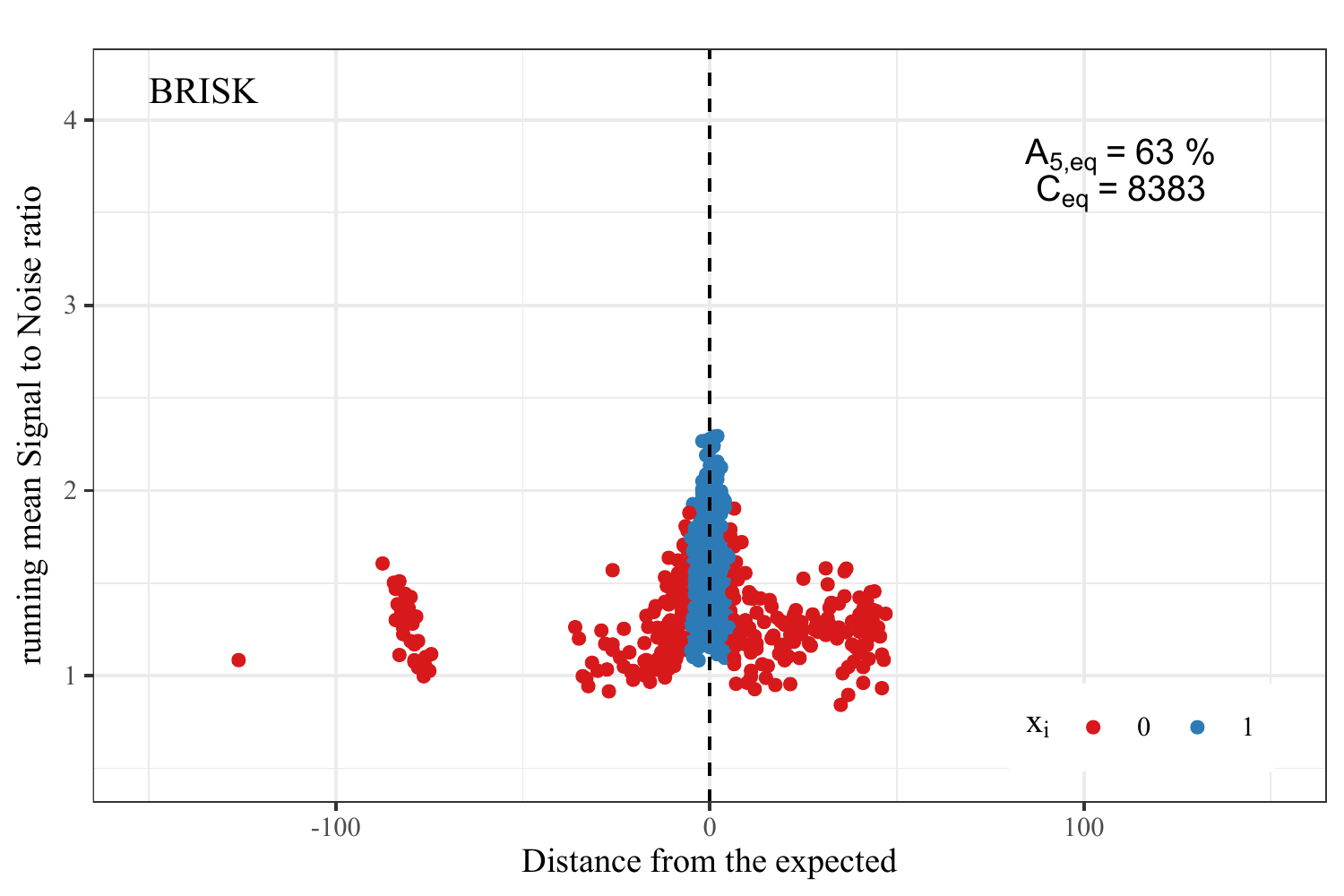}\\
    \includegraphics[height=.125\textheight]{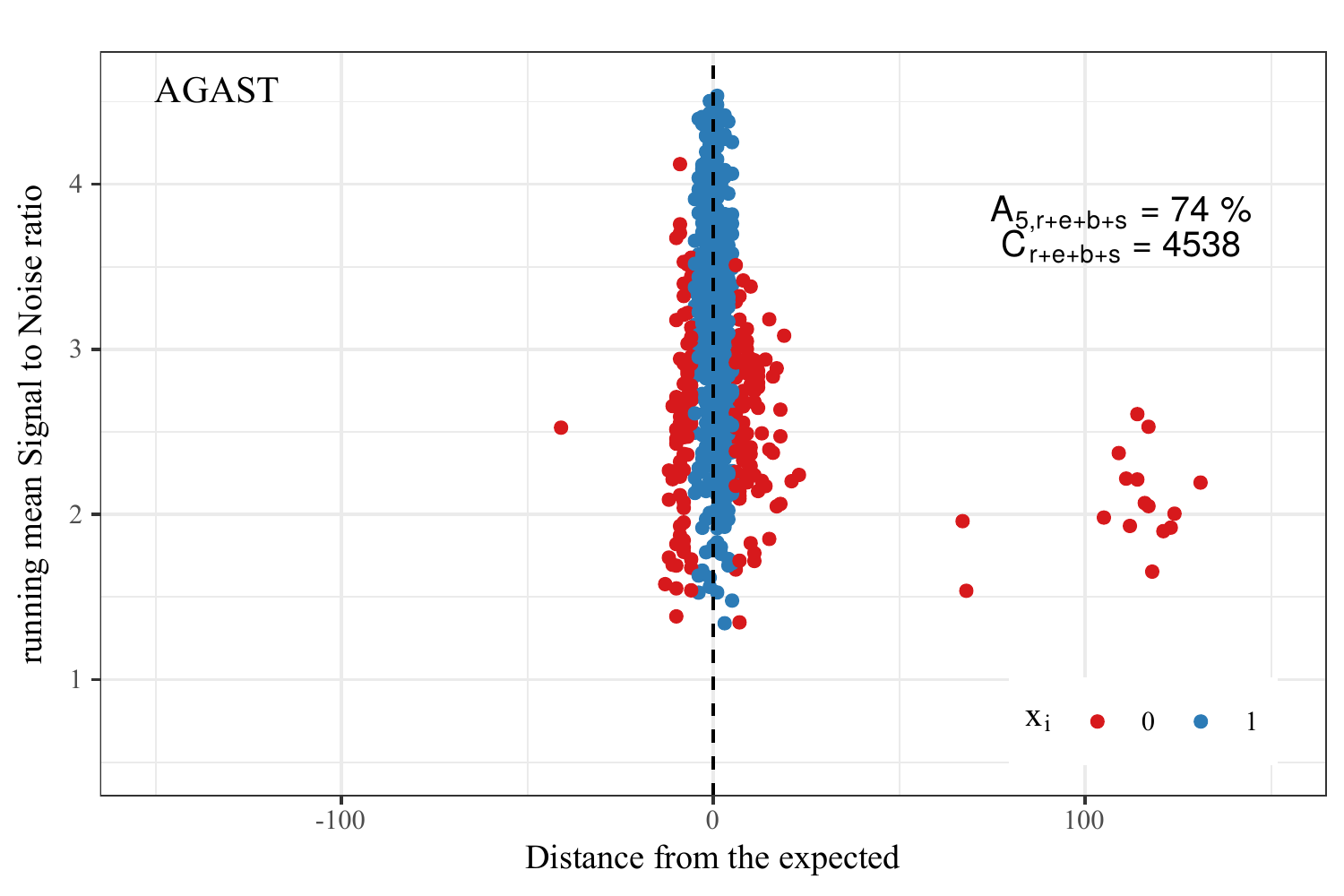}
    \includegraphics[height=.125\textheight]{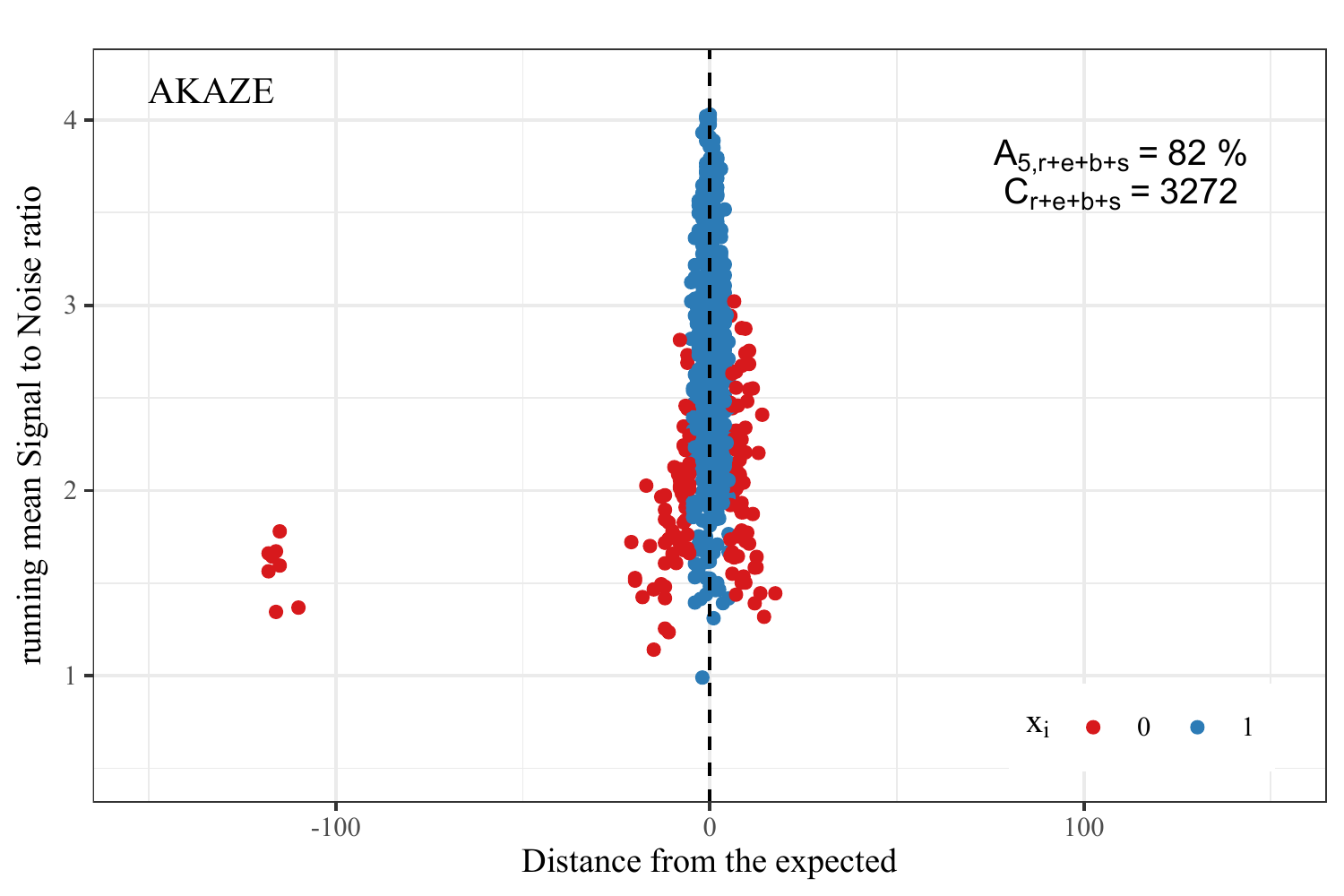}
    \includegraphics[height=.125\textheight]{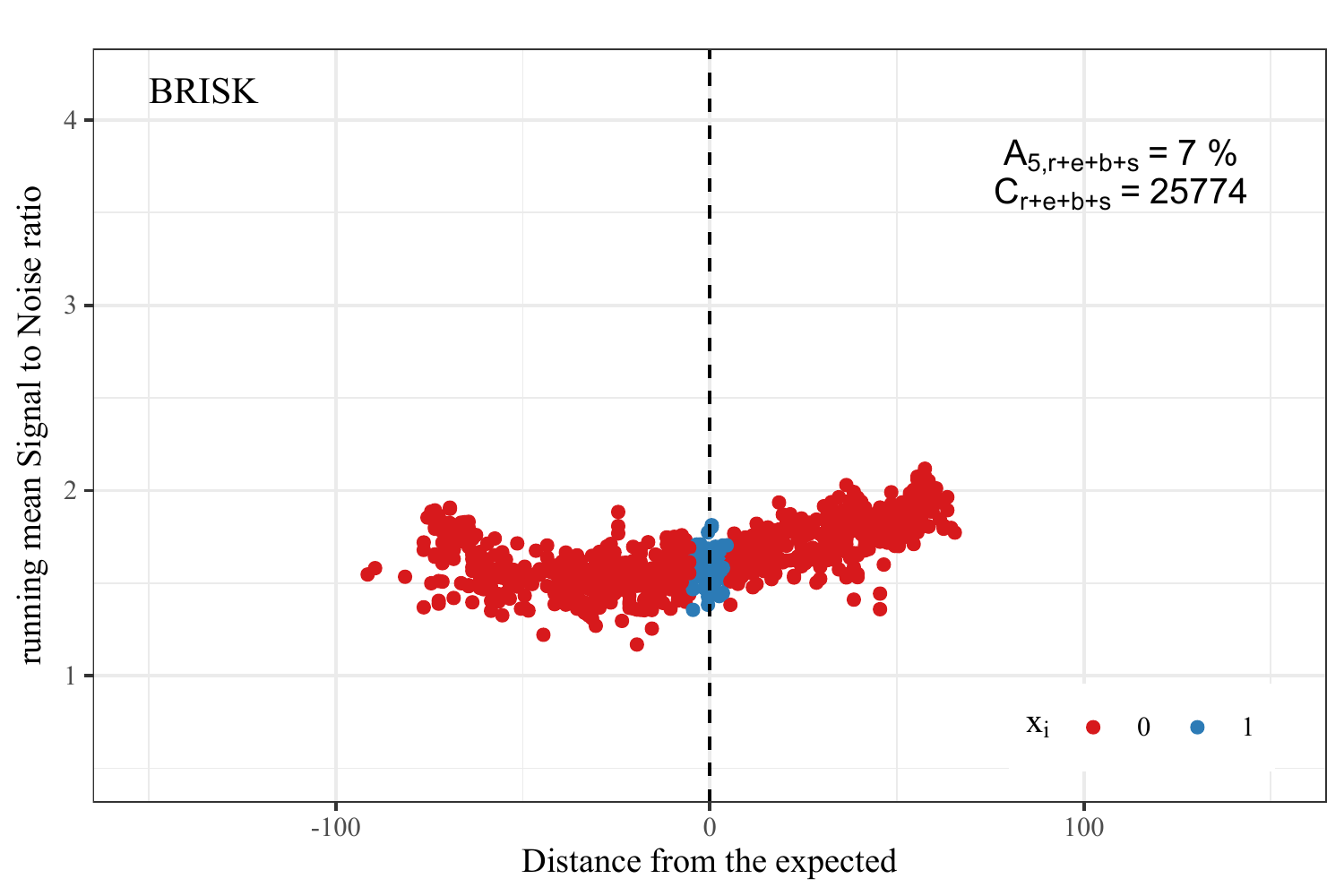}\\
    \caption{\label{fig:accuracy_all}Accuracy $A_{d,c}$ of each used FD method (from the left column to right column: AGAST, AKAZE, BRISK) and the applied image preprocessing (from the first row to last: none; rotation; rotation and skull extraction; rotation and scaling; rotation, skull extraction and scaling; equalisation; rotation, equalisation, skull extraction and scaling. The blue dots correspond to the input images satisfying the condition in Eq.~\ref{eq:accuracy}. The y-axis corresponds to the running mean SNR and the x-axis to the distance from the expected. The vertical dashed line represents the expected image ID. The accuracy $A_{d,c}$ and cumulative distance $C_d$ computed by Eq.~\ref{eq:cum_distance} are stated in each figures.}
\end{figure*}

\begin{figure*}[p]
    \centering
    \includegraphics[height=.125\textheight]{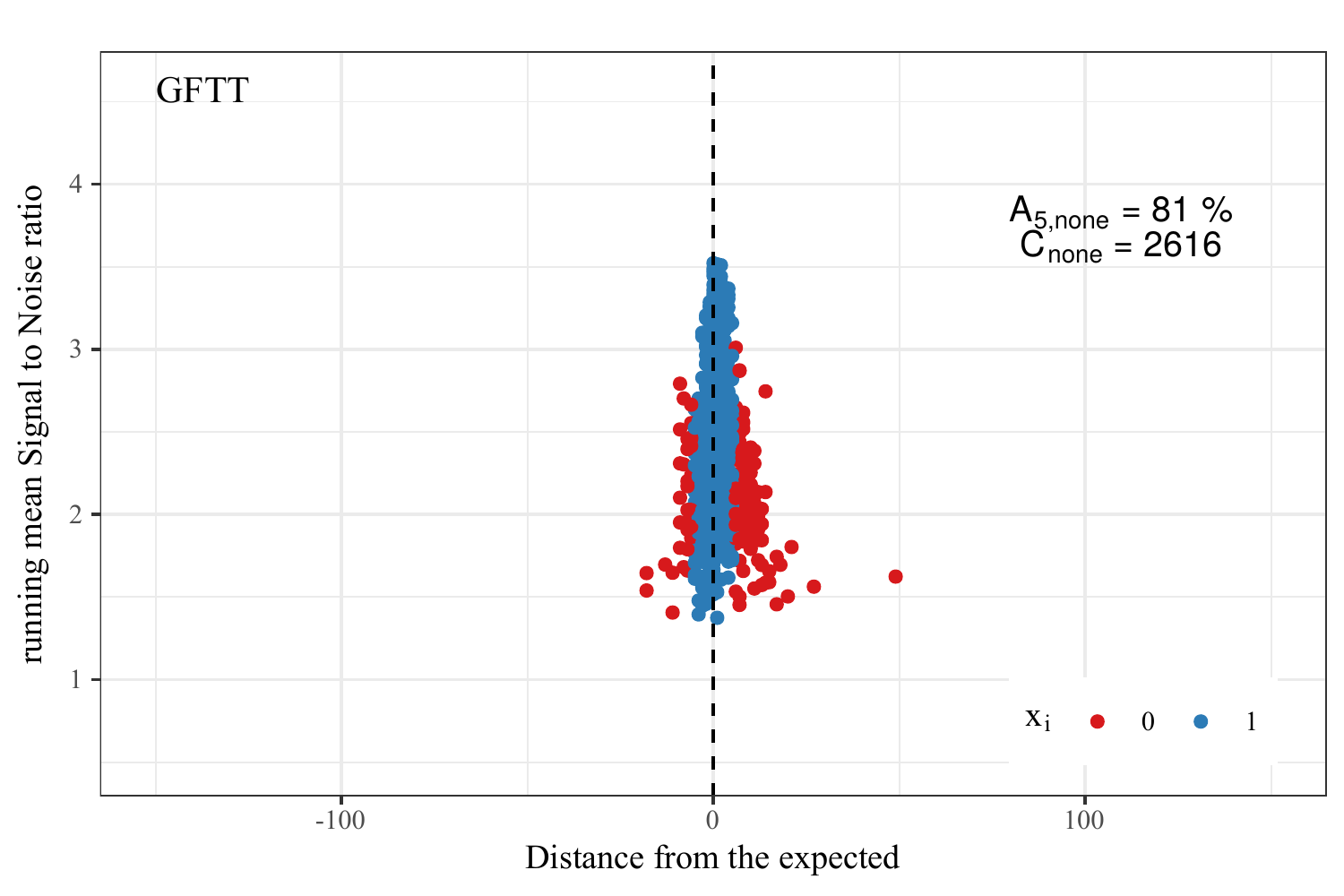}
    \includegraphics[height=.125\textheight]{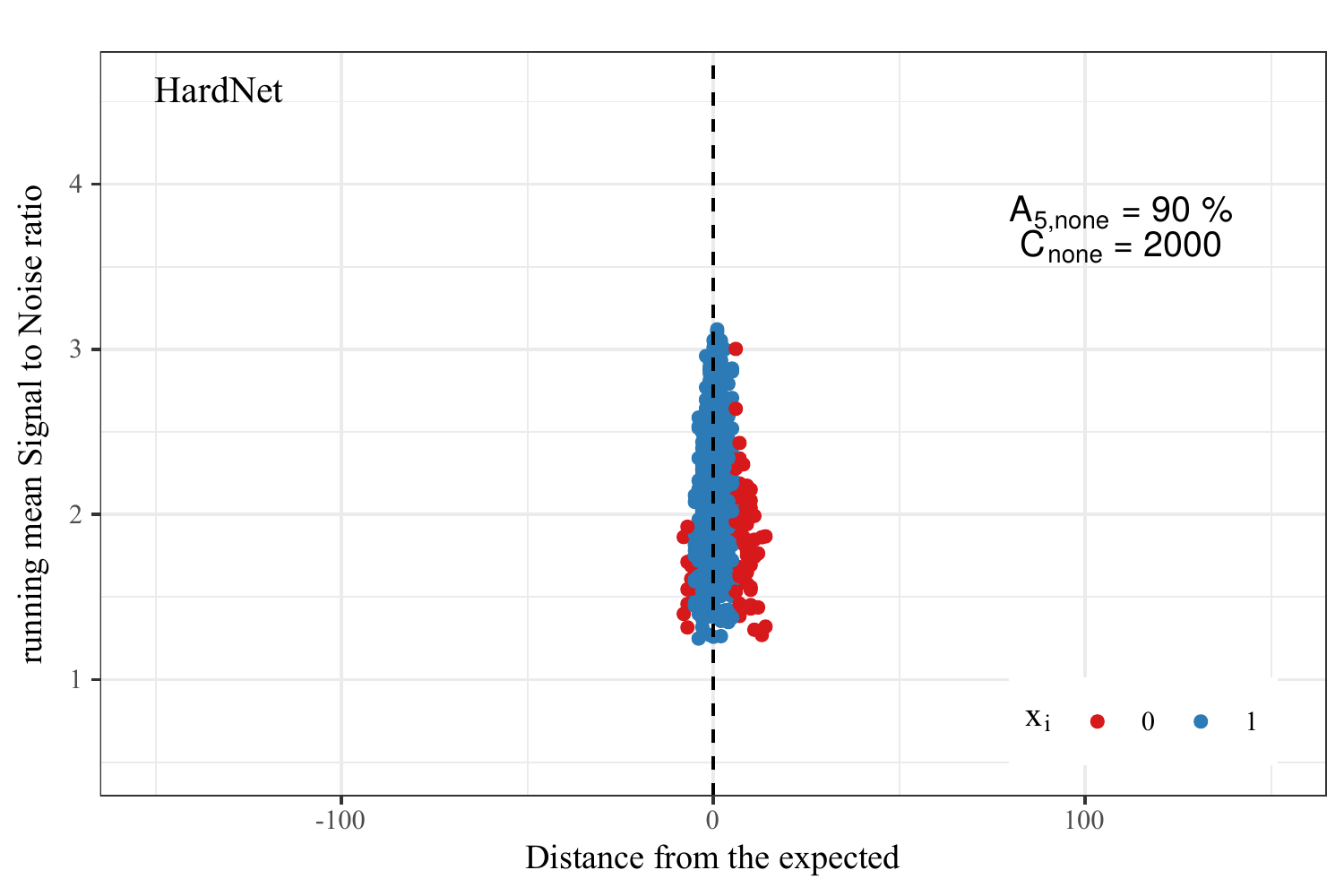}
    \includegraphics[height=.125\textheight]{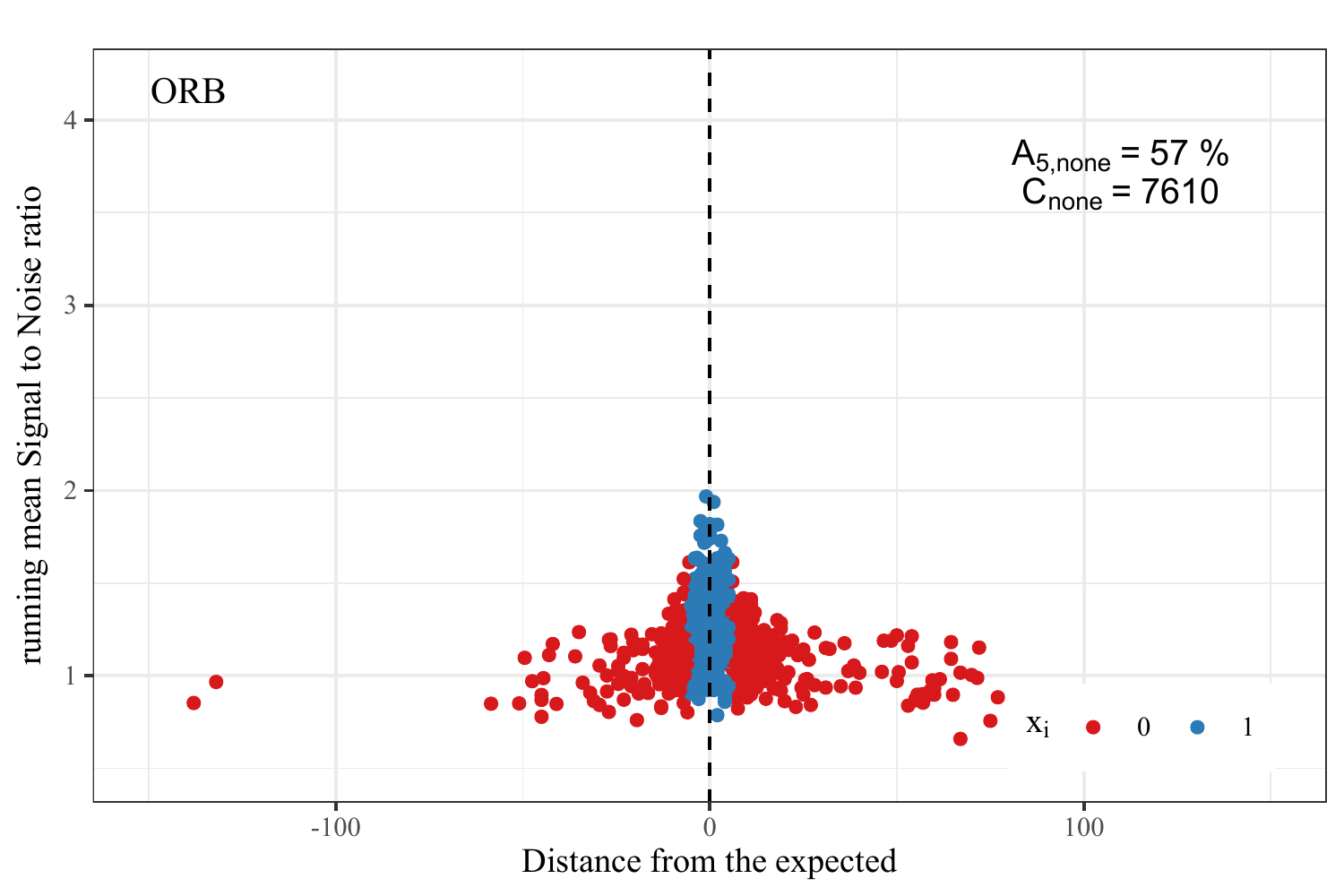}\\
    \includegraphics[height=.125\textheight]{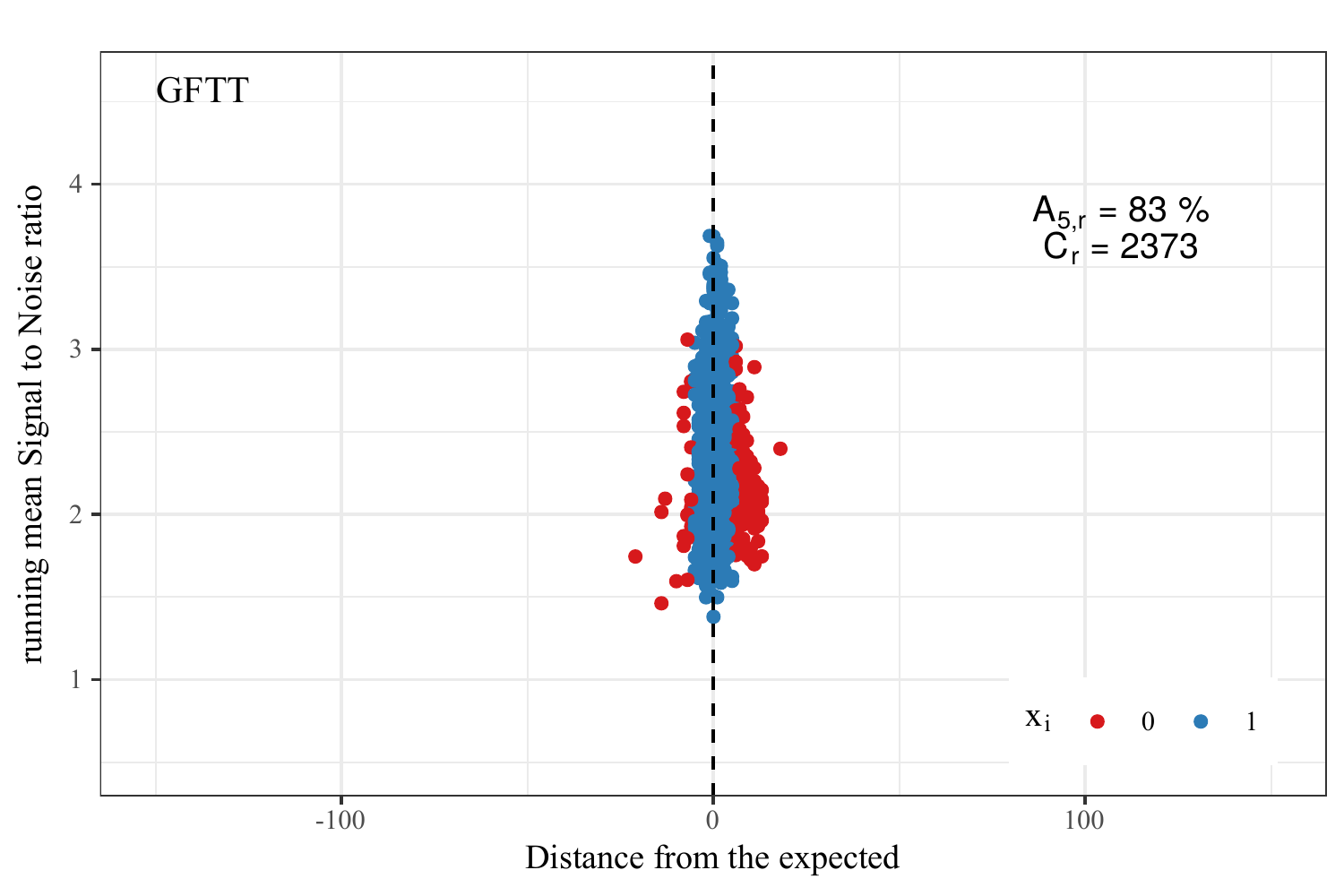}
    \includegraphics[height=.125\textheight]{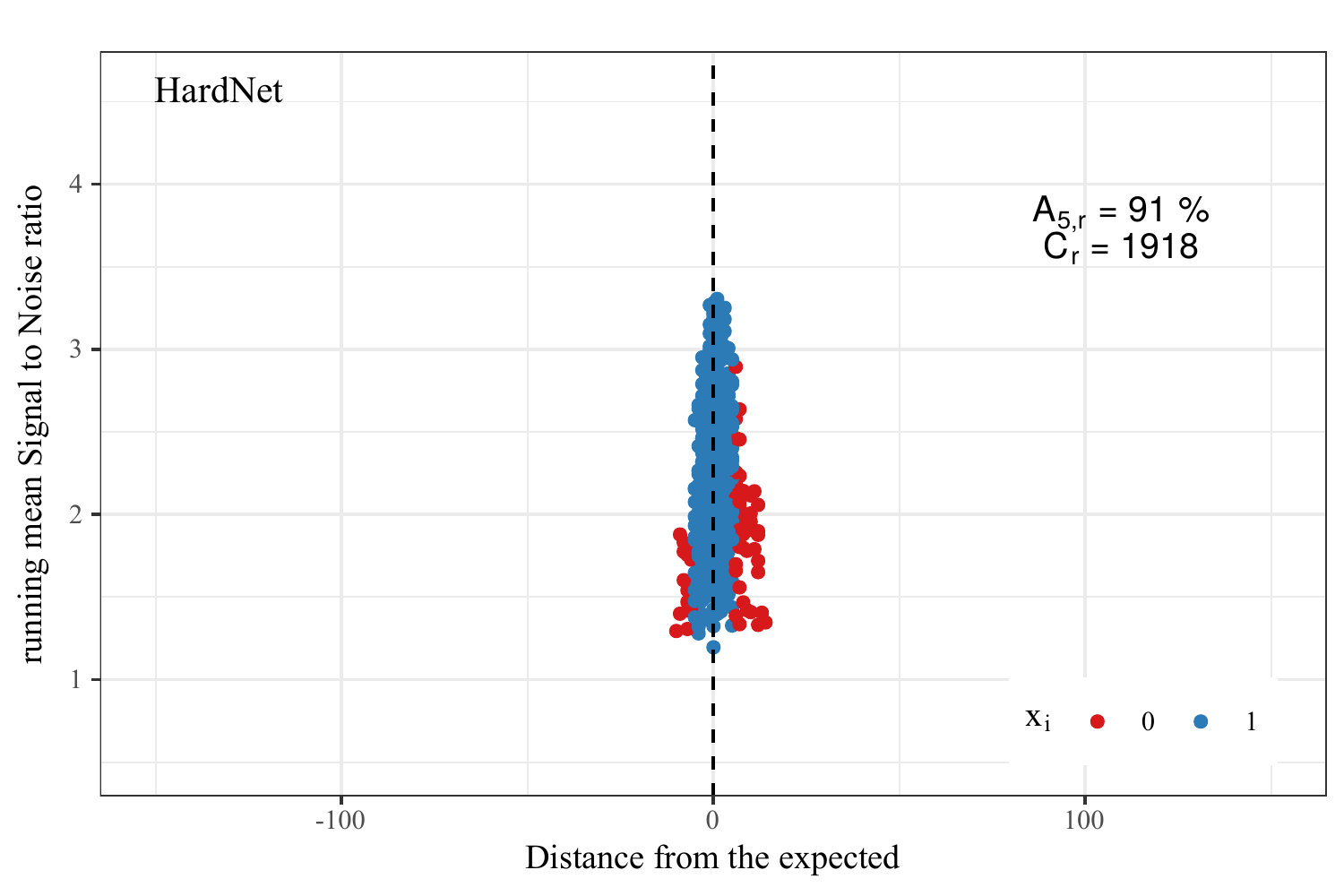}
    \includegraphics[height=.125\textheight]{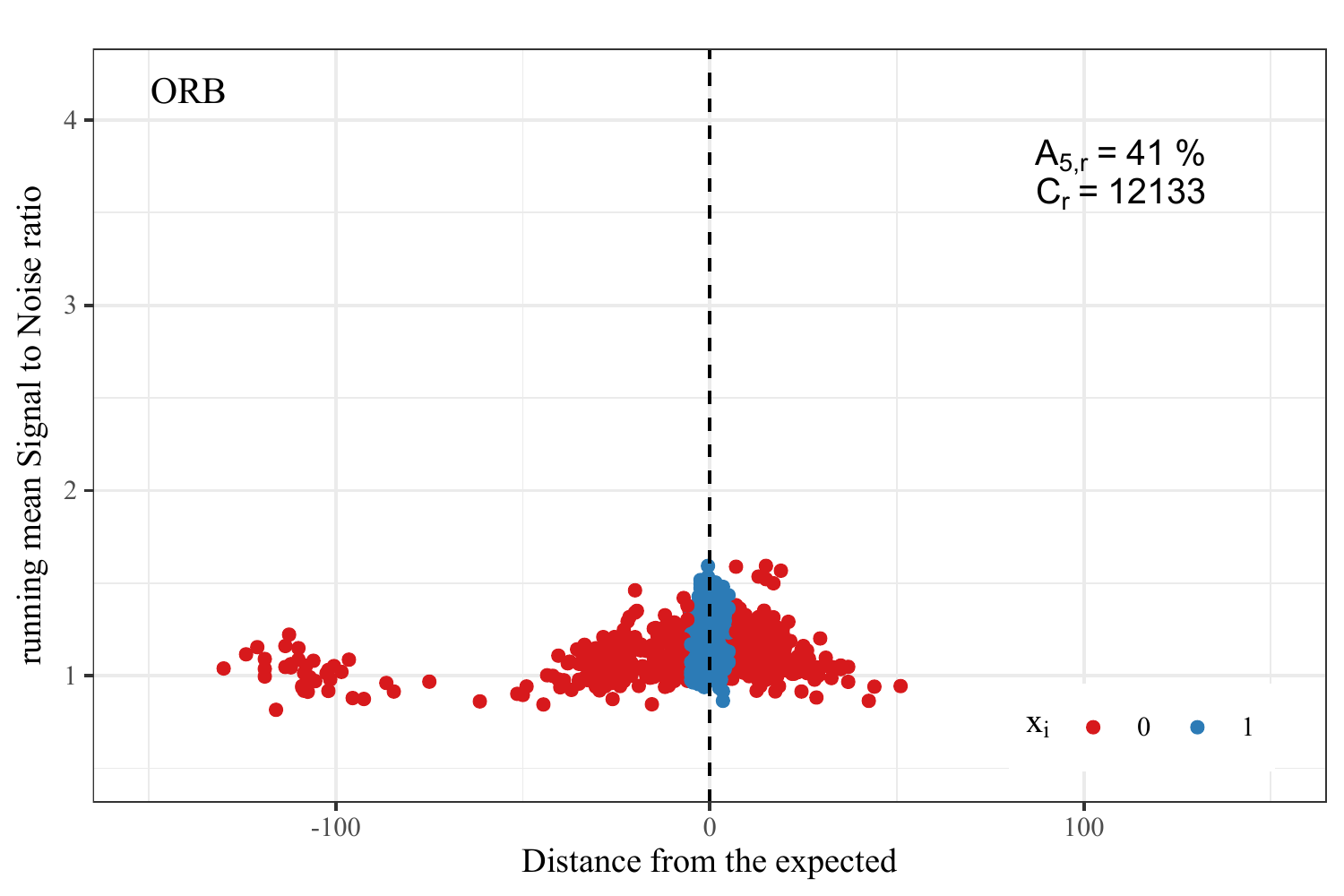}\\
    \includegraphics[height=.125\textheight]{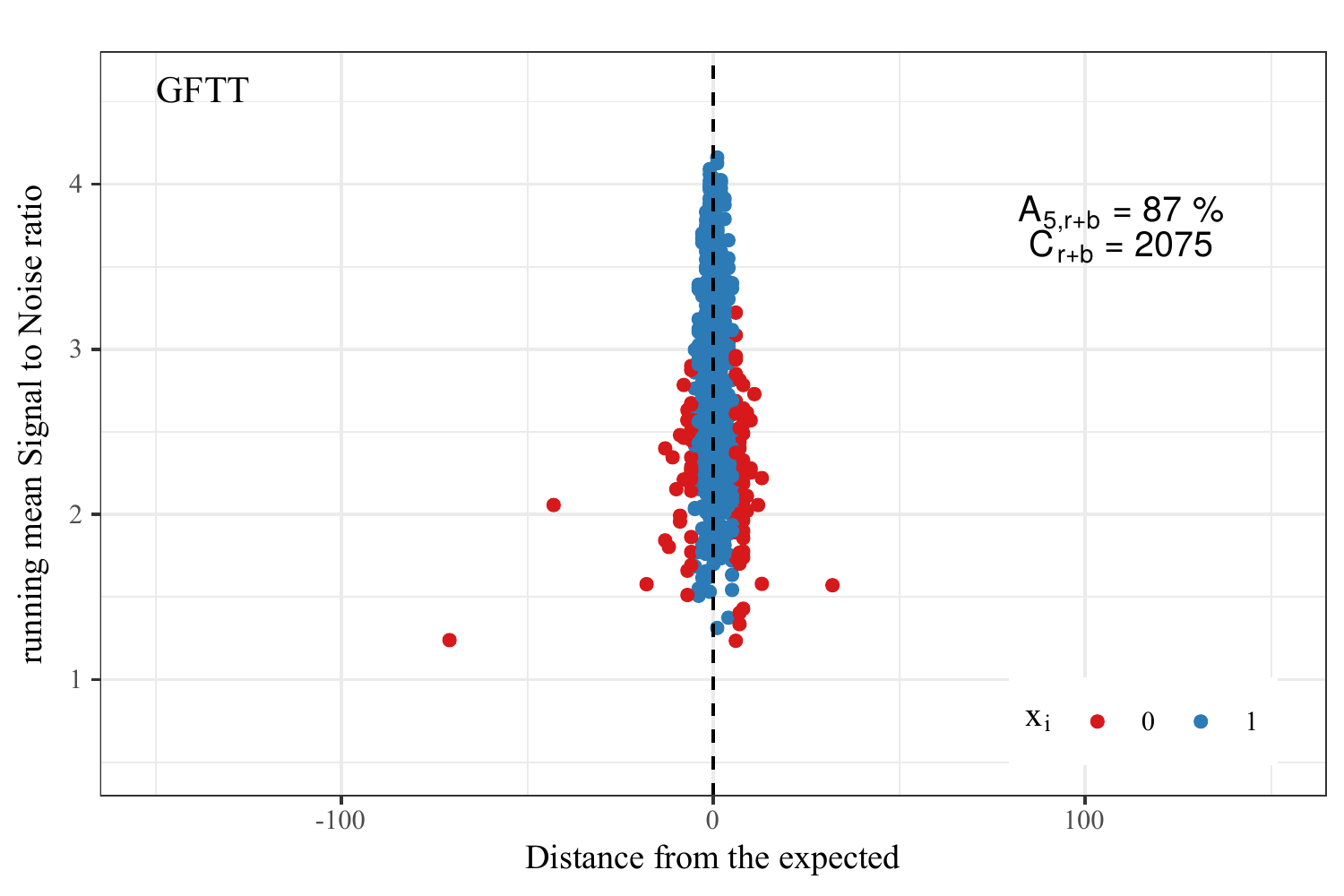}
    \includegraphics[height=.125\textheight]{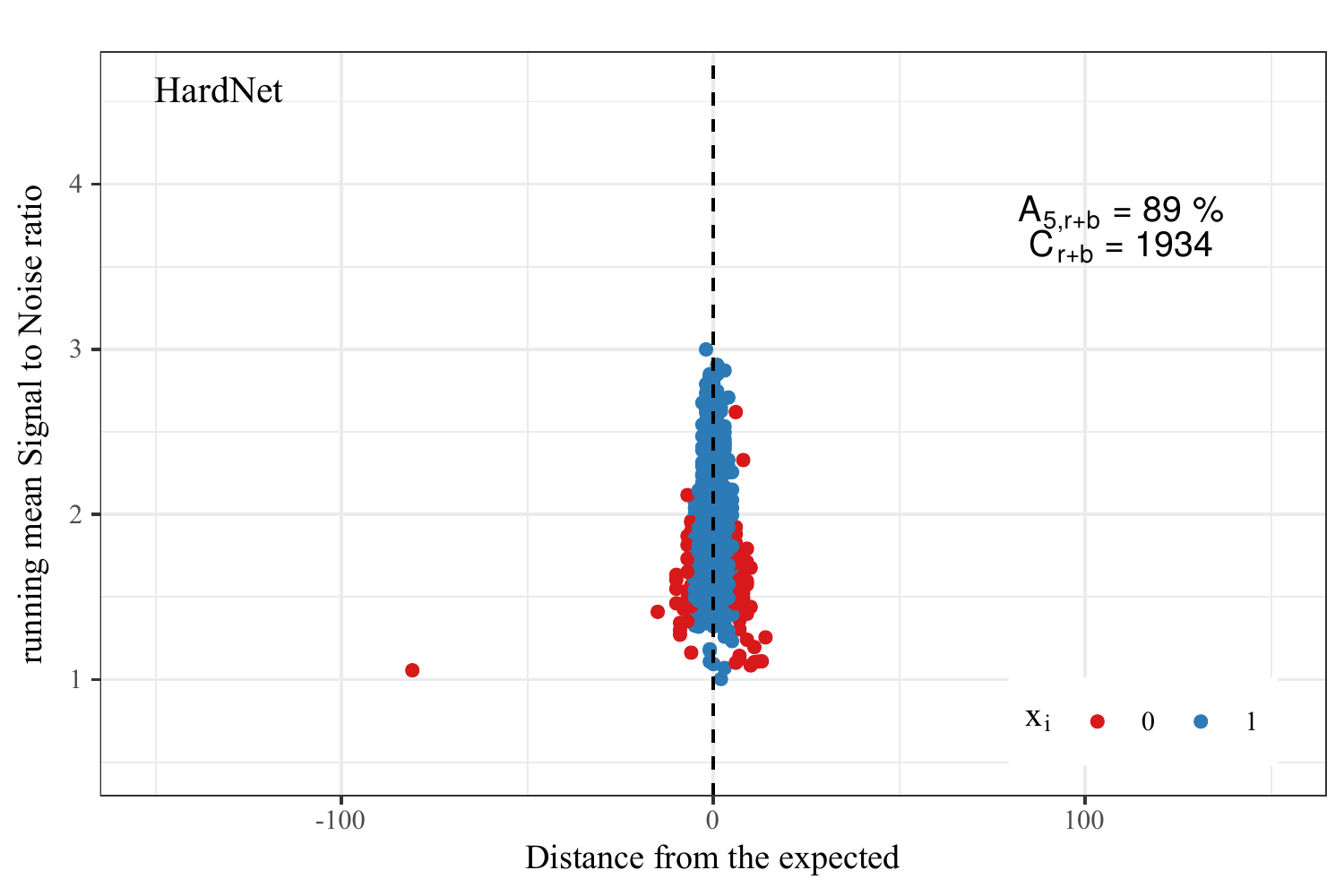}
    \includegraphics[height=.125\textheight]{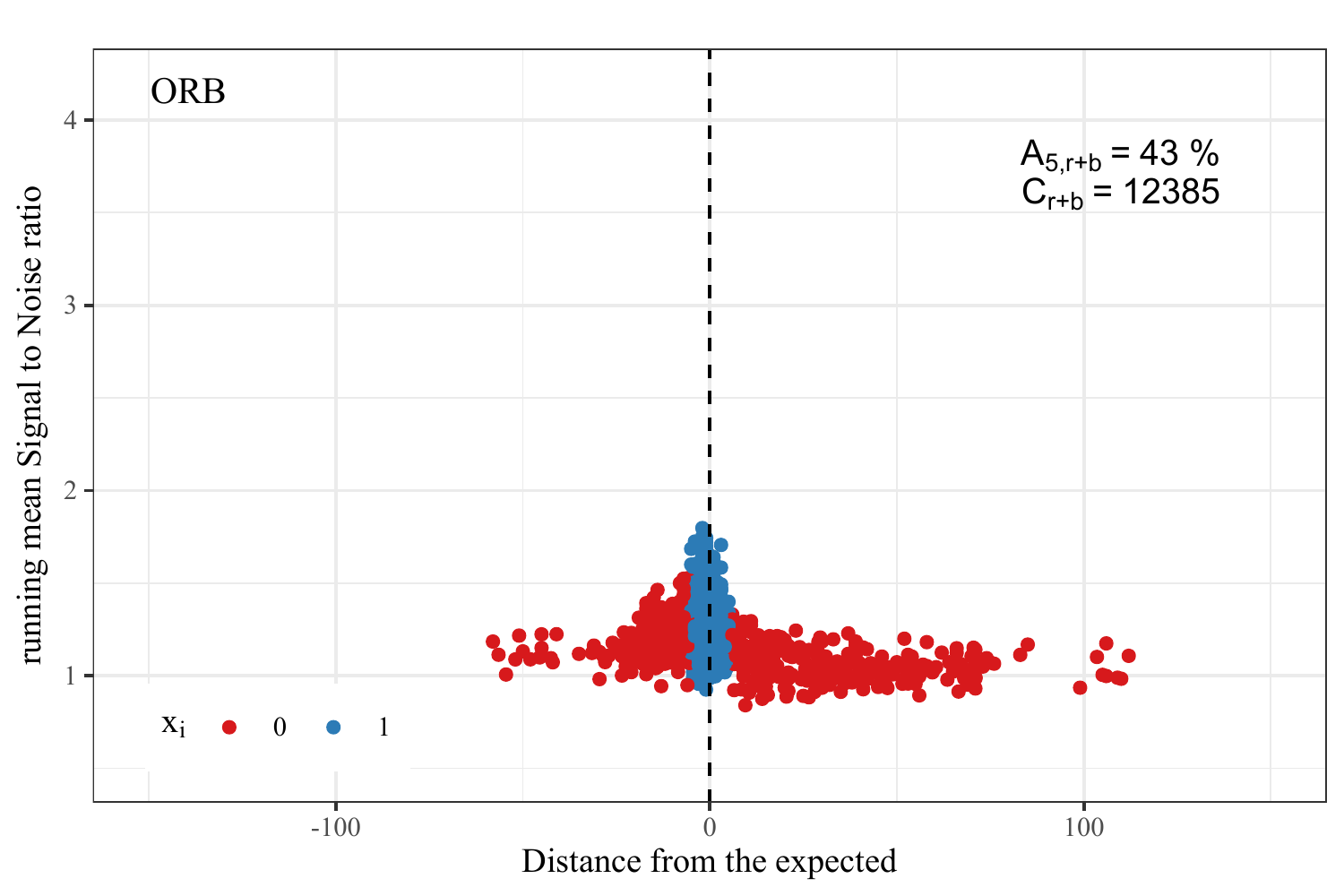}\\
    \includegraphics[height=.125\textheight]{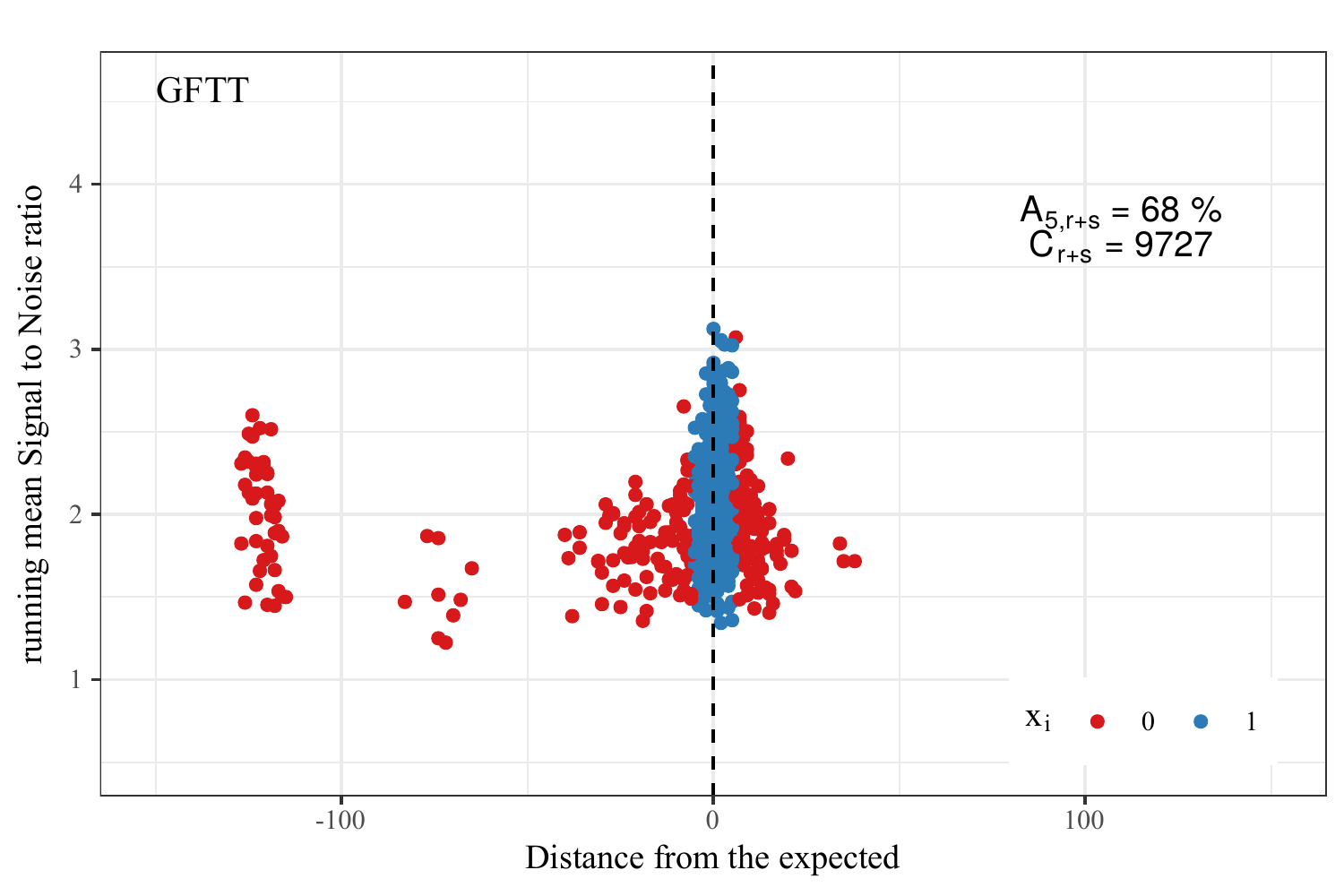}
    \includegraphics[height=.125\textheight]{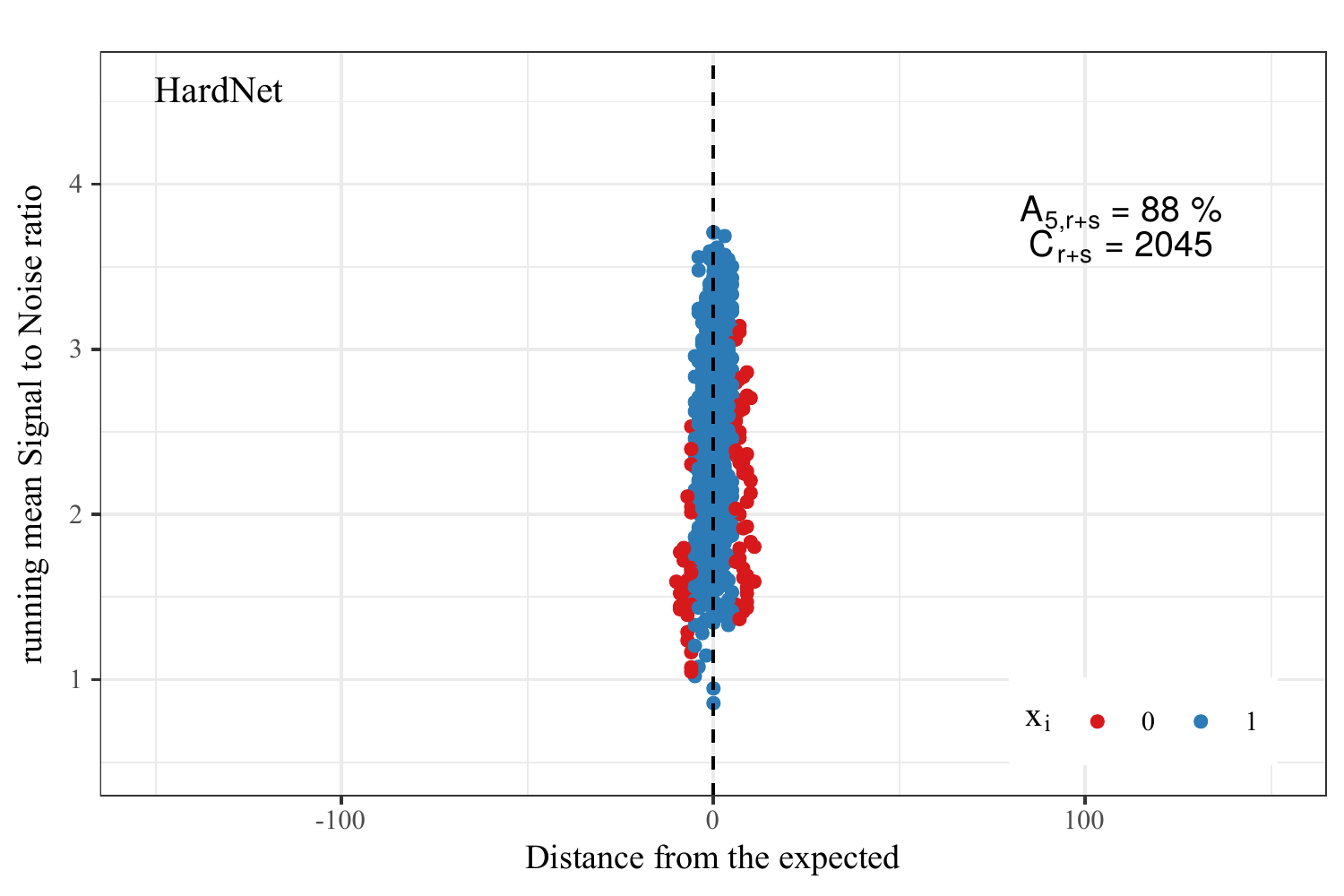}
    \includegraphics[height=.125\textheight]{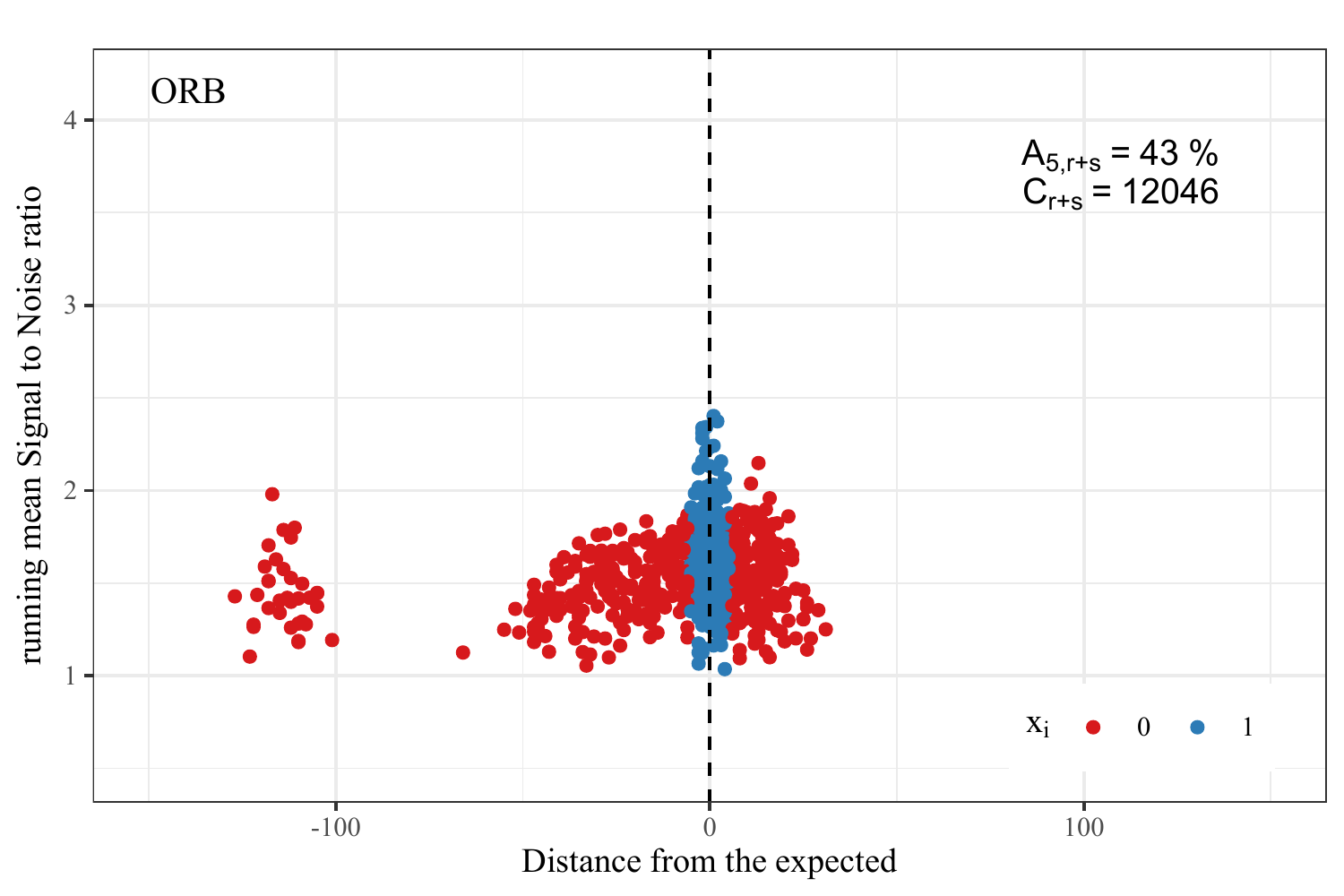}\\
    \includegraphics[height=.125\textheight]{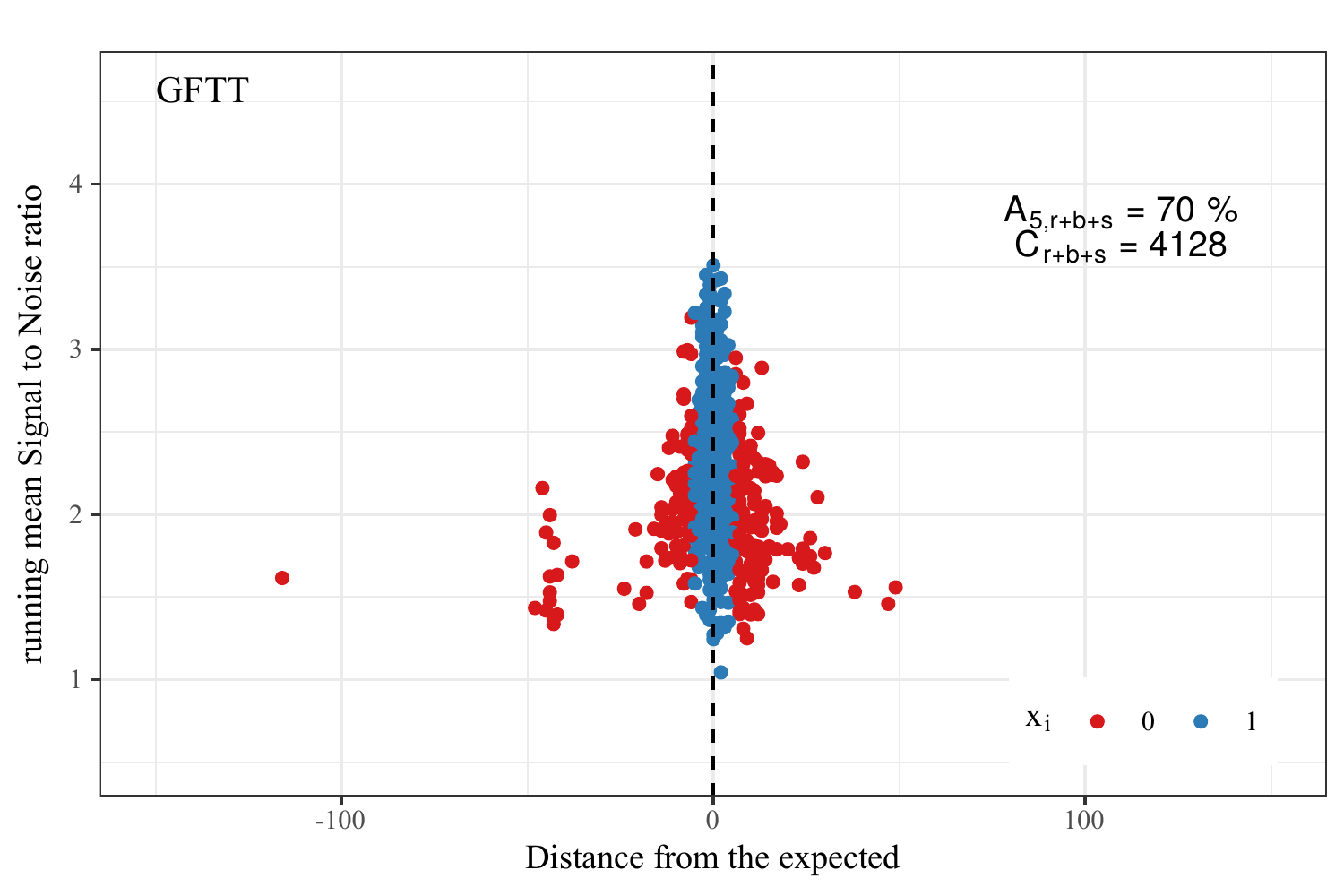}
    \includegraphics[height=.125\textheight]{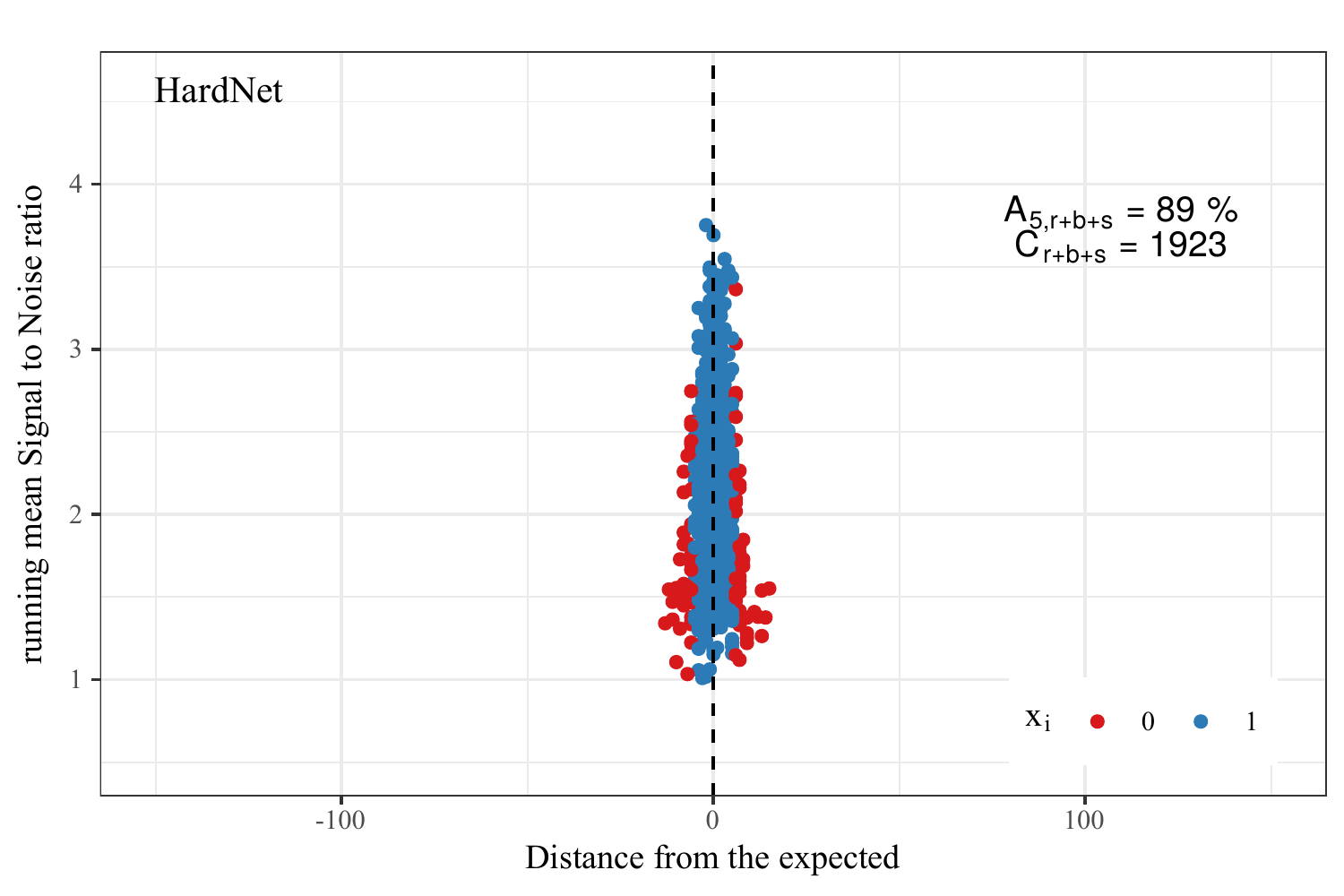}
    \includegraphics[height=.125\textheight]{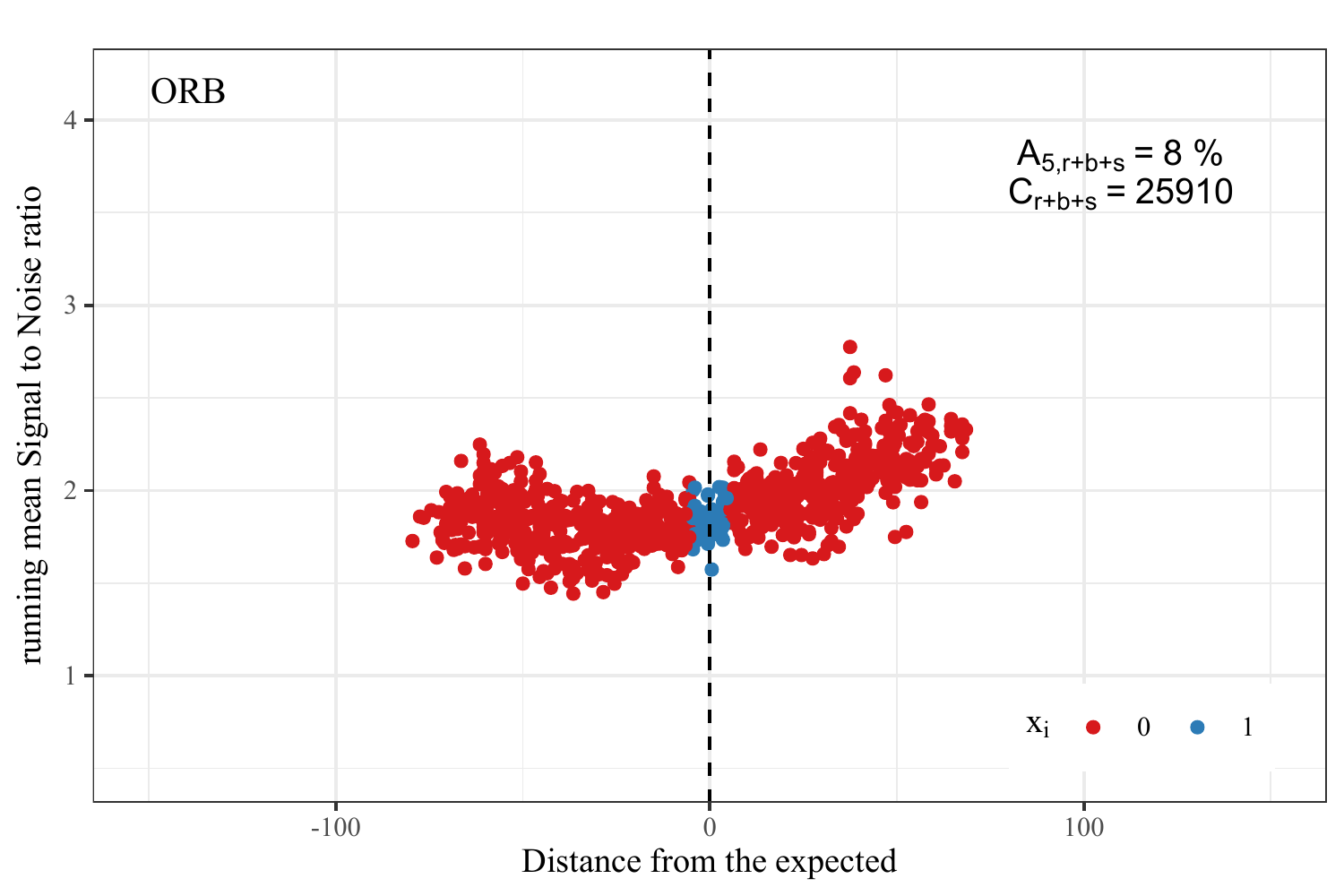}\\
    \includegraphics[height=.125\textheight]{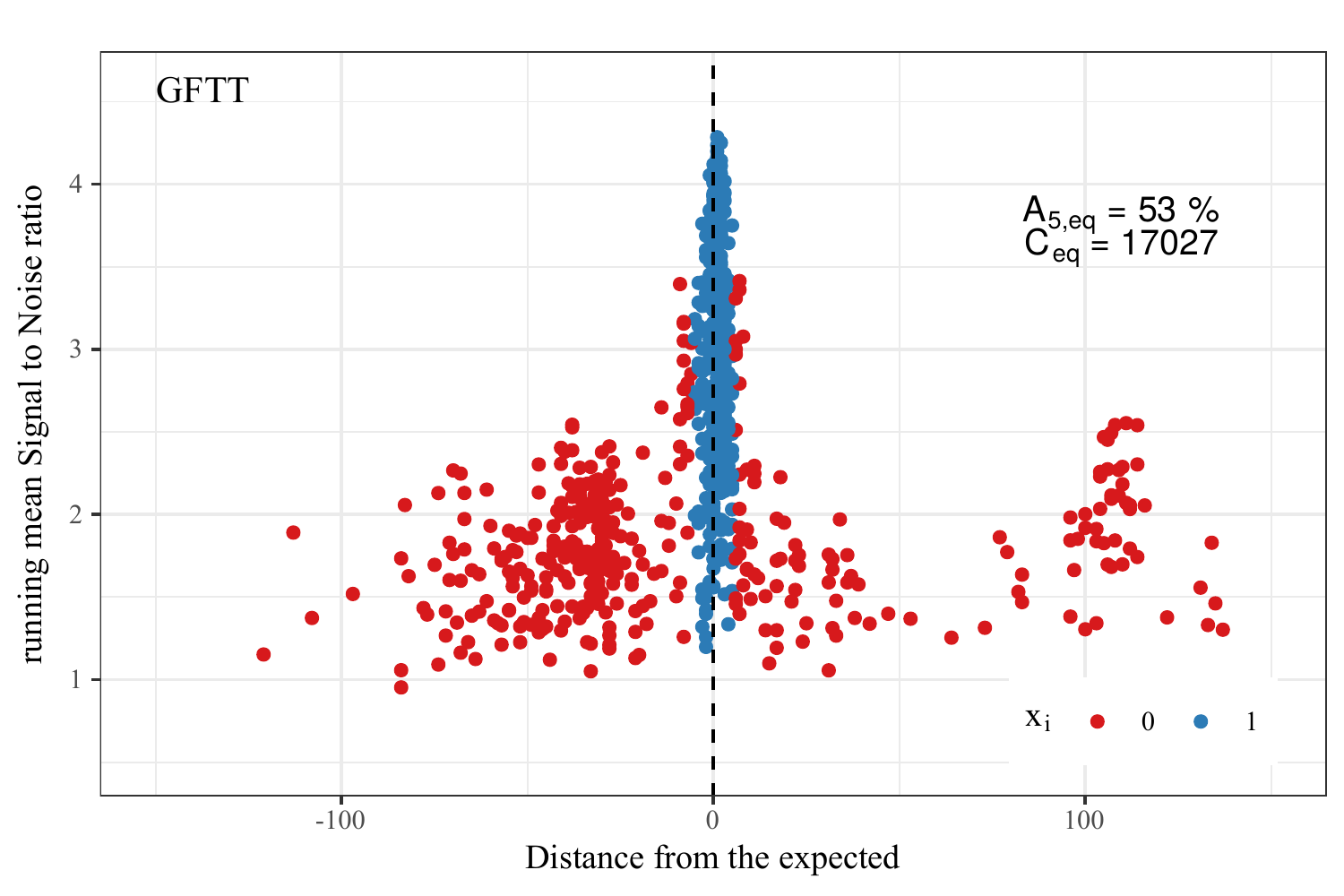}
    \includegraphics[height=.125\textheight]{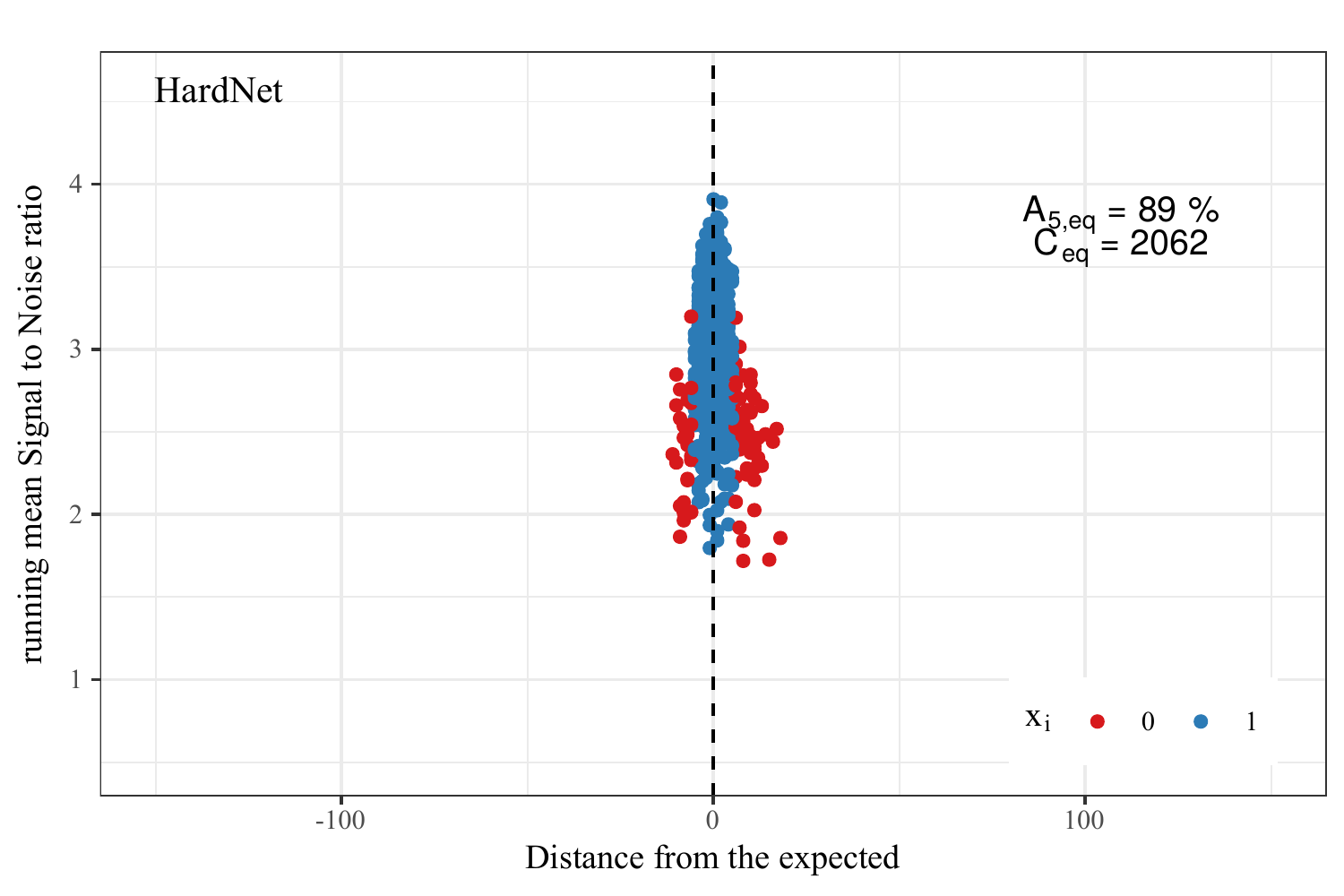}
    \includegraphics[height=.125\textheight]{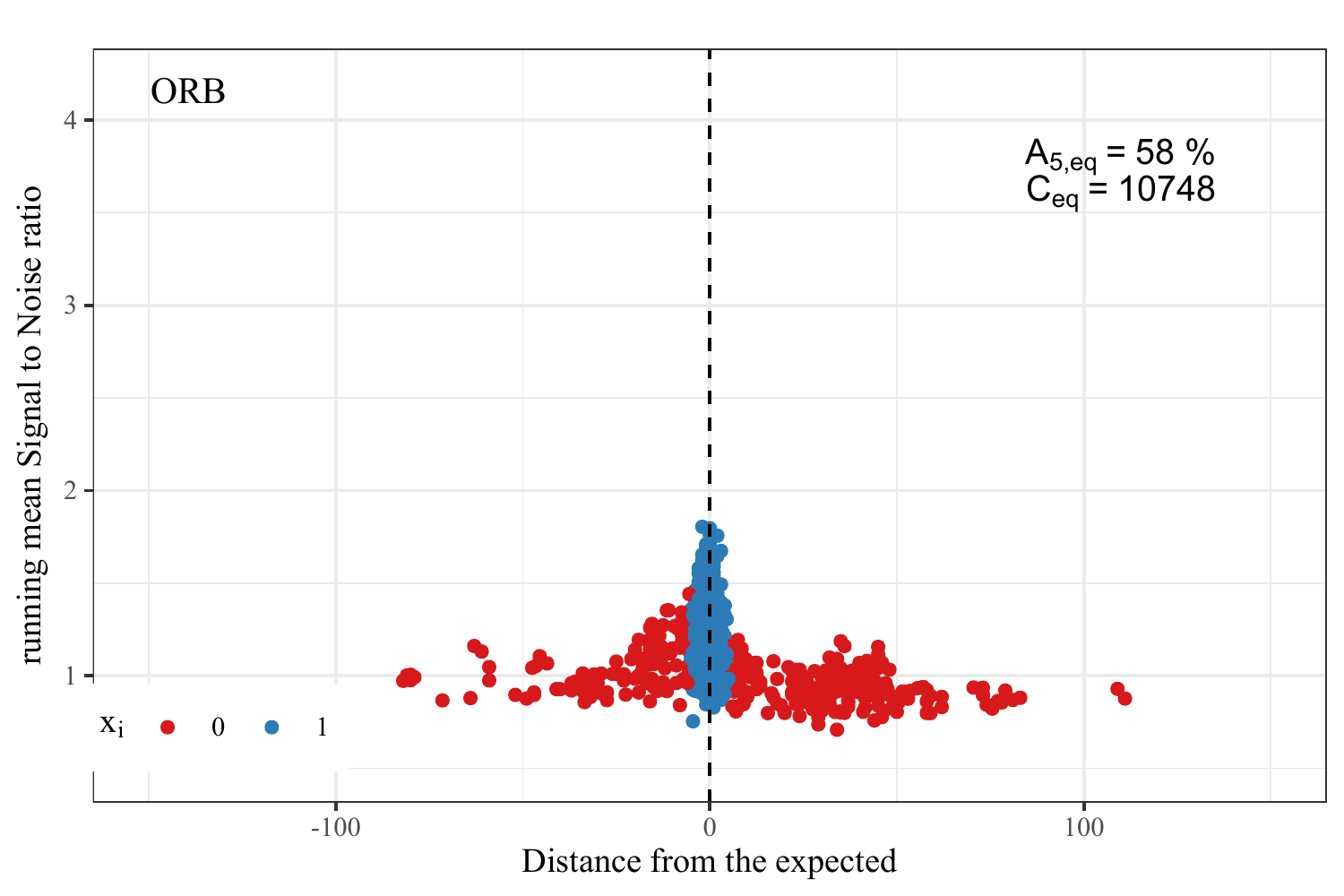}\\
    \includegraphics[height=.125\textheight]{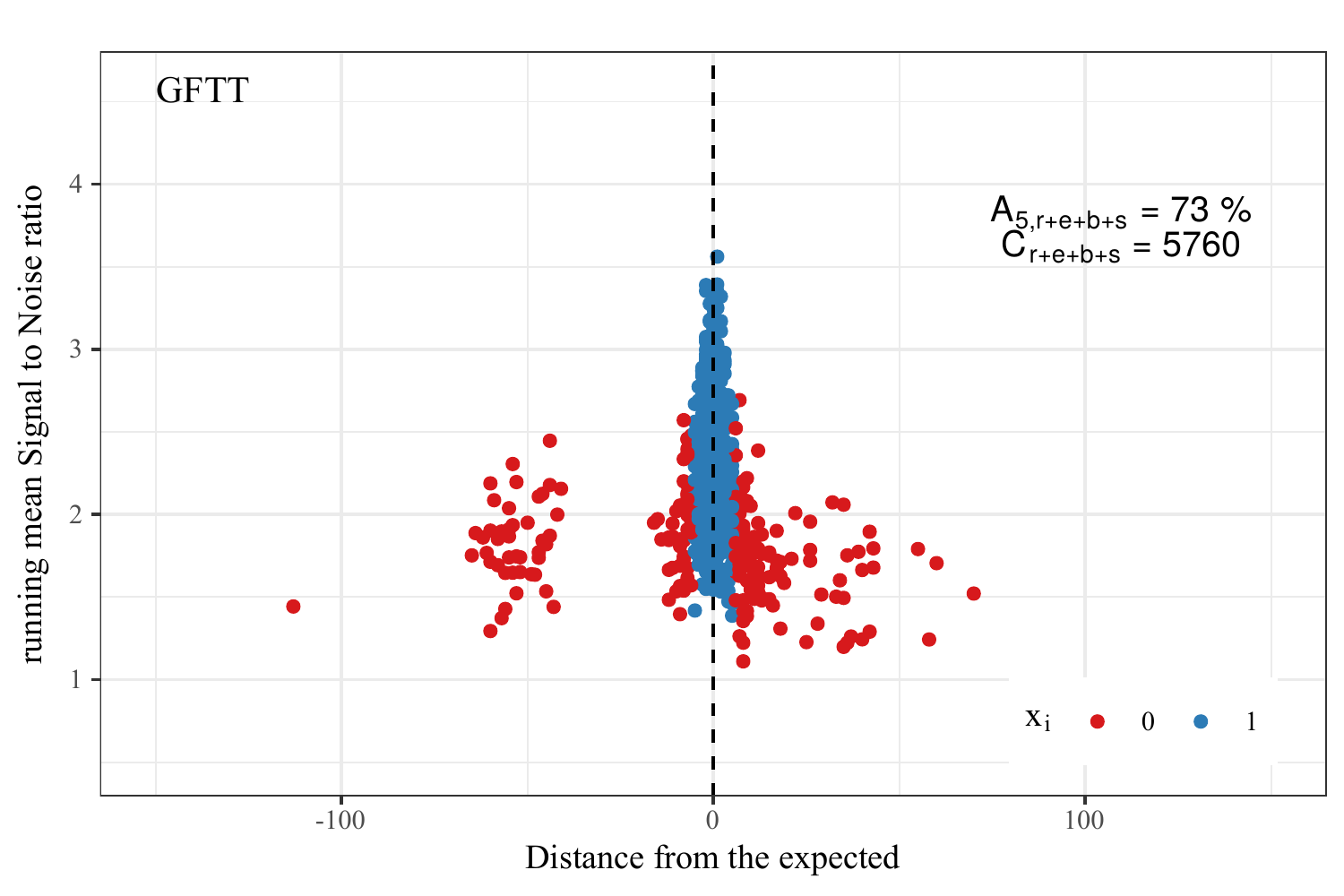}
    \includegraphics[height=.125\textheight]{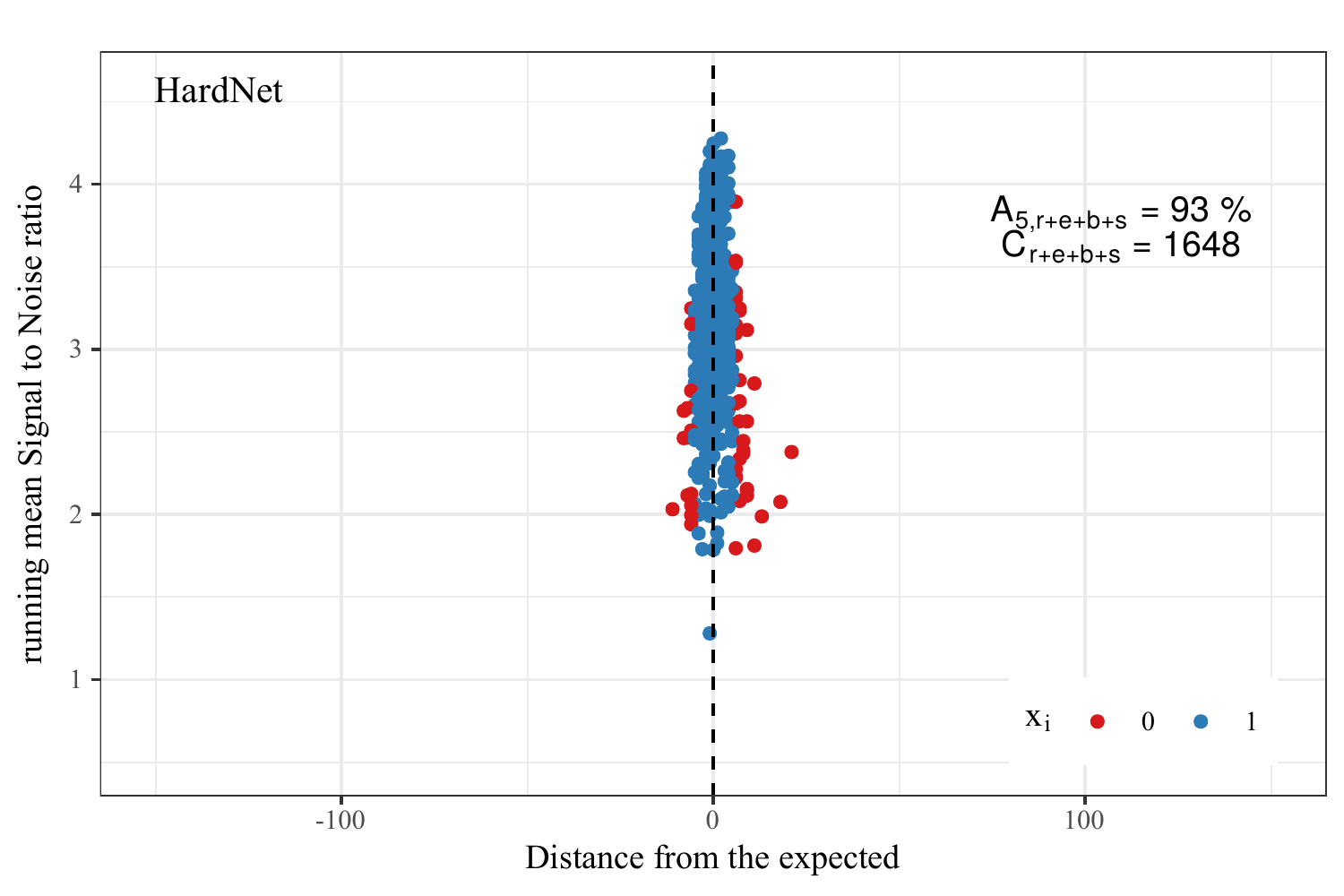}
    \includegraphics[height=.125\textheight]{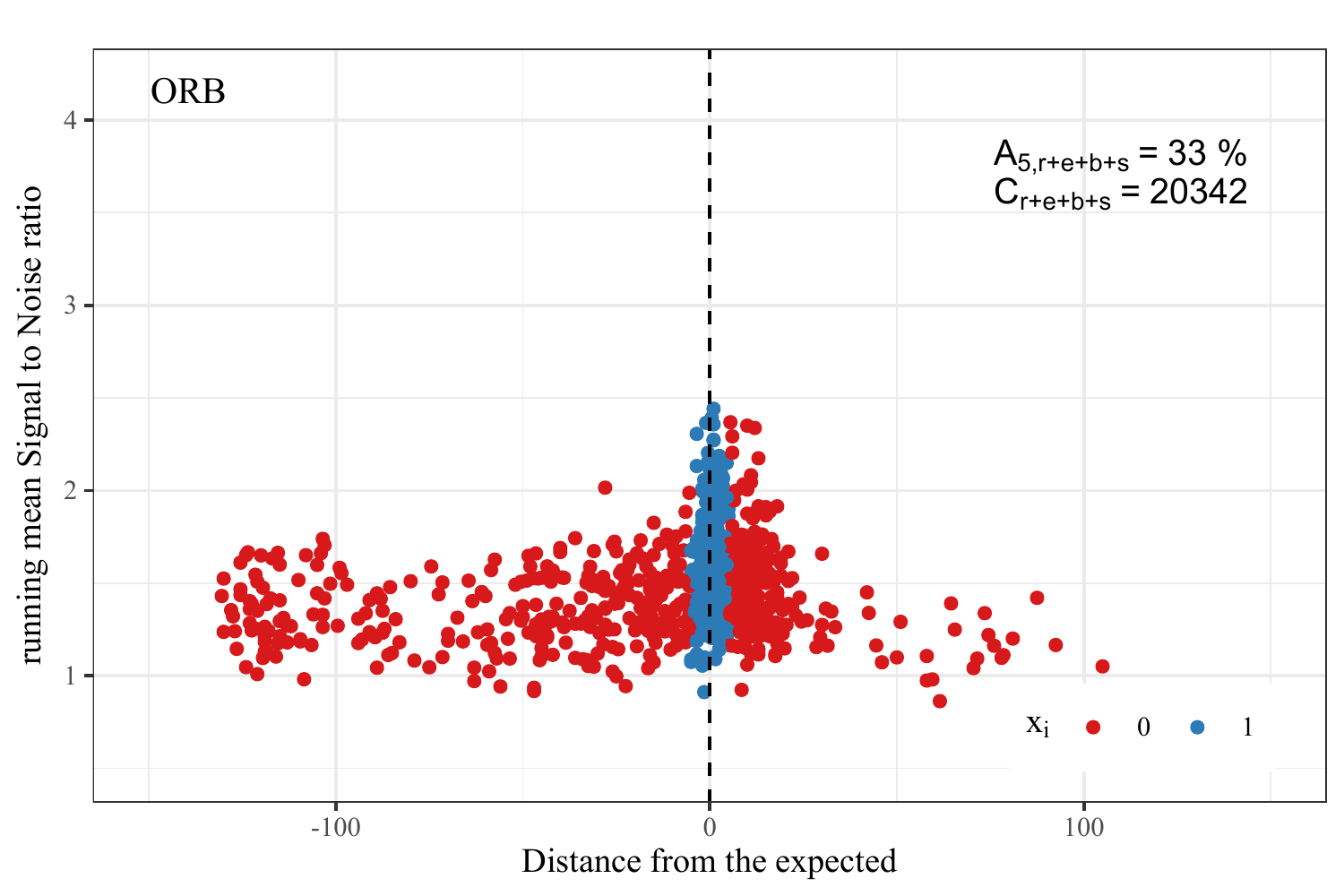}\\
    \caption{\label{fig:accuracy_all2}Accuracy $A_{d,c}$ of each used FD method (from left column to right column: GFTT, HardNet, ORB) and the applied image preprocessing (from first row to last: none; rotation; rotation and skull extraction; rotation and scaling; rotation, skull extraction and scaling; equalisation; rotation, equalisation, skull extraction and scaling. The blue dots correspond to the input images satisfying the condition in Eq.~\ref{eq:accuracy}. The y-axis corresponds to the running mean SNR and the x-axis to the distance from the expected. The vertical dashed line represents the expected image ID. The accuracy $A_{d,c}$ and cumulative distance $C_d$ computed by Eq.~\ref{eq:cum_distance} are stated in each figures.}
\end{figure*}

\begin{figure*}[htbp]
    \centering
    \includegraphics[width=.33\linewidth]{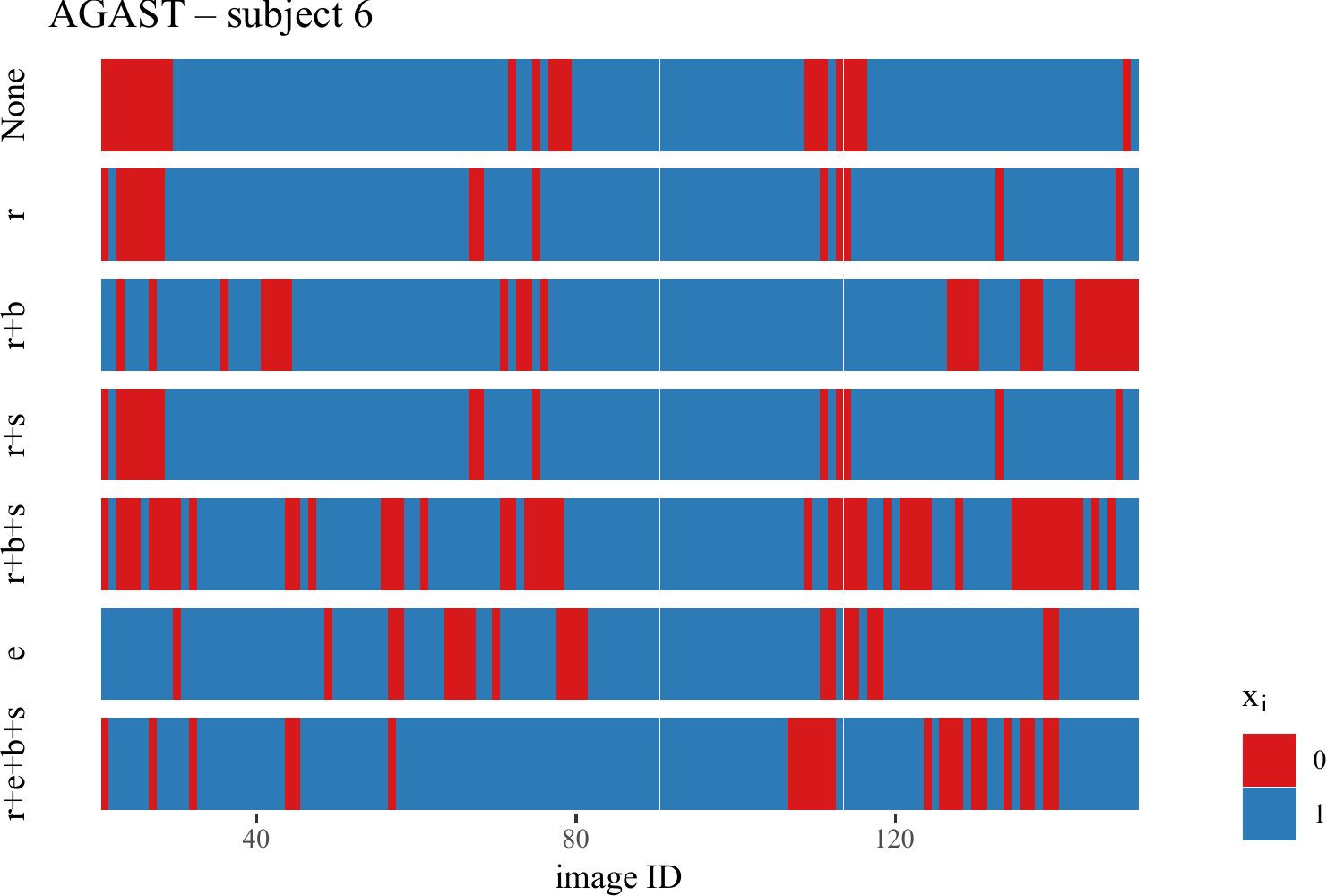}\hfill
    \includegraphics[width=.33\linewidth]{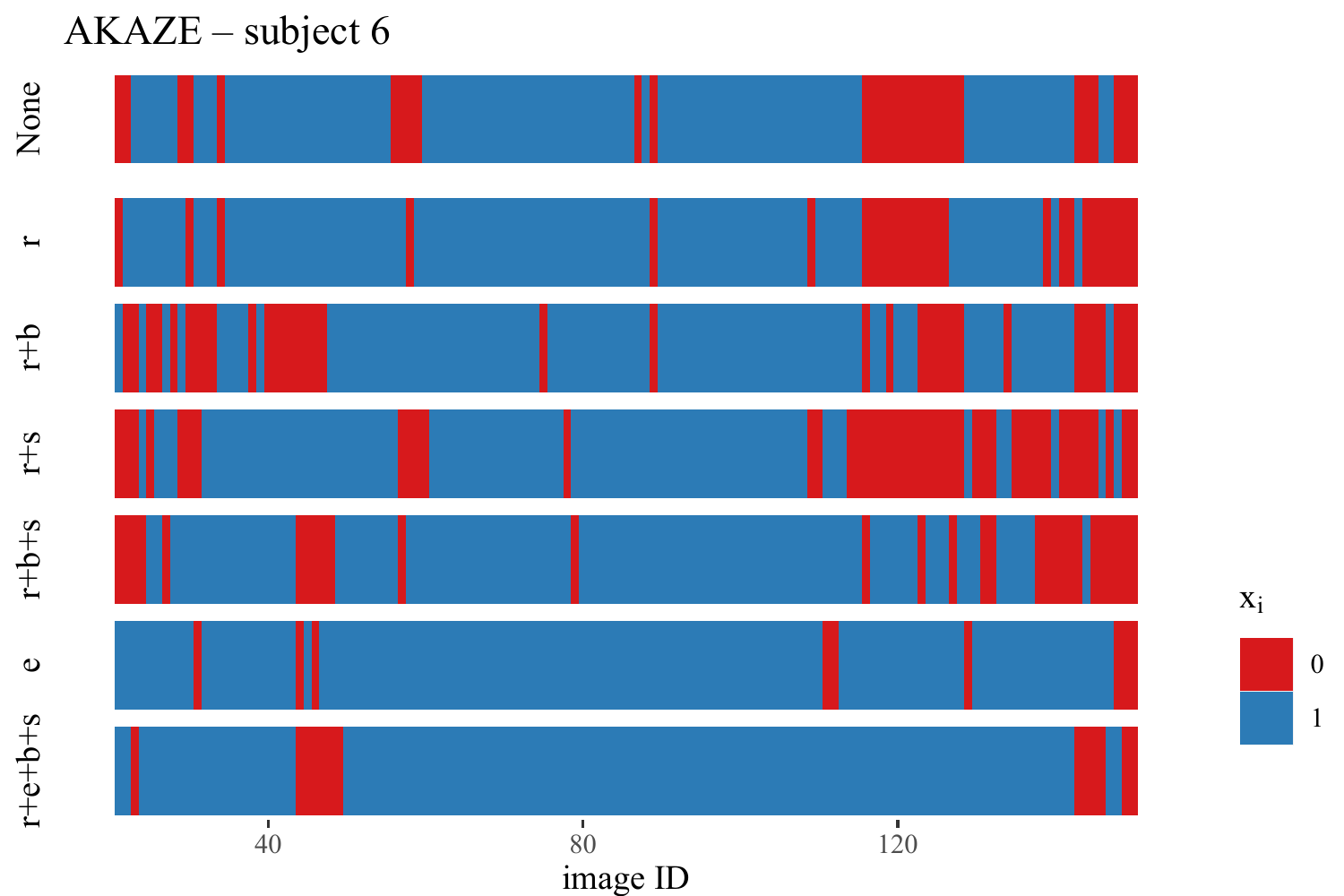}\hfill
    \includegraphics[width=.33\linewidth]{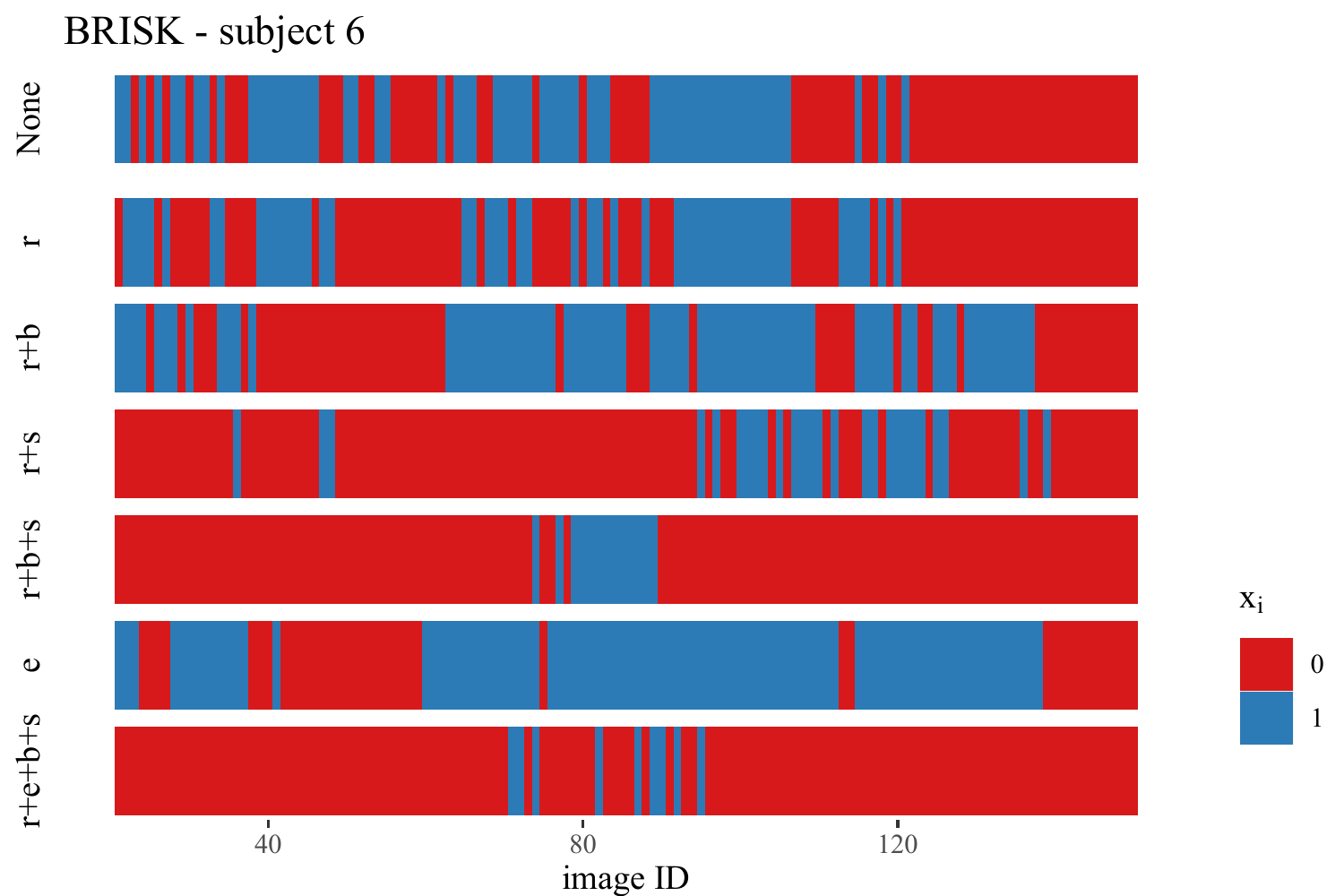}\\
    \includegraphics[width=.33\linewidth]{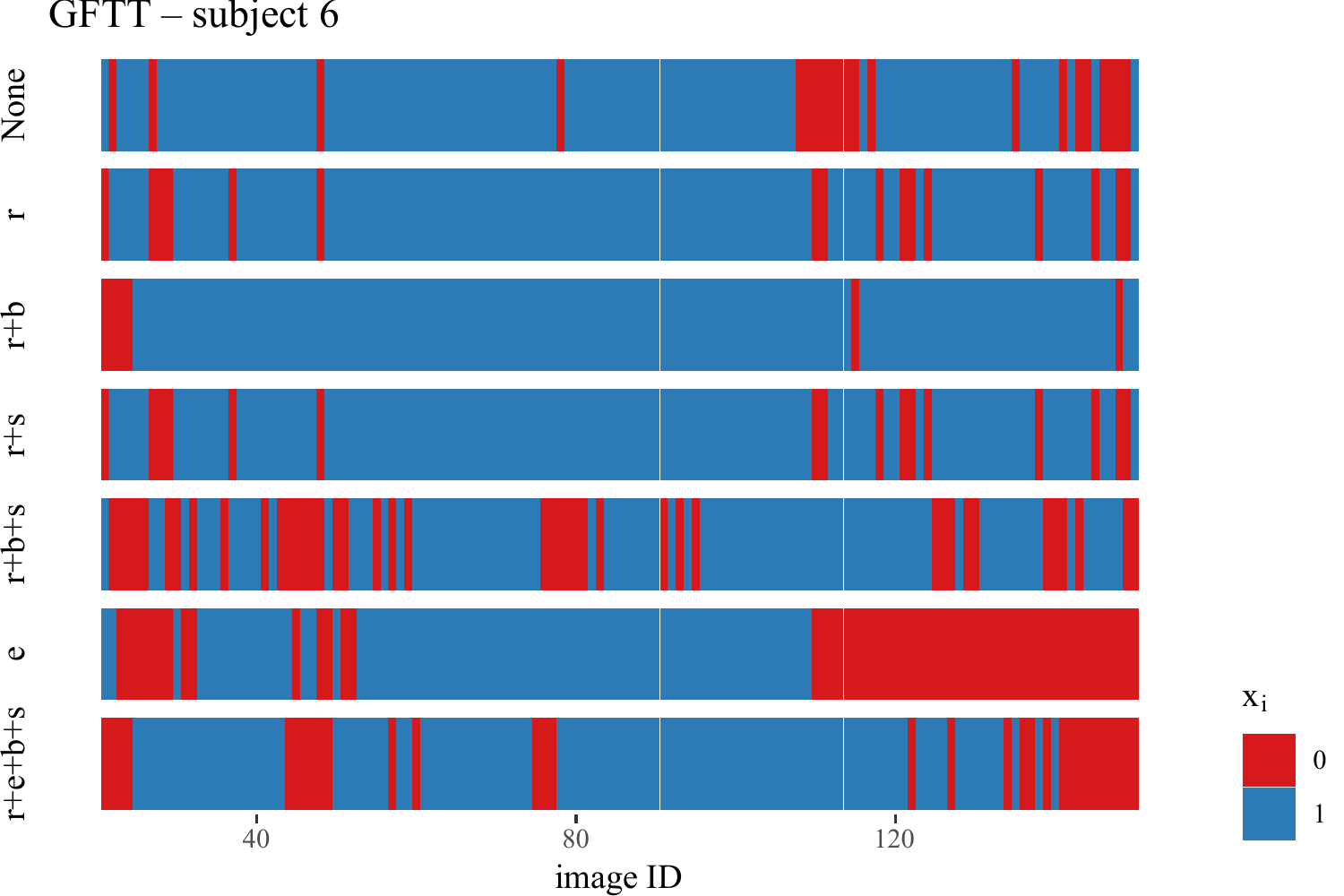}\hfill
    \includegraphics[width=.33\linewidth]{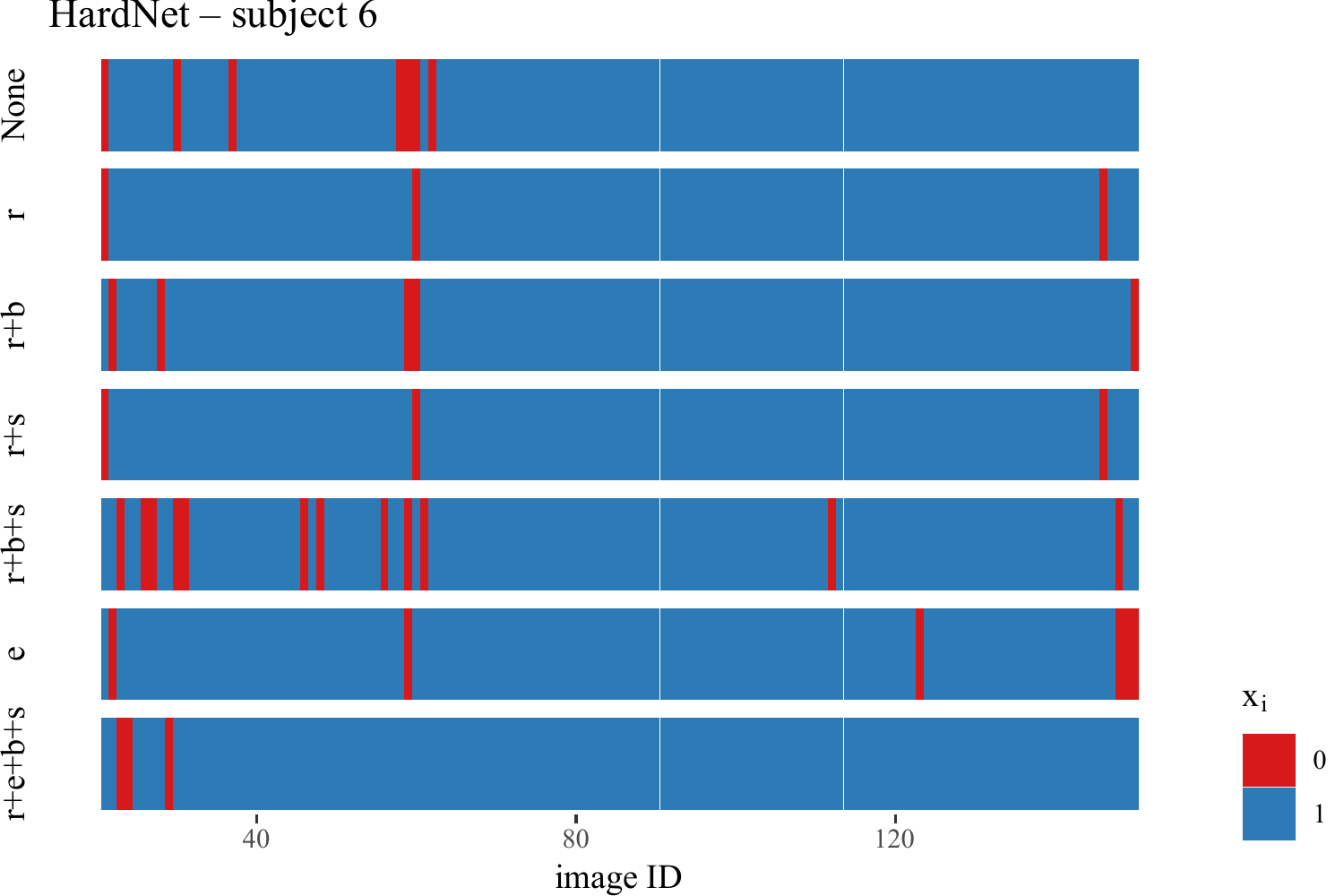}\hfill
    \includegraphics[width=.33\linewidth]{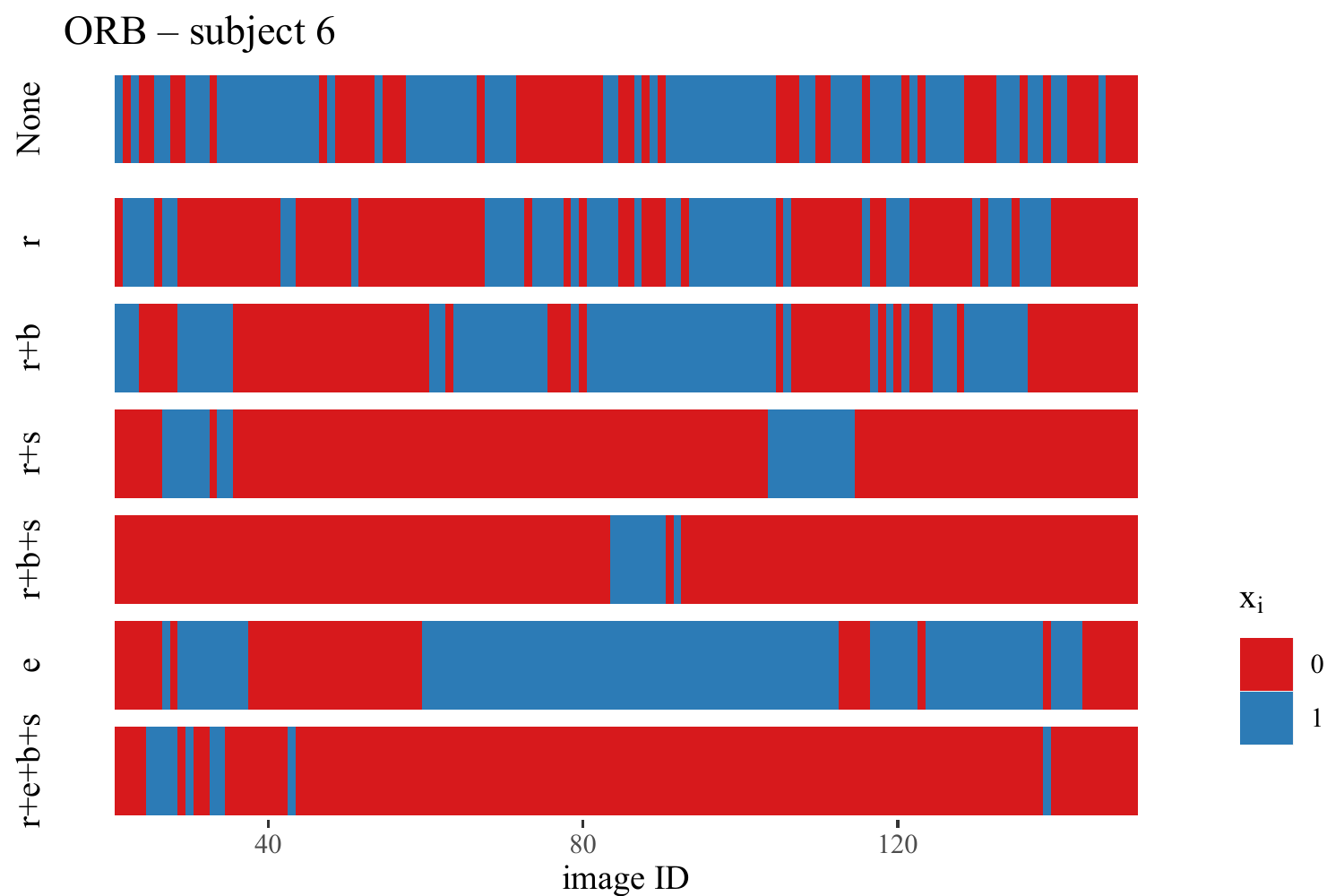}
    \caption{\label{fig:where06}Relation between the condition $x_i$ of accuracy $A_{5,c}$ (Eq.~\ref{eq:accuracy}) and the image ID (from image index 20 up to index 150) with respect to image corrections (rows in the plot) and FD methods. The blue bars indicate that the condition $x_i$ is met, i.e., the found position of the best matching image ID is not more than $c=5$~mm from both sides, whilst the red bars indicate the opposite.}
\end{figure*}

Figures~\ref{fig:accuracy_all},~\ref{fig:accuracy_all2} shows the result of accuracy $A_{5,c}$ (noted in each figure, corresponds with the colour of the dots) and distance of the best matched ID slice and the expected one (x-axis). We show the accuracy $A_{5,c}$ for each FD method used in columns for different preprocessing steps in rows. Namely AGAST, AKAZE and BRISK in Fig.~\ref{fig:accuracy_all}, whilst GFTT, HardNet and ORB in Fig.~\ref{fig:accuracy_all2}. For the preprocessing steps we applied only or a combination of rotation (r), scaling (s), skull extraction (b) and equalisation (e). In each case we also provide the value of cumulative distance $C_c$ (see Eq.~\ref{eq:cum_distance}), where the lower value the better as it shows the spread of the IDs differences across all input images. Finally,  the Fig.~\ref{fig:where06} shows the visualisation whenever the patient 6 MRI slice was correctly matched (blue bar) or not (red bar) with the effect of selected preprocessing steps, i.e. we can define which section of the axial plane have the FD methods problems. The condition $x_i$ (here is equal to 5) is used for the calculation of the accuracy $A_{d,c}$ defined in Eq.~\ref{eq:accuracy} and it corresponds to the allowed range of correctness. In general we see, that the FD methods (except HardNet) have problems with the MRI slices from the top and bottom of the head.

\begin{figure*}[htbp]
    \centering
    \includegraphics[width=\linewidth]{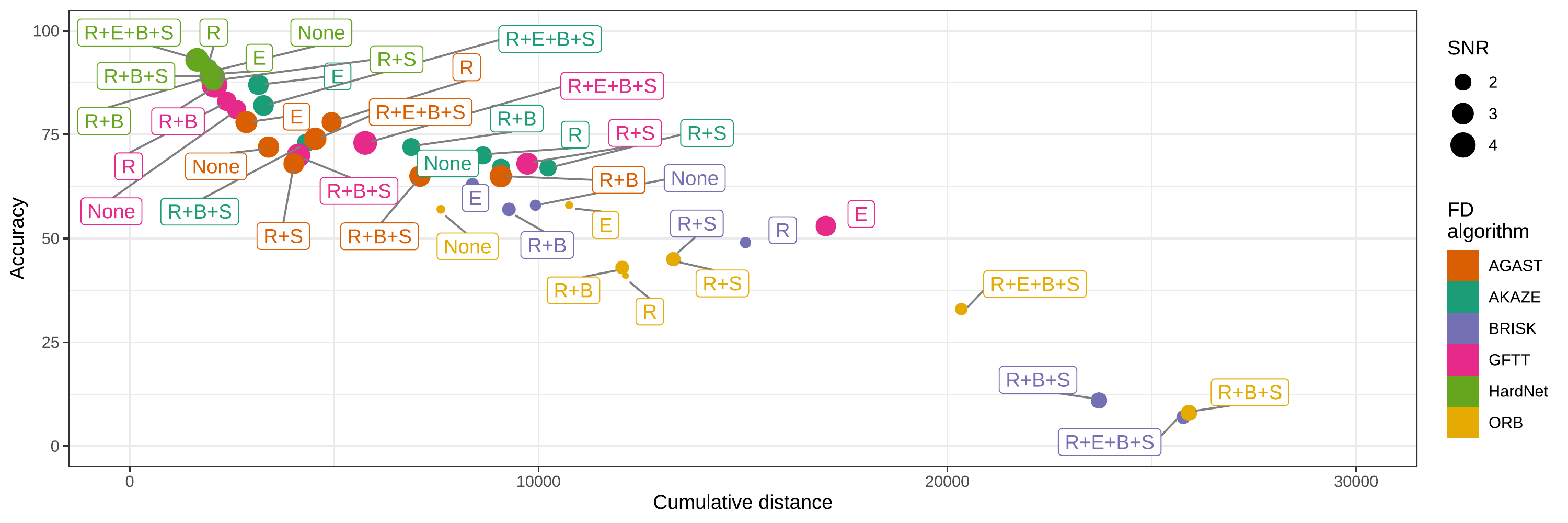}
    \caption{Summary of the FD methods relation between the cumulative distance $C_c$ (Eq.~\ref{eq:cum_distance}) and accuracy $A_{5,c}$ (Eq.~\ref{eq:accuracy}) with respect to each image preprocessing (labels). The point colours correspond to the FD methods while the point size to the running mean SNR.}
    \label{fig:fd_algo_SNR_summary}
\end{figure*}

The overall summary with their mean SNR$_{ij}$ (Eq.~\ref{eq:snr}), accuracy $A_{d,c}$ (Eq.~\ref{eq:accuracy}) and cumulative distance $C_c$ (Eq.~\ref{eq:cum_distance}) for each featured detection and image preprocessing is stated in the Table~\ref{tab:accuracy} and in Fig.~\ref{fig:fd_algo_SNR_summary}. 


The results from the subsection~\ref{sec:robustness} show that, when correcting the input images by preprocessing, we could expect improvements in SNR and accuracy $A_{d,c}$. However, only in the case of AKAZE do we see a significant increase (from the accuracy of 67\% up to 87\%), whilst AGAST, ORB, BRISK and GFTT show rather worse results. This decrease is probably due to the fact that methods are too sensitive to such image degradation/improvements or to the fact that their methods identify key regions with insufficiently unique features to compare MRI images from a different patient. However, the SNR in the case of AGAST and GFTT when a rotation, a skull extraction or both is applied slightly increases. From the tested methods HardNet shows the most consistent behaviour when a preprocessing is applied or not.

Although the HardNet shows quite good results in accuracy $A_{5,c}$ and cumulative distance $C_c$, the SNR value is around 4 at best (AGAST $\sim$4.5, AKAZE $\sim$4, GFTT~$\sim$4.4, BRISK $\sim$2.5, ORB $\sim$2.4), which is relatively low.

\begin{table*}[htbp]
\begin{center}
\caption{\label{tab:accuracy} Summary of the FD methods compared with images from different subjects. In the columns we show the effect of the selected combination of image preprocessing techniques and their influence on the Accuracy $A_{5,c}$ (see Eq.~\ref{eq:accuracy}), cumulative distance $C_c$ (see Eq.~\ref{eq:cum_distance}) and the average of the signal to noise ratio.}
\begin{tabular}{@{}lrrrrrrr@{}}
\toprule
\multicolumn{1}{c}{\multirow[b]{5}{*}{AGAST}} & \multicolumn{7}{c}{Image  preprocessing} \\ \cmidrule(l){2-8} 
\multicolumn{1}{c}{} & none & rotation & rotation\newline & rotation & rotation & equalisation & rotation \\
\multicolumn{1}{c}{} & & & skull ex. & scaling & skull ex. &  & equalisation \\
\multicolumn{1}{c}{} & & & & & scaling & & skull ex. \\
\multicolumn{1}{c}{} & & & & & & & scaling\\ 
\midrule
Accuracy [\%] & 72 & 78 & 65 & 68 & 65 & 78 & 74 \\
Cumulative distance & 3396 & 4939 & 9077 & 4013 & 7099 & 2854 & 4538 \\
mean SNR & 2.93 & 2.64 & 3.19 & 2.78 & 2.94 & 3.12 & 3.2 \\ 
\midrule
\multicolumn{1}{c}{AKAZE} & \multicolumn{7}{c}{} \\
\midrule
Accuracy [\%] & 67 & 70 & 72 & 67 & 73 & 87 & 82 \\
Cumulative distance & 9083 & 8639 & 6888 & 10233 & 4315 & 3149 & 3272 \\
mean SNR & 2.26 & 2.23 & 2.2 & 2.09 & 2.27 & 2.74 & 2.76 \\ 
\midrule
\multicolumn{1}{c}{BRISK} & \multicolumn{7}{c}{} \\
\midrule
Accuracy [\%] & 58 & 49 & 57 & 45 & 11 & 63 & 7 \\
Cumulative distance & 9925 & 15063 & 9276 & 13299 & 23708 & 8383 & 25774 \\
mean SNR & 1.35 & 1.34 & 1.57 & 1.64 & 1.94 & 1.49 & 1.61 \\
\midrule
\multicolumn{1}{c}{GFTT} & \multicolumn{7}{c}{} \\
\midrule
Accuracy [\%] & 81 & 83 & 87 & 68 & 70 & 53 & 73 \\
Cumulative distance & 2616 & 2373 & 2075 & 9727 & 4128 & 17027 & 5760 \\
mean SNR & 2.93 & 2.64 & 3.19 & 2.78 & 2.94 & 3.12 & 3.2 \\
\midrule
\multicolumn{1}{c}{HardNet} & \multicolumn{7}{c}{} \\
\midrule
Accuracy [\%] & 90 & 91 & 89 & 88 & 89 & 89 & 93 \\
Cumulative distance & 2000 & 1918 & 1934 & 2045 & 1923 & 2062 & 1648 \\
mean SNR & 2.24 & 2.32 & 2.02 & 2.56 & 2.25 & 3.11 & 3.53 \\
\midrule
\multicolumn{1}{c}{ORB} & \multicolumn{7}{c}{} \\
\midrule
Accuracy & 57 & 41 & 43 & 45 & 8 & 58 & 33 \\
Cumulative distance & 7610 & 12133 & 12046 & 13299 & 25910 & 10748 & 20342 \\
mean SNR & 1.2 & 1.16 & 1.58 & 1.64 & 1.92 & 1.18 & 1.44 \\
\bottomrule
\end{tabular}
\end{center}
\end{table*}


\subsection{Comparing with Atlas}
\label{sec:atlas}
Nowadays, as the amount of medical image data is growing, such atlases are mostly used as a gold standard validation for quantitative analysis or testing new techniques. In addition, as the atlases are mostly simulated models, they are also independent of the technological aspects of MRI devices, or in general of the used device. Moreover, comparing a set of MRI images of one patient with a simulated or ``ideal'' atlas by using FD methods can bring us a basic awareness of how to create atlases, or what additional information (metadata) is good to keep, which preprocessing steps are required and how to work with them in general.  

Here, we follow the same methodology of testing as in the previous subsection~\ref{sec:accuracy} but for the reference we have used a simulated atlas. Respectively, as the ground truth for our analysis procedure, we downloaded the atlas/phantom generated from BrainWeb: Simulated MRI Volumes for Normal Brain~\citep{Cocosco:1997:BrainWebOI, Kwan-etal:1999:BrainWeb, Kwan:1996:Proc, Collins-etal:1998:BrainWeb}. Currently, the simulated brain database (SBD) offers 3 modalities, 6 slice thicknesses, 6 levels of noise, and 3 levels of intensity non-uniformity, i.e., we chose the one that most closely matches our downloaded data set (1~mm slice thickness, T1 modality, 0~\% noise level, and 0~\% intensity non-uniformity). 

For brevity we show only results for HardNet and GFTT, as they demonstrate the best results in general, however, we provide all results of tested FD methods in Table~\ref{tab:atlas_all}. Similarly, we show only the results for no image preprocessing enhancements and selected ones, specifically the combination of rotation, scaling, equalisation, and skull extraction. The results are summarised in~Fig.~\ref{fig:atlas_akaze}. 

\begin{table*}[h!]
\begin{center}
\caption{\label{tab:atlas_all}Achieved performance of Accuracy $A_{5,c}$ (Eq.~\ref{eq:accuracy}), Cumulative distance $C_c$ (Eq.~\ref{eq:cum_distance}) and running mean SNR of all tested FD methods without and with applied their best image processing with respect of comparison with the atlas.}
\begin{tabular}{clrrr}
\toprule
\multicolumn{5}{c}{Comparing FD method with atlas (BRAINWEB)} \\ \midrule
\multicolumn{1}{l}{FD method} & Im. processing & \multicolumn{1}{l}{Accuracy $A_{5,c}$} & \multicolumn{1}{l}{Cum. distance $C_c$} & \multicolumn{1}{l}{run. mean SNR} \\ \midrule
\multirow{2}{*}{AGAST}     & none   & 18   & 18823   & 2.64   \\
                           & e      & 32   & 21153   & 2.50   \\ \hline
\multirow{2}{*}{AKAZE}     & none   & 20   & 26849   & 1.97   \\
                           & rebs   & 29   & 19888   & 1.22   \\ \hline
\multirow{2}{*}{BRISK}     & none   & 16   & 24196   & 0.93   \\
                           & e      & 15   & 22960   & 1.31   \\ \hline
GFTT                       & none   & 27   & 10361   & 1.83   \\
\multicolumn{1}{l}{}       & rb     & 52   & 8313    & 2.17   \\ \hline
\multirow{2}{*}{HardNet}   & none   & 53   & 5438    & 1.51   \\
                           & rebs   & 52   & 6000    & 1.57   \\ \hline
\multirow{2}{*}{ORB}       & none   & 11   & 26804   & 1.12   \\
                           & e      & 17   & 19560   & 1.29   \\ \bottomrule
\end{tabular}
\end{center}
\end{table*}

Overall, when we compare MRI slices with the atlas all the FD methods show poor results. The best-performing method is HardNet. HardNet accuracy without preprocessing achieved $A_{5,\mathrm{none}}{\sim} 53$\,\% with running mean SNR ${\sim}$1.5 and cumulative distance $\sim$5400. Apply the preprocessing of rotation, brain extraction, scaling and equalization slightly improved only the running SNR  to $\sim$1.57. In terms of accuracy and cumulative distance, it remained similar or slightly worse. AGAST and GFTT also show good results in accuracy ($A_{5,\mathrm{none}}{\sim}$30\,\%, especially when the image preprocessing is applied the accuracy increase significantly (for GFTT $\sim$53\,\% and AGAST $\sim$31\,\%. However, the cumulative distance $C_c$ for both methods is high $\sim$10000--20000, i.e., showing certain uncertainty. AKAZE accuracy without preprocessing is 20\,\% with a running mean SNR $\sim$1.97 and cumulative distance $\sim$26000. The best result is achieved with all preprocessing steps where $A_\mathrm{5,rebs} = 29$\%, $C_\mathrm{rebs}=19888$, and running mean SNR $\sim$1.22. This does not represent a significant improvement. The ORB and BRISK methods achieve similar results with accuracy $A_\mathrm{5,none} = 12$\,\% and $A_\mathrm{5,none} = 16$\,\%, cumulative distance $C_\mathrm{none} = 26804$ and  $C_\mathrm{none} = 24196$, running SNR 1.12 and 0.97. The low value of SNR ($<3$) suggests that matched slices are found due to random occurrence or are found with low confidence liable to noise fluctuations. We suppose, that these results are caused by the different interpretations of the grayscale tissues between the atlas and MRI slices. Therefore, additional data information on grayscale ranges for each component of the brain could be useful for later analysis. 

\subsubsection{Other anatomical planes}
\label{sub:atlas_other_planes}
In the previous sections, we have analysed in detail the behaviour of different FD methods for the axial plane case. In general practice, however, other anatomical planes, the sagittal and the coronal planes are used as well. Therefore, here we provide a brief analysis of the accuracy  $A_{5,c}$ for the two best-performing FD methods HardNet and GFTT. We present the comparison of all three anatomical planes with the atlas.

\begin{figure*}[htbp]
    \centering
    \includegraphics[width=.32\linewidth]{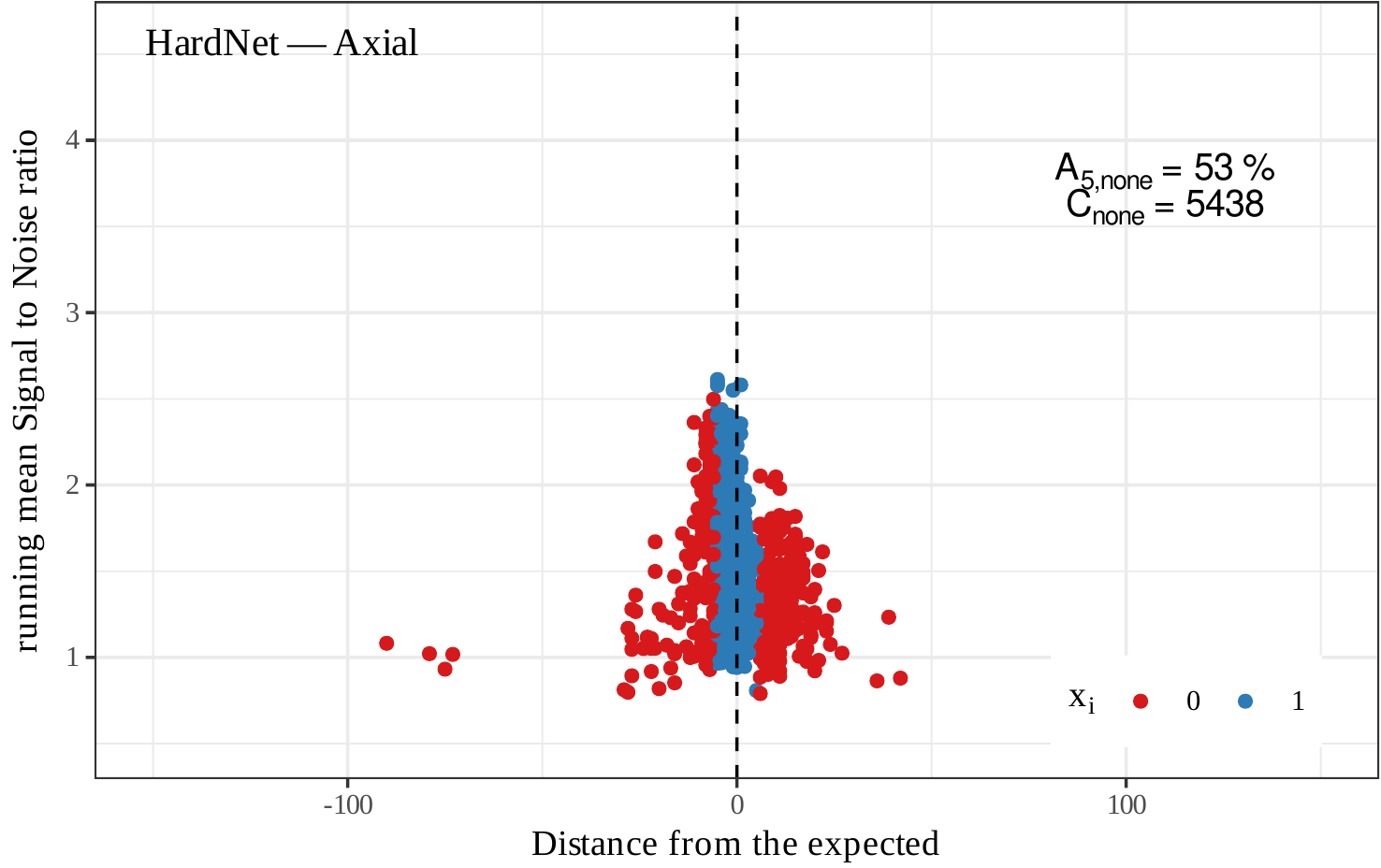}
    \includegraphics[width=.32\linewidth]{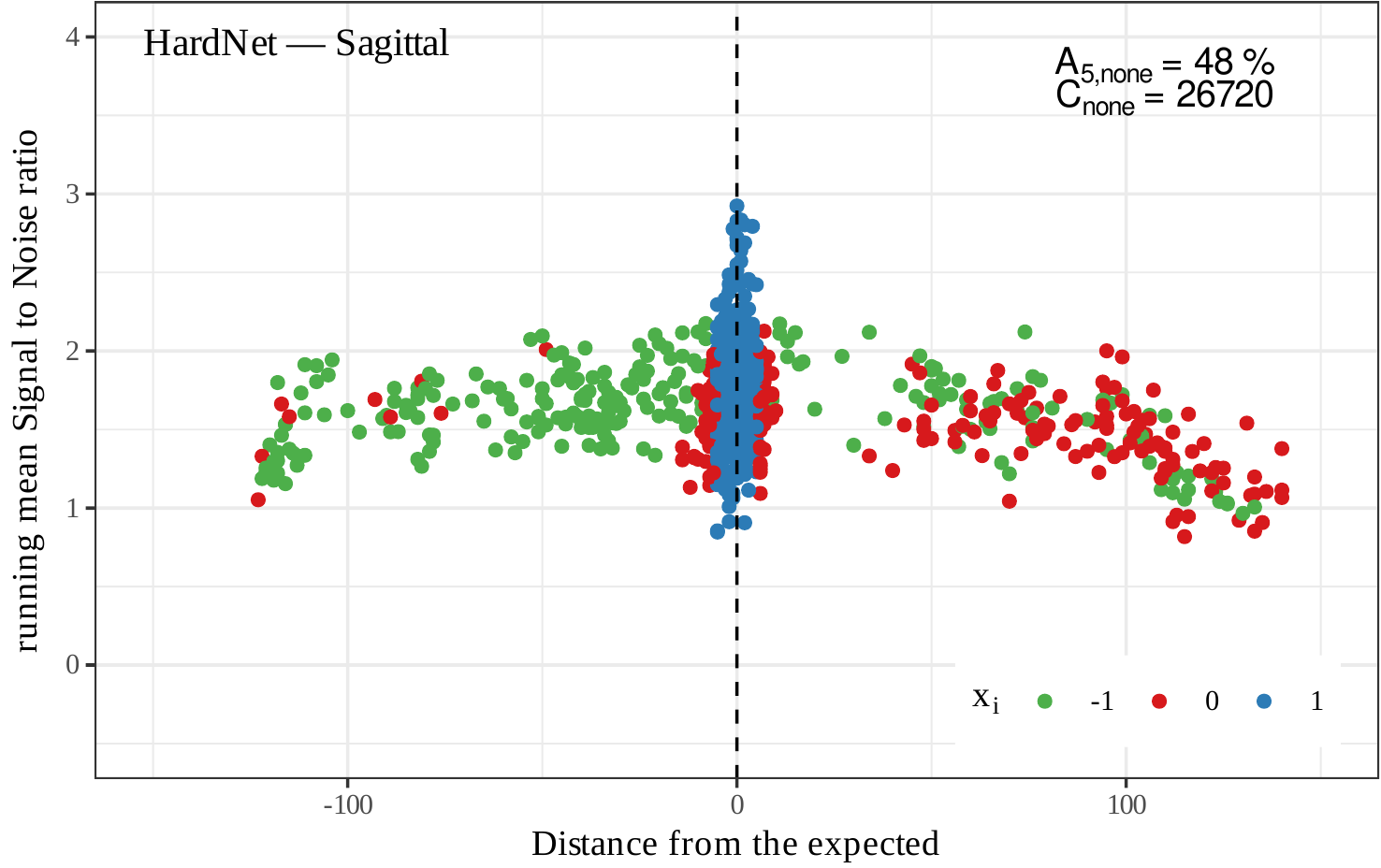}
    \includegraphics[width=.32\linewidth]{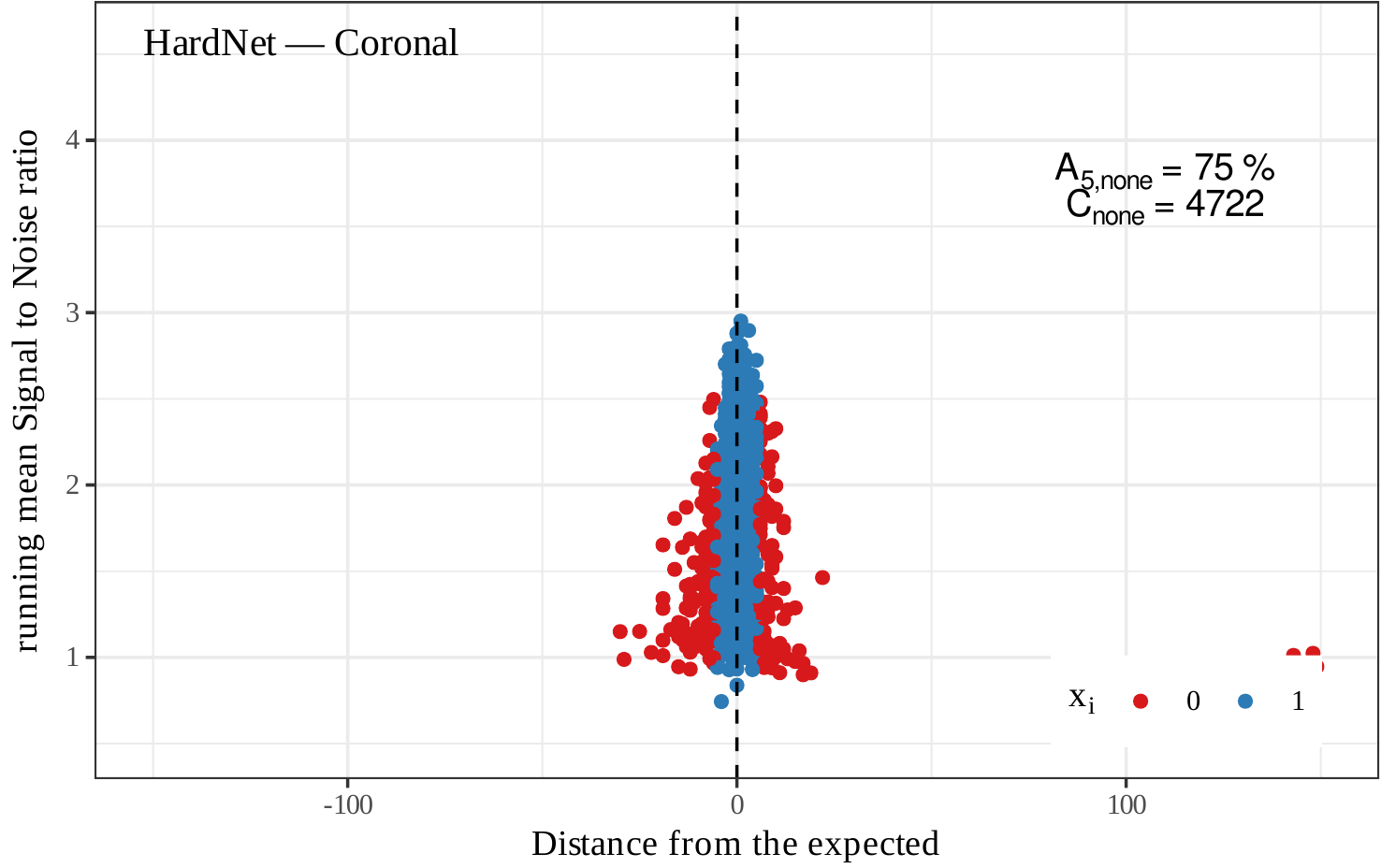}\\
    \includegraphics[width=.32\linewidth]{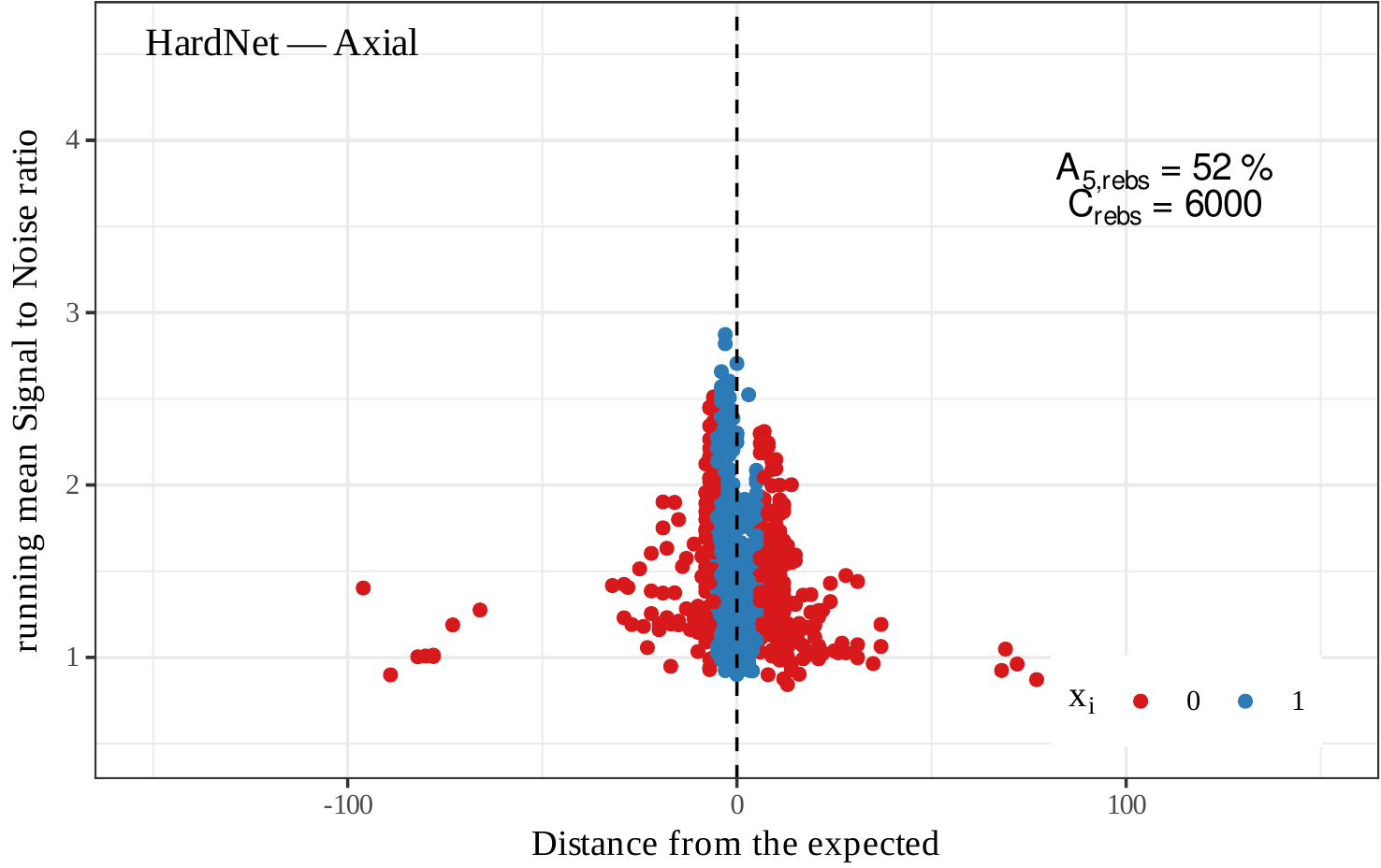}
    \includegraphics[width=.32\linewidth]{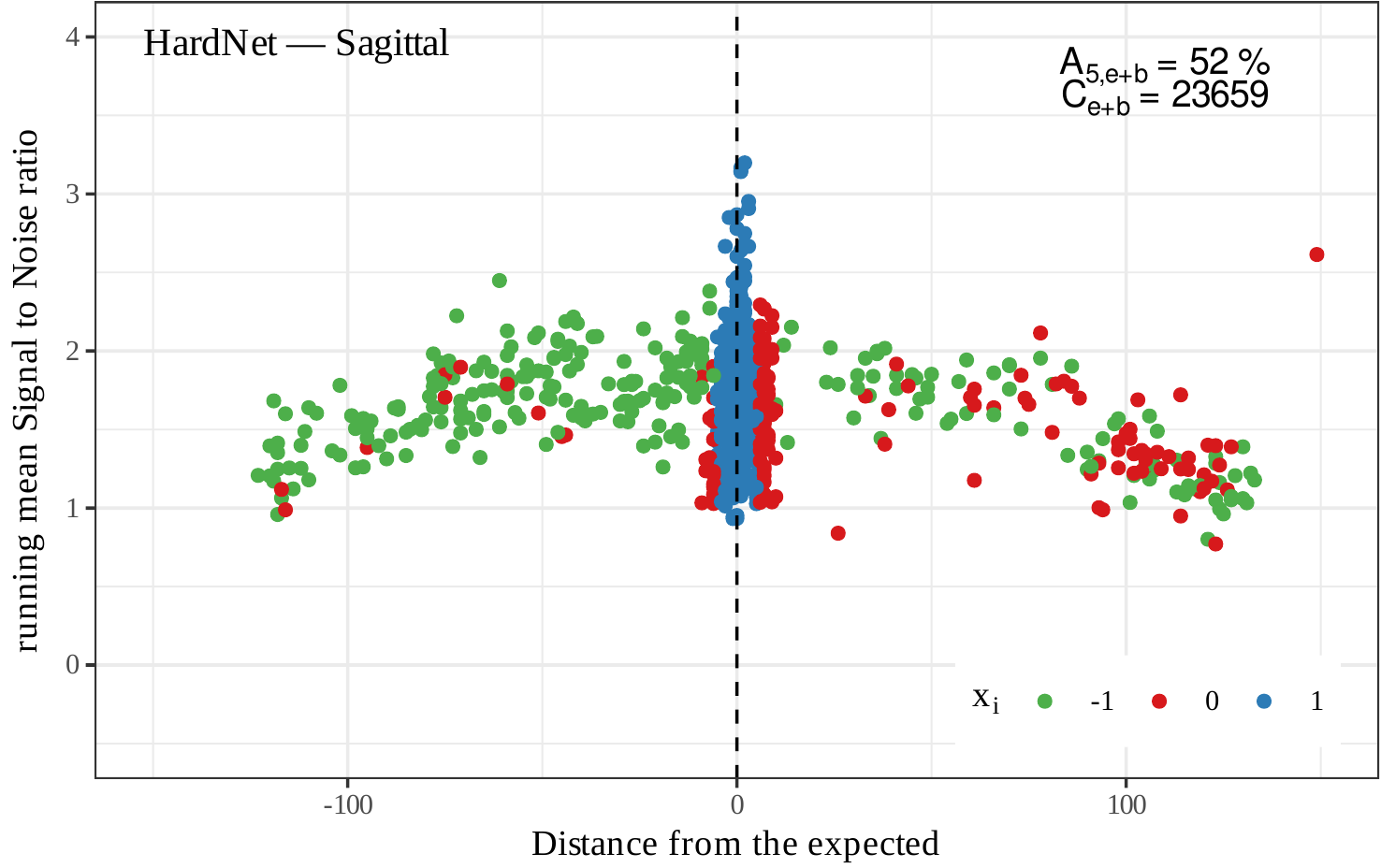}
    \includegraphics[width=.32\linewidth]{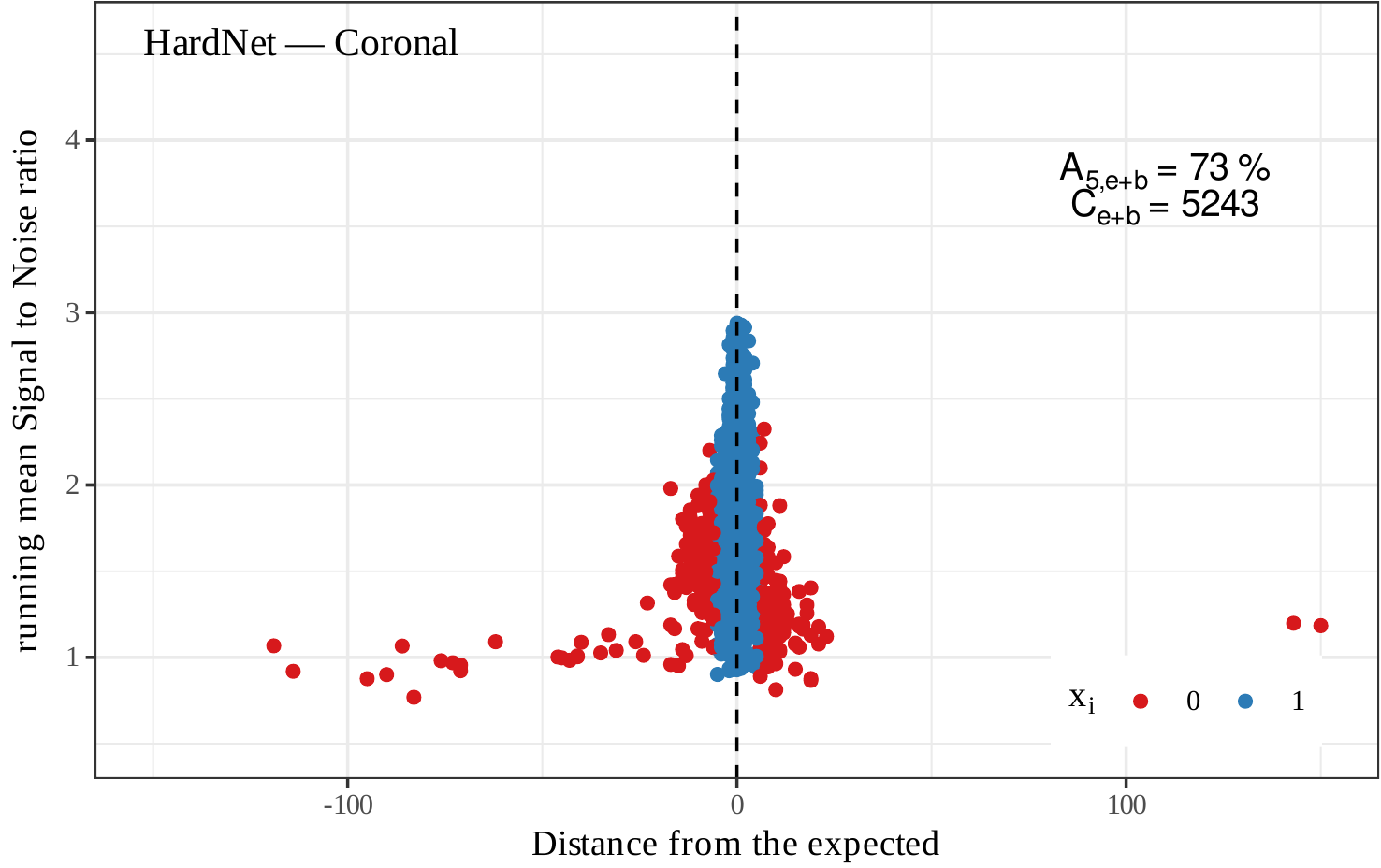}\\[0.2cm]
    \includegraphics[width=.32\linewidth]{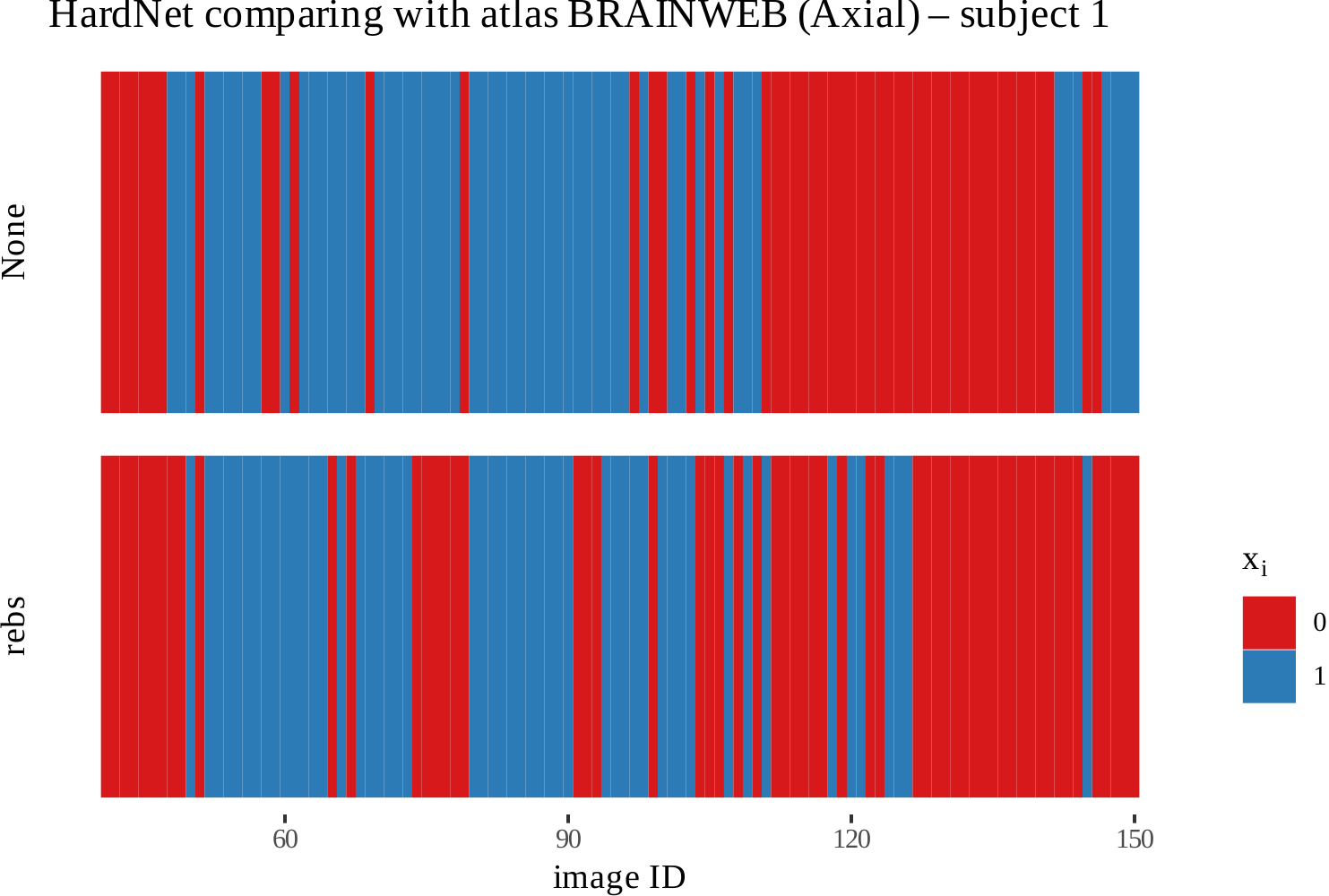}
    \includegraphics[width=.32\linewidth]{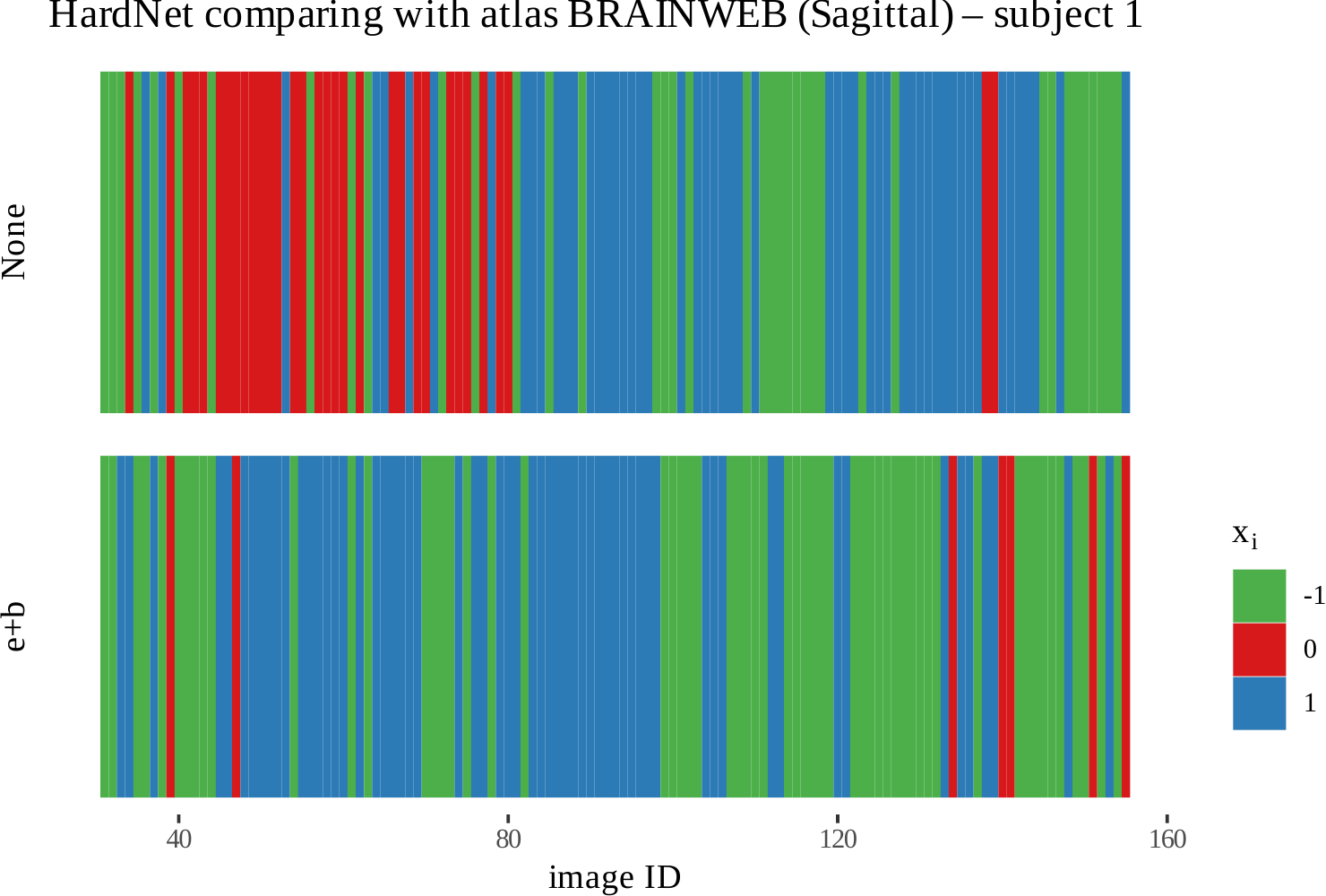}
    \includegraphics[width=.32\linewidth]{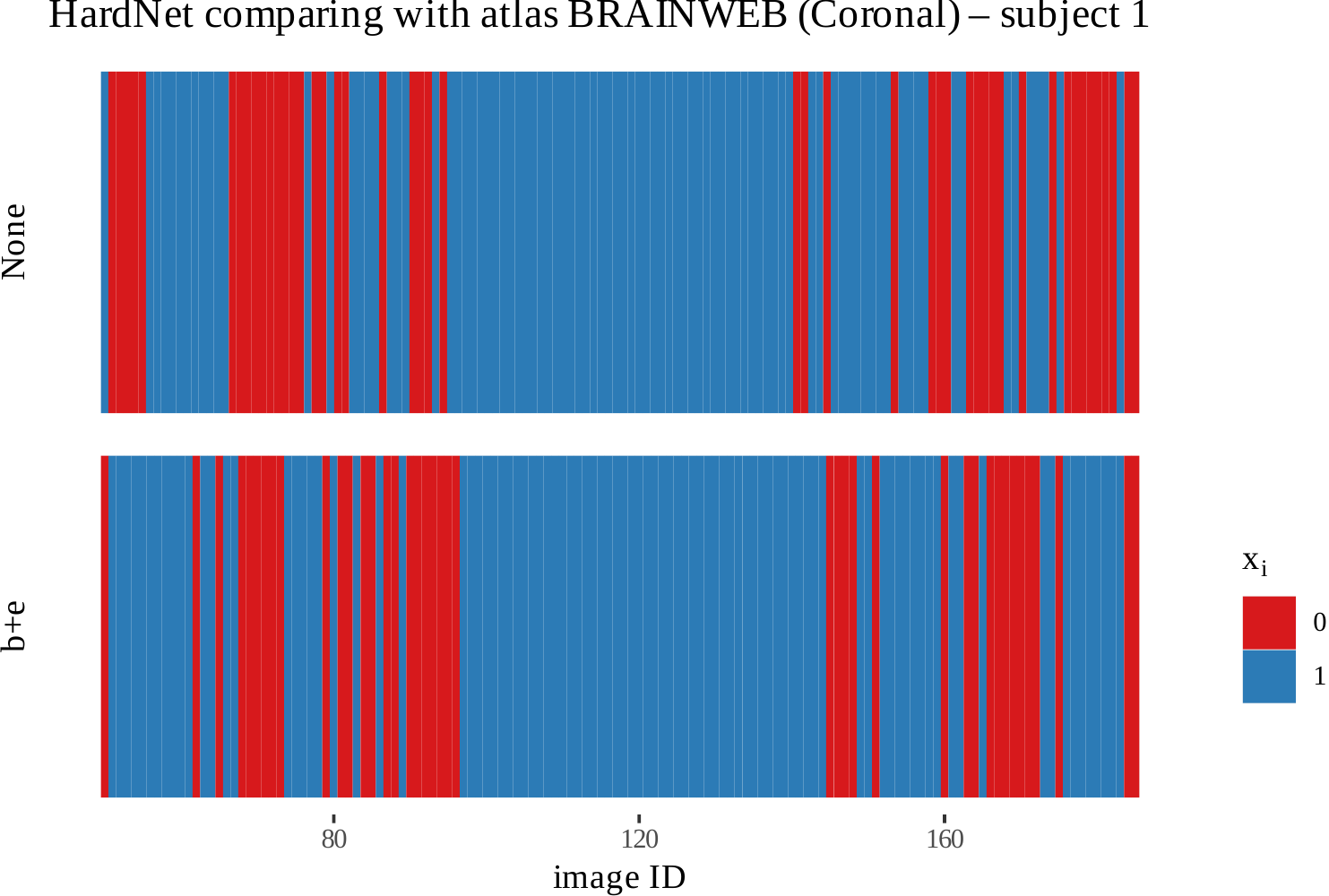}\\
    \includegraphics[width=.32\linewidth]{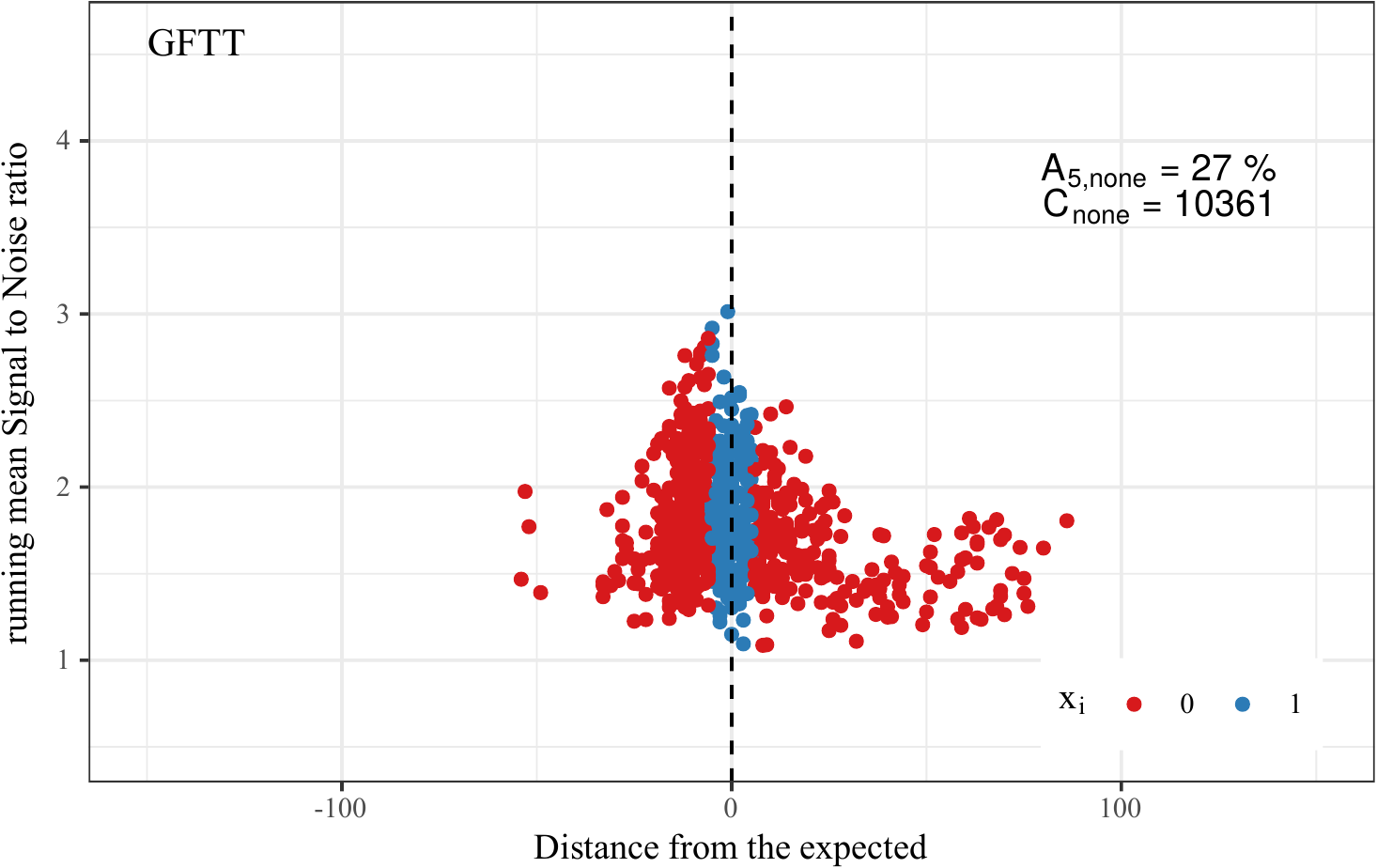}
    \includegraphics[width=.32\linewidth]{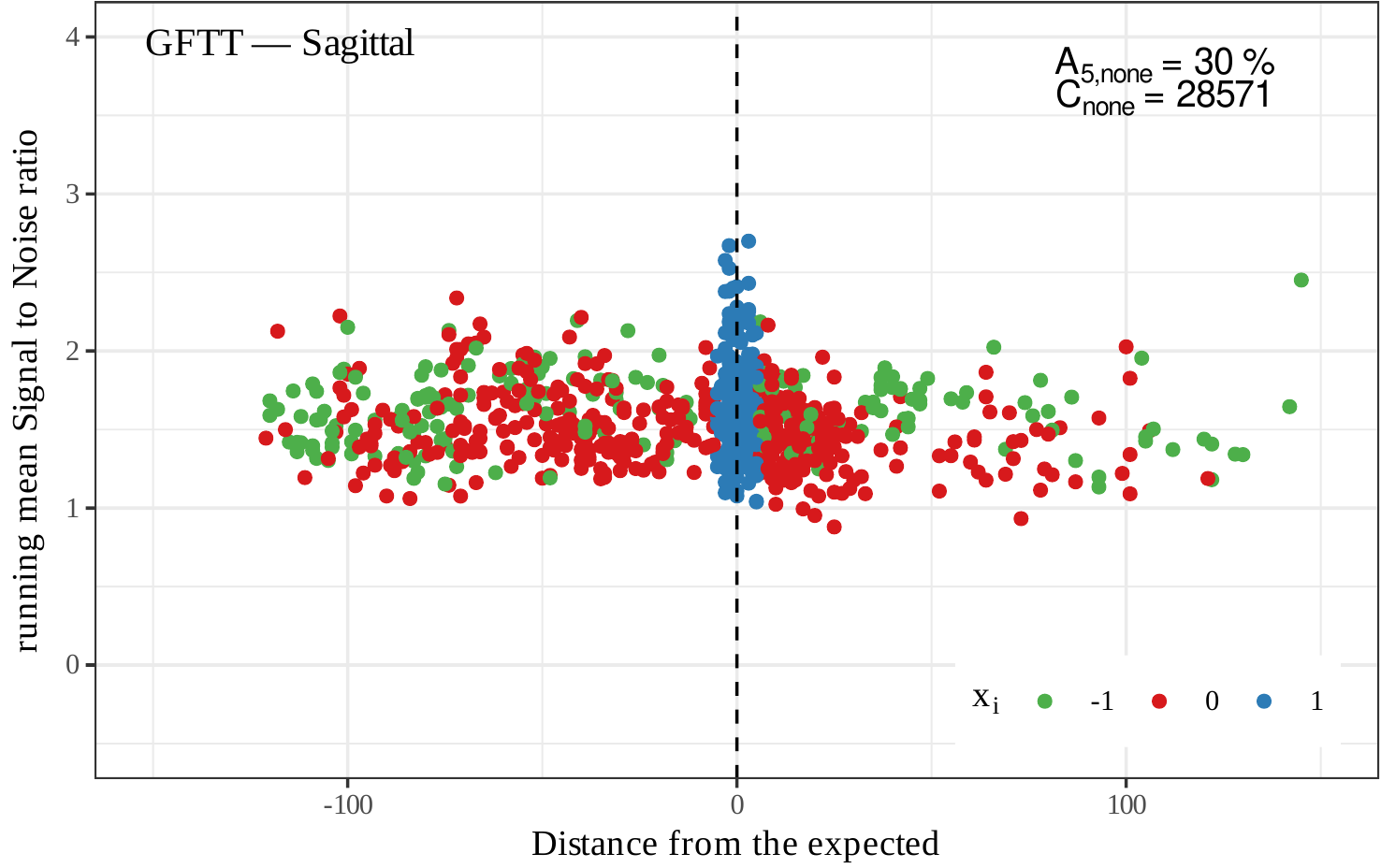}
    \includegraphics[width=.32\linewidth]{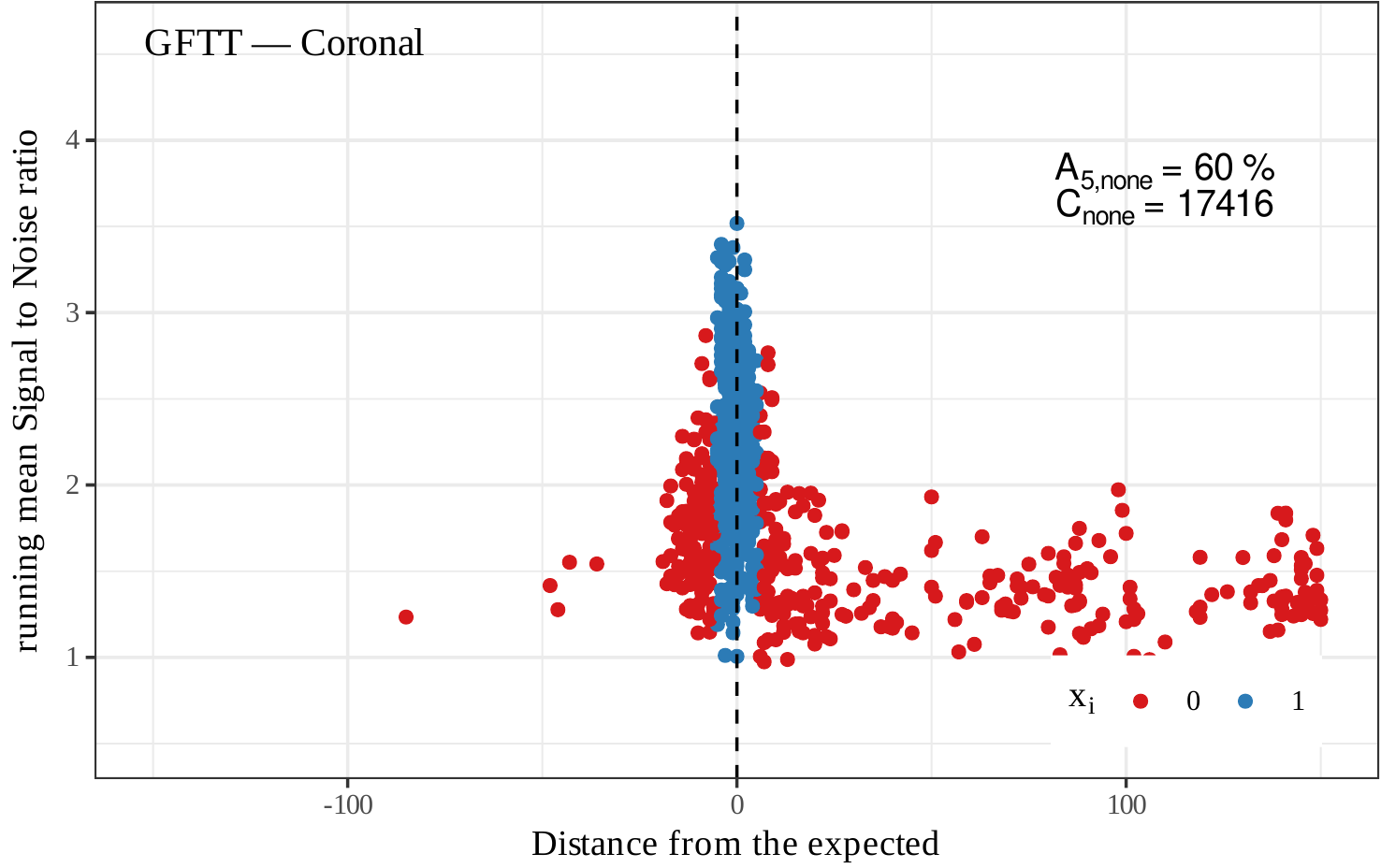}\\
    \includegraphics[width=.32\linewidth]{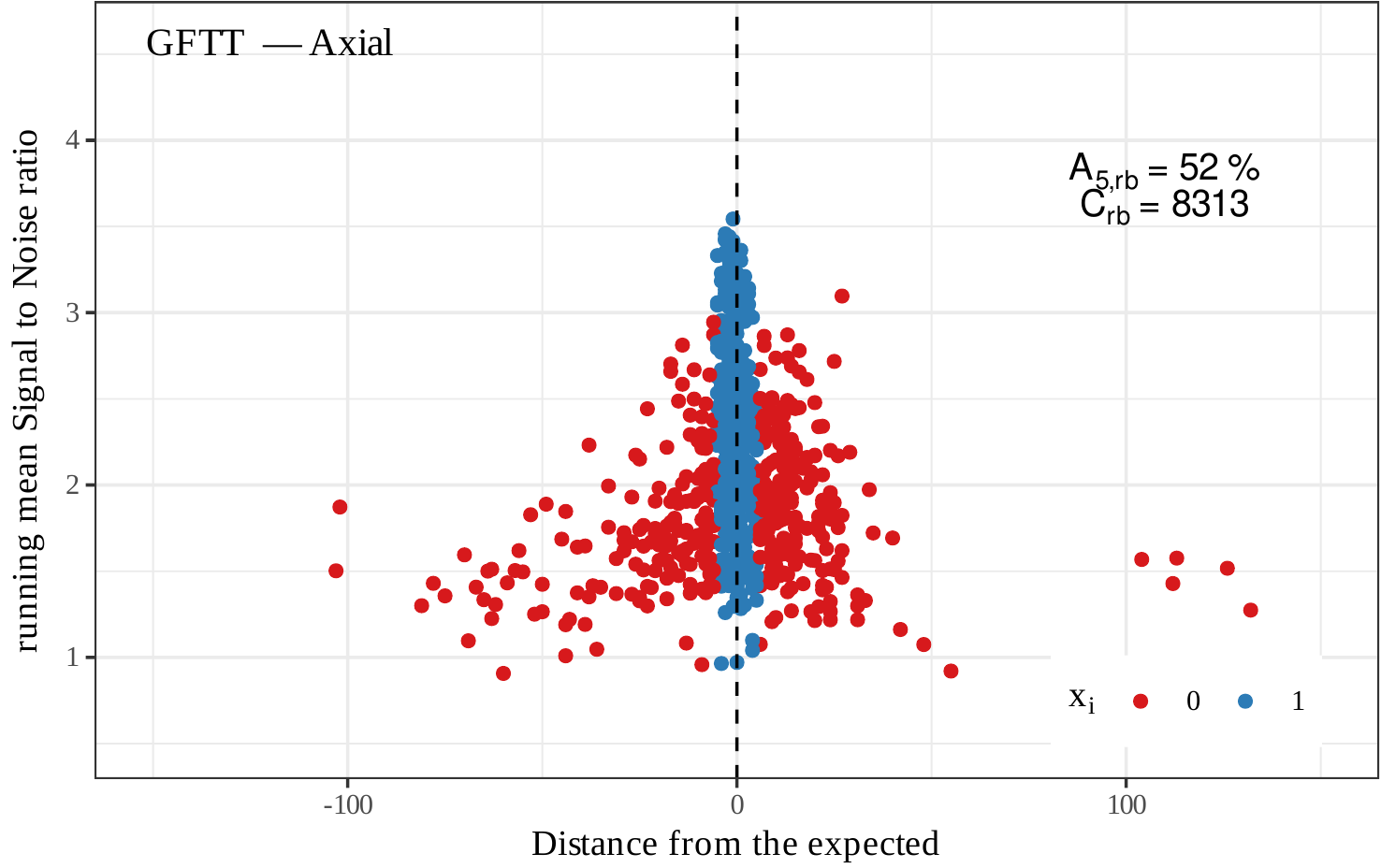}
    \includegraphics[width=.32\linewidth]{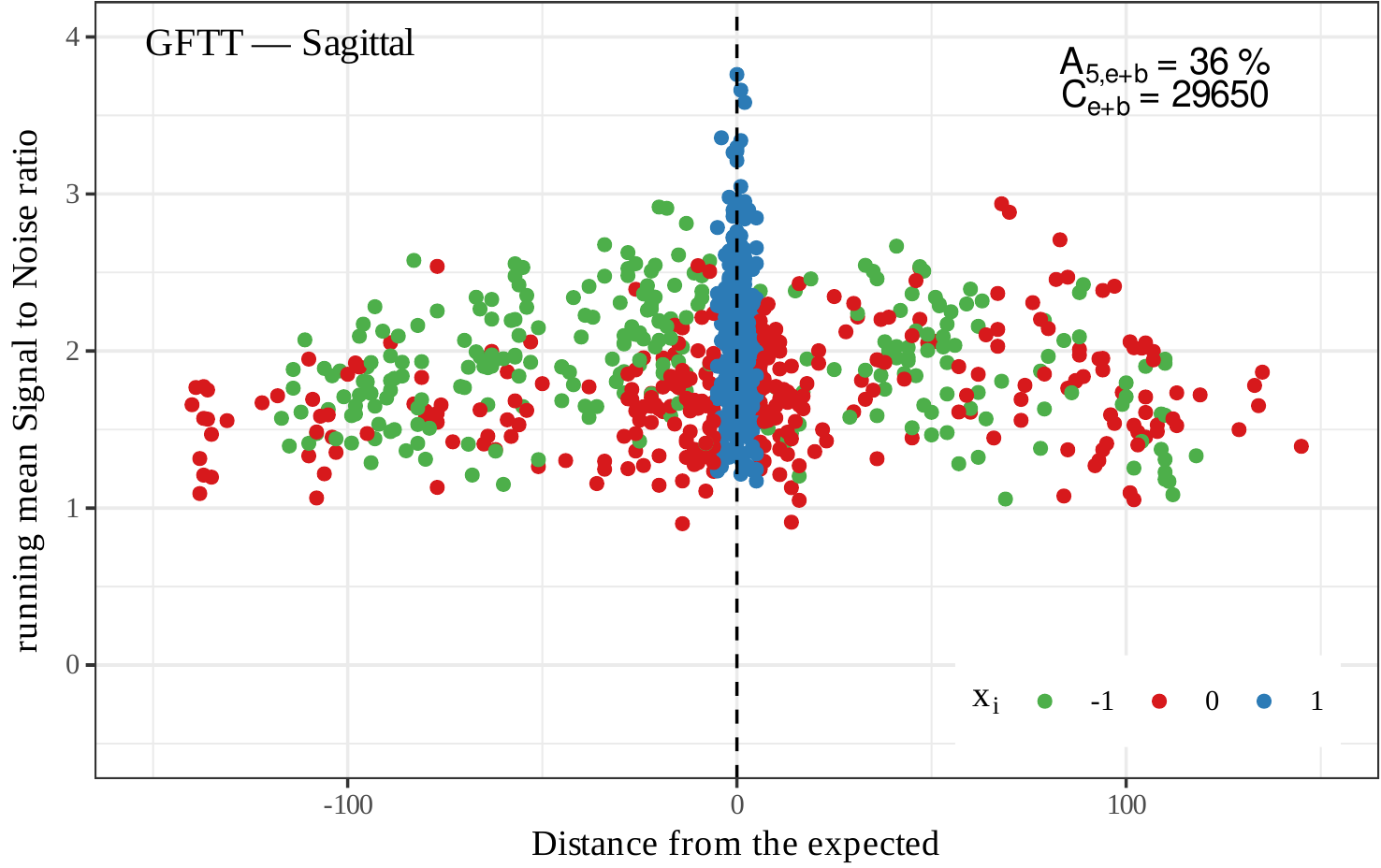}
    \includegraphics[width=.32\linewidth]{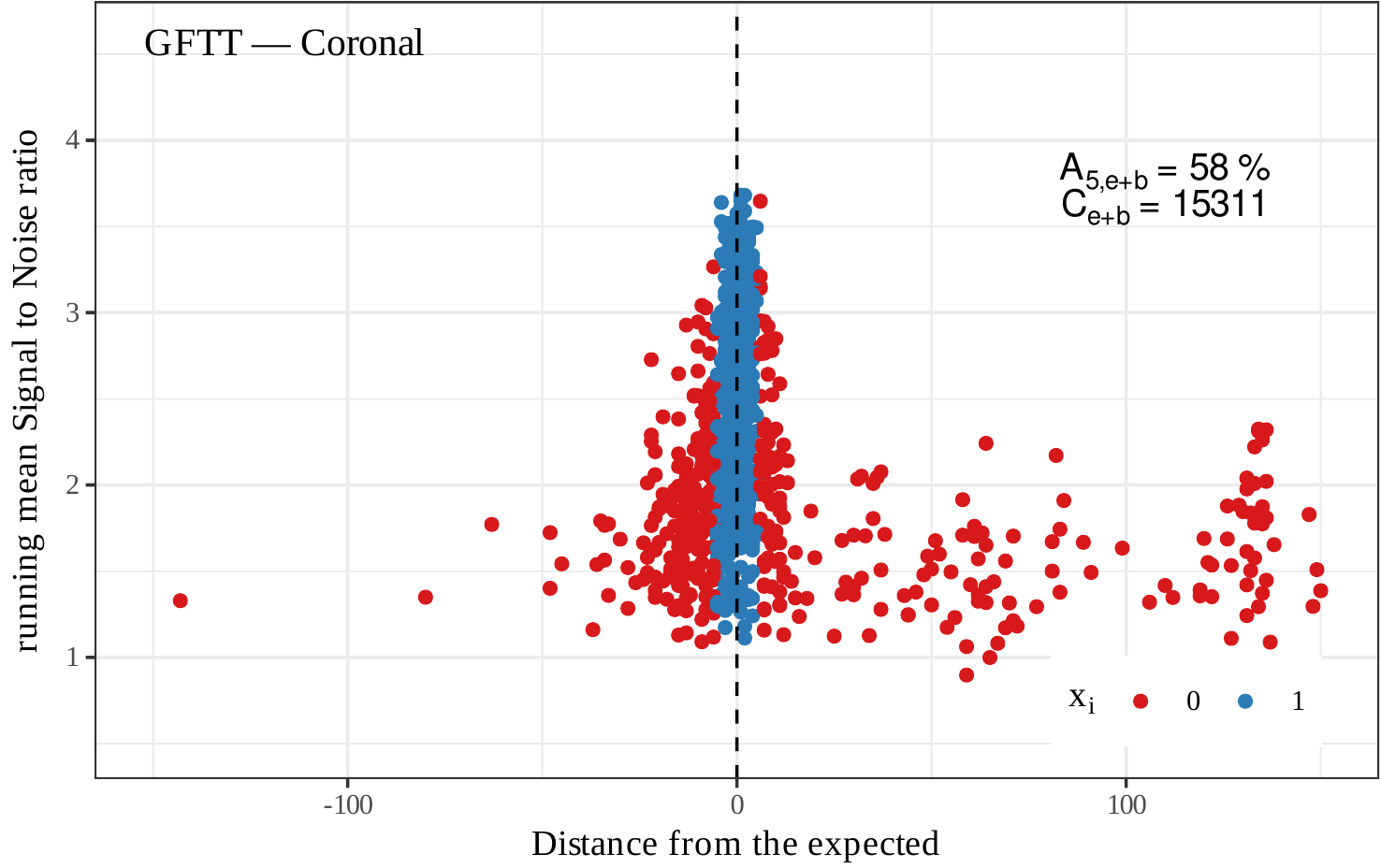}\\[0.2cm]
    \includegraphics[width=.32\linewidth]{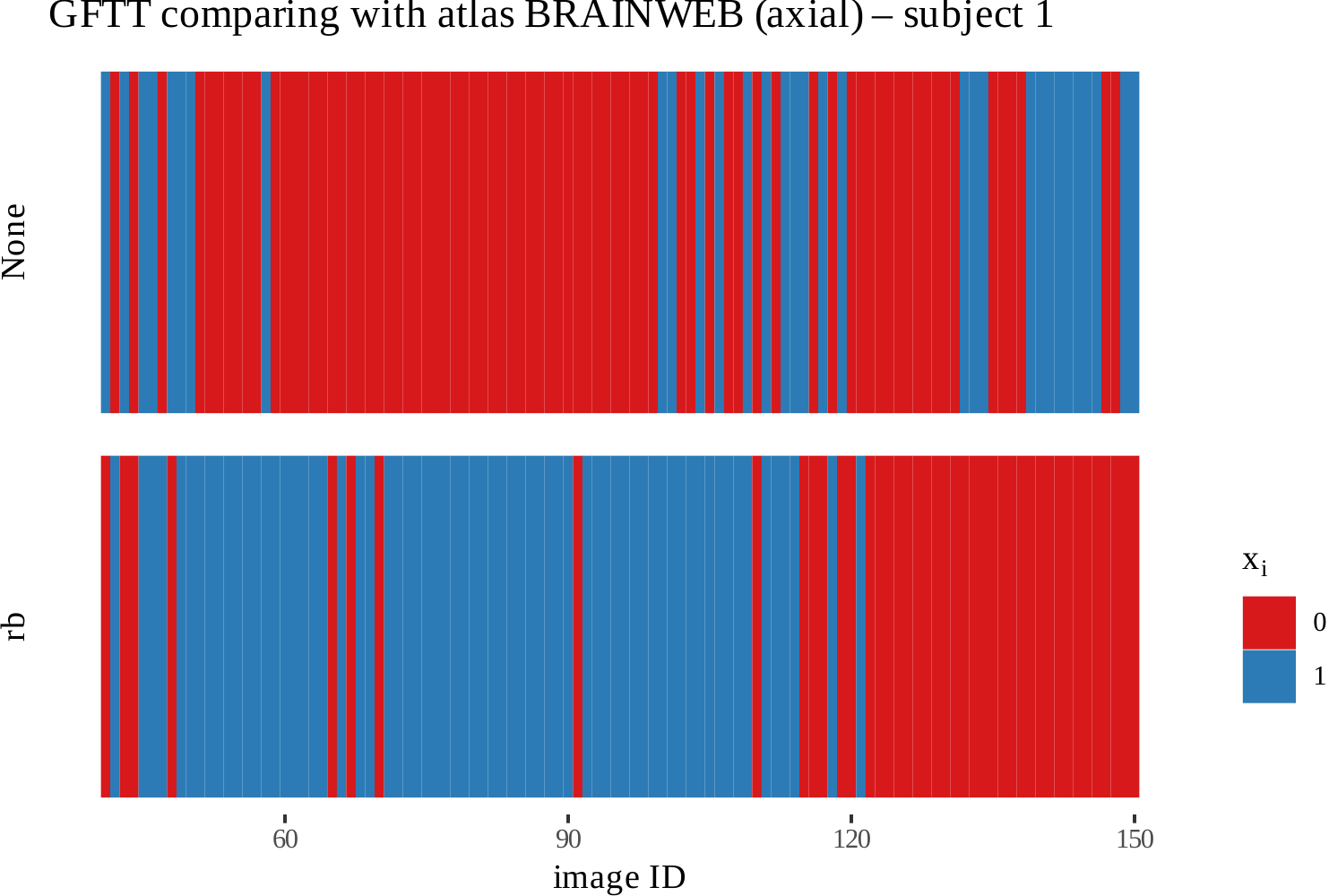}
    \includegraphics[width=.32\linewidth]{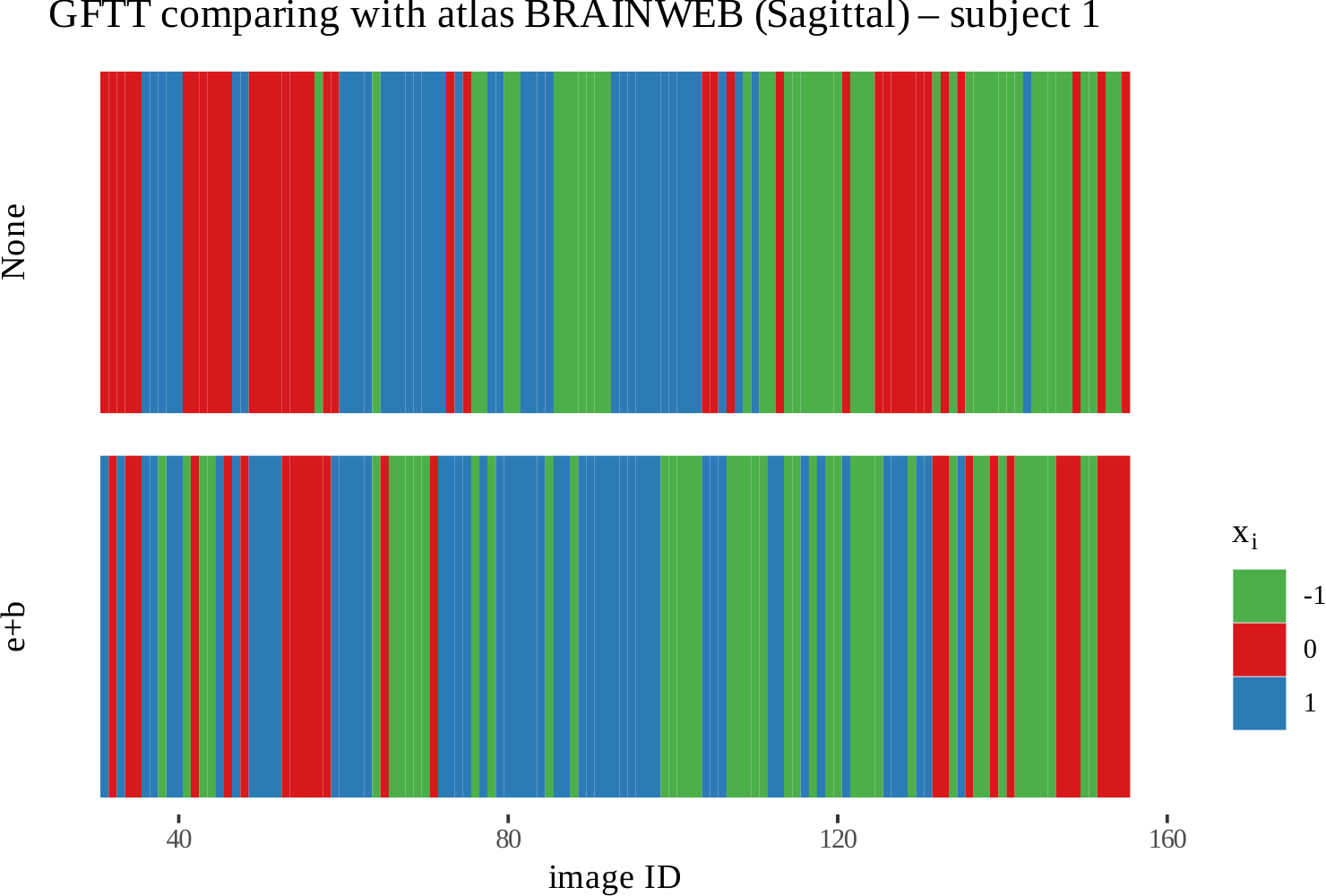}
    \includegraphics[width=.32\linewidth]{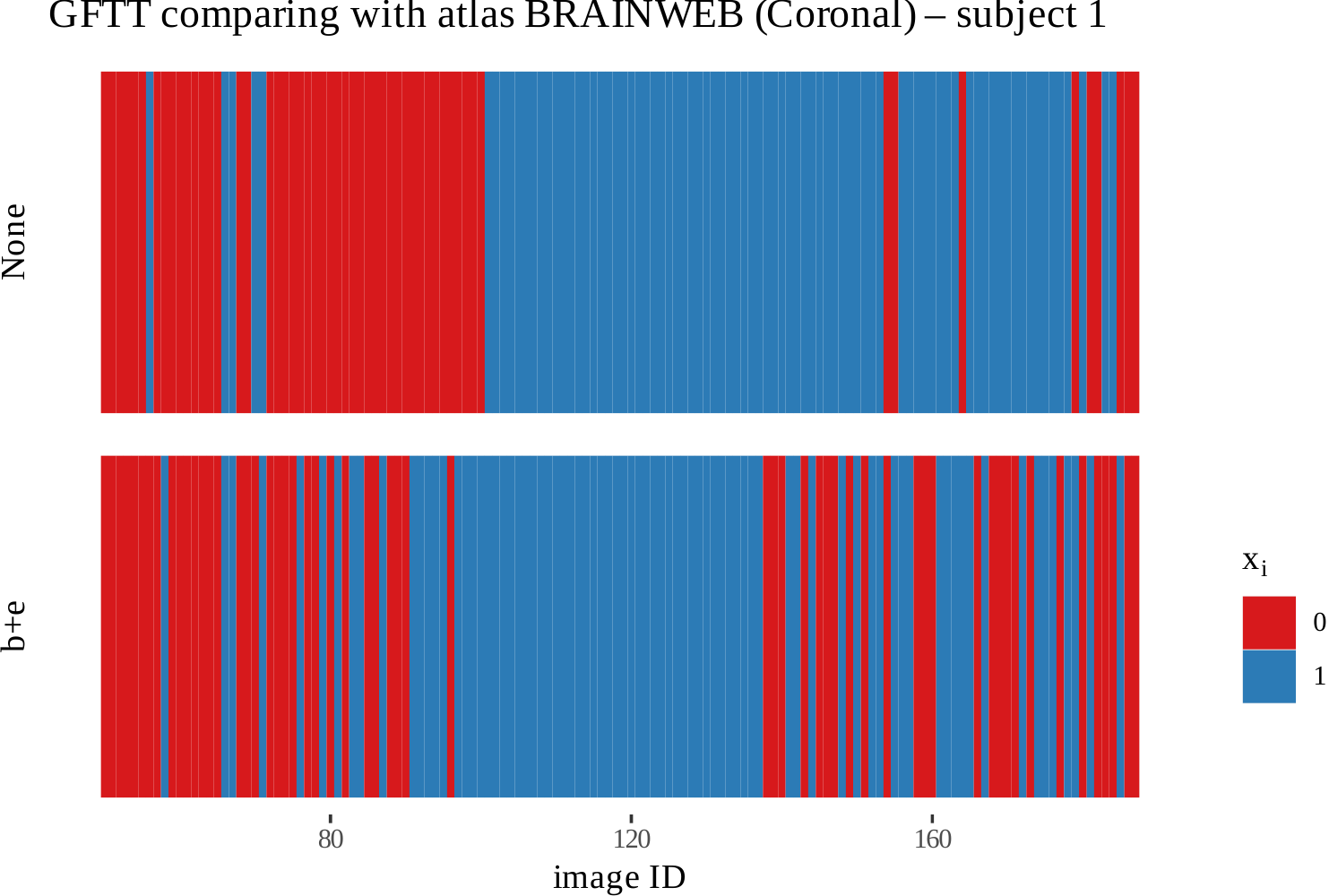}\\
    \caption{Comparison of the achieved accuracy $A_{5,c}$ for HardNet (first three rows) and GFTT  (second three rows) method with Atlas and selected image preprocessing for different anatomical planes: axial, sagittal and coronal (first, second and third column). In the case of the sagittal plane, we also included the green colour representing images with similar, but lower, SNR that were mismatched due to the symmetry of hemispheres. These images, if selected, would have fulfilled the accuracy condition.}
    \label{fig:atlas_akaze}
\end{figure*}

The results for all anatomical planes are summarised in Fig.~\ref{fig:atlas_akaze}. Similarly to the axial plane, the HardNet FD method achieved better results compared to the GFTT method and achieves at best $A_{5,be}{\sim}53\,\%$ accuracy for the sagittal plane and $A_{5,be}{\sim}75\,\%$ for the coronal plane. Whilst GFTT accuracy for the sagittal plane is at best $A_{5,be}{\sim}36\,\%$ and coronal plane $A_{5,be}{\sim}60\,\%$. The cumulative distance $C_c$ in the case of the sagittal plane for both FD methods is similar $\sim$10000--30000. This is due to the fact that the FD method generally fails to distinguish well between the left and the right hemispheres. Although, in general, the two hemispheres are not necessarily equal, the FD methods used gave similar SNR values, and without additional information, we were not able to select the correct image. However, as part of most standards, the orientation and location information is included in the metadata. Therefore we can limit the search for matching images only to the left or right hemispheres, thus avoiding the problem of symmetry.

\begin{figure*}
    \centering
    \includegraphics[width=0.48\linewidth]{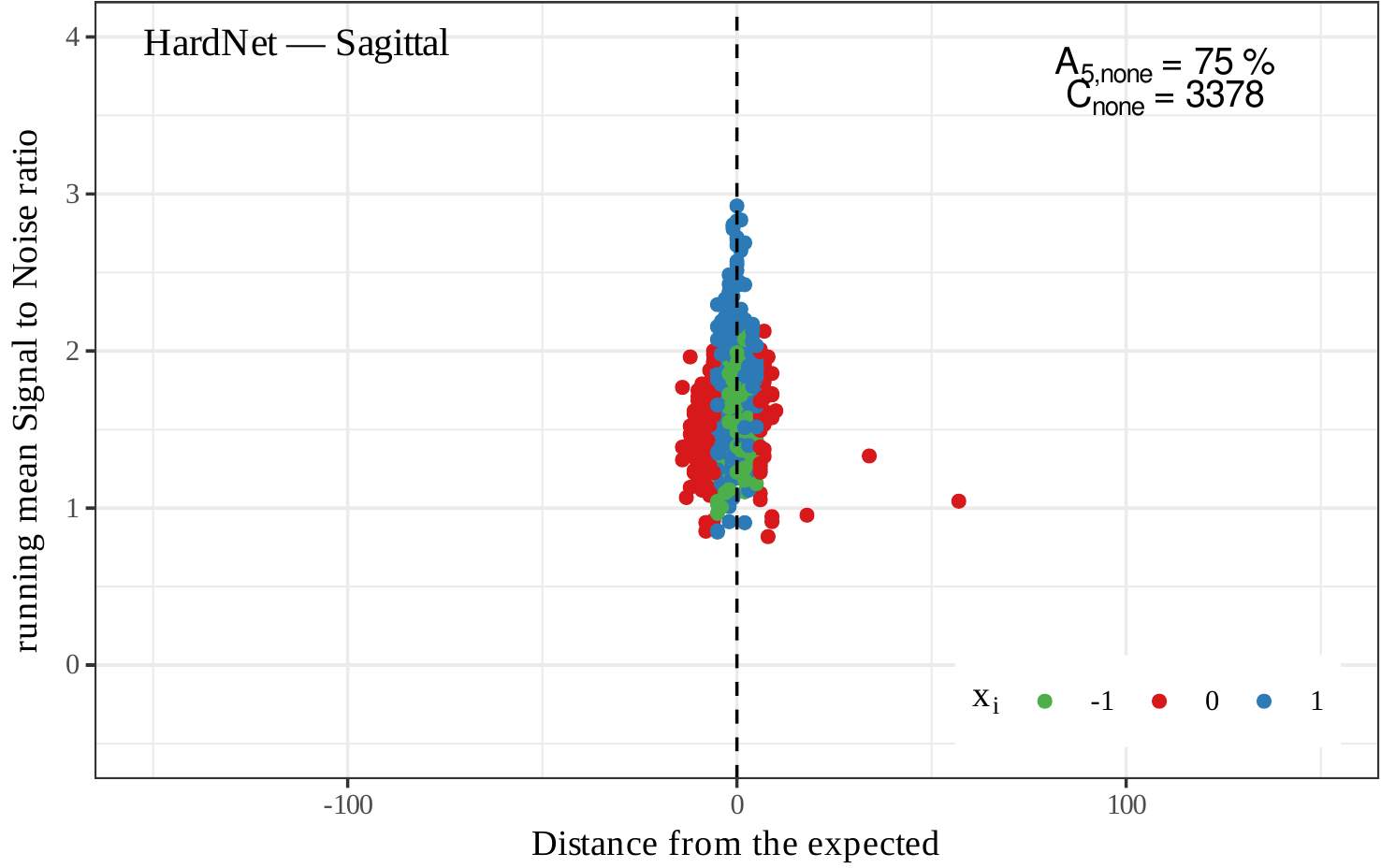}
    \includegraphics[width=0.48\linewidth]{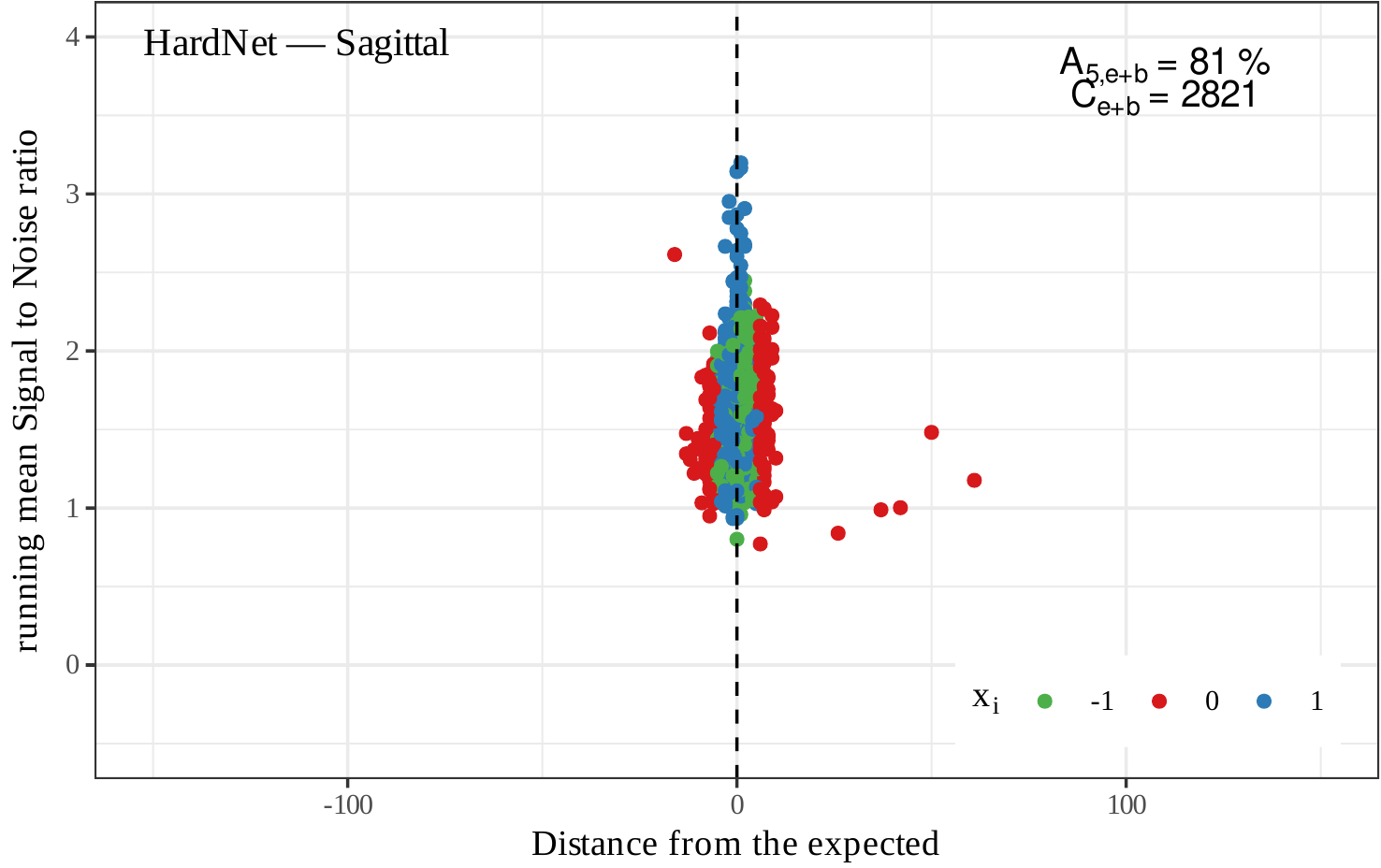}\\
    \includegraphics[width=0.48\linewidth]{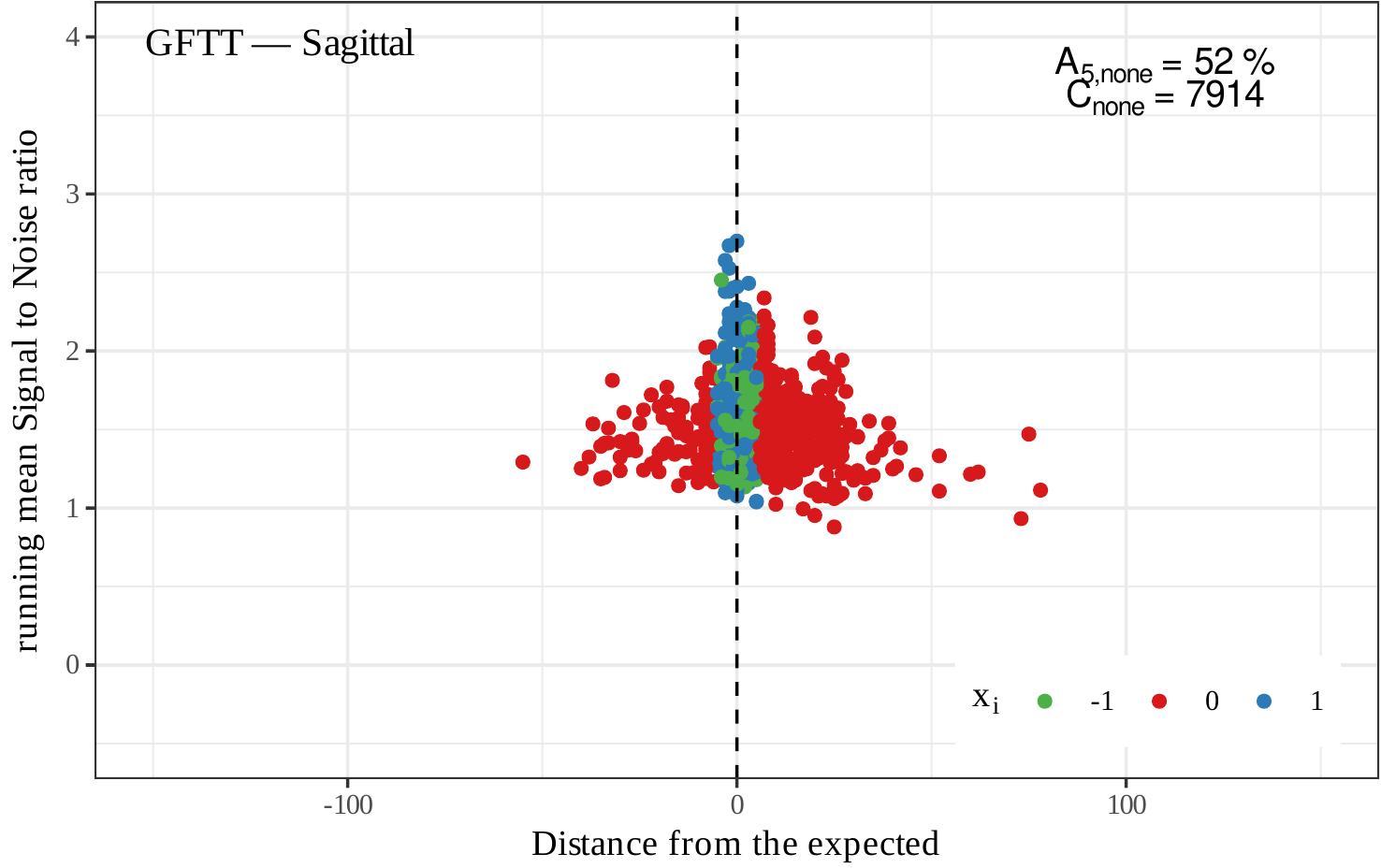}
    \includegraphics[width=0.48\linewidth]{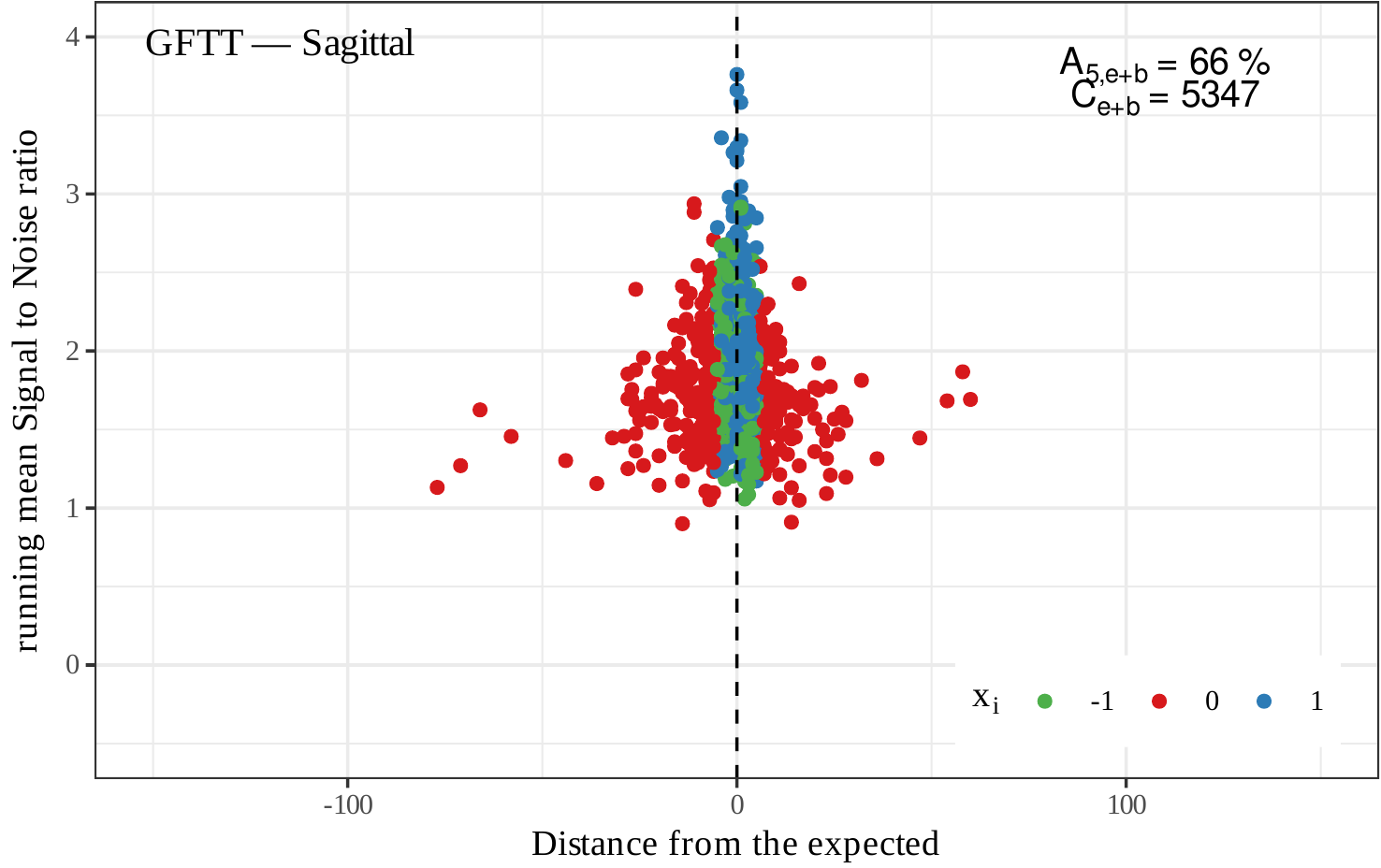}\\  
    \caption{\label{fig:atlas_sagital_sym} The accuracy $A_{5,c}$ for HardNet (first row) and GFTT (second row) methods without preprocessing (left) and with preprocessing (right) for the sagittal plane. Different hemispheres are distinguished by colour. Blue dots for the left and green for the right hemisphere satisfy the accuracy condition $x_i$ ($d=5$) by Eq.~\ref{eq:accuracy}.}
\end{figure*}

If we use the additional information to select the correct hemisphere, we see that the cumulative distance $C_c$ significantly improves to a value ${\sim}2500$--8000, as well as the accuracy $A_{5,c}$, improves up to 85\% for HardNet and 66\% for GFTT. These results are summarised in Fig.~\ref{fig:atlas_sagital_sym}, where the blue/green dots represent the images satisfying the accuracy condition $x_i$ the correct/opposite hemisphere.

In the case of the coronal plane, both FD methods with the atlas show better accuracy than in the case of the axial plane. Specifically, $A_{5,c}{\sim}75\%$ and a cumulative distance $C_c{\sim}5000$ for HardNet, and $A_{5,c}{\sim}60\%$ and $C_c{\sim}15000$ for GFTT.



\section{Conclusion and Future work}
\label{sec:Conclusion}
    We have used six freely available FD methods (FD methods), namely AGAST, AKAZE, BRISK, GFTT, HardNet, and ORB, and evaluated their ability to match MRI brain slices. To that end, we have tested how are selected FD methods invariant to various image degradations (rotation, up-scaling, added Gaussian noise). Then we compared methods based on their capability of matching MRI brain slices of different patients but using the same MRI scanner. Lastly, we have evaluated how well different methods can match MRI scans of a patient with a simulated brain atlas.

Methods for localisation of plain sections of the axial brain by magnetic resonance are one of the basic prerequisites for testing programs for digital image analysis that would detect pathological changes in the image. In order to test these programs for changes in a given brain structure, it is necessary to have methods that localise the given region-of-interest and displays it. Manual selection of these regions for further marking of such structures would be time-consuming. Especially in the case of use in longitudinal studies, where it is necessary to obtain similar information during the same type of examination on a different machine. In clinical practice, these methods of imaging brain structures could be used in combination with programs for digital image analysis, especially in patients with neurodegenerative diseases, when changes in the brain parenchyma occur gradually and only a visual assessment by a doctor is not sufficient to recognize pathological processes.
    
    The method invariance to image degradation was tested on MRI scans of the same patient. The task was to match the selected MRI slice to the same but degraded MRI slice among the rest of the MRI slices of the same patient. The invariance itself was measured as the decrease in signal-to-noise ratio (SNR) when the MRI slice was matched with an unmodified and degraded image. From tested methods the most invariant was GFTT with decreases of 5\% for rotation, 15\% for upscaling, and 12\% for Gaussian noise, followed by AKAZE (10\%, 25\%, 10\%) and AGAST (3\%, 55\%, 14\%). The method used in the case of HardNet showed a gain in SNR (+27\%, +27\%, +24\%). Although, there is a decrease in the number of matched points the level of noise decreased even further, therefore HardNet is one of the best methods regarding invariance to image degradation. Although it is worth noting that during our testing of HardNet, we found out that the number of matching points changed by 1--2 points when the test was run on the same sets again. The BRISK method demonstrated an average 50\% decrease for all applied degradations thus being most sensitive to image degradation from selected methods. ORB exhibited on average 35\% reduction in SNR with a 30\% decrease for rotation and Gaussian noise and a 40\% decrease for up-scaling. This shows that for MRI slices BRISK and ORB are not well suited.
    
    When FD methods were compared based on the ability to match all MRI slices from a patient to their counterparts from different patients provided on the same machine then the best method is HardNet which managed to correctly match 93\% of the images, followed by AKAZE and GFTT both matched 87\% of the images, then AGAST with 78\%. BRISK and ORB managed to match 63\% and 58\% of images correctly. However the confidence in these matches is arguably still low. For the leading FD methods, the SNR for most of the matches was below SNR$=4$, whilst for ORB and BRISK the most matched with a value of SNR$=2.5$.
    
    Comparing MRI slices with simulated brain atlas failed for all tested methods. The best result was achieved with HardNet where only 53\% of images were correctly matched with low confidence. The reason for this mismatch is probably that the grayscale values of interest are too different and may prevent the FD method to select ``unique'' points of interest, thus prevent us from matching the correct images. As one solution a dynamical histogram renormalization may be needed based on the raw (e.g. DICOM) data format, or in the case of atlases a defined map of grayscale values ranges. 
    
    As future work, testing the potential FD method would be useful to increase the tests of robustness with a mapping of grayscale ranges as well as include other free available FD methods or, at best, to develop a specific FD method suited for medical images from the beginning. From the tests we provided, such FD method needs to locate points of interest similarly like AKAZE, which features are generally detected in the form of blobs instead as corners (for example BRISK, ORB). It could even lead to the interpretation of the medical images as a whole or a larger part of it. Another possibility would be to tune internally the available FD methods. However, as would be helpful in the future to compare or search slices from different modalities. The tuning could help only specific modalities, and for other it will need to be tuned again. 
    
    Also would be interesting the creation of a custom atlas suited for image indexing using FD methods should be considered. Such atlas would need to add more information relevant to FD methods and could be an indicator for other types of atlases in the future. Also, it would be interesting to run a similar summary on different images than the brain.
\section{Acknowledgement}
The authors would like to express their gratitude to the Research Centre for Theoretical Physics and Astrophysics, Institute of Physics, Silesian University in Opava for institutional support. This work was supported by the Ministry of Health of the Czech Republic grants number NV-19-04-00270, NV-19-04-00362 and NU21-09-00357, and Palack\'{y} University grant number JG\_2019\_004.
\bibliographystyle{abbrvnat}
\bibliography{fd_in_med_arxiv}
\end{document}